\documentclass[prc,nofootinbib,longbibliography]{revtex4-1}
\usepackage{slashed}
\usepackage{multirow,array}
\usepackage{color}
\usepackage{mathtools}
\usepackage{mathrsfs}
\usepackage{amssymb}

\newcommand{\beq}{\begin{eqnarray}}
\newcommand{\eeq}{\end{eqnarray}}
\newcommand{\be}{\begin{eqnarray*}}
\newcommand{\ee}{\end{eqnarray*}}

\begin{document}

\title{Fluid dynamic propagation of initial baryon number perturbations on a Bjorken flow background}

\author{Stefan Floerchinger,}
\email{stefan.floerchinger@cern.ch}
\affiliation{Physics Department, Theory Unit, CERN, CH-1211 Gen\`eve 23, Switzerland}
\author{Mauricio Martinez}
\email{martinezguerrero.1@osu.edu}
\affiliation{Department of Physics, The Ohio State University, Columbus, Ohio 43210, USA}

\begin{abstract}
Baryon number density perturbations offer a possible route to experimentally measure baryon number susceptibilities and heat conductivity of the quark gluon plasma. We study the fluid dynamical evolution of local and event-by-event fluctuations of baryon number density, flow velocity and energy density on top of a (generalized) Bjorken expansion. To that end we use a background-fluctuation splitting and a Bessel-Fourier decomposition for the fluctuating part of the fluid dynamical fields with respect to the azimuthal angle, the radius in the transverse plane and rapidity. We examine how the time evolution of linear perturbations depends on the equation of state as well as on shear viscosity, bulk viscosity and heat conductivity for modes with different azimuthal, radial and rapidity wave numbers. Finally we discuss how this information is accessible to experiments in terms of the transverse and rapidity dependence of correlation functions for baryonic particles in high energy nuclear collisions.
\end{abstract}



\maketitle

%%%%%%%%%%%%%%%%%%%%%%%%%%%%%%%%%%%%%%%
\section{Introduction}
\label{sec:intro}
%%%%%%%%%%%%%%%%%%%%%%%%%%%%%%%%%%%%%%% 
One of the most important goals of the experimental program of high energy nuclear collisions is to determine the transport and thermodynamical properties of QCD as a function of temperature $T$ and baryon chemical potential $\mu$. During the past few decades, the experimental data measured at the Relativistic Heavy Ion Collider (RHIC) at the Brookhaven National Laboratory and the Large Hadron Collider (LHC) at CERN in Geneva, has shown collective behavior of the QCD matter created after the collision of heavy nuclei at high energies~\cite{ALICE:2011ab,Aamodt:2011by,Chatrchyan:2012wg,Aad:2013xma,Adamczyk:2013waa,Adare:2011tg,Alver:2008zza}. The low momentum region of the transverse hadron spectra and the two particle correlation functions are well described by relativistic viscous fluid dynamics with a very small value of the shear viscosity over entropy ratio~\footnote{See  Refs.~\cite{Heinz:2013th,Gale:2013da} for the recent developments in relativistic hydrodynamics.}. These results have been taken as evidence for the production of an almost perfect liquid, a strongly coupled quark gluon plasma. 

The hydrodynamic modeling of heavy ion collisions solves on an event-by-event basis the relativistic fluid equations corresponding to energy-momentum conservation laws together with the so called constitutive relations for the shear viscous tensor and bulk pressure. Within this approach, little attention has been paid to the possible role of the baryon density $n$ and/or baryon chemical potential $\mu$. At high energies, this is justified because $n$ and $\mu$ are very small, at least in the midrapidity region. However, interesting physics could be probed by investigating event-by-event fluctuations in the local baryon number density.

Baryon number fluctuations have been mainly discussed in the context of heavy ion collisions at lower energy where larger values of $\mu$ can be realized. Interesting features of the QCD phase diagram can emerge there~\cite{Fukushima:2010bq}. Different effective models have predicted the existence of a first-order phase boundary that separates hadronic matter from the quark gluon plasma at larger values of the baryon chemical potential. This boundary comes to an end at some critical values of the temperature $T_c$ and baryon chemical potential $\mu_c$. Right now there is no conclusive evidence for the location of a critical point in the $T-\mu$ plane from lattice QCD calculations at finite baryon density~\cite{Ding:2015ona}. 

On the other hand, in heavy ion collisions it has been proposed to study second and higher order cumulants of particle multiplicity distributions as a function of  the center of mass energy~$\sqrt{s}$~\cite{Gazdzicki:1992ri,Stodolsky:1995ds,Shuryak:1997yj,Mrowczynski:1997kz,Stephanov:1998dy,Stephanov:1999zu,Berdnikov:1999ph,
Jeon:2000wg,Asakawa:2000wh,Voloshin:2001ei,Pruneau:2002yf,Hatta:2003wn,Nonaka:2004pg,Aziz:2004qu,Stephanov:2008qz,Schuster:2009jv,Athanasiou:2010kw,Ling:2013ksb,Albright:2015uua,Mukherjee:2015swa}. From thermodynamic considerations, it is expected that these moments scale with the correlation length which is expected to become large near the QCD critical point~\cite{Stephanov:1998dy,Stephanov:1999zu,Berdnikov:1999ph,Stephanov:2008qz,Athanasiou:2010kw,Stephanov:2011pb}. Possible signs of the critical point have been measured at RHIC but at present these do not provide a conclusive evidence~\cite{Adamczyk:2013dal,Aggarwal:2010wy,Abelev:2009bw,Sarkar:2013iaa}. If the expanding fireball of nuclear matter passes through a critical region (close to a critical point), one can extract information about the equation of state and the critical behavior of transport coefficients from the particle spectra formed at the freeze-out surface. It is important to determine whether the possible signatures of the critical point can survive the entire evolution of the expanding fireball.
 
In the fluid dynamic framework, different aspects of the evolution of the fireball can change the pattern expected from purely thermodynamic considerations. Thermodynamic fluctuations are in principle part of a fluid dynamic description, at least in an extended sense where one accounts also for noise. Fluctuations evolve in time and space during the expansion of the fireball and thus, these are indeed effected by the equation of state and specially the transport coefficients such as the viscosities and conductivities. Close to equilibrium, there is also a deep theoretical connection between thermodynamic fluctuations in fluid dynamic fields and dissipative transport properties as stated by the fluctuation-dissipation theorem; see, e.\ g.,\ Refs.\ \cite{landau1980statistical,Kovtun:2012rj,Kapusta:2011gt}. In the vicinity of the critical point, heat conductivity $\kappa$ as well as the shear and bulk viscosities $\eta$ and $\zeta$ show critical behavior~\cite{Hohenberg:1977ym,strathmann1995,Berg1988,Kapusta:2012zb}. 

Besides genuine thermodynamic fluctuations (or noise), there is another possible source of fluctuations in the fluid dynamic approach to heavy ion collisions. These are the fluctuations already present in the initial state when the fluid dynamic treatment becomes valid. Their origin can be either the substructure of the colliding nuclei or the far-from-equilibrium dynamics preceding a fluid dynamic regime. This kind of initial state perturbation is particularly important for energy and/or entropy density. Fluctuations in the geometric distribution of nucleons within a nucleus lead to initial density perturbations which - after a fluid dynamical evolution - determine the spectrum of harmonic flow coefficients and, for example the form of the two-particle correlation function (``the ridge'') in heavy ion collisions. 

In a very similar way to fluctuations in the initial energy density, one can also expect, for example from a Glauber-type description of the initial state, initial fluctuations in the baryon number density. Indeed, baryon number density carried by protons and neutrons is presumably not distributed homogeneously within a nucleus and fluctuates locally and from event to event. In addition, the baryons and anti-baryons produced by pair production directly after the collision are subject to some local and event-by-event fluctuations~\cite{DMS}. 

In order to discriminate the effects associated to the thermodynamic fluctuations from the initial state fluctuations, it is necessary to understand their space-time evolution. In the present work we will concentrate mainly on the dynamics of initial state fluctuations although parts of our formalism are relevant also for the evolution of thermodynamic fluctuations. Initial state fluctuations are interesting on their own. For instance, the evolution of the initial perturbations of energy density depends on the viscosities, in particular shear viscosity. In a similar way, the evolution of baryon number density depends on heat conductivity (in the Landau frame one may see heat conductivity equivalently as baryon number diffusion). If one has a theoretical understanding of initial state perturbations in baryon number density and their fluid dynamic evolution, it is possible to study their consequences for particle spectra at freeze-out. Provided possible signals are large enough to be seen within the constraints set by finite statistics, there could be a possibility to constrain the heat conductivity of the quark gluon plasma from experimental data. This would be very interesting for not only low energy collision experiments which aim at exploring the QCD phase diagram, but also at RHIC and LHC energies where baryon number diffusion could be another characteristic of the quark-gluon plasma.

As a first step in this direction we study here the  fluid dynamic propagation of local and event-by-event fluctuations of the baryon number density, flow velocity and energy density. These fluctuations propagate on top of a hydrodynamical background which for simplicity, we consider to be described by Bjorken's model~\cite{Bjorken:1982qr} (which includes finite baryon number density). In order to study the fluid dynamic propagation of perturbations we use a background-fluctuation splitting and a Bessel-Fourier decomposition for the fluctuating part of the fluid fields~\cite{Florchinger:2011qf,Floerchinger:2013rya,Floerchinger:2013vua,Floerchinger:2013tya,Floerchinger:2013hza,Floerchinger:2014fta,Brouzakis:2014gka,Floerchinger:2014yqa}. We derive the evolution equations of the linear fluctuations and solve them for different initial conditions, values of the transport coefficients and equation of state.

This work is organized as follows. In Sec.~\ref{sec:fluid} we review briefly the theory of relativistic fluid dynamics at finite chemical potential putting emphasis on the role of the equation of state and current estimates of the transport coefficients in the strong and weakly coupling regimes. The main features of the temporal evolution of the background fields are discussed in Sec.~\ref{sec:Bjo}. In Sec.~\ref{sec:fluc+Bjo} we formulate the theory of linear perturbations on top of this evolving background and discuss numerical solutions. In Sec.~\ref{sec:corrFunct} we draw some conclusions for a potential experimental observable, the correlation function of net baryon number as a function of azimuthal angles and rapidity. General conclusions are presented in Sec.\ \ref{sec:concl}. 
Some technical details of our calculations are presented in Appendixes~\ref{app:ThermodynamicRelations} and~\ref{app:Linearizedrelativisticfluiddynamics}, respectively. 

%%%%%%%%%%%%%%%%%%%%%%%%%%%%%%%%%%%%%%%
\section{Relativistic fluid dynamics with a globally conserved charge}
\label{sec:fluid}
%%%%%%%%%%%%%%%%%%%%%%%%%%%%%%%%%%%%%%%

We consider a relativistic fluid with one globally conserved quantum number current (baryonic number for our purposes). The energy-momentum tensor and number current are
\begin{equation}
\begin{split}
T^{\mu\nu}&=\epsilon \, u^\mu u^\nu + (p+\pi_\text{bulk})\Delta^{\mu\nu}+\pi^{\mu\nu}\,,\\
N^\mu&= n \, u^\mu+\nu^\mu.
\end{split}
\label{eq:erergymomentumCurrent}
\end{equation}
Here, $\epsilon$ is the energy density, $u^\mu$ is the fluid velocity, $\pi^{\mu\nu}$ is the shear stress tensor, $\pi_\text{bulk}$ is the bulk viscous pressure, $n$ is the particle density and $\nu^\mu$ is the particle diffusion current. We choose the signature of the metric $g_{\mu\nu}$ to be $(-,+,+,+)$ and the projector orthogonal to the fluid velocity is
\begin{equation}
\Delta^{\mu\nu} = g^{\mu\nu} + u^\mu u^\nu.
\end{equation}
The fluid velocity is normalized to $u_\mu u^\mu = -1$.
We work in the Landau frame where the fluid velocity is chosen such that $u_\mu T^{\mu\nu} = - \epsilon \, u^\nu$. The shear stress tensor is transverse to the fluid velocity,
\begin{equation}
u_\mu \pi^{\mu\nu} = 0.
\end{equation}
The shear stress tensor is also symmetric and traceless. The particle number density is defined by $n=-u_\mu N^\mu$ such that the diffusion current is orthogonal to the fluid velocity, $u_\mu \nu^\mu=0$.

It is clear that an arbitrary (symmetric) energy-momentum tensor $T^{\mu\nu}$ (with a time-like eigenvector) and current $N^\mu$ can be written in the above form. The decomposition becomes unique by requiring that the pressure $p$ is related to the energy density $\epsilon$ and the baryon density $n$ by the same relation as in thermodynamic equilibrium, i.e. by an equation of state $p=p(\epsilon,n)$ . 

The evolution of the energy momentum-tensor and the particle current are constrained by the conservation equations
\begin{equation}
\begin{split}
\nabla_\mu T^{\mu\nu}&=0,\\
\nabla_\mu N^\mu &=0,
\end{split}
\label{eq:conservationlaws}
\end{equation}
where $\nabla_\mu$ denotes the covariant derivative. In this general form the conservation equations hold also in curved space-time but we are interested here in curvilinear systems defined in Minkowski space without taking into consideration the gravitational field. From Eqs.\ \eqref{eq:erergymomentumCurrent} and \eqref{eq:conservationlaws} one obtains the evolution equations for the energy density, fluid velocity and particle density
\begin{equation}
\begin{split}
D \epsilon + (\epsilon + p + \pi_\text{bulk}) \nabla_\mu u^\mu + \pi^{\mu\nu} \nabla_\mu u_\nu & =0,\\
(\epsilon + p + \pi_\text{bulk} ) \, D u^\nu  + \Delta^{\nu\mu} \, \partial_\mu (p + \pi_\text{bulk}) +\Delta^\nu\,_\alpha \nabla_\mu \pi^{\mu\alpha}&=0,\\
D n +n \nabla_\mu u^\mu +\nabla_\mu \nu^\mu &=0\,.
\end{split}
\label{eq:eveqnsgeneral}
\end{equation}
Here we have introduced the comoving derivative defined as $ D= u^\mu \nabla_\mu$. 

To close the evolution equations \eqref{eq:eveqnsgeneral} one needs expressions for $\pi_\text{bulk}$, $\pi^{\mu\nu}$ and $\nu^\mu$. Within the formalism of  fluid dynamics one writes these objects as a derivative expansion in terms of the fluid velocity $u^\mu$ and thermodynamic variables $\epsilon$, $n$. In the present work we concentrate for simplicity on the first order of this expansion. One should keep in mind that terms of second order are expected to improve the results quantitatively and are in general needed for an acceptable causal structure and linear stability \cite{Hiscock:1983zz,Pu:2009fj}. 

The constitutive relation for the shear stress is
\begin{equation}
\label{eq:NSshear}
\pi^{\mu\nu} = - 2 \eta \, \sigma^{\mu\nu}= - 2 \eta  \left[ \frac{1}{2} \Delta^{\mu\alpha} \Delta^{\nu\beta} + \frac{1}{2} \Delta^{\mu\beta} \Delta^{\nu\alpha} - \frac{1}{3} \Delta^{\mu\nu} \Delta^{\alpha\beta} \right] \nabla_\alpha u_\beta,
\end{equation}
where $\eta$ is the shear viscosity transport coefficient. The bulk viscous pressure is obtained from the following expression
\begin{equation}
\label{eq:NSbulk}
\pi_\text{bulk} = - \zeta \, \theta  = - \zeta \, \nabla_\mu u^\mu,
\end{equation}
where $\zeta$ is the bulk viscosity and $\theta$ is the expansion scalar. Finally, the particle diffusion current is
\beq
\label{eq:dissNS}
\nu^\alpha &=& - \kappa \left[ \frac{n T}{\epsilon + p} \right]^2 \iota^\alpha = - \kappa \left[ \frac{n T}{\epsilon + p} \right]^2 \Delta^{\alpha\beta}\partial_{\beta}\left(\frac{\mu}{T}\right)\,.
\eeq
where $\kappa$ is the heat conductivity. In the last equation we have introduced the chemical potential $\mu$, which is conjugate to the baryon density $n$, and the temperature $T$.

In summary, the hydrodynamic equations at this stage involve the fluid velocity $u^\mu$ (with three independent components), the energy density $\epsilon$, pressure $p$, baryon density $n$, baryon chemical potential $\mu$, temperature $T$ as well as the shear viscosity $\eta$, bulk viscosity $\zeta$ and the thermal conductivity $\kappa$. Only two thermodynamic variables are independent and they also determine the transport properties $\eta$, $\zeta$ and $\kappa$. In a non-equilibrium situation only energy density $\epsilon= u_\mu u_\nu T^{\mu\nu}$ and baryon number density $n=-u_\mu N^\mu$ are directly related to the physical energy-momentum tensor $T^{\mu\nu}$ and number current $N^\mu$. All other thermodynamic variables are defined indirectly via their relation to $\epsilon$ and $n$ in thermal equilibrium.

For the practical calculations one is in principle free to use any set of independent thermodynamic variables.  The form of Eqs.~\eqref{eq:eveqnsgeneral} suggests the use of the energy density $\epsilon$ and baryon density $n$. However, because most microscopic calculations are done in the grand canonical ensemble, the thermodynamic equation of state and the transport coefficients are usually obtained as a function of the temperature $T$ and chemical potential $\mu$, for example $p=p(T,\mu)$. Thus, it can be advantageous to use $T$ and $\mu$ as independent variables in fluid dynamics, as well. This avoids the inversion of functions which can be numerically difficult. One should keep in mind that $T$ and $\mu$ in a non-equilibrium situation are defined via their relation to $\epsilon$ and $n$. Eq.~\ \eqref{eq:eveqnsgeneral} can be transformed using thermodynamic relations compiled in Appendix~\ref{app:ThermodynamicRelations}. The evolution equation for energy density becomes
\begin{equation}
\label{eq:appAeq4}
\left[ T \frac{\partial^2 p}{\partial T^2} + \mu \frac{\partial^2 p}{\partial T\partial\mu} \right] D T 
+ \left[ T \frac{\partial^2 p}{\partial T\partial \mu} + \mu \frac{\partial^2 p}{\partial \mu^2} \right] D\mu + (\epsilon + p)\, \theta - 2 \eta \; \sigma_{\alpha\beta} \sigma^{\alpha \beta} - \zeta \; \theta^2 = 0.
\end{equation}
where we have now used the constitutive relations \eqref{eq:NSshear} and \eqref{eq:NSbulk}. The evolution equation for the fluid velocity is now of the form
\begin{equation}
(\epsilon + p)\, D u^\nu +  \Delta^{\nu\alpha}(s \, \partial_{\alpha} T + n\, \partial_{\alpha} \mu)- \Delta^{\nu}_{\;\; \alpha}{} \nabla_\beta \left( 2\, \eta\, \sigma^{\alpha\beta} + \zeta \,\Delta^{\alpha\beta} \, \nabla_\gamma u^\gamma \right)=0\,,
\label{eq:appAeq5}
\end{equation}
and finally, the particle number conservation law becomes
\begin{equation}
\begin{split}
& \frac{\partial^2 p}{\partial T\partial\mu} D T + \frac{\partial^2 p}{\partial \mu^2} D \mu + n\,\theta 
+ \nabla_\alpha \nu^\alpha=0\,.
\end{split}
\label{eq:appAeq6}
\end{equation}
Note that Eqs.\ \eqref{eq:appAeq4} and \eqref{eq:appAeq6} form a linear system of equations that can be solved for $D T= u^\alpha \partial_\alpha T$ and $D \mu = u^\alpha \partial_\alpha \mu$ as long as 
\begin{equation}
\frac{\partial^2 p}{\partial T^2}\frac{\partial^2 p}{\partial\mu^2} - \left(\frac{\partial^2 p}{\partial T \partial\mu}\right)^2 \neq 0.
\end{equation}

To solve the fluid dynamic equations we will use a background-fluctuation splitting. To this end we write the fluid dynamic fields as
\begin{equation}
u^\mu = \bar u^\mu + \delta u^\mu, \quad\quad \epsilon = \bar \epsilon + \delta \epsilon,
\end{equation}
and similar for the other fields. We are interested in perturbations $\delta u^\mu$, $\delta\epsilon$ etc.\ that are small enough so that only linear terms in the evolution equations need to be kept. The background fields $\bar u^\mu$, $\bar \epsilon$ etc.\ satisfy the fluid dynamic equations \eqref{eq:eveqnsgeneral} while the perturbations satisfy linear equations that depend on the background solution. We derive these linearized equations for arbitrary background fields in Appendix \ref{app:Linearizedrelativisticfluiddynamics}. The structure of the linearized equations permits us to simply use $\delta \epsilon$, $\delta n$ and three independent components of the fluid velocity as variables (the fourth component of the fluid velocity follows from the constraint $\bar u^\mu \delta u_\mu = 0$). However, all the background-dependent thermodynamic quantities can be expressed in terms of $\bar T$ and $\bar \mu$. Useful thermodynamic relations for this purpose are compiled in Appendix \ref{app:ThermodynamicRelations}. 

In the rest of this section we briefly discuss some simple parametrization of the thermodynamic equations of state $p(T,\mu)$ and transport properties $\eta(T,\mu)$, $\zeta(T,\mu)$, $\kappa(T,\mu)$. We emphasize that our formalism can be used for an arbitrary form of these functions once these have been determined from a particular microscopic description.

%%%%%%%%%%%%%%%%%%%%%%%%%%%%%%%%%%%%%%%%%%%%%%%%%%%%%%%%%%%%%%%%%%%%%%%%%%%%%%%%%%%%%%%%%%%%%%%%%%%%%%%%
\subsection{Equation of state}
\label{subsec:eos}
%%%%%%%%%%%%%%%%%%%%%%%%%%%%%%%%%%%%%%%%%%%%%%%%%%%%%%%%%%%%%%%%%%%%%%%%%%%%%%%%%%%%%%%%%%%%%%%%%%%%%%%%

The fluid hydrodynamical equations require an equation of state (EOS) $p(T,\mu)$ as an input. In principle, the equation of state can be calculated from the inherent quantum field theory associated to a particular system but this is a formidable task. In recent years there have been important advances to determine analytically and numerically the thermodynamical properties of QCD at high temperatures and chemical potential by considering effective thermal field theories~\cite{Bazavov:2014pvz,Borsanyi:2013bia,
Haque:2012my,Haque:2014rua,Haque:2013sja,Vuorinen:2003fs,deForcrand:2010ys} while in the low temperature and chemical potential regimes one expects that a non-interacting hadron resonance gas provides a reasonably good approximation~\cite{Huovinen:2009yb}. 

At intermediate temperatures, non-perturbative methods are needed to describe the transition which separates the hadronic, confined phase and the quark-gluon plasma (QGP) phase. While several studies of lattice QCD simulations are available at the moment at vanishing chemical potential $\mu=0$~\footnote{For a recent review of the lattice QCD studies we refer to the reader to Ref.~\cite{Ding:2015ona}.}, at $\mu>0$ lattice simulations are not possible due to the sign problem. However, different alternatives have been studied in order to circumvent this problem such as reweighting~\cite{Csikor:2004ik}, Taylor-expansion in $\mu$~\cite{Fodor:2002km,Allton:2003vx,Allton:2005gk,Bernard:2007nm,Basak:2009uv,DeTar:2010xm,Borsanyi:2012cr,Hegde:2014wga}, analytic continuation from imaginary $\mu$~\cite{Takaishi:2010kc}, the density of states method, or using the canonical ensemble. Of course, each of these methods have their advantages and disadvantages.

One of the main goals in the analysis of fluid dynamic fluctuations and their propagation is to provide a phenomenological determination of the equation of state (or at least of some of its properties). In the derivation of the evolution equations for the background and fluctuating fields we shall keep the equation of state $p(T,\mu)$ unspecified as far as possible in analytic expressions. For some numerical calculations and illustrations we use the simplest possible case, a non-interacting gas of $N_F$ massless quarks that come in $N_C$ colors and $N_C^2-1$ gluons,
\beq
\label{eq:eos}
p(T,\mu)= \frac{1}{4!}a_1\,T^4+\frac{1}{4}a_2\,T^2\mu^2+\frac{1}{4!}a_3\,\mu^4\,,\label{eq:idealEOS}
\eeq
where we use the abbreviations
\begin{equation}
\begin{split}
a_1&=\frac{8\pi^2}{15} %4!\frac{\pi^2}{45}
\Biggl(N_C^2-1+\frac{7}{4}N_C N_F\Biggr)\, ,\\
a_2&=%4 \frac{N_C N_F}{6\cdot 9}\,,\\
\frac{2 N_C N_F}{27}\,,\\
a_3&=%4! \frac{N_C N_F}{12\pi^2\cdot 81 }.
\frac{2 N_C N_F}{81\pi^2}.
\end{split}
\label{eq:EOSa}
\end{equation}
The baryon chemical potential $\mu$ measures the net baryon density of the system. In our convention, quarks carry baryon number charge $1/3$ and anti-quarks $-1/3$. 

Corrections to the ideal EOS arise as a consequence of interactions and the breaking of conformal invariance by dimensional transmutation and non-zero quark masses. They are most important at low temperatures. We follow here the Wuppertal collaboration which has parametrized the QCD equation of state for finite chemical potential in terms of a Taylor expansion~\cite{Borsanyi:2012cr}. The leading order expression for the trace anomaly or QCD interaction measure $I(T,\mu)=\epsilon(T,\mu)-3p(T,\mu)$ is 
\begin{equation}
\label{eq:intmeas}
\frac{I(T,\mu)}{T^4}=\frac{I(T,0)}{T^4}+\frac{\mu^2}{2T}\frac{\partial \chi_2 (T)}{\partial T}
\end{equation}
where $I(T,0)$ is the interaction measured at $\mu=0$ and $\chi_2(T)$ is the leading-order Taylor coefficient. Both terms, $I(T,0)$ and $\chi_2 (T)$, can be parametrized analytically as~\cite{Borsanyi:2012cr,Borsanyi:2010cj}
\begin{subequations}
\label{eq:lattpar}
\begin{align}
\label{eq:lattmeas-nomu}
\frac{I(T,0)}{T^4}&=e^{-h_1/t-h_2/t^2}\,\left[h_0+\frac{f_0\left(\tanh(f_1\,t+f_2)+1\right)}{1+g_1\,t+g_2\,t^2}
\right]\,,\\
\label{eq:taycoffmu}
\chi_2(T)&=e^{-h_3/t-h_4/t^2}\,f_3\,\left(\tanh(f_4\,t+f_5)+1\right)
\end{align}
\end{subequations}
where $t=T/(0.2$ GeV). For $N_f=2+1$ flavors of quarks with physical masses and finite baryon chemical potential $\mu$ the parameters in Eq.~\eqref{eq:lattpar} are $h_0 = 0.1396$, $h_1 = −0.1800$, $h_2 = 0.0350$, $f_0 = 2.76$, $f_1 = 6.79$, $f_2 = −5.29$, $g_1 = −0.47$,
$g_2 = 1.04$, $h_3=-0.5022$, $h_4=0.5950$, $f_3=0.0940$, $f_4=6.3290$ and $f_5=-4.8303$~\cite{Borsanyi:2012cr,Borsanyi:2010cj}. The pressure at finite $\mu$ is given by
\begin{equation}
\label{eq:EOSnonid}
\frac{p(T,\mu)}{T^4}=\frac{p(T,0)}{T^4} +\frac{1}{2}\frac{\mu^2}{T^2}\chi_2\,.
\end{equation}
At $\mu=0$ the relation between the pressure and the trace anomaly~\eqref{eq:lattmeas-nomu} is 
\begin{equation}
\frac{p(T,0)}{T^4}=\int_0^T\,dT'\,\frac{I(T',0)}{T'^5}\,. 
\end{equation}
All other thermodynamic quantities can be derived from $p(T,\mu)$ using the standard relations (compiled in appendix \ref{app:ThermodynamicRelations}).
The equation of state~\eqref{eq:EOSnonid} with the above parametrization is valid for small chemical potentials $\mu/T<3$ in the temperature window $0\,<\,T\,<\,400$ MeV. We will use Eq.~\eqref{eq:EOSnonid} to study the influence of the EOS for the dynamics of the background fluid dynamic fields.
 
%%%%%%%%%%%%%%%%%%%%%%%%%%%%%%%%%%%%%%%%%%%%%%%%%%%%%%%%%%%%%%%%%%%%%%%%%%%%%%%%%%%%%%%%%%%%%%%%%%%%%%%%
\subsection{Transport coefficients}
\label{subsec:transcoeff}
%%%%%%%%%%%%%%%%%%%%%%%%%%%%%%%%%%%%%%%%%%%%%%%%%%%%%%%%%%%%%%%%%%%%%%%%%%%%%%%%%%%%%%%%%%%%%%%%%%%%%%%%
%
\begin{table}
\begin{center}
\begin{tabular}{ | c|| c | c | c |}
\hline
{\bf Transport coefficient } & 
{\bf  Weakly-coupled QCD } & 
{\bf Strongly-coupled theories } \\ \hline
\multirow{2}{*} {$\eta$} & \multirow{2}{*} {$ k\,\frac{T^3}{g^4 \log(1/g)}$} & \multirow{2}{*} {$\frac{s(T,\mu)}{4 \pi}$}\\ 
&  &  \\ \hline
\multirow{2}{*} {$\zeta$} & \multirow{2}{*} {$15\,\eta(T)\left(\frac{1}{3}-c_s^2(T)\right)^2$} & \multirow{2}{*}{$2\,\eta(T,\mu)\left(\frac{1}{3}-c_s^2(T,\mu)\right)$} \\ 
&  &  \\ \hline
\multirow{2}{*} {$\kappa$} & $\sim\mu^2/g^4$ for $\mu\gg T$ &   \multirow{2}{*} {$8\pi^2\,\frac{T}{\mu^2}\eta (T,\mu)$}\\ 
& $\sim T^4/(g^4\mu^2)$for $\mu\ll T$ &  \\ 
\hline
\end{tabular}
\caption{Estimated values of the shear viscosity and different parametrizations for the bulk viscosities and heat conductivity for weakly-coupled QCD~\cite{Arnold:2000dr,Arnold:2003zc,Arnold:2006fz,Heiselberg:1993cr,Danielewicz:1984ww} and 
strongly coupled theories with holographic duals~\cite{Policastro:2001yc,Son:2006em,Buchel:2005cv}. See text for discussion.}
\label{Table1}
\end{center}
\end{table}
In addition to the thermodynamic equation of state, the fluid dynamical description needs as an input transport coefficients. These can either be determined experimentally, or, if a microscopic underlying theory is known, they can at least in principle be calculated as a function of the thermodynamic variables via Kubo relations. In this section we briefly summarize the current theoretical knowledge for the shear  and bulk viscosities and thermal conductivity of QCD and related theories, both in weakly and strongly coupled regimes\footnote{A more detailed discussion of the properties of the transport coefficients discussed in this work can be found in Ref.~\cite{Schafer:2009dj}.}.  

%%%%%%%%%%%%%%%%%%%%%%%%%%%%%%%%%%%%%%%%%%%%%%%%%%%%%%%%%%%%%%%%%%%%%%%%%%%%%%%%%%%%%%%%%%%%%%%%%%%%%%%%
\subsection{Weak coupling regime}
\label{subsec:weak}
%%%%%%%%%%%%%%%%%%%%%%%%%%%%%%%%%%%%%%%%%%%%%%%%%%%%%%%%%%%%%%%%%%%%%%%%%%%%%%%%%%%%%%%%%%%%%%%%%%%%%%%%
When the interaction strength is small, effective thermal field theory methods allows us to calculate the transport coefficients. For weakly coupled QCD in the high temperature and vanishing chemical potential regime, the leading logarithmic result for the shear viscosity is~\cite{Arnold:2000dr,Arnold:2003zc,Arnold:2006fz}
\begin{equation}
\label{w-shear}
\eta(T)= k\frac{T^3}{g^4\log(1/g)}\,,
\end{equation}
where $g$ is the strong coupling constant. In the previous expression $k$ is a constant that depends on the number of fermions species~\cite{Arnold:2000dr,Arnold:2003zc,Arnold:2006fz}. 
Arnold {\itshape et al.} showed that at leading log accuracy and for high temperatures with vanishing chemical potential there is an approximate scaling between the shear ($\eta$) and bulk ($\zeta$) viscosities for weakly coupled QCD~\cite{Arnold:2006fz}
\beq
\label{scalw-bulk-shear}
\zeta (T)\approx 15\eta(T)\left(\frac{1}{3}-c_s^2(T)\right)^2\,,
\eeq
where $c_s^2=d p/d\epsilon$ is the speed of sound. A similar expression was first derived by Weinberg  for a gas of photons~\cite{Weinberg:1971mx}. To date there is no complete leading logarithmic calculation of the heat conductivity $\kappa(T,\mu)$ and so far only two estimates of $\kappa(T,\mu)$ have been provided in the literature for different kinematic regions of the $T-\mu$ plane~\cite{Heiselberg:1993cr,Danielewicz:1984ww}
\beq
\label{eq:kappaw}
\kappa (T,\mu)=
\begin{cases}
 F(T,m_D)\,\mu^2/g^4, & \text{for $\mu\gg T$}, \\
C\,T^4/(g^4\,\mu^2), & \text{for $\mu\ll T$}. 
\end{cases}
\eeq
where $F(T,m_D)$ is a function that depends on the temperature and the Debye screening mass $m_D$ (see Ref.~\cite{Heiselberg:1993cr} for details). In the case of small chemical potential, the proportionality constant $C$ depends on the number of flavors and the gauge group~\cite{Danielewicz:1984ww}. In the limit where $\mu\to 0$ the heat conductivity $\kappa\sim \mu^{-2}$ is divergent. However, the particle diffusion current~\eqref{eq:dissNS} remains finite~\cite{Danielewicz:1984ww}. In the context of relativistic kinetic theory, some general expressions for the transport coefficients with constant cross section or within the relaxation time approximation have been derived recently~\cite{Denicol:2012es,Denicol:2012cn,Molnar:2013lta,Denicol:2014vaa,
Jaiswal:2014isa,Florkowski:2015lra}. However, these calculations do not take into account the quantum screening effects of the QCD plasma.

Despite relatively large uncertainties, experimental results indicate that the value of the shear viscosity over the entropy ratio $\eta/s$ is smaller than the one calculated from weakly coupled QCD~\eqref{w-shear}
~\cite{Heinz:2013th,Gale:2013da}. For the case of the bulk viscosity the situation is less clear: the uncertainties in its experimental determination are even larger (see Ref.~\cite{Noronha-Hostler:2013gga} and references therein). In addition, there are no experimental constraints for the value of heat conductivity in high energy-nuclear collisions so far.

%%%%%%%%%%%%%%%%%%%%%%%%%%%%%%%%%%%%%%%%%%%%%%%%%%%%%%%%%%%%%%%%%%%%%%%%%%%%%%%%%%%%%%%%%%%%%%%%%%%%%%%%
\subsection{Strong coupling regime}
\label{subsec:strong}
%%%%%%%%%%%%%%%%%%%%%%%%%%%%%%%%%%%%%%%%%%%%%%%%%%%%%%%%%%%%%%%%%%%%%%%%%%%%%%%%%%%%%%%%%%%%%%%%%%%%%%%%
From the previous discussion it is clear that at this moment perturbative QCD calculations of the transport coefficients are not completely under control for all the possible physical values of the temperature and chemical potential. On the other side, there are certain classes of strongly interacting theories where transport coefficients can be determined for almost all values of $T$ and $\mu$. These are field theories with known gravitational duals where the computations can be done via the anti-de Sitter/conformal field theory (AdS/CFT) correspondence. Despite the fact that those theories are not equivalent to QCD, they share some qualitative aspects with it and thus, these theories might provide some guidance in the regimes where pQCD calculations are not reliable. We take here a pragmatical approach and consider the estimates of the transport coefficients based on holographic calculations as toy models which allow us to study the propagation of perturbations in fluid dynamic fields. For large t'Hooft coupling and for ${\mathcal N}$= 4 SYM theory, holographic methods give the well known result~\cite{Policastro:2001yc},
\begin{equation}
\frac{\eta(T,\mu)}{s(T,\mu)}= \frac{1}{4\pi}\,.
\label{eq:etas}
\end{equation}
This result holds also for any holographic theory at sufficiently large coupling and number of colors as long as the theory is spatially isotropic. This relation for $\eta/s$ holds even in the presence of non-zero chemical potential~\cite{Son:2006em}. Initially this result was conjectured to be an universal lower bound but today there is evidence showing that this relation does not hold in general~\cite{Brigante:2007nu,Brigante:2008gz,Kats:2007mq,Natsuume:2007ty,Buchel:2008vz,Erdmenger:2010xm,Rebhan:2011vd,Critelli:2014kra}. Incidentally, the value of the shear viscosity extracted from experiments in high energy nuclear collisions is closer to the one predicted for strongly coupled theories~\eqref{eq:etas} than the one calculated in weakly coupled QCD~\eqref{w-shear} (see Ref.~\cite{Heinz:2013th} for a recent review). 

The shear viscosity has also been calculated for pure Yang-Mills theory using lattice gauge theory for specific values of temperature \cite{Meyer:2007ic,Meyer:2009jp}. The estimated values for $\eta/s$ are somewhat above the AdS/CFT values. Similarly, $\eta/s$ as a function of temperature for vanishing baryon chemical potential has also been estimated for Yang-Mills theory as well as QCD by using diagrammatic functional relations and gluon spectral functions obtained by numerical analytic continuation from Euclidean quantum field theory \cite{Christiansen:2014ypa,Haas:2013hpa}. The minimal value for QCD was found to be $\eta/s\approx 0.17$ at temperature $T\approx 1.3 \, T_c$. 

For holographic theories that deviate from conformal behavior the bulk viscosity has also been calculated~\cite{Buchel:2005cv}\footnote{We pointed out to the reader that Eq.~\eqref{eq:bulk-visc} was derived by means of the gauge/gravity duality in Ref.~\cite{Buchel:2005cv} for a specific model. Other non conformal field theories~\cite{Benincasa:2005iv} where the duality holds provide some modifications to the parametrization given by Eq.~\eqref{eq:bulk-visc}.}
\beq
\label{eq:bulk-visc}
\zeta(T,\mu)=2\eta(T,\mu) \left(\frac{1}{3}-c_s^2(T,\mu)\right)\,.
\eeq
As in the case of the shear viscosity value~\eqref{eq:etas} this relation holds for certain theories with finite chemical potential~\cite{Buchel:2010gd} but it is not an universal bound~\cite{Buchel:2011uj}. By comparing the scalings between $\zeta$ and $\eta$, Eqs.~\eqref{eq:bulk-visc} and~\eqref{scalw-bulk-shear}, one observes that they differ in the strong and weak coupling regime. This mismatch between both parametrizations is currently not understood. In the case of the thermal conductivity $\kappa$, the calculations for strongly coupled plasmas with finite chemical potential give the following result~\cite{Son:2006em}
\begin{equation}
\label{eq:thercond}
\kappa(T,\mu)=8\pi^2\frac{T}{\mu^2}\eta(T,\mu)\,,
\end{equation}
which is an analog of the Wiedemann-Franz law~\cite{Landau}\footnote{The relation~\eqref{eq:thercond} was derived originally for a conformal holographic theory. However, this expression does not hold for non-conformal systems within the AdS/CFT correspondence~\cite{Rougemont:2015ona}.}. As in the weakly coupled case~\eqref{eq:kappaw}, the heat conductivity is divergent $\sim\mu^{-2}$  while the particle diffusion current~\eqref{eq:dissNS} is finite. Recently the temperature-dependence of the first and second order transport coefficients have been studied in a particular holographic model~\cite{Finazzo:2014cna}. 

We summarize the discussion presented in this section in Table ~\ref{Table1}, where we show the estimates of the transport coefficients in both strong and weak coupling regimes. Mainly for reasons of simplicity, we shall concentrate here on the parametrizations of the transport coefficients in the strong coupling regime Eqs.~\eqref{eq:etas},~\eqref{eq:bulk-visc} and~\eqref{eq:thercond} for our numerical calculatons. Another advantage of using the parametrization of strongly coupled theories is that both transport coefficients, the bulk viscosity $\zeta$ and the heat conductivity $\kappa$, are proportional to the shear viscosity $\eta$ and thus, one can not only study the effect of the dissipative corrections but also one can investigate the `weak' and `strong' regimes by varying the values of $\eta/s$. We keep the functions $\eta(T,\mu)$, $\zeta(T,\mu)$, $\kappa(T,\mu)$ unspecified as far as possible in our analytic calculations.

%%%%%%%%%%%%%%%%%%%%%%%%%%%%%%%%%%%%%%%%%%%%%%%%%%%%%%%%%%%%%%%%%%%%%%%%%%%%%%%%%%%%%%%%%%%%%%%%%%%%%%%%
\section{Bjorken boost invariant solution}
\label{sec:Bjo}
%%%%%%%%%%%%%%%%%%%%%%%%%%%%%%%%%%%%%%%%%%%%%%%%%%%%%%%%%%%%%%%%%%%%%%%%%%%%%%%%%%%%%%%%%%%%%%%%%%%%%%%%
 
In this section we study the solutions of the fluid dynamical equations for a quark-gluon plasma undergoing boost invariant longitudinal expansion. We assume translational and rotational symmetry in the transverse plane and arrive at a simple model for the early stages of a heavy ion collision first studied by Bjorken~\cite{Bjorken:1982qr}. Our analysis is extended to the case where there is a non-vanishing baryon number density. The relatively simple homogeneous solutions will also serve as a background for a more elaborate discussion of perturbations around it in Sec.~\ref{sec:fluc+Bjo}. 

It is convenient to change from  Cartesian coordinates $x^\mu=(t,x_1,x_2,x_3)$ to the Milne coordinates $(\tau,r,\phi,\eta)$ where $\tau=\sqrt{t^2-x_3^2}$ is the longitudinal proper time, $\eta=\text{arctanh}(x_3/t)$ is the longitudinal (space) rapidity and $r$ and $\phi$ are the usual polar coordinates in the transverse plane. The metric in the Milne coordinates is $g_{\mu\nu}=\text{diag}(-1,1,r^2,\tau^2)$. The main advantage of using these coordinate systems is that the symmetries of the Bjorken solution are explicitly manifest. Specifically, the symmetry group $\text{ISO}(2)\otimes \text{SO}(1,1)\otimes \text{Z}_2$ consists of translations and rotations in the transverse plane, longitudinal boosts $\eta\to\eta+\Delta\eta$ and reflections $\eta \to - \eta$~\cite{Gubser:2010ze}. The Bjorken flow velocity $u^\mu=(1,0,0,0)$ is the only invariant unit vector and the symmetry also implies that all fluid dynamic fields depend only on the longitudinal proper time $\tau$~\cite{Bjorken:1982qr}.

From Eqs.~\eqref{eq:eveqnsgeneral} one finds that the evolution equations for energy density and particle number density are~
\begin{equation}
\begin{split}
\partial_\tau \epsilon + (\epsilon + p) \frac{1}{\tau} - \left(\tfrac{4}{3}\eta + \zeta \right) \frac{1}{\tau^2} = 0,\\
\partial_\tau n + n \frac{1}{\tau} = 0.
\end{split}
\label{eq:BjorkenExpEN}
\end{equation}
We have used here the Christoffel symbols of the Milne coordinate system. The non-vanishing ones are $\Gamma^\eta_{\tau \eta} = \Gamma^\eta_{\eta \tau} = 1/\tau$, $\Gamma^{\tau}_{\eta\eta} = \tau$, $\Gamma^\phi_{r \phi} = \Gamma^\phi_{\phi r} = 1/r$, $\Gamma^r_{\phi\phi} = - r$. The shear tensor defined in Eq.~\eqref{eq:NSshear} becomes $\sigma^{\mu\nu}=\text{diag}\left(0,-\frac{1}{3\tau}, -\frac{1}{3\tau r^2},\frac{2}{3\tau^3}\right)$ with $\sigma_{\mu\nu}\sigma^{\mu\nu} = \frac{2}{3\tau^2}$. The expansion scalar is $\theta= \frac{1}{\tau}$ and the projector orthogonal to the fluid velocity is $\Delta^\mu_{\;\;\nu} = \text{diag}(0,1,1,1)$. The particle diffusion current $\nu^\mu$~\eqref{eq:dissNS} is a vector orthogonal to $u^\mu$ and therefore vanishes exactly for the Bjorken flow.

While the particle number density is simply diluted by the one-dimensional expansion, the evolution of energy density in \eqref{eq:BjorkenExpEN} contains an additional loss term from the thermodynamic work done by the expansion and a gain term from shear and bulk viscous effects. After the variable change to $T$ and $\mu$ eq.\ \eqref{eq:BjorkenExpEN} becomes
\begin{equation}
\begin{split}
\partial_\tau T & + \frac{ - \frac{n}{\tau} \frac{\partial^2 p}{\partial T \partial\mu} + \frac{s}{\tau} \left( 1 - \frac{4\eta/3+\zeta}{s T \tau} \right) \frac{\partial^2 p}{\partial \mu^2}  }{\frac{\partial^2 p}{\partial T^2}\frac{\partial^2 p}{\partial\mu^2}- \left(\frac{\partial^2 p}{\partial T \partial\mu}\right)^2} = 0,\\
\partial_\tau \mu & + \frac{\frac{n}{\tau} \frac{\partial^2 p}{\partial T^2} -  \frac{s}{\tau}\left( 1 - \frac{4\eta/3+\zeta}{s T \tau} \right) \frac{\partial^2 p}{\partial T \partial\mu} }{\frac{\partial^2 p}{\partial T^2}\frac{\partial^2 p}{\partial\mu^2}- \left(\frac{\partial^2 p}{\partial T \partial\mu}\right)^2} = 0.
\end{split}
\label{eq:BjorkenEOM}
\end{equation}
We observe that the size of viscous corrections to an isentropic expansion is determined by the parameter
\begin{equation}
\gamma=\frac{4\eta/3+\zeta}{s T \tau}.
\label{eq:viscosityfactor}
\end{equation}
Formally, the gradient expansion underlying viscous fluid dynamics can be used for $\gamma\ll 1$. Note that for a given thermodynamic equation of state $p(T,\mu)$ and viscosities $\eta(T,\mu)$, $\zeta(T,\mu)$ one can solve the two coupled ordinary differential equations \eqref{eq:BjorkenEOM}.

In the remainder of this section we discuss as a simple illustrative example the equation of state of an ideal gas of massless quarks and gluons in Eq.\ \eqref{eq:idealEOS}. The evolution equations \eqref{eq:BjorkenEOM} for the temperature $T$ and chemical potential $\mu$ become
\begin{equation}
\begin{split}
\partial_\tau T + \frac{1}{3\tau} T - \left(\frac{\gamma T}{\tau}\right) \frac{\frac{1}{3}a_1 a_2 T^4 + (\frac{1}{3}a_1 a_3 + a_2^2) T^2 \mu^2 + a_2 a_3 \mu^4}{a_1 a_2 T^4 + (a_1 a_3 - 3 a_2^2) T^2 \mu^2 + a_2 a_3 \mu^4} = 0,\\
\partial_\tau \mu + \frac{1}{3\tau} \mu + \left(\frac{\gamma T}{\tau}\right) \frac{\frac{2}{3}a_1 a_2 T^3\mu + 2 a_2^2 T \mu^3}{a_1 a_2 T^4 + (a_1 a_3 - 3 a_2^2) T^2 \mu^2 + a_2 a_3 \mu^4} = 0,
\end{split}
\label{eq:BjorkenEvolutionTmuIdealEOS}
\end{equation}
where the coefficients $a_1$, $a_2$ and $a_3$ are given in Eq.~\eqref{eq:EOSa}. Note that we use conventions
where $\mu$ is the chemical potential for baryons, and the chemical potential for quarks is $\mu_q=\mu/3$.

Let us first discuss some interesting limiting cases of Eqs.~\eqref{eq:BjorkenEvolutionTmuIdealEOS}:
\begin{enumerate}
\item \textbf{Ideal fluid dynamic expansion}. When shear and bulk viscosities vanish, $\eta=\zeta=0$, the temperature and the chemical potential decouple from each other. This allows us to solve Eqs.\ \eqref{eq:BjorkenEvolutionTmuIdealEOS} exactly, which gives
\begin{subequations}
\label{eq:scalingTmu}
\begin{align}
\label{eq:scalingT}
T(\tau) &= T(\tau_0) \left(\frac{\tau_0}{\tau}\right)^{1/3}, \\ 
\label{eq:scalingmu}
\mu(\tau) &= \mu(\tau_0) \left(\frac{\tau_0}{\tau}\right)^{1/3}. 
\end{align}
\end{subequations}
The scaling solution of the temperature is not modified by the presence of the chemical potential and it coincides with the well known result found by Bjorken \cite{Bjorken:1982qr}.

\item {\bf Vanishing chemical potential}. The point with $\mu =0$ corresponds to a (partial) fixed point of the evolution equations~\eqref{eq:BjorkenEvolutionTmuIdealEOS} with extended symmetry (baryon number parity).
The evolution equation for temperature becomes
\beq
\label{eq:bjoTnomu}
\partial_\tau T+\frac{T}{3\tau}(1-\gamma)=0.
\eeq
where $\gamma$ is given by Eq.~\eqref{eq:viscosityfactor}. For vanishing bulk viscosity, $\zeta=0$, and constant ratio $\eta/s$, the exact solution to the previous equation is~\cite{Kouno:1989ps,Muronga:2001zk,Muronga:2003ta,Baier:2006um}
\beq
\label{eq:bjosolTnomu}
T(\tau)=T(\tau_0)\,\left(\frac{\tau_0}{\tau}\right)^{1/3}\left[1+\frac{2}{3\tau_0 T(\tau_0)}\frac{\eta}{s}\left(1-\left(\frac{\tau_0}{\tau}\right)^{2/3}\right)\right]\,.
\eeq
Viscous corrections are relevant only at early times where velocity gradients are large while at late times these are suppressed and thus, $T(\tau)\sim\tau^{-1/3}$. 

\item {\bf Small chemical potential}. For $ \mu/T\ll 1$ the dynamics of $T$ is approximmately determined by Eq.~\eqref{eq:bjoTnomu} while the evolution equation for $\mu$ is
\begin{equation}
\partial_\tau \mu + \frac{\mu}{3\tau} \left( 1 + 2 \gamma \right) = 0.
\label{eq:evolutionmusmallmu}
\end{equation}
The viscous effects (encoded in the parameter $\gamma$) have the tendency to accelerate the decrease of $\mu$ due to the expansion. This is in contrast to the temperature where viscosity has the opposite effect. To lowest order in $\eta/s$, the solution of \eqref{eq:evolutionmusmallmu} is
\beq
\label{eq:bjosolsmallmu}
\mu(\tau)=\mu(\tau_0)\,\left(\frac{\tau_0}{\tau}\right)^{1/3}\left[1-\frac{4}{3\tau_0 T(\tau_0)}\frac{\eta}{s}\left(1-\left(\frac{\tau_0}{\tau}\right)^{2/3}\right)\right]\,.
\eeq

\item {\bf Small temperature}. For $ T/\mu\ll 1$ the evolution equation for the chemical potential is the one of eq.~\eqref{eq:scalingmu} with a simple scaling solution. For the temperature we obtain to lowest order in $T/\mu$
\begin{equation}
\label{eq:smallT}
\partial_\tau T + \frac{T}{3\tau}(1 - 3\gamma)=0,
\end{equation}
which has a solution similar to Eq.\ \eqref{eq:bjosolTnomu} when $\eta/s$ and $\zeta/s$ have constant values. If one chooses $T(\tau_0) = 0$ as initial condition the solution to Eq.~\eqref{eq:smallT} becomes
\begin{equation}
T(\tau) = \frac{4\eta + 3 \zeta}{2s} \left( \frac{1}{\tau_0^{2/3}\tau^{1/3}} - \frac{1}{\tau}\right).
\label{eq:solsmallT}
\end{equation}
Even if the temperature vanishes initially, the system is heated up due to shear and bulk dissipative effects. In contrast to $\mu=0$, vanishing temperature $T=0$ does not correspond to a (partial) fixed point of the evolution.
\end{enumerate}

\begin{figure}
\includegraphics[width=1\textwidth]{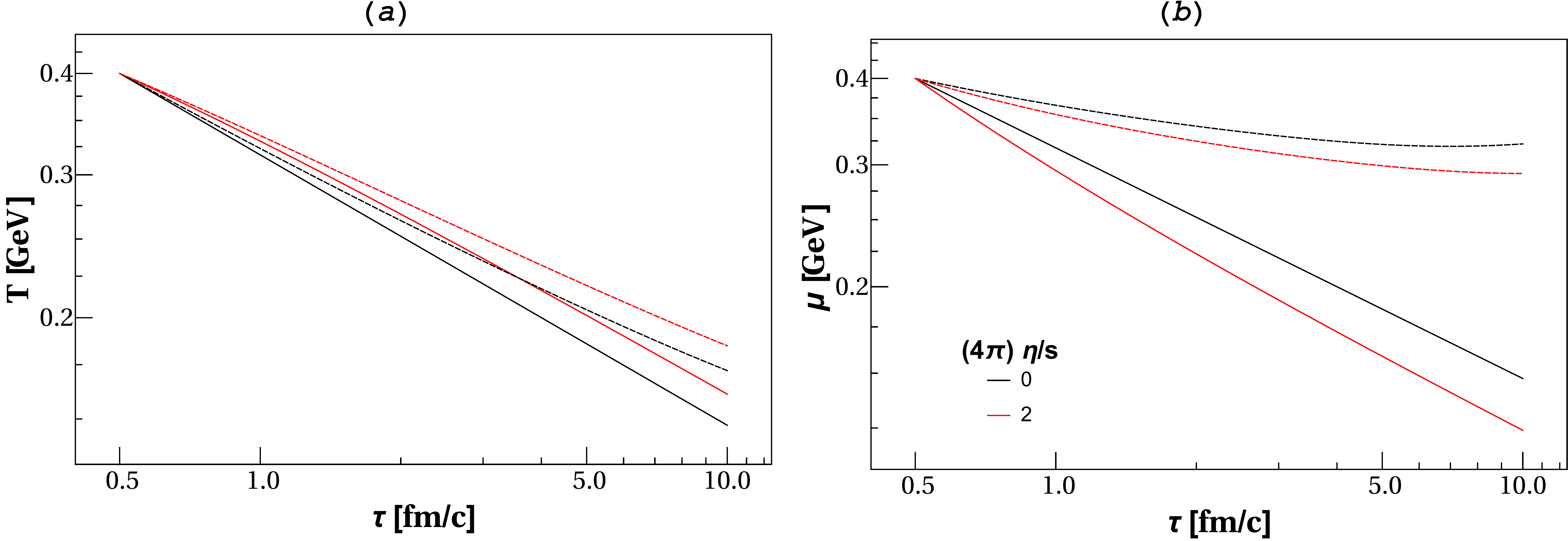}
\caption{(Color online) Log-log plot of the time evolution of (a) temperature (left panel) and (b) chemical potential (right panel) for the ideal EOS~\eqref{eq:idealEOS} (solid lines) and the lattice-based EOS~\eqref{eq:EOSnonid} (dashed lines). We choose here $\eta/s =0$ (black line) and $\eta/s=2/(4\pi)$ (red line). For the initial conditions we select $T(\tau_0)=\mu(\tau_0) = 0.4 \text{ GeV}$ and $\tau_0= 0.5 \text{ fm/c}$.}
\label{fig1}
\end{figure}

Let us now consider the evolution equations \eqref{eq:BjorkenEvolutionTmuIdealEOS} in the general case where we find their solution numerically. In Fig.~\ref{fig1} we show the time evolution of the temperature (left panel) and chemical potential (right panel) for different constant values of $\eta/s=0$ and $\eta/s=2/(4\pi)$ (black and red lines respectively) and two different parametrizations of the equation of state:  the ideal EOS~\eqref{eq:idealEOS} (solid lines) and the lattice-based EOS~\eqref{eq:EOSnonid} (dashed lines). The initial values at time $\tau_0=0.5\text{ fm/c}$ are taken to be $T(\tau_0) = 0.4 \text{ GeV}$ and $\mu(\tau_0)= 0.4 \text{ GeV}$. For the ideal EOS the coefficients $a_1$, $a_2$ and $a_3$ are taken according to Eq.~\eqref{eq:EOSa} with $N_C=N_F=3$. The bulk viscosity~\eqref{eq:bulk-visc} vanishes exactly for the ideal EOS~\eqref{eq:idealEOS} but it becomes a function of the temperature and chemical potential for the lattice-based EOS~\eqref{eq:EOSnonid}.

First we discuss the properties of the numerical solutions of Eqs.~\eqref{eq:BjorkenEOM} for the ideal (and massless) EOS~\eqref{eq:idealEOS}. For both variables, $T$ and $\mu$, the effect of viscous corrections are more relevant during the early stages of the expansion while at late times their effects are negligible as expected. In the left panel of Fig.~\ref{fig1} we see that the viscosity reduces the effect of the longitudinal expansion on the temperature. This is simply the expected heating by dissipative effects. At the final time $\tau_f=10 \,\text{fm/c}$ the temperature is larger by values of the order of $10\%$ for $\eta/s=2/(4\pi)$ compared to the ideal fluid expansion. For the chemical potential we find that the inclusion of dissipative corrections has the opposite effect, i.e. the chemical potentials decrease faster in the viscous case. This is clearly seen in the right panel of Fig.~\ref{fig1} when comparing the final values of the chemical potential $\mu (\tau_f)$. The changes with respect to the ideal fluid expansion are also somewhat larger, of the order of $15\%$ for $\eta/s=2/(4\pi)$.

When using the lattice-based EOS~\eqref{eq:EOSnonid} we find that the numerical solutions of Eqs.~\eqref{eq:BjorkenEOM} for $T$ and $\mu$ are qualitatively similar to the ones obtained from the ideal EOS during the early stages of the evolution. As a function of time, the temperature is always decreasing and the dissipative corrections are larger at early times than at late times. The chemical potential decreases faster for larger values of the shear viscosity. For the lattice EOS, the changes induced by the dissipative corrections are on the order of $8-15\%$. 

Interestingly, the evolution of $\mu$ with time differs substantially between the two choices for the equation of state. In the right panel of Fig.~\ref{fig1} one observes that the decrease with time is much weaker for the lattice EOS than for the ideal EOS. At the freeze-out time  $\tau_f=10$ fm/c and for vanishing $\eta/s$, one has $\mu(\tau_f)\approx$ 0.29 GeV for the lattice EOS while $\mu(\tau_f)\approx$ 0.12 GeV for the ideal EOS. The difference between those values increases slightly for finite values of $\eta/s$. Moreover, at late times $\mu$ increases slowly (and somewhat more for larger values of $\eta/s$). Our numerical results show also that when using the lattice EOS the values of the temperature are somewhat larger than for the ideal EOS specially at late times.

%--------------------------Fig. 1------------------------------------ 
%----------------------------------------------------------------------------------------------------
%--------------------------Fig. 2------------------------------------ 
\begin{figure}
\begin{center}
\includegraphics[width=1\textwidth]{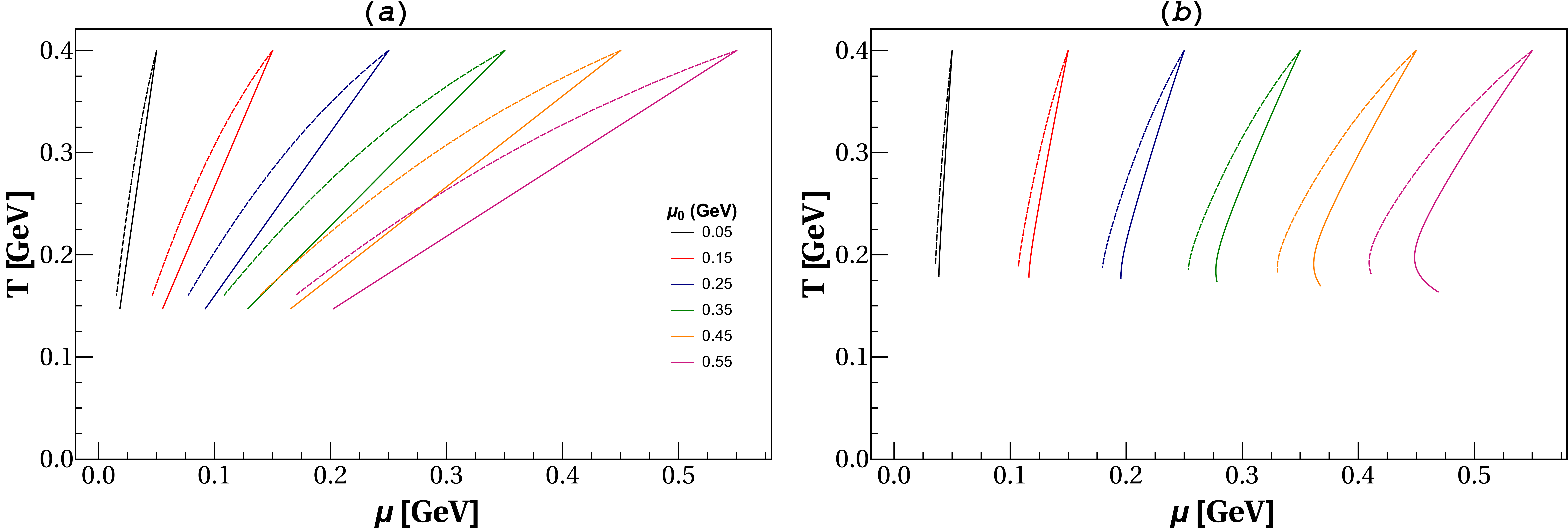}
\end{center}
\caption{(Color online) Flow trajectory of Bjorken expansions in the $\mu$-$T$-plane, initialized for (a) ideal EOS~\eqref{eq:idealEOS} (left panel) and (b) lattice-based EOS~\eqref{eq:EOSnonid}  (right panel). For the initial conditions we choose $\tau_0=0.5 \text{ fm/c}$, $T(\tau_0) = 0.4 \text{ GeV}$ and different values of $\mu(\tau_0)=\{0.05,0.15,0.25,0.35,0.45,0.55\}$ GeV. In both panels we compare the viscous effects by choosing $\eta/s=2/(4\pi)$ (dashed lines) to the the case of vanishing viscosity, $\eta/s=0$ (solid lines). All lines end at fixed final time $\tau_f = 10 \text{ fm/c}$. Note that we use conventions where $\mu$ is the chemical potential for baryons,
the chemical potential for quarks is $\mu_q=\mu/3$.}
\label{fig2}
\end{figure}
%----------------------------------------------------------------------------------------------------

In Fig.\ \ref{fig2} we show the Bjorken flow trajectories in the plane of chemical potential $\mu$ and temperature $T$ for the ideal EOS~\eqref{eq:idealEOS} (left panel) and the lattice EOS (right panel). For the initial conditions we choose $\tau_0=0.5$ fm/c, $T(\tau_0)=0.4$ GeV and different values of $\mu(\tau_0)=\{0.05,0.15,0.25,0.35,0.45,0.55\}$ GeV. For both equations of state we vary the shear viscosity to entropy $\eta/s=2/(4\pi)$ (dashed lines) and $\eta/s=0$ (solid lines). All trajectories end at fixed final time $\tau=10\text{ fm/c}$. 

For the ideal EOS (left panel of Fig.~\ref{fig2}) we observe that the viscosity weakens the effect of the expansion on the temperature $T$ while it does the opposite for the chemical potential $\mu$ and thus the trajectories end at larger values of $T$ and smaller vales of $\mu$ for non-zero $\eta/s$. This is in agreement with the previous discussion of the temporal evolution of $T$ and $\mu$. 
For the lattice EOS (right panel of Fig.~\ref{fig2}) we observe similar trajectories for small initial values of $\mu(\tau_0)$. For larger values of $\mu(\tau_0)$, the trajectories start to bend towards larger values of $\mu$ while they continue to decrease towards lower values of $T$. This behavior is understood from the previous discussion, as well. 

In summary, the time-evolution of temperature and chemical potential for a Bjorken expansion is given by Eqs.\ \eqref{eq:BjorkenEOM} for an arbitrary EOS. The evolution of $\mu$ as a function of time is quite sensitive to the choice of the EOS. The effect of the viscosity is relatively small. This is actually expected for the homogeneous background while we expect more prominent dissipative effects for non-homogeneous perturbations around it.~\footnote{The effect of shear viscosity is also sizable for the transverse expansion (radial flow) and for elliptic flow~\cite{Song:2007ux}.} We turn to those in the next section.

%%%%%%%%%%%%%%%%%%%%%%%%%%%%%%%%%%%%%%%%%%%%%%%%%%%%%%%%%%%%%%%%%%%%%%%%%%%%%%%%%%%%%%%%%%%%%%%%%%%%%%%%
\section{Fluctuations around Bjorken flow}
\label{sec:fluc+Bjo}
%%%%%%%%%%%%%%%%%%%%%%%%%%%%%%%%%%%%%%%%%%%%%%%%%%%%%%%%%%%%%%%%%%%%%%%%%%%%%%%%%%%%%%%%%%%%%%%%%%%%%%%%

After having studied the solution of the hydrodynamic evolution equations with Bjorken boost invariance and transverse translational symmetries we study now the evolution of fluctuations or deviations from that solution. We will concentrate here on deviations that are small enough in magnitude to describe their evolution by linearized evolution equations. In other words, we write the fluid dynamic fields as
\begin{equation}
u^\mu = \bar u^\mu + \delta u^\mu, \quad \quad \epsilon = \bar \epsilon + \delta \epsilon, \quad\quad n= \bar n + \delta n,
\end{equation}
where $\bar u^\mu$, $\bar \epsilon$, $\bar n$ is the Bjorken-type solution discussed in the previous section. The linearized evolution equations for the perturbations $\delta u^\mu$, $\delta \epsilon$, $\delta n$ are discussed for a generic background solution and arbitrary coordinate system in Appendix \ref{app:Linearizedrelativisticfluiddynamics}. If one specializes to the Bjorken background and the coordinate system $(\tau,r,\phi,\eta)$, the independent fluid dynamic fields are in the first order formalism $\delta \epsilon$, $\delta n$, $\delta u^r$, $\delta u^\phi$ and $\delta u^\eta$. (We take the background fluid velocity $\bar u^\mu$ and the full fluid velocity $u^\mu=\bar u^\mu + \delta u^\mu$ to be normalized, $u^\mu u_\mu = \bar u^\mu \bar u_\mu = -1$, such that one has $\delta u^\tau=0$ at linear order in perturbations). 
Equation \eqref{eq:B3} yields the following equation for the perturbation in energy density (each hydrodynamical fluctuating field depends on ($\tau,r,\phi,\eta$) which we suppress for better readability)
\begin{equation}
\begin{split}
\partial_\tau \delta \epsilon & + \left[ \frac{1}{\tau}+\frac{1}{\tau} \left( \frac{\partial p}{\partial \epsilon} \right)_n - \frac{1}{\tau^2} \left( \frac{\partial \zeta}{\partial \epsilon} \right)_n - \frac{4}{3\tau^2} \left( \frac{\partial \eta}{\partial \epsilon} \right)_n \right] \delta\epsilon\\
&+ \left[  \frac{1}{\tau} \left( \frac{\partial p}{\partial n} \right)_\epsilon - \frac{1}{\tau^2} \left( \frac{\partial \zeta}{\partial n} \right)_\epsilon - \frac{4}{3\tau^2} \left( \frac{\partial \eta}{\partial n} \right)_\epsilon \right] \delta n \\
&+ \left[ \bar\epsilon + \bar p - \frac{2}{\tau}\bar \zeta + \frac{4}{3\tau} \bar \eta \right] \left( \partial_r \delta u^r + \frac{1}{r} \delta u^r + \partial_\phi \delta u^\phi + \partial_\eta \delta u^\eta \right) - \frac{4}{\tau} \bar \eta \; \partial_\eta \delta u^\eta = 0.
\end{split}
\label{eq:LinEvEqEnergy}
\end{equation}
The thermodynamic derivatives like $(\partial p/\partial \epsilon)_n$, etc., are to be evaluated here on the background solution and similarly the transport coefficients and their derivatives. The evolution equation for the perturbation in baryon number density is
\begin{equation}
\begin{split}
\partial_\tau \delta n  & + \frac{1}{\tau} \delta n + \left[ \bar n - \bar\kappa \left[\frac{\bar n\bar T}{\bar \epsilon + \bar p}\right]^2 \partial_\tau\left(\frac{\bar\mu}{\bar T}\right) \right] \left( \partial_r \delta u^r + \frac{1}{r} u^r + \partial_\phi \delta u^\phi + \partial_\eta \delta u^\eta \right) \\
& - \bar\kappa \left[\frac{\bar n\bar T}{\bar \epsilon + \bar p}\right]^2 \left( \frac{\partial(\mu/T)}{\partial\epsilon} \right)_n \left( \partial_r^2 + \frac{1}{r}\partial_r  + \frac{1}{r^2} \partial_\phi^2 + \frac{1}{\tau^2} \partial_\eta^2\right) \delta\epsilon \\
& - \bar\kappa \left[\frac{\bar n\bar T}{\bar \epsilon + \bar p}\right]^2 \left( \frac{\partial(\mu/T)}{\partial n} \right)_\epsilon \left( \partial_r^2 + \frac{1}{r}\partial_r  + \frac{1}{r^2} \partial_\phi^2 + \frac{1}{\tau^2} \partial_\eta^2\right) \delta n = 0.
\end{split}
\label{eq:LinEvEqDensity}
\end{equation}
The derivative operator of second order that appears in the last two lines in front of $\delta \epsilon$ and $\delta n$, respectively, is the Laplace operator in the spatial coordinates $r$, $\phi$ and $\eta$. 

The fluid velocity in the radial direction is determined by the following evolution equation
\begin{equation}
\begin{split}
& \left( \bar \epsilon + \bar p - \frac{1}{\tau}\bar \zeta + \frac{2 }{3\tau}\bar \eta \right) \partial_\tau \delta u^r + \left[ \partial_\tau \bar p - \frac{1}{\tau}\partial_\tau\bar \zeta +\frac{1}{\tau^2} \bar \zeta + \frac{2}{3\tau} \partial_\tau \bar \eta + \frac{4}{3\tau^2} \bar \eta \right] \delta u^r \\
& + \left[ \left( \frac{\partial p}{\partial \epsilon} \right)_n - \frac{1}{\tau}\left( \frac{\partial \zeta}{\partial \epsilon} \right)_n + \frac{2}{3\tau} \left( \frac{\partial \eta}{\partial \epsilon} \right)_n \right] \partial_r \delta \epsilon 
+ \left[ \left( \frac{\partial p}{\partial n} \right)_\epsilon - \frac{1}{\tau}\left( \frac{\partial \zeta}{\partial n} \right)_\epsilon + \frac{2}{3\tau} \left( \frac{\partial \eta}{\partial n} \right)_\epsilon \right] \partial_r \delta n \\
& - \bar \zeta \left[ \left(\partial_r^2  + \frac{1}{r} \partial_r  - \frac{1}{r^2} \right) \delta u^r + \partial_r \partial_\phi \delta u^\phi + \partial_r \partial_\eta \delta u^\eta \right]\\
& - \bar \eta \left[ \left(\frac{4}{3} \partial_r^2 + \frac{4}{3r} \partial_r - \frac{4}{3 r^2} + \frac{1}{r^2} \partial_\phi^2  + \frac{1}{\tau^2} \partial_\eta^2 \right)\delta u^r + \left(\frac{1}{3}\partial_r \partial_\phi  - \frac{2}{r}\partial_\phi \right)\delta u^\phi + \frac{1}{3} \partial_r \partial_\eta \delta u^\eta\right] = 0 ,
\end{split}
\label{eq:LinEvEqRadialVelocity}
\end{equation}
the one in the azimuthal direction by
\begin{equation}
\begin{split}
& \left( \bar \epsilon + \bar p - \frac{1}{\tau}\bar \zeta + \frac{2 }{3\tau}\bar \eta \right) \partial_\tau \delta u^\phi + \left[ \partial_\tau \bar p - \frac{1}{\tau}\partial_\tau\bar \zeta +\frac{1}{\tau^2} \bar \zeta + \frac{2}{3\tau} \partial_\tau \bar \eta + \frac{4}{3\tau^2} \bar \eta \right] \delta u^\phi \\
& + \left[ \left( \frac{\partial p}{\partial \epsilon} \right)_n - \frac{1}{\tau}\left( \frac{\partial \zeta}{\partial \epsilon} \right)_n + \frac{2}{3\tau} \left( \frac{\partial \eta}{\partial \epsilon} \right)_n \right] \frac{1}{r^2} \partial_\phi \delta \epsilon 
+ \left[ \left( \frac{\partial p}{\partial n} \right)_\epsilon - \frac{1}{\tau}\left( \frac{\partial \zeta}{\partial n} \right)_\epsilon + \frac{2}{3\tau} \left( \frac{\partial \eta}{\partial n} \right)_\epsilon \right] \frac{1}{r^2}\partial_\phi \delta n \\
& - \bar \zeta \left[ \left( \frac{1}{r^2} \partial_r \partial_\phi+ \frac{1}{r^3} \partial_\phi \right) \delta u^r + \frac{1}{r^2}\partial_\phi^2 \delta u^\phi + \frac{1}{r^2}\partial_\phi \partial_\eta \delta u^\eta \right]\\
& - \bar \eta \left[ \left(\frac{1}{3r^2} \partial_r\partial_\phi + \frac{7}{3r^3} \partial_\phi \right)\delta u^r + \left(\partial_r^2  + \frac{3}{r}\partial_r + \frac{4}{3r^2} \partial_\phi^2 + \frac{1}{\tau^2} \partial_\eta^2 \right)\delta u^\phi + \frac{1}{3 r^2} \partial_\phi \partial_\eta \delta u^\eta\right] = 0 ,
\end{split}
\label{eq:LinEvEqAzimuthalVelocity}
\end{equation}
and finally the fluid velocity component in the rapidity direction is governed by
\begin{equation}
\begin{split}
& \left( \bar \epsilon + \bar p - \frac{1}{\tau}\bar \zeta - \frac{4}{3\tau}\bar \eta \right) \partial_\tau \delta u^\eta 
+ \left[ \partial_\tau \bar p + \frac{2}{\tau}(\bar \epsilon + \bar p)  - \frac{1}{\tau}\partial_\tau\bar \zeta +\frac{1}{\tau^2} \bar \zeta - \frac{4}{3\tau} \partial_\tau \bar \eta - \frac{4}{3\tau^2} \bar \eta \right] \delta u^\eta \\
& + \left[ \left( \frac{\partial p}{\partial \epsilon} \right)_n - \frac{1}{\tau}\left( \frac{\partial \zeta}{\partial \epsilon} \right)_n - \frac{4}{3\tau} \left( \frac{\partial \eta}{\partial \epsilon} \right)_n \right] \frac{1}{\tau^2} \partial_\eta \delta \epsilon 
+ \left[ \left( \frac{\partial p}{\partial n} \right)_\epsilon - \frac{1}{\tau}\left( \frac{\partial \zeta}{\partial n} \right)_\epsilon - \frac{4}{3\tau} \left( \frac{\partial \eta}{\partial n} \right)_\epsilon \right] \frac{1}{\tau^2}\partial_\eta \delta n \\
& - \bar \zeta \left[ \left( \frac{1}{\tau^2} \partial_r \partial_\eta+ \frac{1}{\tau^2 r} \partial_\eta \right) \delta u^r + \frac{1}{\tau^2}\partial_\phi \partial_\eta \delta u^\phi + \frac{1}{\tau^2}\partial_\eta^2 \delta u^\eta \right]\\
& - \bar \eta \left[ \left(\frac{1}{3\tau^2} \partial_r\partial_\eta + \frac{1}{3\tau^2 r} \partial_\eta \right)\delta u^r + \frac{1}{3 \tau^2} \partial_\phi\partial_\eta \delta u^\phi + \left(\partial_r^2  + \frac{1}{r}\partial_r + \frac{1}{r^2} \partial_\phi^2 + \frac{4}{3\tau^2} \partial_\eta^2 \right)\delta u^\eta \right] = 0 .
\end{split}
\label{eq:LinEvEqRapidityVelocity}
\end{equation}
Equations~\eqref{eq:LinEvEqEnergy} - \eqref{eq:LinEvEqRapidityVelocity} are hyperbolic coupled linear differential equations for the variables $\delta \epsilon$, $\delta n$, $\delta u^r$, $\delta u^\phi$ and $\delta u^\eta$: They contain only first order derivatives with respect to the time coordinate $\tau$ but up to second order derivatives with respect to the spatial coordinates $r$, $\phi$ and $\eta$. In the second order gradient expansion the equations would be elliptical but also contain more degrees of freedom and transport coefficients. 

In order to analyze the differential equations \eqref{eq:LinEvEqEnergy} - \eqref{eq:LinEvEqRapidityVelocity} it is convenient to use a Bessel-Fourier transformation. For the perturbation in energy density this reads
\begin{equation}
\begin{split}
\delta \epsilon(\tau,r,\phi,\eta) = \int_0^\infty dk \, k \sum_{m=-\infty}^\infty \int \frac{dq}{2\pi} \, \delta\epsilon(\tau,k,m,q) \, e^{i(m\phi + q\eta)} J_m(k r) ,
\end{split}
\label{eq:BesselFourierExpansionEnergy}
\end{equation}
with inverse relation
\begin{equation}
\begin{split}
\delta \epsilon(\tau,k,m,q) = \int_0^\infty dr \, r \frac{1}{2\pi}\int_0^{2\pi} d\phi  \int d\eta \; \delta\epsilon(\tau,r,\phi,\eta) \, e^{-i(m\phi + q\eta)} J_m(k r) .
\end{split}
\end{equation}
Since $\delta\epsilon(\tau,r,\phi,\eta)\in \mathbb{R}$ and $J_{-m}(k r) = (-1)^m J_m(k r) $ one has 
\begin{equation}
\delta\epsilon^{*}(\tau,k,m,q) = (-1)^m \delta\epsilon(\tau,k,-m,-q).
\end{equation}
For the baryon number density fluctuation $\delta n$ and the rapidity component of the fluid velocity $\delta u^\eta$ one can use the same expansion. For the fluid velocity components $\delta u^r $ and $\delta u^\phi$ we write instead
\begin{equation}
\begin{split}
\delta u^r(\tau,r,\phi,\eta) & = \frac{1}{\sqrt{2}} \left[\delta u^-(\tau,r,\phi,\eta) + \delta u^+(\tau,r,\phi,\eta) \right], \\\quad \delta u^\phi(\tau,r,\phi,\eta) & = \frac{i}{r \sqrt{2}} \left[ \delta u^-(\tau,r,\phi,\eta) - \delta u^+(\tau,r,\phi,\eta) \right], 
\end{split}
\end{equation}
with $ \delta u^{+*}(\tau,r,\phi,\eta) = \delta u^-(\tau,r,\phi,\eta)$. We expand $\delta u^-(\tau,r,\phi,\eta)$ and $\delta u^+(\tau,r,\phi,\eta)$ similar to Eq.\ \eqref{eq:BesselFourierExpansionEnergy} but replace $J_{m}(kr)$ by $J_{m-1}(kr)$ and $J_{m+1}(kr)$, respectively. The reality constraint becomes
\begin{equation}
\delta u^{+*}(\tau,k,m,q) = (-1)^{m+1}\delta u^{-}(\tau,k,-m,-q).
\end{equation}
In terms of the Bessel-Fourier transformed variables one can easily perform the spatial derivatives in Eqs.\ \eqref{eq:LinEvEqEnergy} - \eqref{eq:LinEvEqRapidityVelocity}. To that end it is useful to use the relations
\begin{equation}
\begin{split}
\frac{m}{r} J_m(kr) & = \frac{k}{2} \left[ J_{m-1}(kr) + J_{m+1}(kr) \right] ,\\
\frac{\partial}{\partial r} J_m(kr) & = \frac{k}{2} \left[ J_{m-1}(kr) - J_{m+1}(kr) \right] .
\end{split}
\label{eq:derRelationBessel}
\end{equation}

\begin{figure}
\begin{center}
\includegraphics[width=0.6\textwidth]{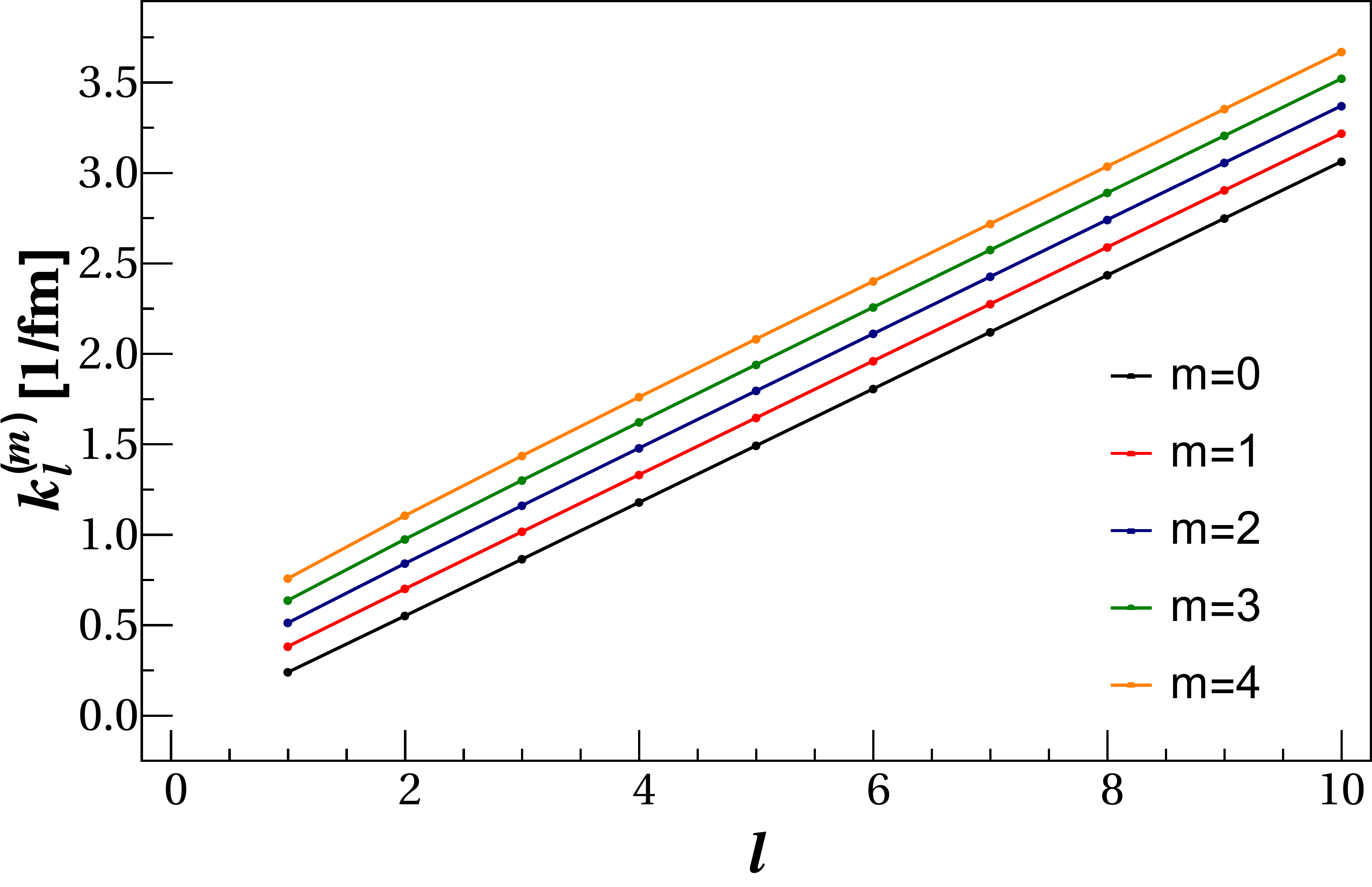}
\end{center}
\caption{(Color online) Wavenumber $k^{(m)}_l = z^{(m)}_l/R$ as a function of the discrete radial wavenumber $l$ and for different values of the azimuthal wavenumber $m=1$ (lowest curve) to $m=4$ (uppermost curve). These values arise for the boundary condition $\delta \epsilon = 0$ at $r=R$ and we choose $R=10 \, \text{fm}$ for definiteness. The plot shows that $k$ increases with increasing values of both $l$ and $m$, corresponding to finer spatial resolution.}
\label{fig:wavenumbers}
\end{figure}

The Bessel expansion we use in Eqs.\ \eqref{eq:BesselFourierExpansionEnergy} contains an integral over all (positive) values of $k$. This expansion, also known as the Hankel transformation, is appropriate for functions on the open interval $r\in (0,\infty)$. More realistically, the energy distribution in a heavy ion collision is non-zero only on a compact interval $(0,R)$ with some radius $R$ that depends on time during the expansion of the fireball and it is of the order of $R\sim 10 \,\text{fm}$. On such a compact interval the Bessel expansion becomes discrete, in the sense that the integral over $k$ is replaced by a sum over a discrete subset. For example, the boundary condition $\delta \epsilon = 0$ at $r=R$ leads to the values $k^{(m)}_l=z^{(m)}_l/R$ where the $z^{(m)}_l$ are the $l$'th zero crossings of the Bessel function $J_m(z)$. To relatively good approximation $z^{(m)}_l$ is linear in $m$ (for fixed $l$). In Fig.\ \ref{fig:wavenumbers} we illustrate the resulting values for $k^{(m)}_l$ as a function of the discrete radial wavenumber $l$ and for different values of $m$.
%--------------------------Fig. 3------------------------------------
%----------------------------------------------------------------------------------------------------
More generally, one might use an expansion based on $J_m\left(z^{(m)}\rho(r)\right)$ where $\rho(r)$ is a monotonous function into the interval $(0,1)$ and a particularly useful choice for $\rho(r)$ is discussed in Appendix A of Ref.\ \cite{Floerchinger:2014fta}.

The evolution equation for the perturbation in energy density, Eq.\ \eqref{eq:LinEvEqEnergy} becomes in Bessel-Fourier space (all perturbation functions have now the argument $(\tau,k,m,q)$ that we suppress for better readability)
\begin{equation}
\begin{split}
\partial_\tau \delta \epsilon & + \left[ \frac{1}{\tau}+\frac{1}{\tau} \left( \frac{\partial p}{\partial \epsilon} \right)_n - \frac{1}{\tau^2} \left( \frac{\partial \zeta}{\partial \epsilon} \right)_n - \frac{4}{3\tau^2} \left( \frac{\partial \eta}{\partial \epsilon} \right)_n \right] \delta\epsilon\\
&+ \left[  \frac{1}{\tau} \left( \frac{\partial p}{\partial n} \right)_\epsilon - \frac{1}{\tau^2} \left( \frac{\partial \zeta}{\partial n} \right)_\epsilon - \frac{4}{3\tau^2} \left( \frac{\partial \eta}{\partial n} \right)_\epsilon \right] \delta n \\
&+ \left[ \bar\epsilon + \bar p - \frac{2}{\tau}\bar \zeta + \frac{4}{3\tau} \bar \eta \right] \left( \frac{k}{\sqrt{2}} \left(\delta u^+ - \delta u^- \right) + i q \, \delta u^\eta \right) - \frac{4}{\tau} \bar \eta \; i q\, \delta u^\eta = 0.
\end{split}
\label{eq:LinEvEqEnergyBesselFourier}
\end{equation}
Similarly, the evolution equation for the perturbation in baryon number density becomes
\begin{equation}
\begin{split}
\partial_\tau \delta n  & + \frac{1}{\tau} \delta n + \left[ \bar n - \bar\kappa \left[\frac{\bar n\bar T}{\bar \epsilon + \bar p}\right]^2 \partial_\tau\left(\frac{\bar\mu}{\bar T}\right) \right] \left( \frac{k}{\sqrt{2}} \left(\delta u^+ - \delta u^- \right) + i q \, \delta u^\eta \right) \\
& + \bar\kappa \left[\frac{\bar n\bar T}{\bar \epsilon + \bar p}\right]^2 \left( \frac{\partial(\mu/T)}{\partial\epsilon} \right)_n \left( k^2 + \frac{q^2}{\tau^2}\right) \delta\epsilon \\
& + \bar\kappa \left[\frac{\bar n\bar T}{\bar \epsilon + \bar p}\right]^2 \left( \frac{\partial(\mu/T)}{\partial n} \right)_\epsilon \left( k^2 + \frac{q^2}{\tau^2}\right) \delta n = 0.
\end{split}
\label{eq:LinEvEqDensityBesselFourier}
\end{equation}
Let us now turn to the perturbations in the fluid velocity. Equations \eqref{eq:LinEvEqRadialVelocity} and \eqref{eq:LinEvEqAzimuthalVelocity} lead to the following equations for $\delta u^+$ and $\delta u^-$ in Bessel-Fourier space
 \begin{equation}
 \begin{split}
 & \left( \bar \epsilon + \bar p - \frac{1}{\tau}\bar \zeta + \frac{2 }{3\tau}\bar \eta \right) \partial_\tau \delta u^\pm + \left[ \partial_\tau \bar p - \frac{1}{\tau}\partial_\tau\bar \zeta +\frac{1}{\tau^2} \bar \zeta + \frac{2}{3\tau} \partial_\tau \bar \eta + \frac{4}{3\tau^2} \bar \eta \right] \delta u^\pm \\
& \mp \left[ \left( \frac{\partial p}{\partial \epsilon} \right)_n - \frac{1}{\tau}\left( \frac{\partial \zeta}{\partial \epsilon} \right)_n + \frac{2}{3\tau} \left( \frac{\partial \eta}{\partial \epsilon} \right)_n \right] \frac{k}{\sqrt{2}} \delta \epsilon 
\mp \left[ \left( \frac{\partial p}{\partial n} \right)_\epsilon - \frac{1}{\tau}\left( \frac{\partial \zeta}{\partial n} \right)_\epsilon + \frac{2}{3\tau} \left( \frac{\partial \eta}{\partial n} \right)_\epsilon \right] \frac{k}{\sqrt{2}} \delta n \\
& + \left[ \frac{1}{2} \bar \zeta k^2 + \frac{7}{6} \bar \eta k^2 +\bar \eta \frac{q^2}{\tau^2} \right] \delta u^\pm - \left[ \frac{1}{2} \bar \zeta k^2 + \frac{1}{6} \bar \eta k^2 \right] \delta u^\mp \pm i \left[ \frac{\bar \zeta k q}{\sqrt{2}} + \frac{\bar \eta k q}{\sqrt{2}}  \right] \delta u^\eta = 0 ,
 \end{split}
 \label{eq:LinEvEqPlusMinusVelocityBesselFourier}
 \end{equation}
and for the rapidity component we find from Eq.\ \eqref{eq:LinEvEqRapidityVelocity}
\begin{equation}
\begin{split}
& \left( \bar \epsilon + \bar p - \frac{1}{\tau}\bar \zeta - \frac{4}{3\tau}\bar \eta \right) \partial_\tau \delta u^\eta 
+ \left[ \partial_\tau \bar p + \frac{2}{\tau}(\bar \epsilon + \bar p)  - \frac{1}{\tau}\partial_\tau\bar \zeta +\frac{1}{\tau^2} \bar \zeta - \frac{4}{3\tau} \partial_\tau \bar \eta - \frac{4}{3\tau^2} \bar \eta \right] \delta u^\eta \\
& + \left[ \left( \frac{\partial p}{\partial \epsilon} \right)_n - \frac{1}{\tau}\left( \frac{\partial \zeta}{\partial \epsilon} \right)_n + \frac{2}{3\tau} \left( \frac{\partial \eta}{\partial \epsilon} \right)_n \right] \frac{i q}{\tau^2} \delta \epsilon 
+ \left[ \left( \frac{\partial p}{\partial n} \right)_\epsilon - \frac{1}{\tau}\left( \frac{\partial \zeta}{\partial n} \right)_\epsilon + \frac{2}{3\tau} \left( \frac{\partial \eta}{\partial n} \right)_\epsilon \right] \frac{i q}{\tau^2} \delta n \\
& - \left(\bar \zeta + \frac{1}{3}\bar \eta \right) \frac{iq k}{\tau^2\sqrt{2}} \left(\delta u^+ - \delta u^- \right) + \left[ \left( \bar \zeta + \frac{4}{3}\bar \eta \right) \frac{q^2}{\tau^2} + \bar \eta k^2 \right] \delta u^\eta = 0.
\end{split}
\label{eq:LinEvEqRapidityVelocityBesselFourier}
\end{equation}
Note that eqs.\ \eqref{eq:LinEvEqEnergyBesselFourier} - \eqref{eq:LinEvEqRapidityVelocityBesselFourier} are now coupled ordinary differential equations. All spatial derivatives have become algebraic and one can directly integrate for the time dependent perturbations $\delta\epsilon(\tau,k,m,q)$ etc. To construct such a solution one needs as an input the background or Bjorken solution for $\bar T(\tau)$ and $\bar\mu(\tau)$ as well as the relations that express all other thermodynamic densities ($\bar \epsilon$, $\bar p$, $\bar n$ etc.), transport coefficients ($\bar \zeta$, $\bar \eta$, $\bar \kappa$) and derivatives ($(\partial p / \partial \epsilon)_n$, $(\partial p / \partial n)_\epsilon$, $(\partial \zeta/\partial \epsilon)_n$ etc.) in terms of the independent thermodynamic variables $\bar T$ and $\bar \mu$.

Let us first discuss some limiting cases of Eqs.\ \eqref{eq:LinEvEqEnergyBesselFourier} - \eqref{eq:LinEvEqRapidityVelocityBesselFourier} with extended symmetries. 

%%%%%%%%%%%%%%%%%%%%%%%%%%%%%%%%%%%%%%%%%%%%%%%%%%%%%%%%%%%%%%%%%%%%%%%%%%%%%%%%%%%%%%%%%%
\subsection{Statistical baryon number conjugation symmetry}
\label{subsec:statbar}
%%%%%%%%%%%%%%%%%%%%%%%%%%%%%%%%%%%%%%%%%%%%%%%%%%%%%%%%%%%%%%%%%%%%%%%%%%%%%%%%%%%%%%%%%%
If the baryon number density vanishes in the background solution, i.\ e.\ $\bar n = \bar \mu=0$, one has an extended symmetry namely baryon-anti-baryon or baryon number conjugation symmetry corresponding to , $n\to - n$. Odd derivatives such as $(\partial p/\partial n)_\epsilon$ or $(\partial \eta/\partial n)_\epsilon$ have to vanish and one finds that $\delta n$ decouples from the equations for $\delta \epsilon$ in Eq.\ \eqref{eq:LinEvEqEnergyBesselFourier} and the perturbations of fluid velocity in Eqs.\ \eqref{eq:LinEvEqPlusMinusVelocityBesselFourier} and~\eqref{eq:LinEvEqRapidityVelocityBesselFourier}. However, this does not imply that $\delta n$ has to vanish as well. Locally and event-by-event one may have a non-zero baryon number density. The evolution equation for this perturbation is obtained from Eq.\ \eqref{eq:LinEvEqDensityBesselFourier} as
\begin{equation}
\partial_\tau \delta n + \frac{1}{\tau} \delta n + \bar\kappa \left[\frac{\bar n\bar T}{\bar \epsilon + \bar p}\right]^2 \left( \frac{\partial(\mu/T)}{\partial n} \right)_\epsilon \left( k^2 + \frac{q^2}{\tau^2}\right) \delta n = 0.
\label{eq:LinEvEqDensityBesselFourierBaryonConjugationSymmetry}
\end{equation}
The second term on the left hand side accounts simply for the dilution due to the longitudinal expansion while the third term is a diffusion term due to heat conductivity. Note that  $\bar \kappa$ is expected to be singular in the limit $\bar n\to 0$ in such a way that the combination of terms that multiplies $(k^2+\frac{q^2}{\tau^2}) \, \delta n$ remains finite~\cite{Danielewicz:1984ww}. Therefore, the diffusion term indeed plays a role for the evolution of perturbations $\delta n$.

Equation \eqref{eq:LinEvEqDensityBesselFourierBaryonConjugationSymmetry} can be directly integrated and its solution reads as
\begin{equation}
\delta n(\tau,k,m,q) = \left( \frac{\tau_0}{\tau} \right) \exp\left[-k^2 I_1(\tau,\tau_0) - q^2 I_2(\tau,\tau_0) \right] \delta n(\tau_0,k,m,q) ,
\label{eq:baryonfluctlongitudinal}
\end{equation}
where the integrals
\begin{equation}
\begin{split}
I_1(\tau,\tau_0) & = \int_{\tau_0}^\tau  d\tau^\prime \, \bar\kappa \left[\frac{\bar n\bar T}{\bar \epsilon + \bar p}\right]^2 \left( \frac{\partial(\mu/T)}{\partial n} \right)_\epsilon, \\
I_2(\tau,\tau_0) & = \int_{\tau_0}^\tau  d\tau^\prime \, \frac{1}{\tau^{\prime 2}} \, \bar\kappa \left[\frac{\bar n\bar T}{\bar \epsilon + \bar p}\right]^2 \left( \frac{\partial(\mu/T)}{\partial n} \right)_\epsilon ,
\end{split}
\end{equation}
depend on the heat conductivity and thermodynamic quantities on the background Bjorken solution. While the integral $I_1$ is typically dominated by late times $\tau$ (for example for the ideal thermodynamic equation of state \eqref{eq:idealEOS}, heat conductivity of the form \eqref{eq:kappaw} and Bjorken expansion as in Eq.\ \eqref{eq:bjoTnomu}), the integral $I_2$ is dominated by early times $\tau \approx \tau_0$. Moreover, for fast thermalization $\tau_0 \to 0$ one has formally $I_2 \to \infty$ such that in reality it might be rather large. Modes with $q \neq 0$ are therefore strongly damped by dissipative effects of heat conductivity.

The evolution equations for the perturbations in energy density $\delta \epsilon$ and fluid velocity are independent of $\delta n$. Their solution has already been discussed in a similar setup in Ref.\ \cite{Florchinger:2011qf}. 

%%%%%%%%%%%%%%%%%%%%%%%%%%%%%%%%%%%%%%%%%%%%%%%%%%%%%%%%%%%%%%%%%%%%%%%%%%%%%%%%%%%%%%%%%%%
\subsection{Exact Bjorken boost symmetry} 
\label{subsec:exactbjo}
%%%%%%%%%%%%%%%%%%%%%%%%%%%%%%%%%%%%%%%%%%%%%%%%%%%%%%%%%%%%%%%%%%%%%%%%%%%%%%%%%%%%%%%%%%%
%--------------------------Fig. 4------------------------------------
\begin{figure}[h]
\begin{centering}
\begin{tabular}{c c}
\begin{picture}(180,150)
\put(0,8){\includegraphics[scale=0.195]{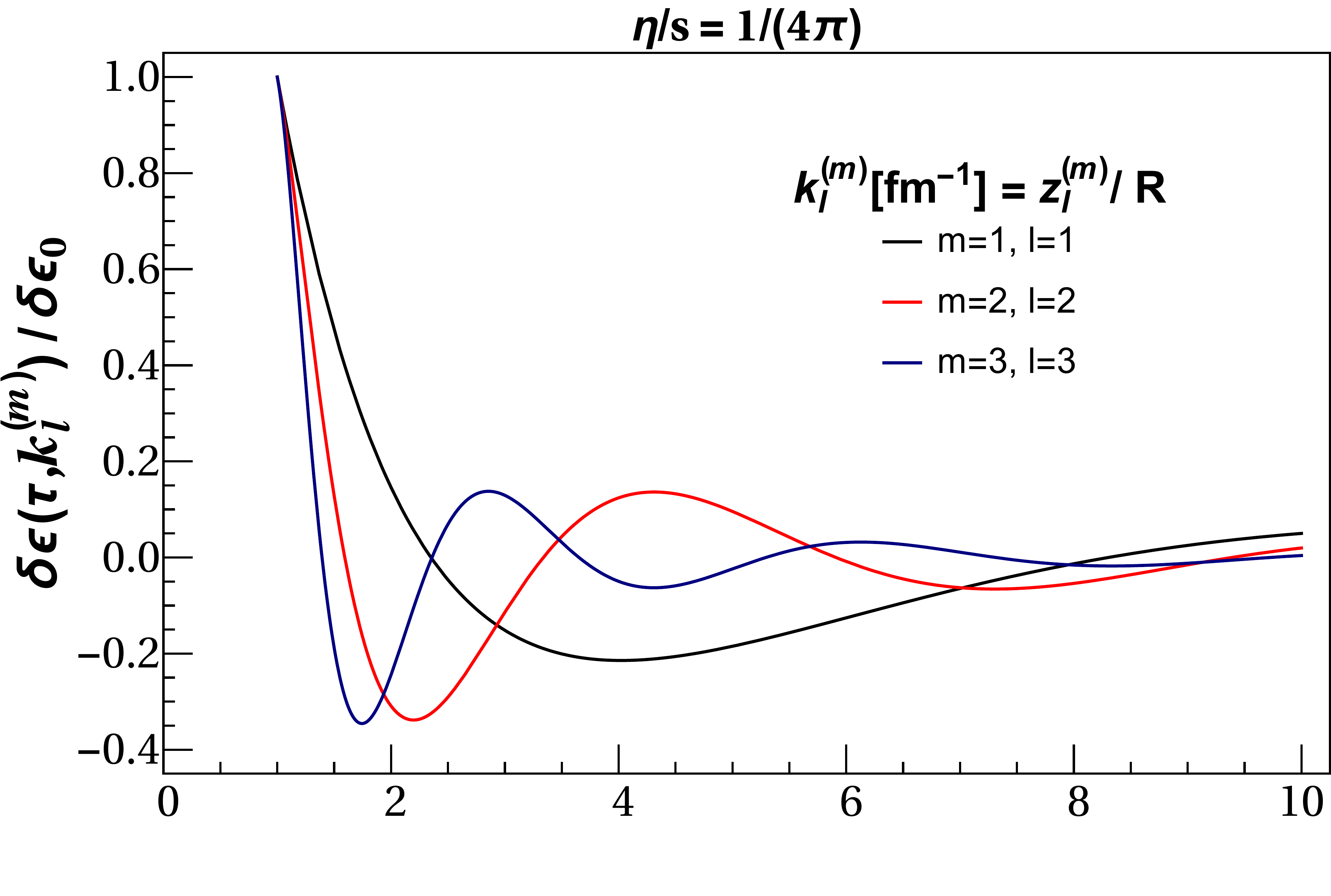}} 
\end{picture}
&
\begin{picture}(450,180)
\put(0,8){\includegraphics[scale=0.139]{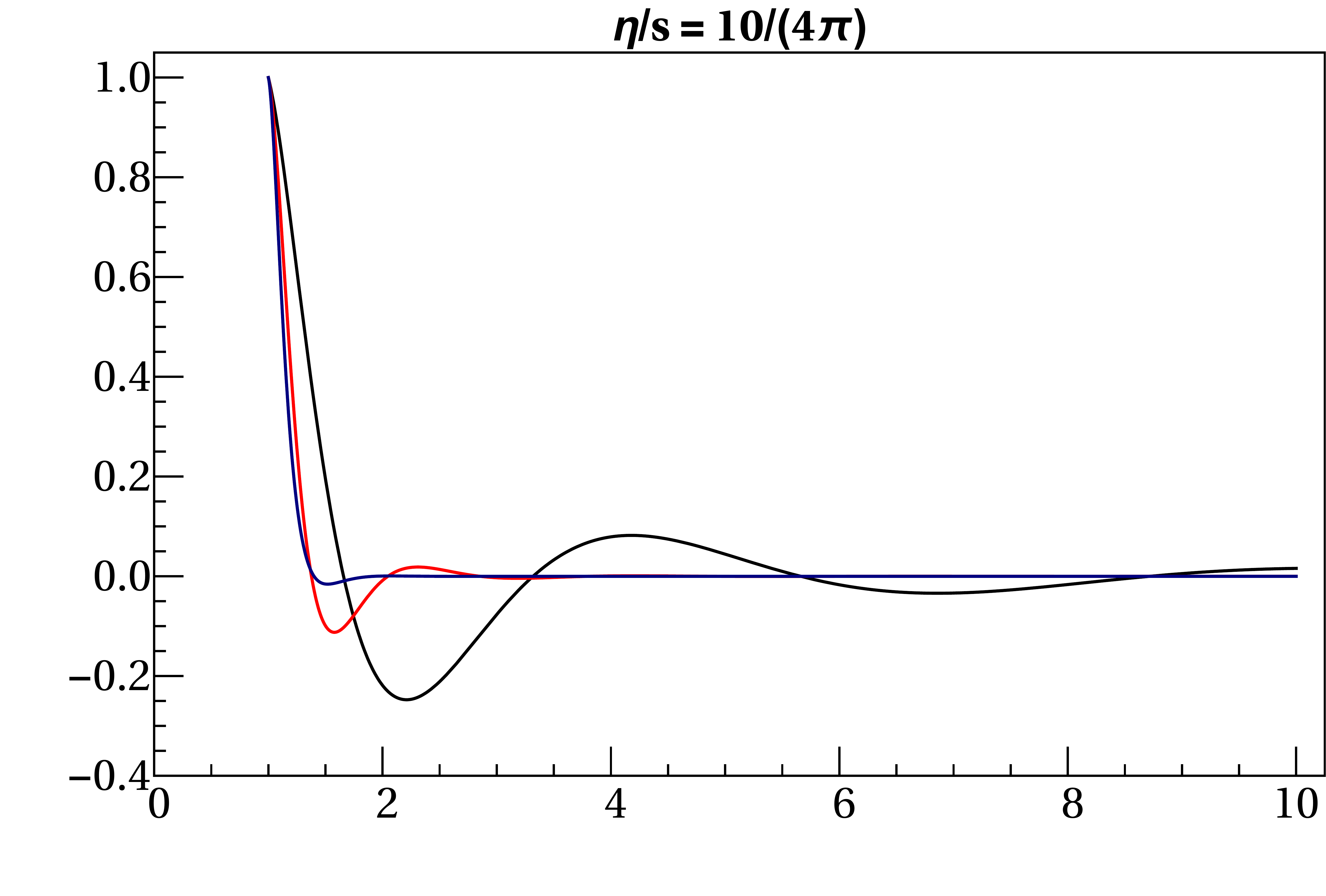}}
\end{picture}
 \\
\begin{picture}(233,150) 
\put(0,8){\includegraphics[scale=0.147]{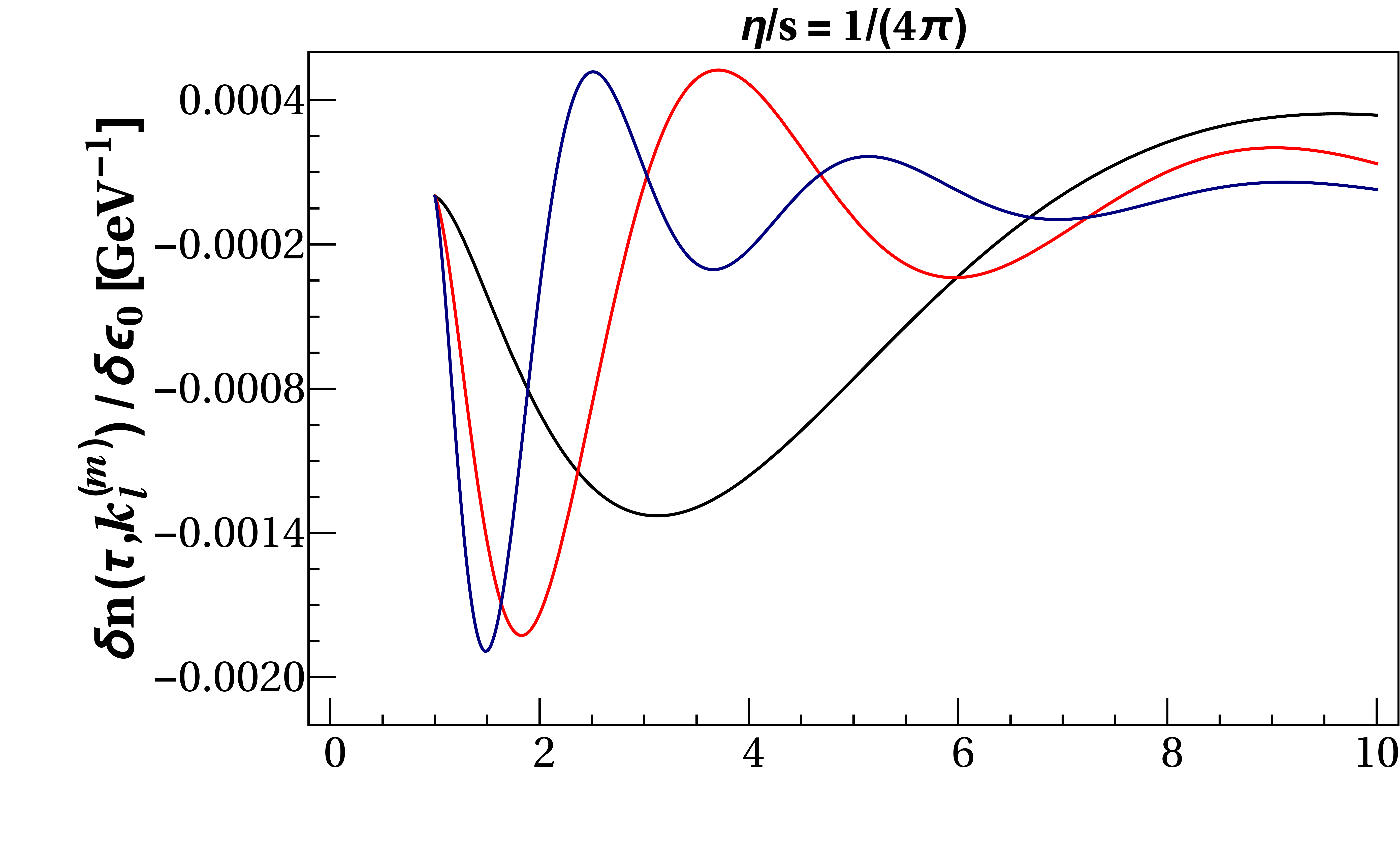} }
\end{picture}
&
\begin{picture}(465,150) 
\put(0,8){\includegraphics[scale=0.135]{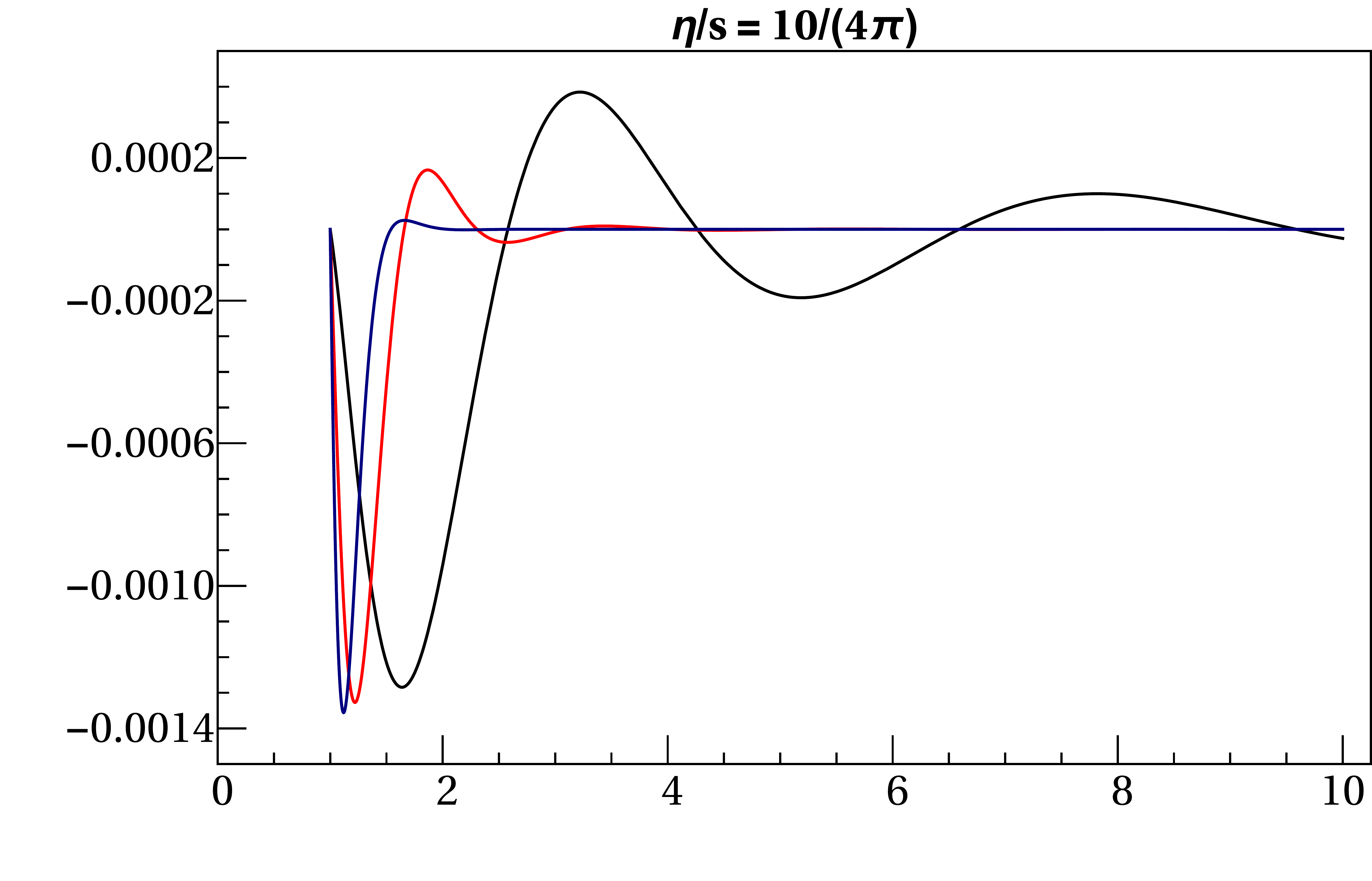}}
\end{picture}
\\
\begin{picture}(212,150) 
\put(0,8){\includegraphics[scale=0.144]{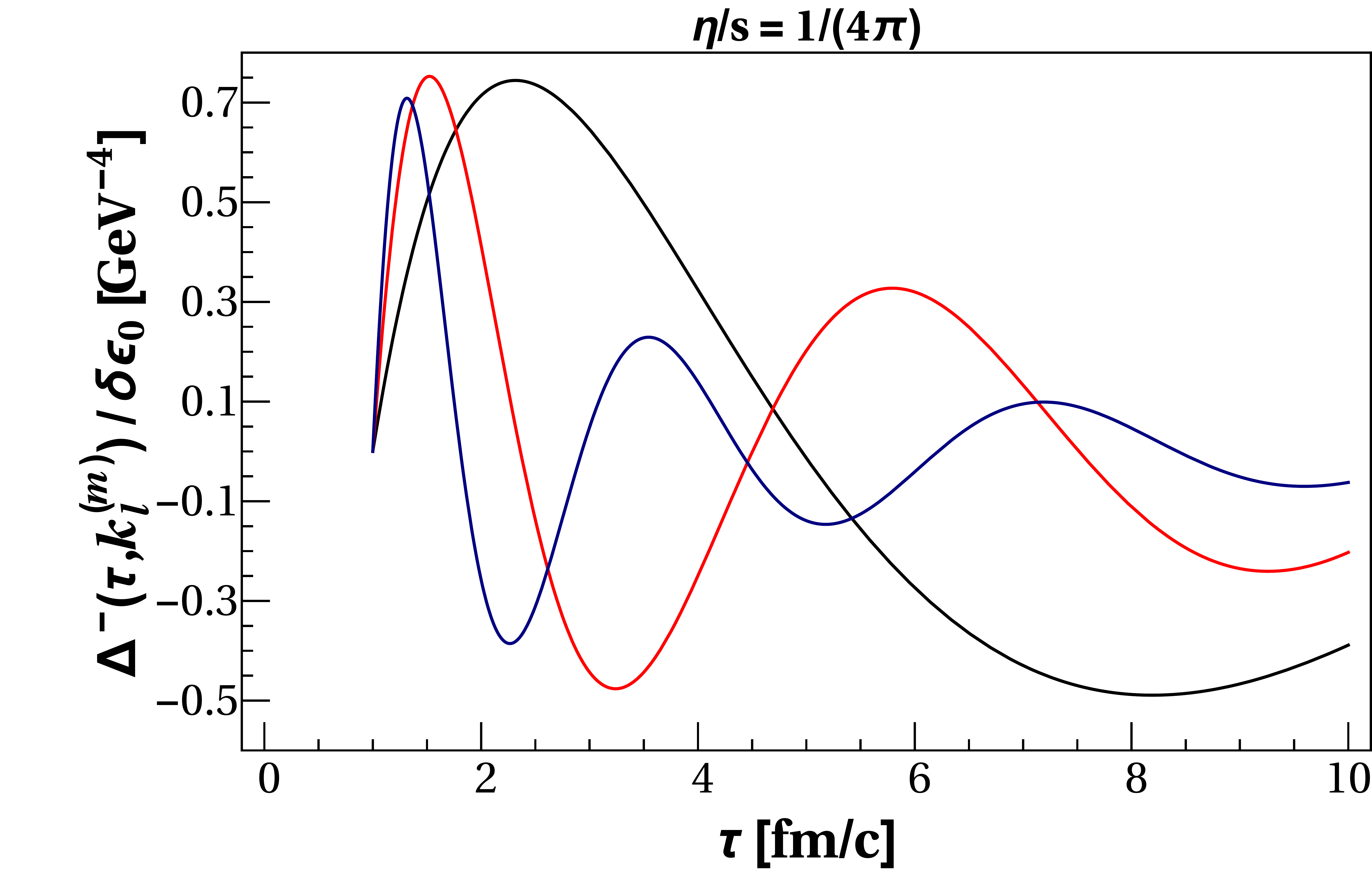}}
\end{picture}
&
\begin{picture}(445,150) 
\put(0,8){\includegraphics[scale=0.138]{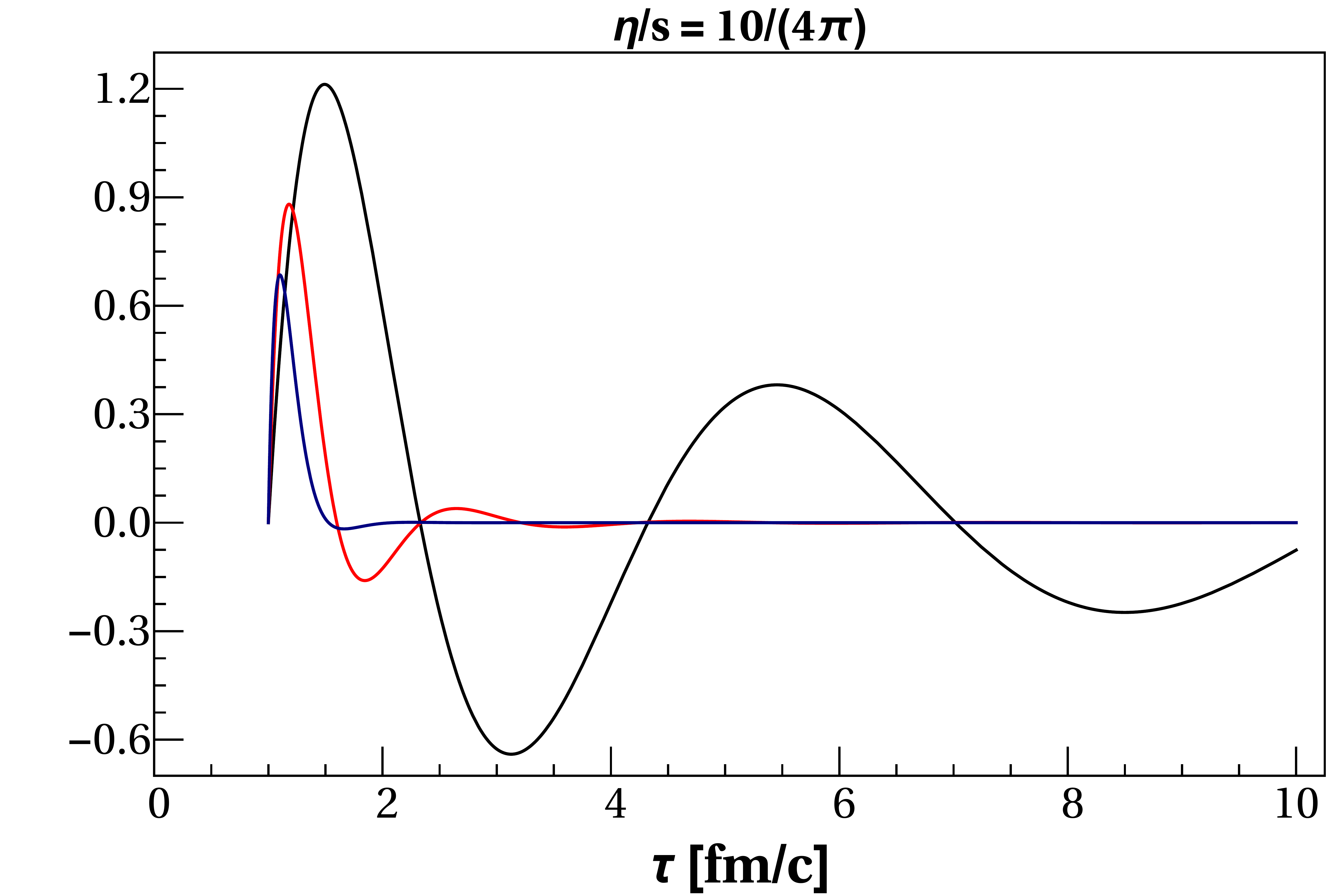}}
\end{picture} 
\end{tabular}
\end{centering}
\caption{(Color online) Evolution of perturbations in energy density, baryon number density and fluid velocity with exact Bjorken boost symmetry ($q = 0$, $\delta u^\eta=0$) for different values of the azimuthal wavenumber $m$ and radial wavenumber $l$. For $R=10$ fm/c one has $k_1^{(1)}=0.38 \, \text{fm}^{-1}$ (black curves), $k_2^{(2)}=0.84 \, \text{fm}^{-1}$ (red curves) and $k_3^{(3)}=1.30 \, \text{fm}^{-1}$ (blue curves). We compare two different values of the ratio of shear viscosity to entropy density $\eta/s=1/(4\pi)$ (left column) and (b) $\eta/s=10/(4\pi)$ (right column). Heat conductivity is related to this by eq.\ \eqref{eq:thercond}. We use $T_0=$ 0.5 GeV, $\mu_0=$0.05 GeV, $\tau_0$= 1 fm/c, $\tau_f$=10 fm/c and for the initial values of the hydrodynamic fluctuations we choose $\delta\epsilon(\tau_0)\neq 0$, $\delta n(\tau_0)=\delta u^+(\tau_0)=\delta u^-(\tau_0)$=0. We denote $\Delta^-=u^+-u^-$ (thus, $\Delta^-_0=u^+_0-u^-_0=0$). See text for further details.} 
\label{F4}
\end{figure}

%----------------------------------------------------------------------------------------------------
%--------------------------Fig. 5------------------------------------
\begin{figure}[h]
\begin{centering}
\begin{tabular}{c c}
\begin{picture}(190,150)
\put(0,8){\includegraphics[scale=0.195]{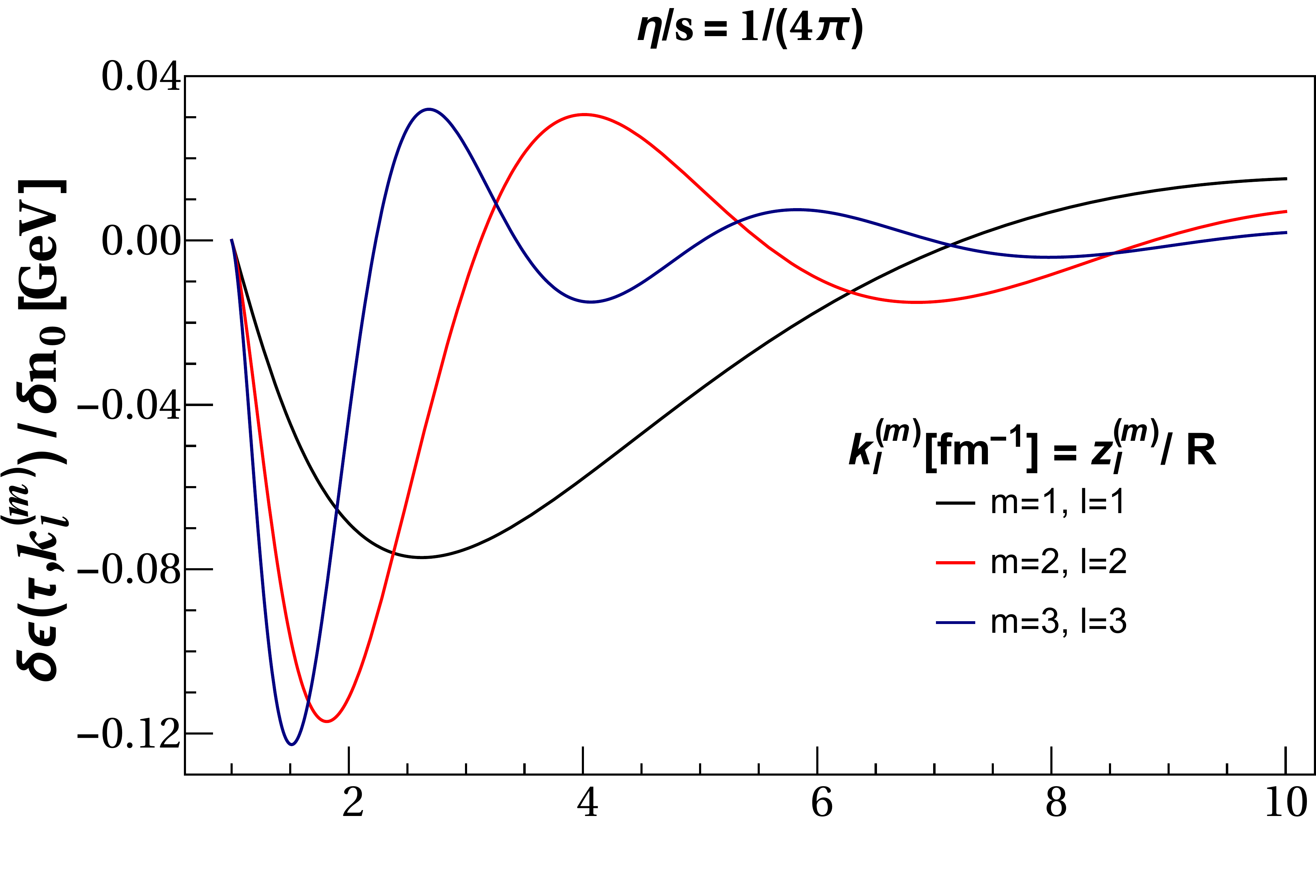}} 
\end{picture}
&
\begin{picture}(450,180)
\put(0,8){\includegraphics[scale=0.139]{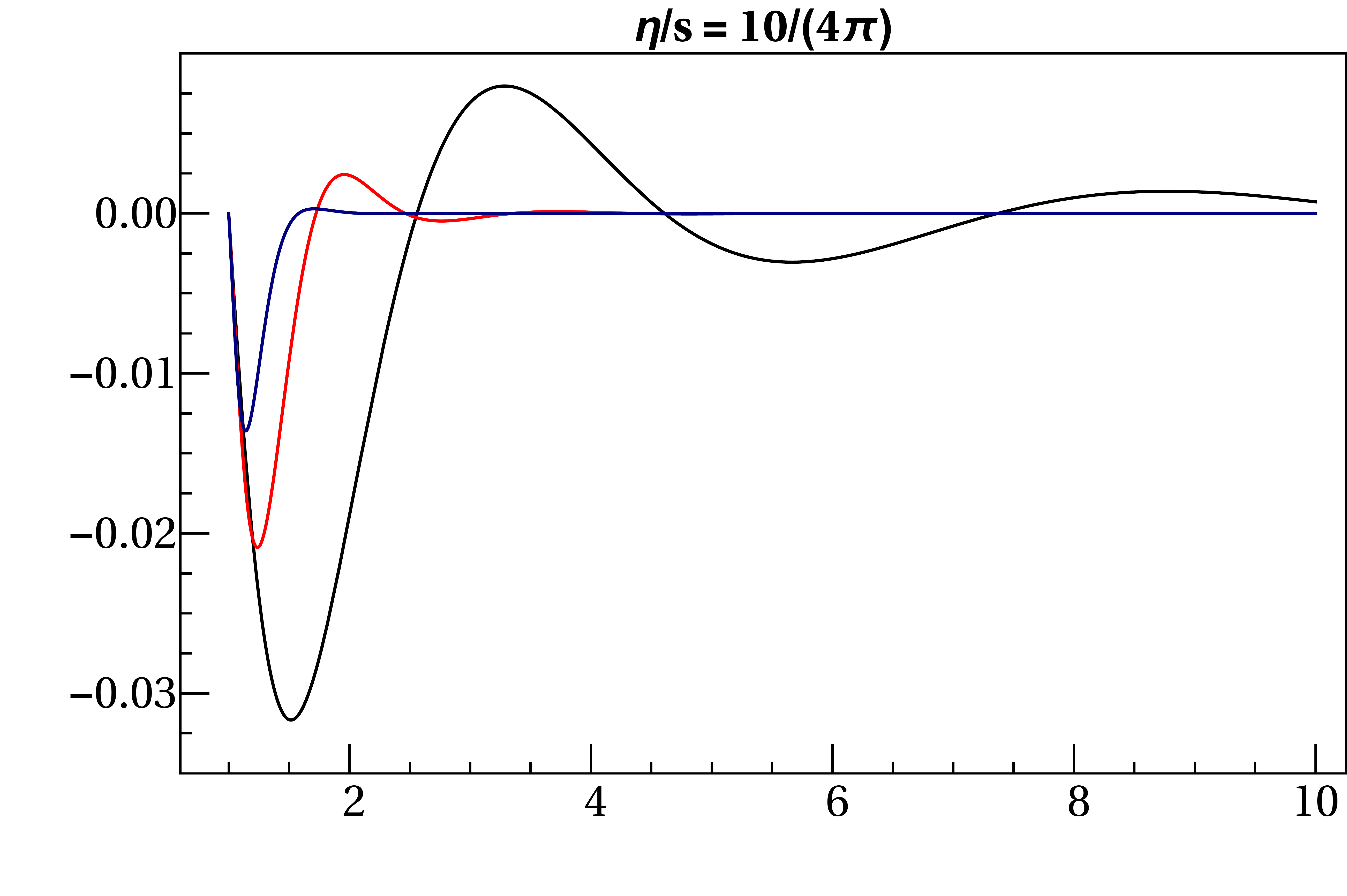}}
\end{picture}
 \\
\begin{picture}(175,150) 
\put(0,8){\includegraphics[scale=0.135]{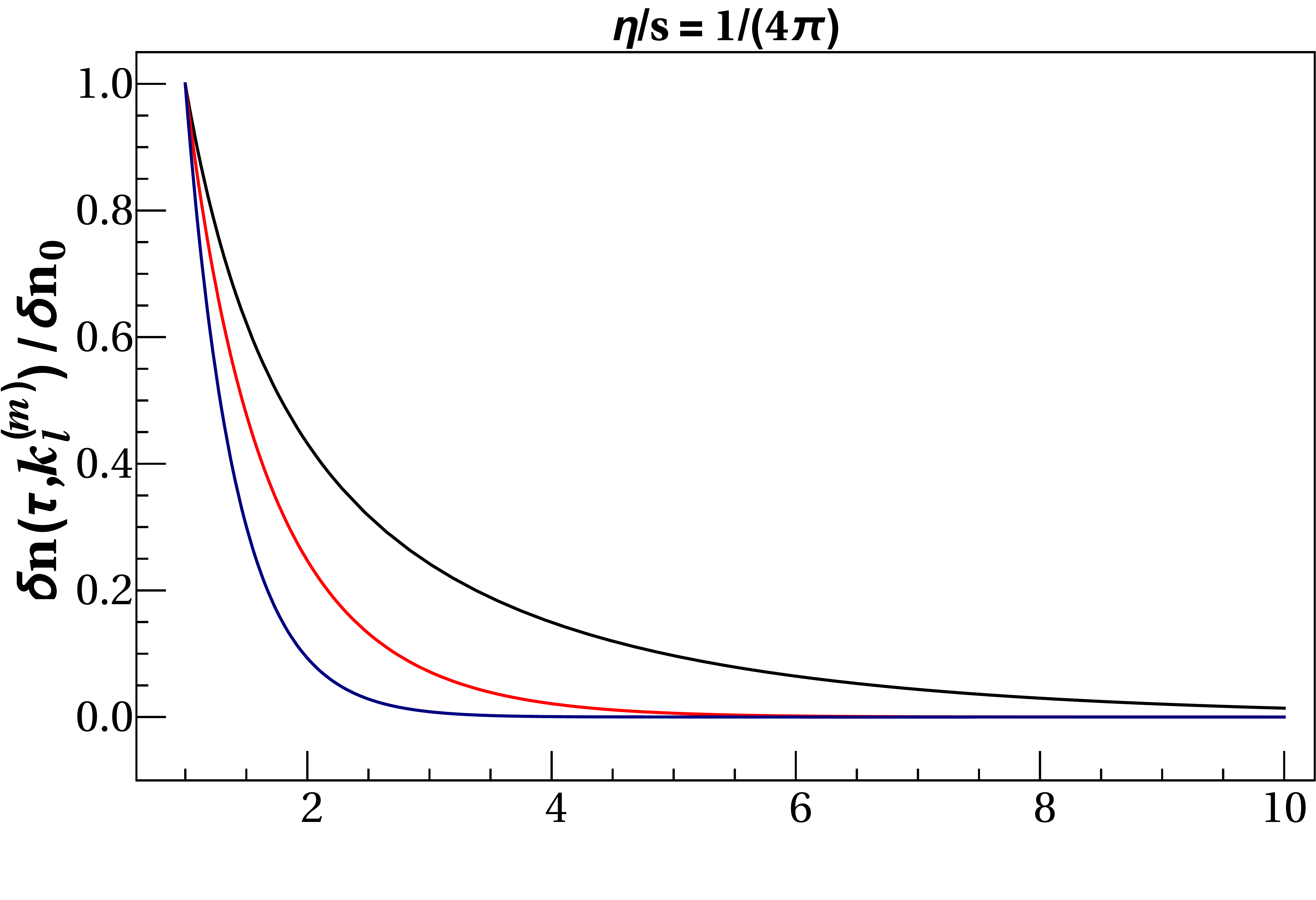} }
\end{picture}
&
\begin{picture}(430,150) 
\put(0,8){\includegraphics[scale=0.132]{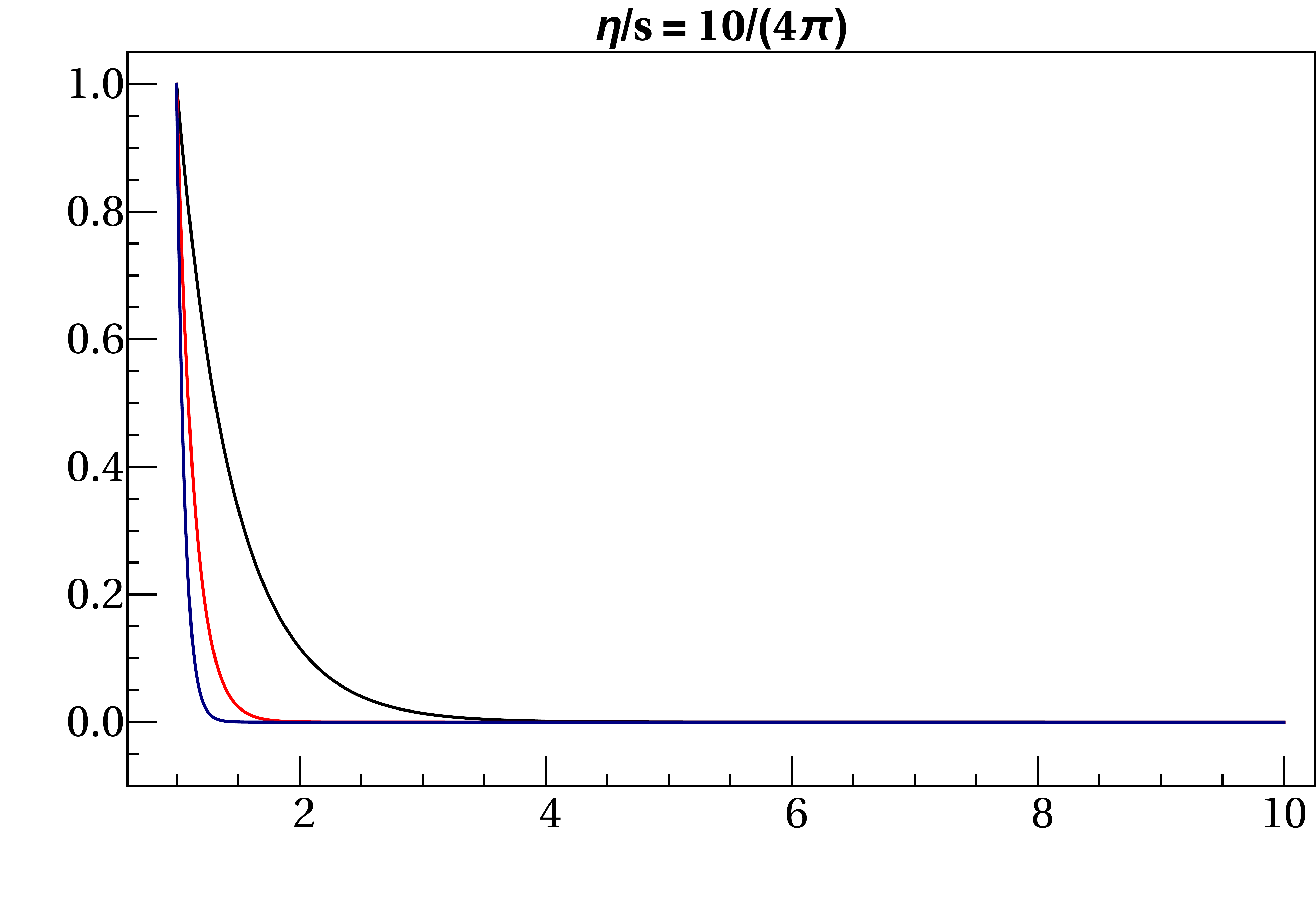}}
\end{picture}
\\
\begin{picture}(225,150) 
\put(0,8){\includegraphics[scale=0.151]{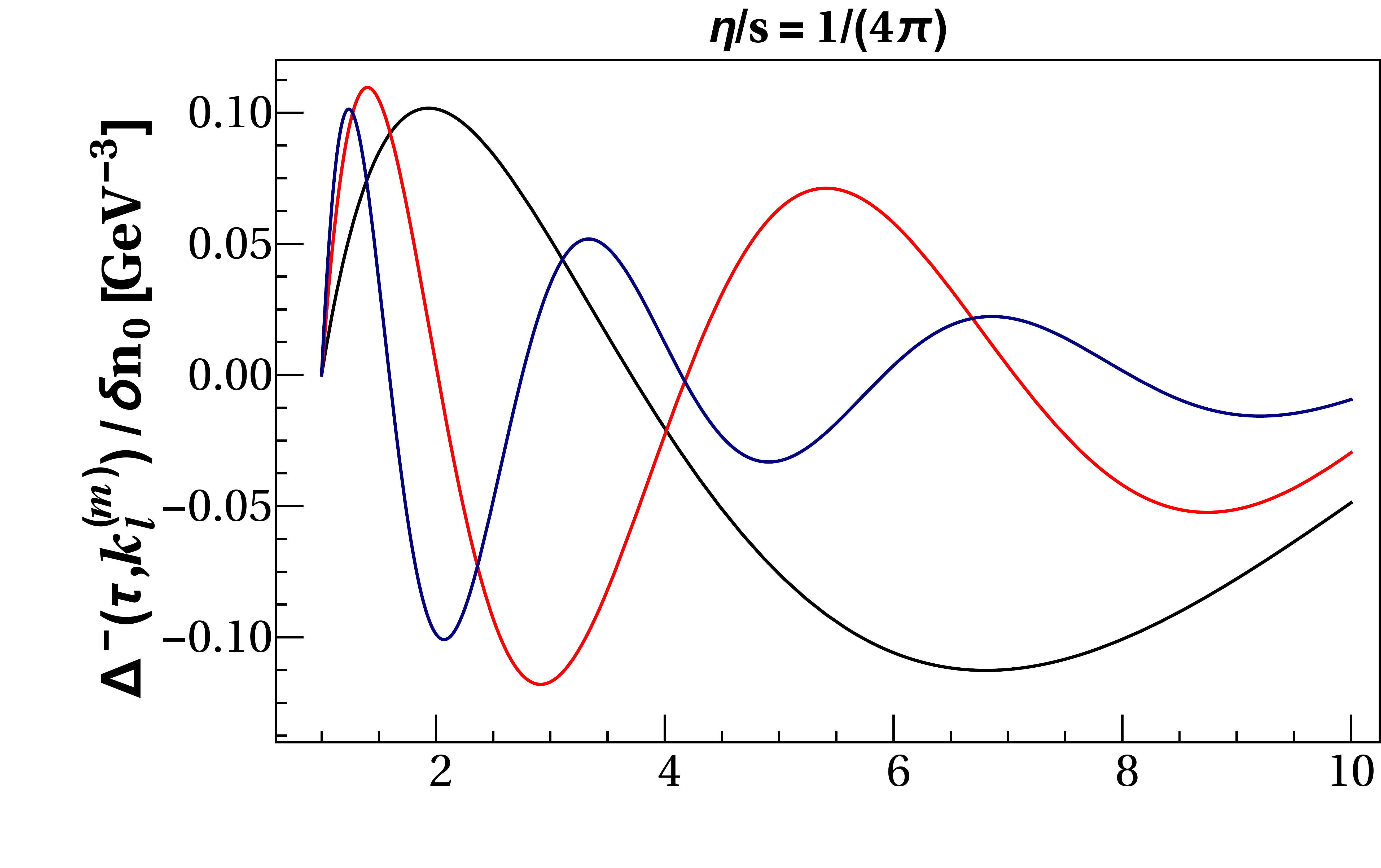}}
\end{picture}
&
\begin{picture}(445,150) 
\put(0,8){\includegraphics[scale=0.1385]{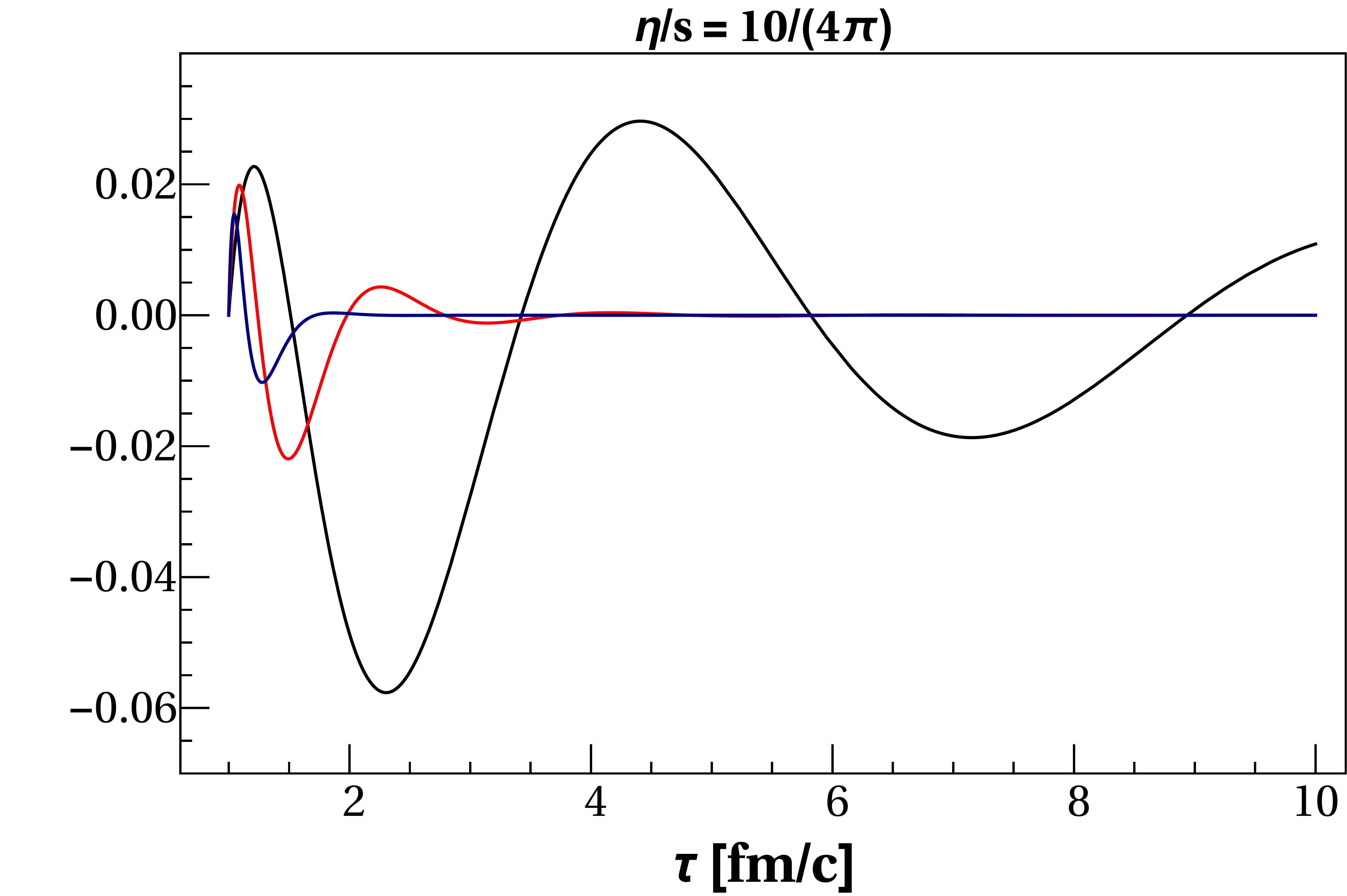}}
\end{picture} 
\end{tabular}
\end{centering}
\caption{(Color online) Same as Fig.~\ref{F4} but for different initial values of the fluid perturbations: $\delta n(\tau_0) \ne 0$, $\delta\epsilon(\tau_0)=\delta u^+(\tau_0)=\delta u^-(\tau_0)$=0 (thus, $\Delta^-_0=u^+_0-u^-_0=0$). See text for further details.} 
\label{F5}
\end{figure}
%----------------------------------------------------------------------------------------------------
%--------------------------Fig. 6------------------------------------
\begin{figure}[h]
\begin{centering}
\begin{tabular}{c c}
\begin{picture}(170,100)
\put(0,8){\includegraphics[scale=0.175]{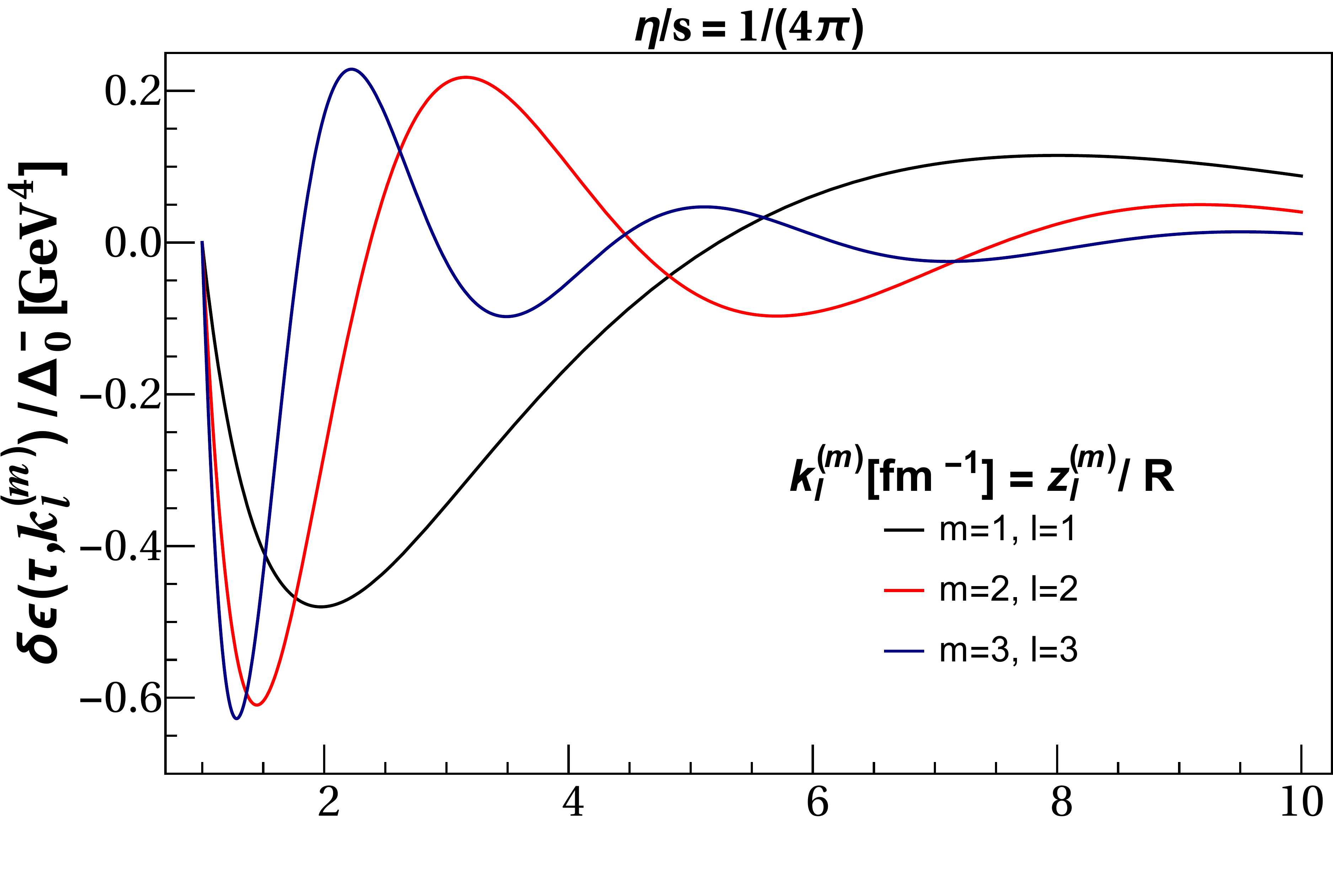}} 
\end{picture}
&
\begin{picture}(705,130)
\put(0,8){\includegraphics[scale=0.127]{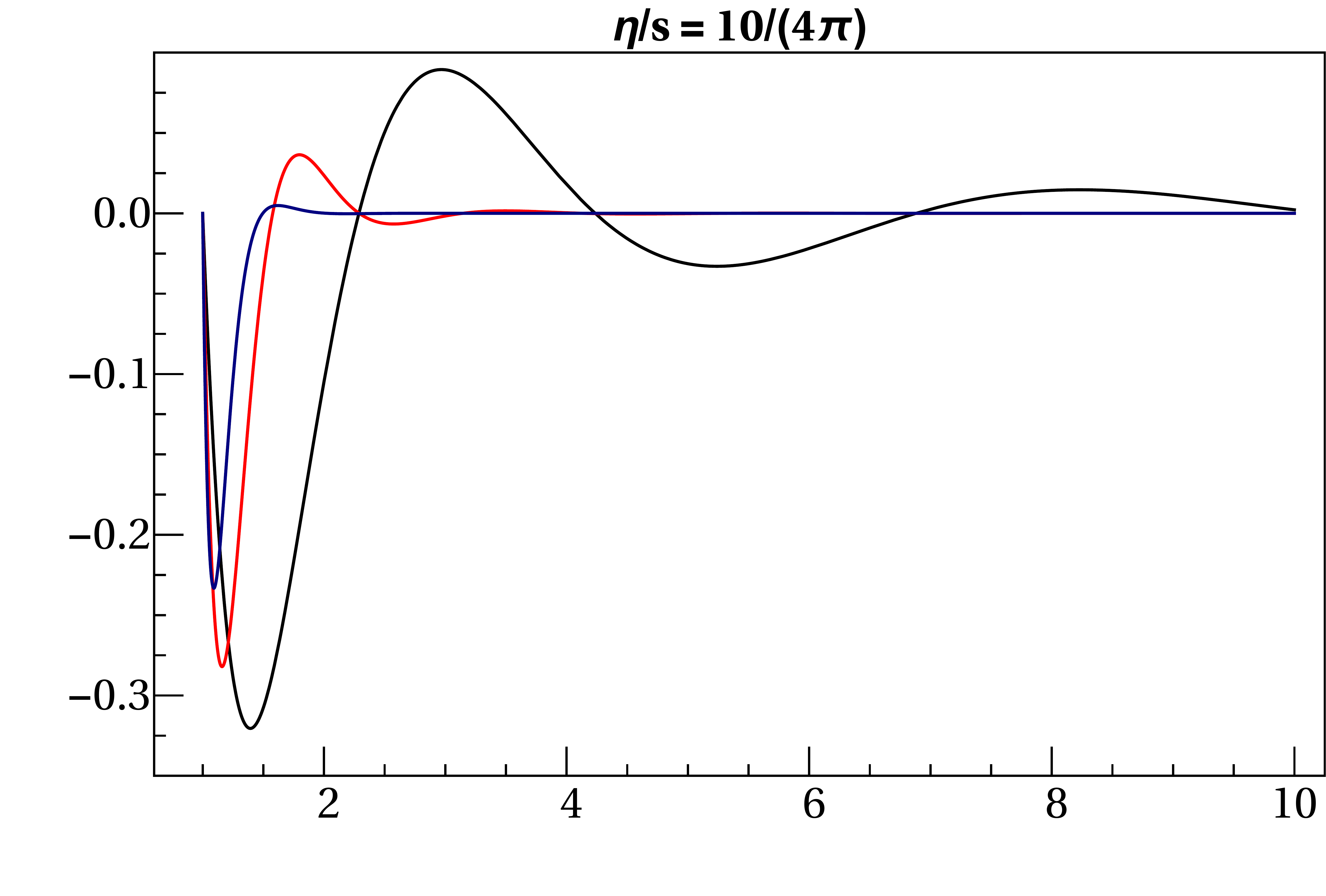}}
\end{picture}
 \\
\begin{picture}(230,100) 
\put(0,8){\includegraphics[scale=0.148]{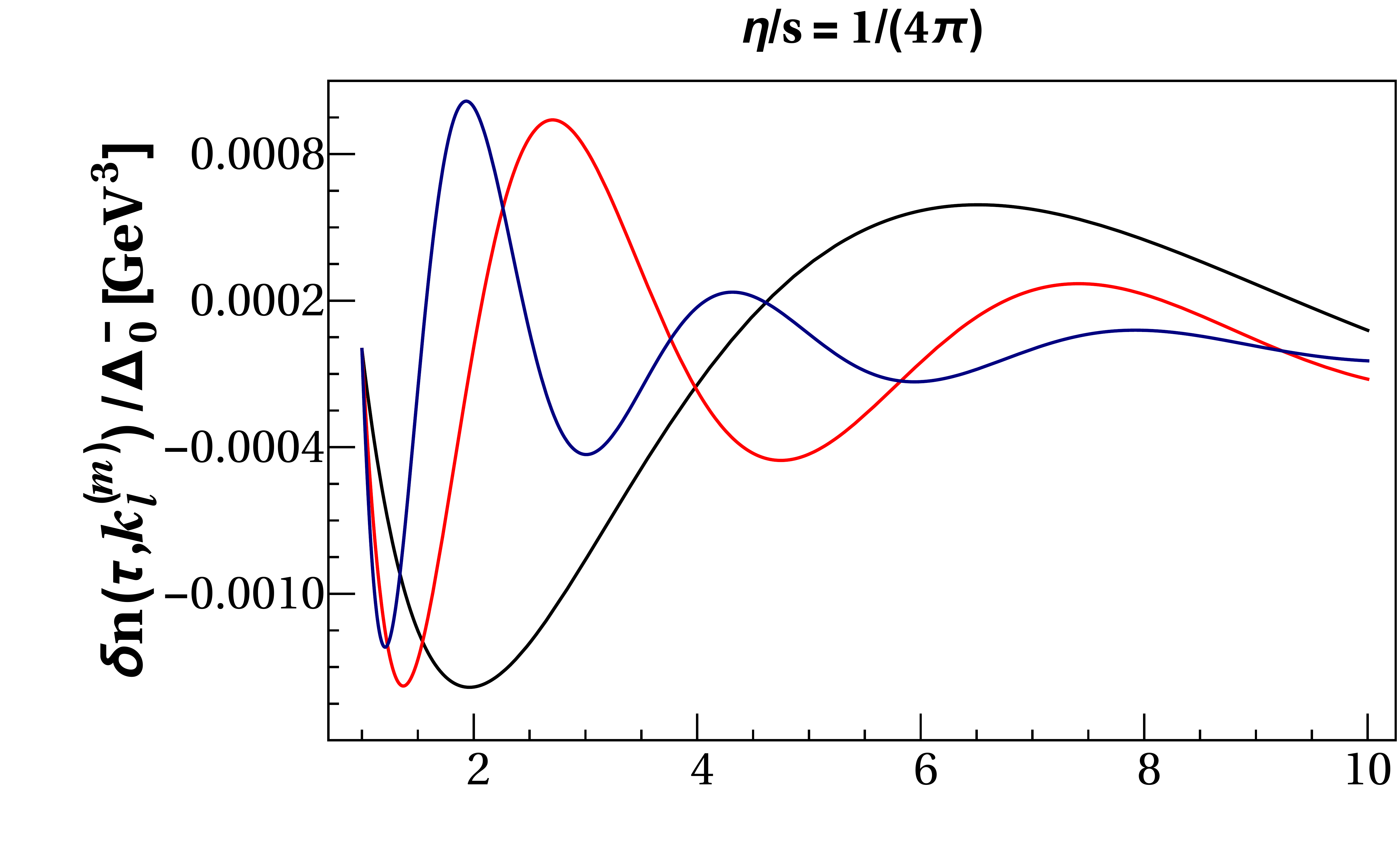} }
\end{picture}
&
\begin{picture}(730,140) 
\put(0,8){\includegraphics[scale=0.135]{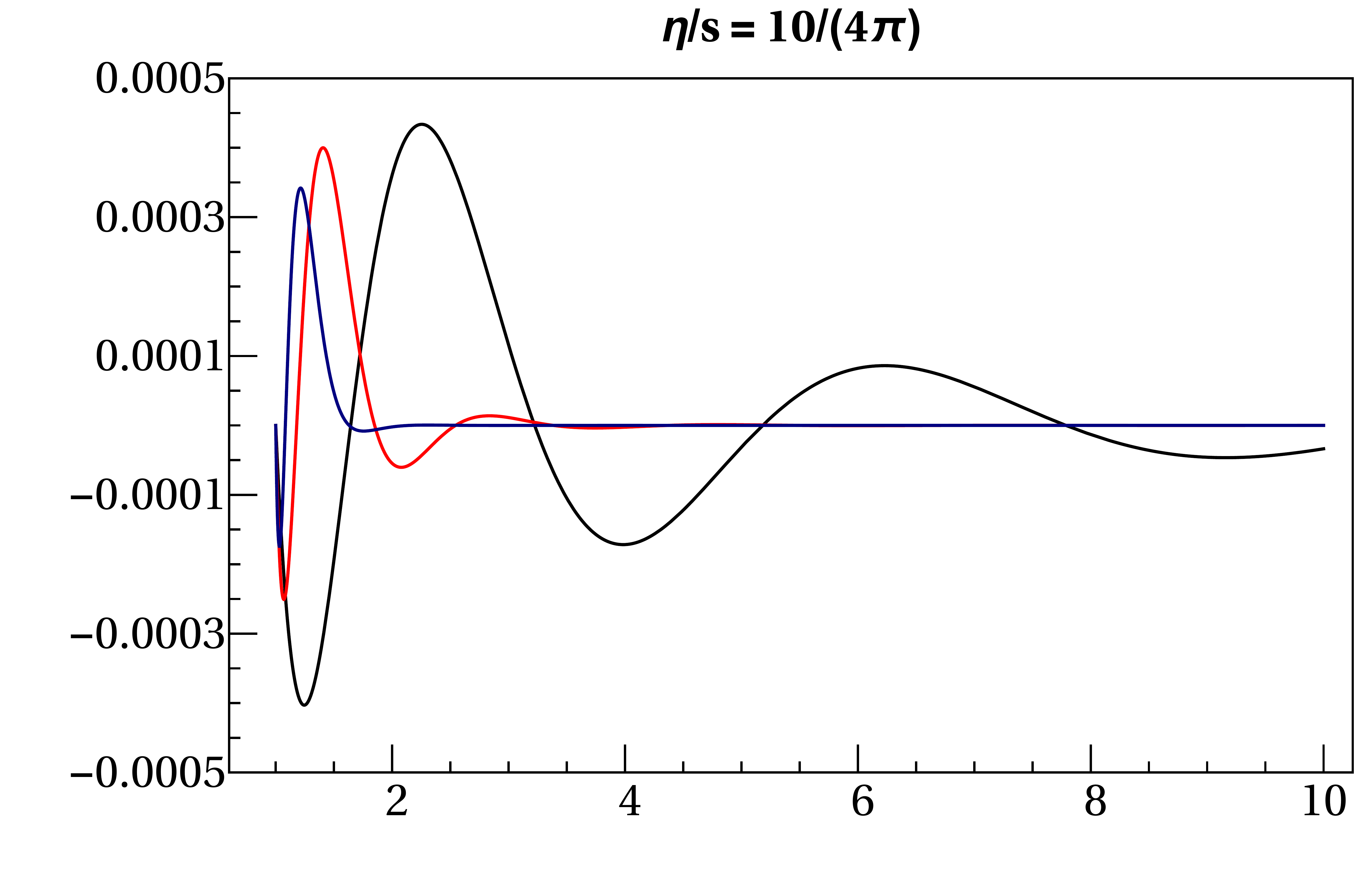}}
\end{picture}
\\
\begin{picture}(180,130) 
\put(0,8){\includegraphics[scale=0.132]{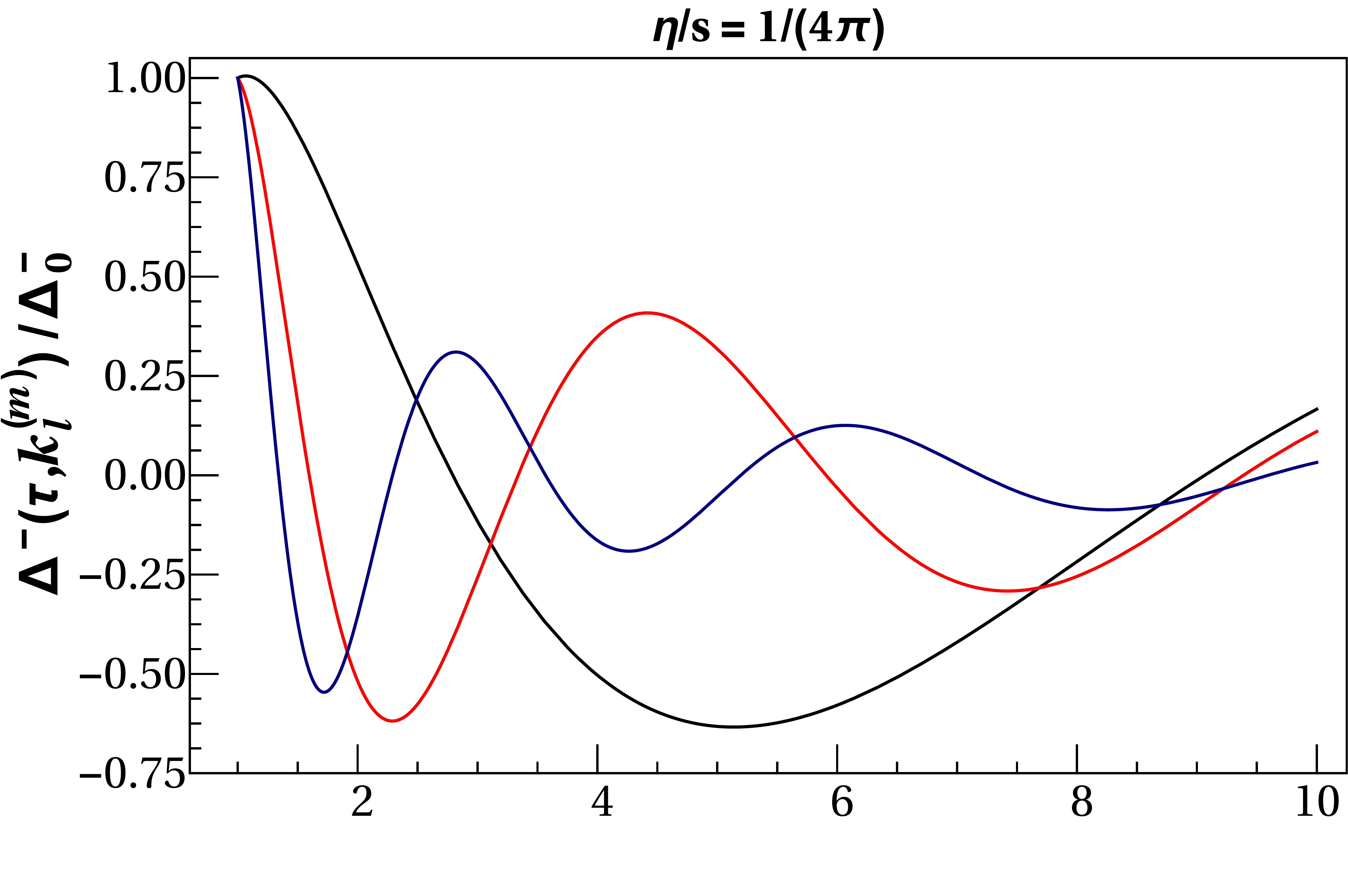}}
\end{picture}
&
\begin{picture}(708,130) 
\put(0,8){\includegraphics[scale=0.128]{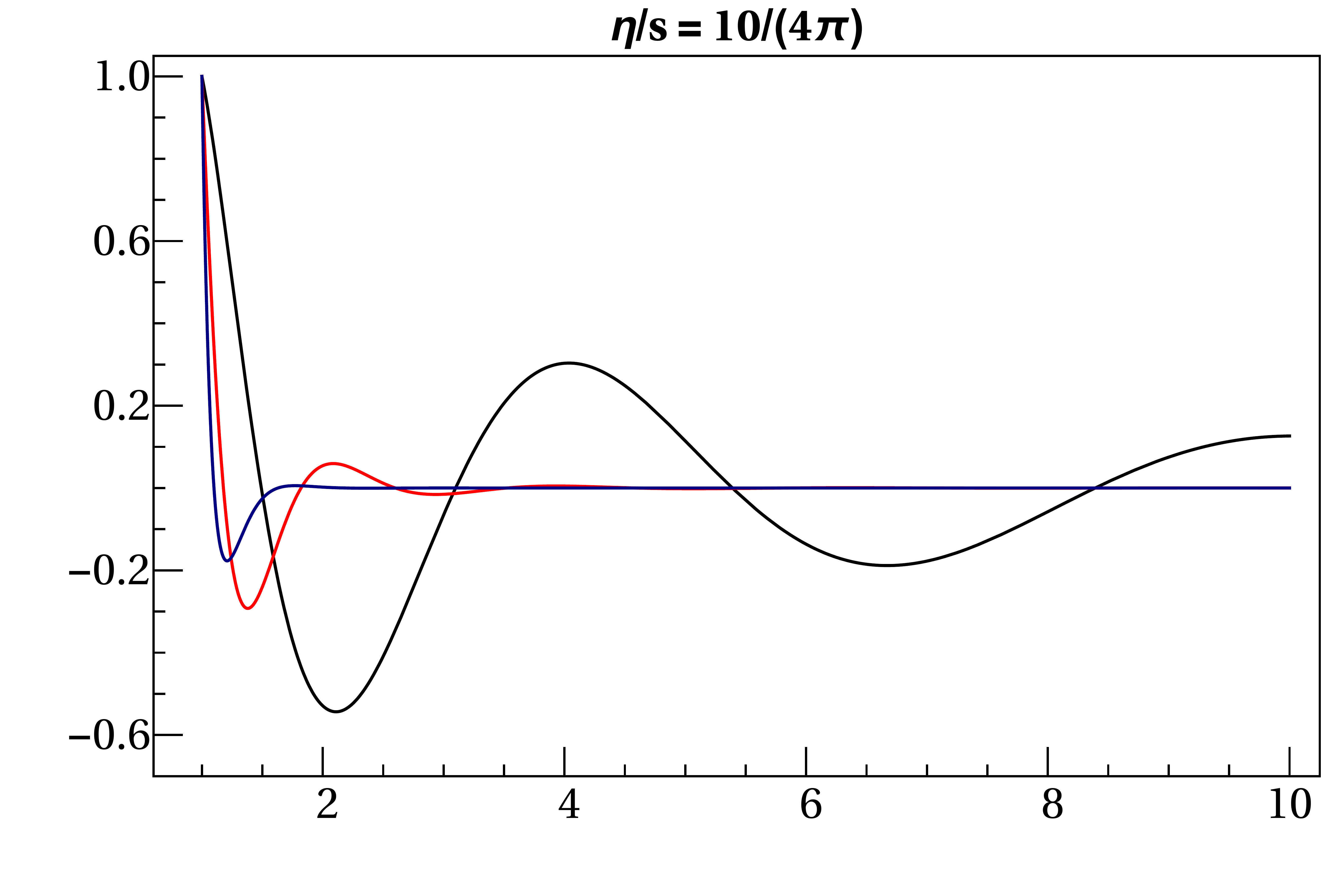}}
\end{picture} 
\\
\begin{picture}(175,130) 
\put(0,8){\includegraphics[scale=0.13]{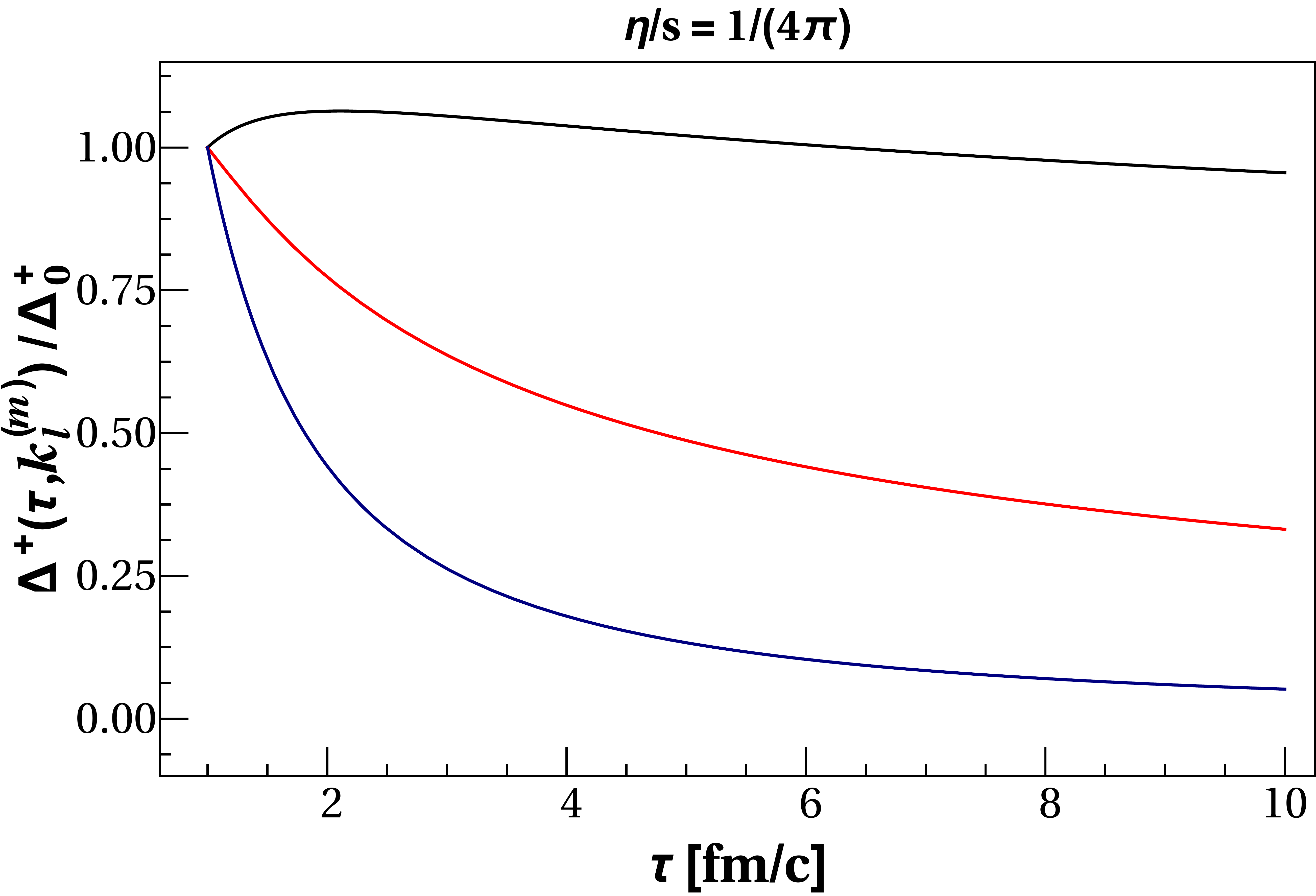} }
\end{picture}
&
\begin{picture}(700,130) 
\put(0,8){\includegraphics[scale=0.125]{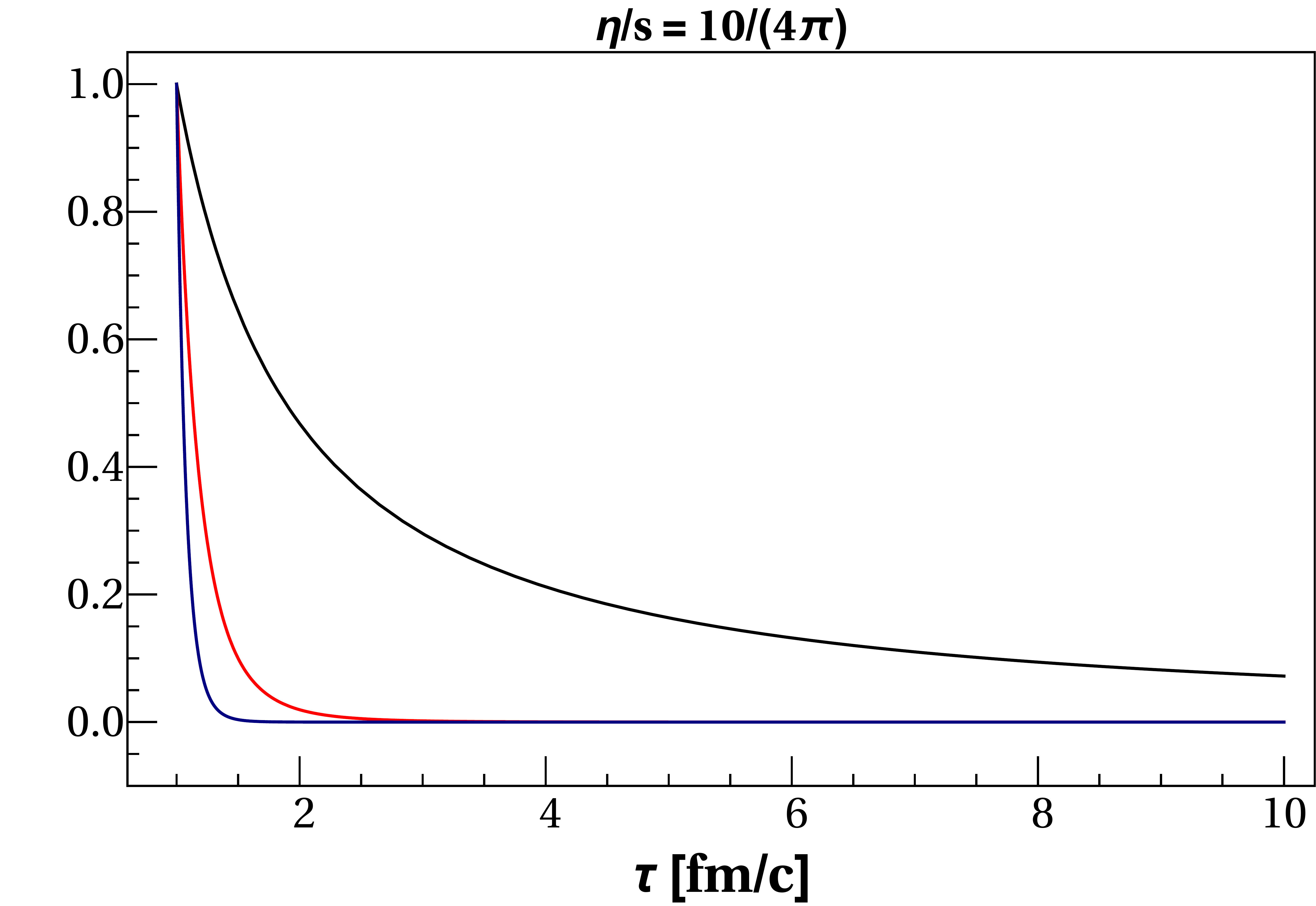}}
\end{picture}
\end{tabular}
\end{centering}
\caption{(Color online) Same as Fig.~\ref{F4} but for different initial values of the fluid perturbations: $\delta u^+(\tau_0)$= 0.4, $\delta u^-(\tau_0)$= 0.3 (thus $\Delta^-_0=u^+_0-u^-_0=0.1$ and $\Delta^+_0=u^+_0+u^-_0=0.7$), $\delta\epsilon(\tau_0)=\delta n(\tau_0)=$0. See text for further details.} 
\label{F6}
\end{figure}
%----------------------------------------------------------------------------------------------------
%--------------------------Fig. 7------------------------------------
\begin{figure}[h]
\begin{centering}
\begin{tabular}{c c }
\begin{picture}(215,150)
\put(0,8){\includegraphics[scale=0.128]{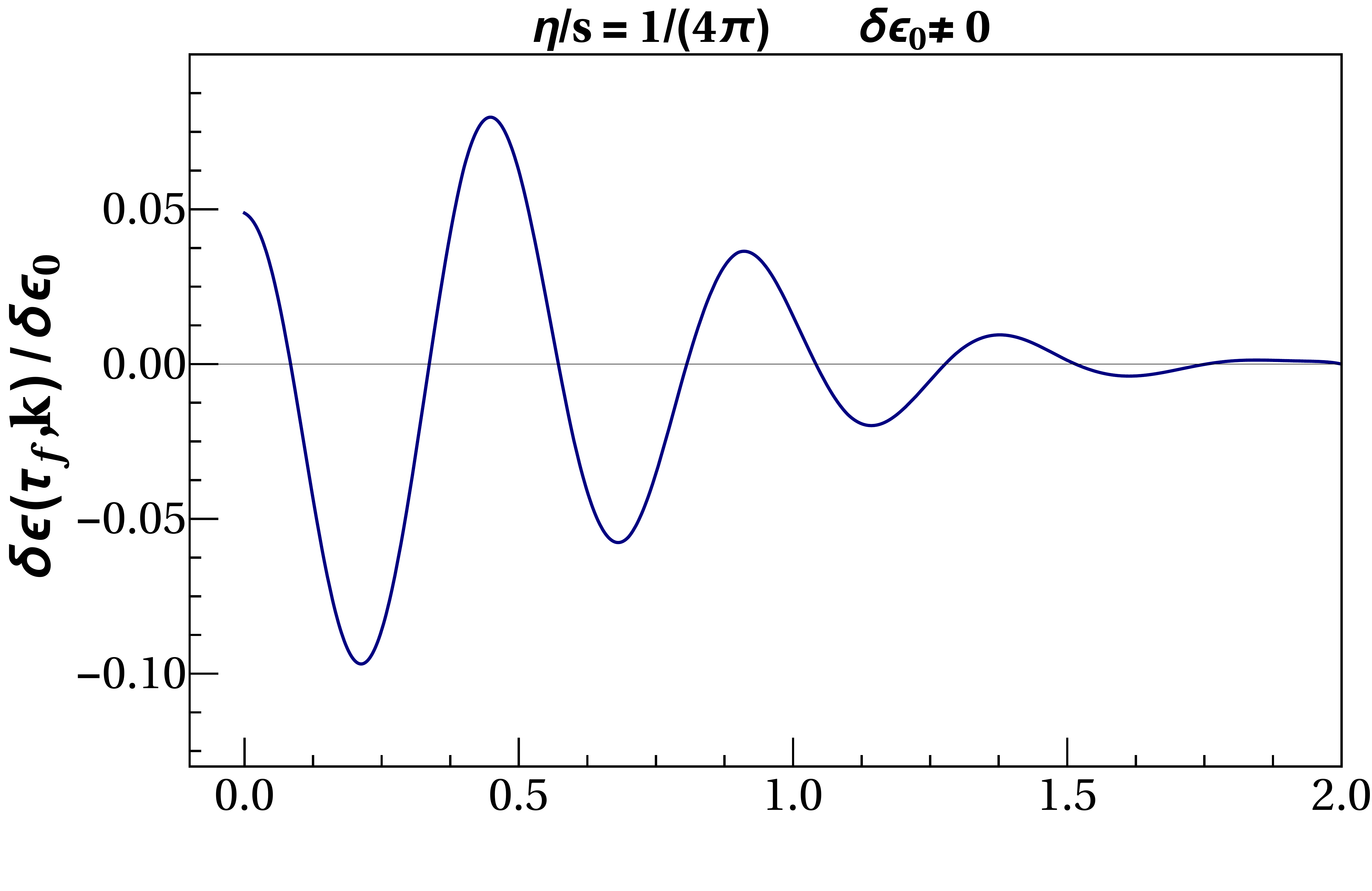}} 
\end{picture}
&
\begin{picture}(410,150)
\put(0,8){\includegraphics[scale=0.145]{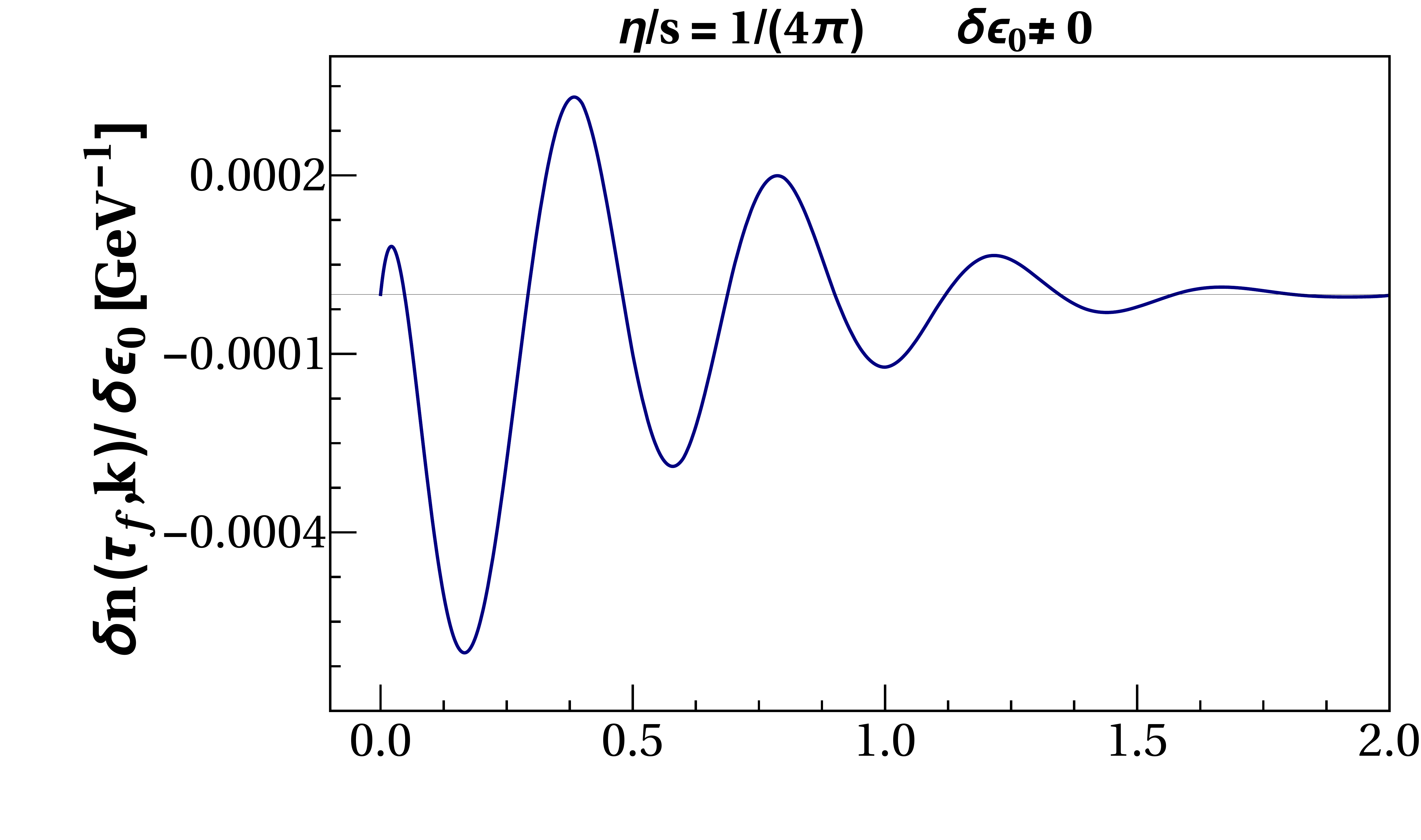}}
\end{picture}
 \\
\begin{picture}(250,130) 
\put(0,8){\includegraphics[scale=0.14]{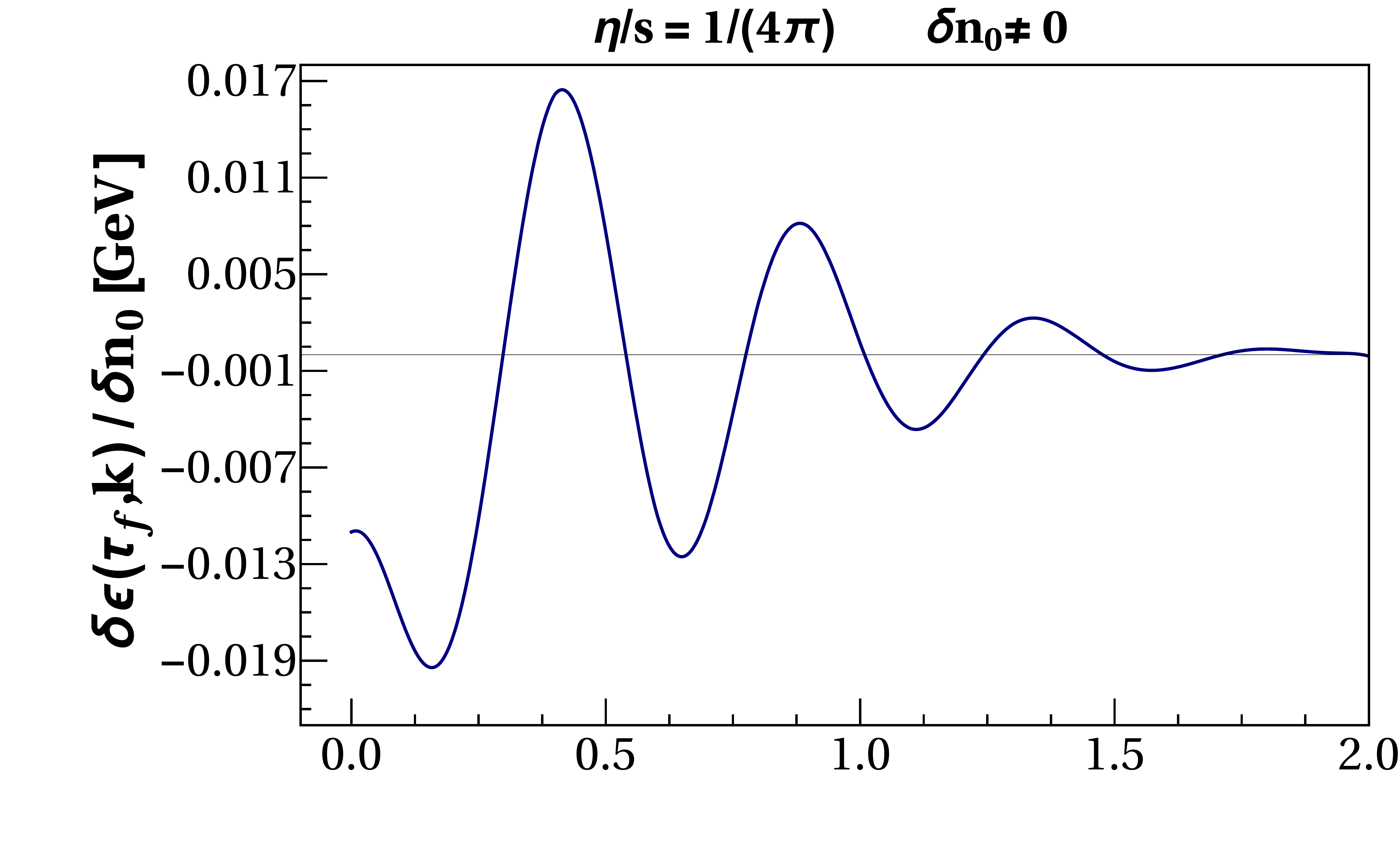} }
\end{picture}
&
\begin{picture}(350,130) 
\put(0,8){\includegraphics[scale=0.125]{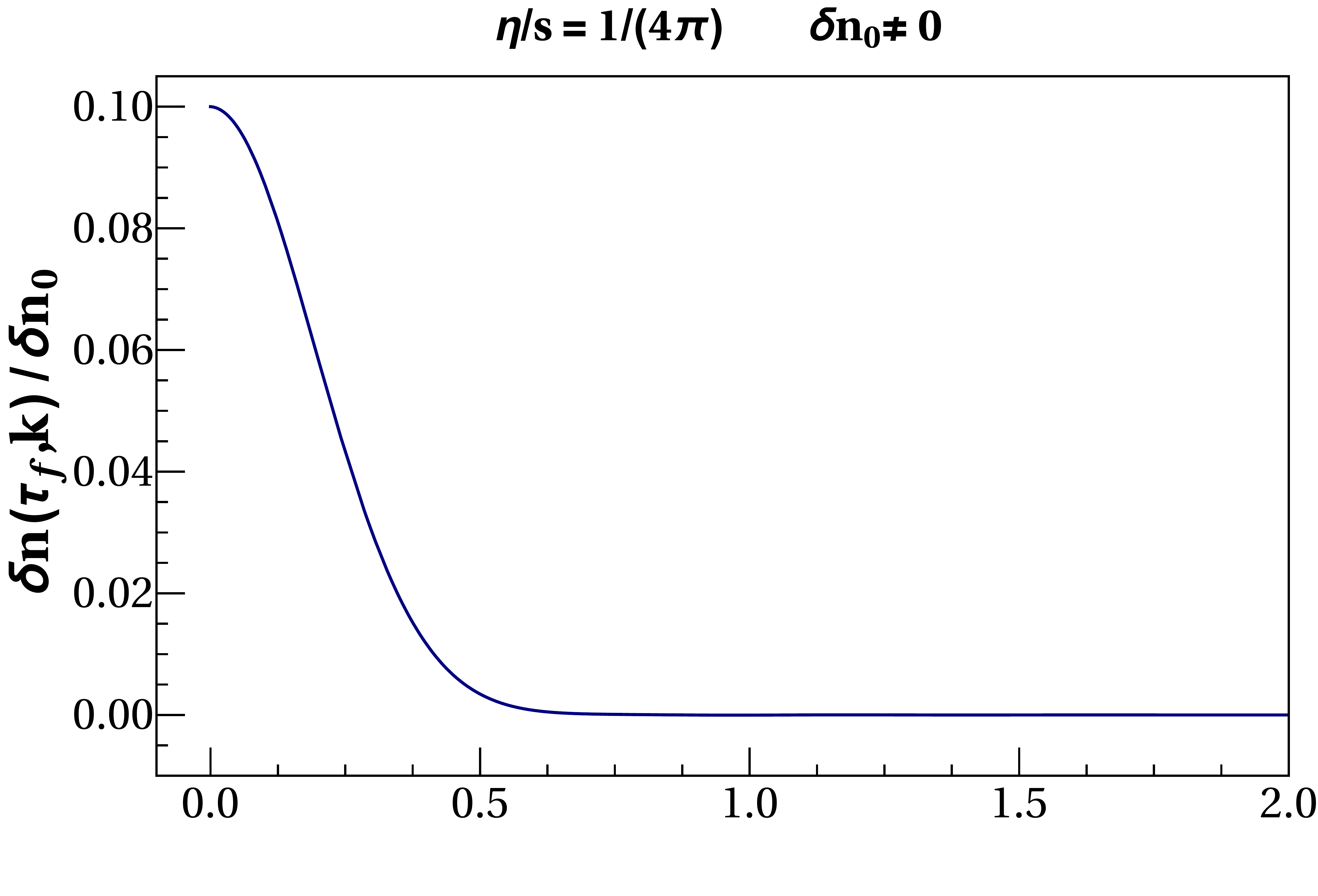}}
\end{picture}
\\
\begin{picture}(240,130) 
\put(0,8){\includegraphics[scale=0.135]{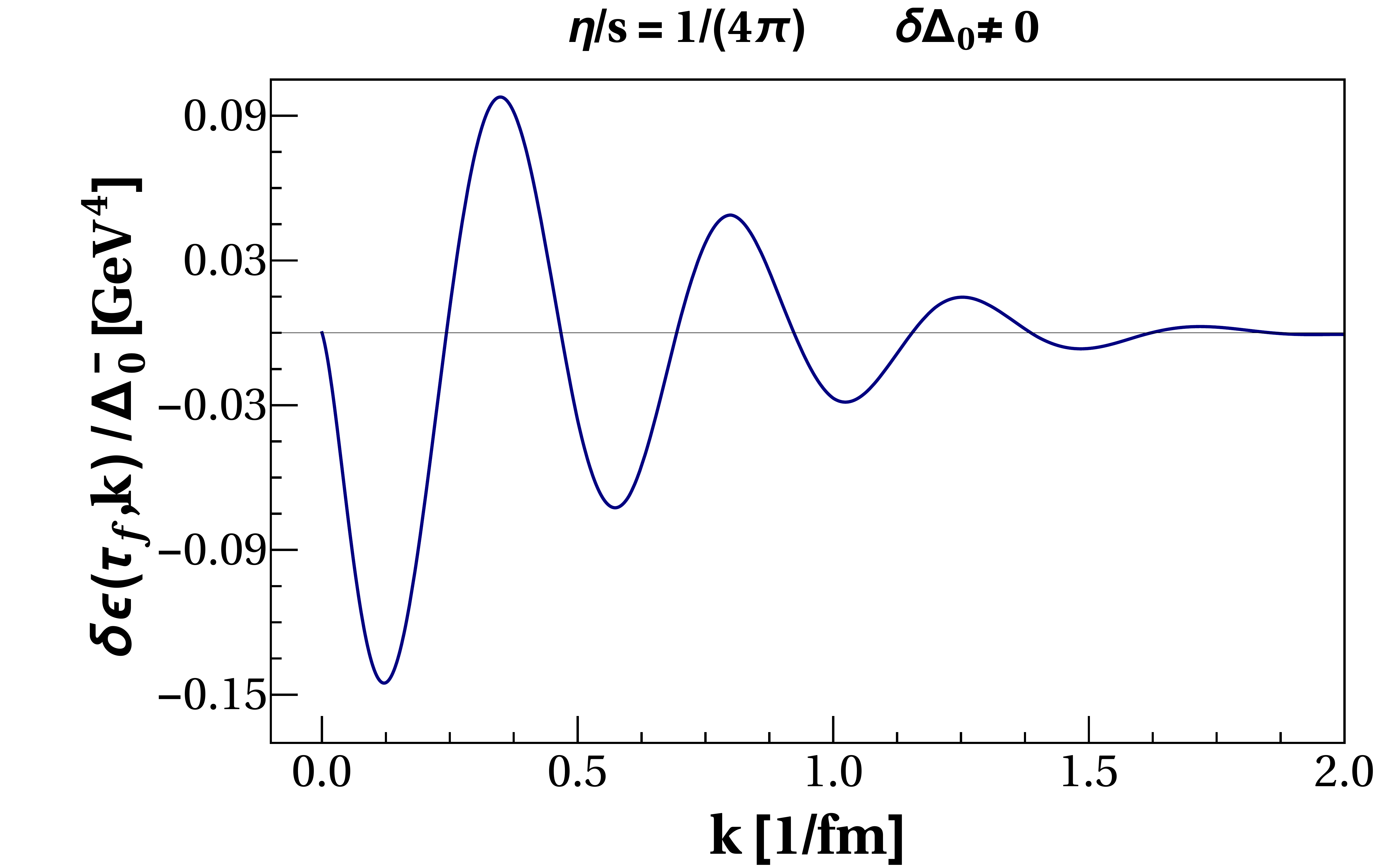}}
\end{picture}
&
\begin{picture}(410,130) 
\put(0,8){\includegraphics[scale=0.145]{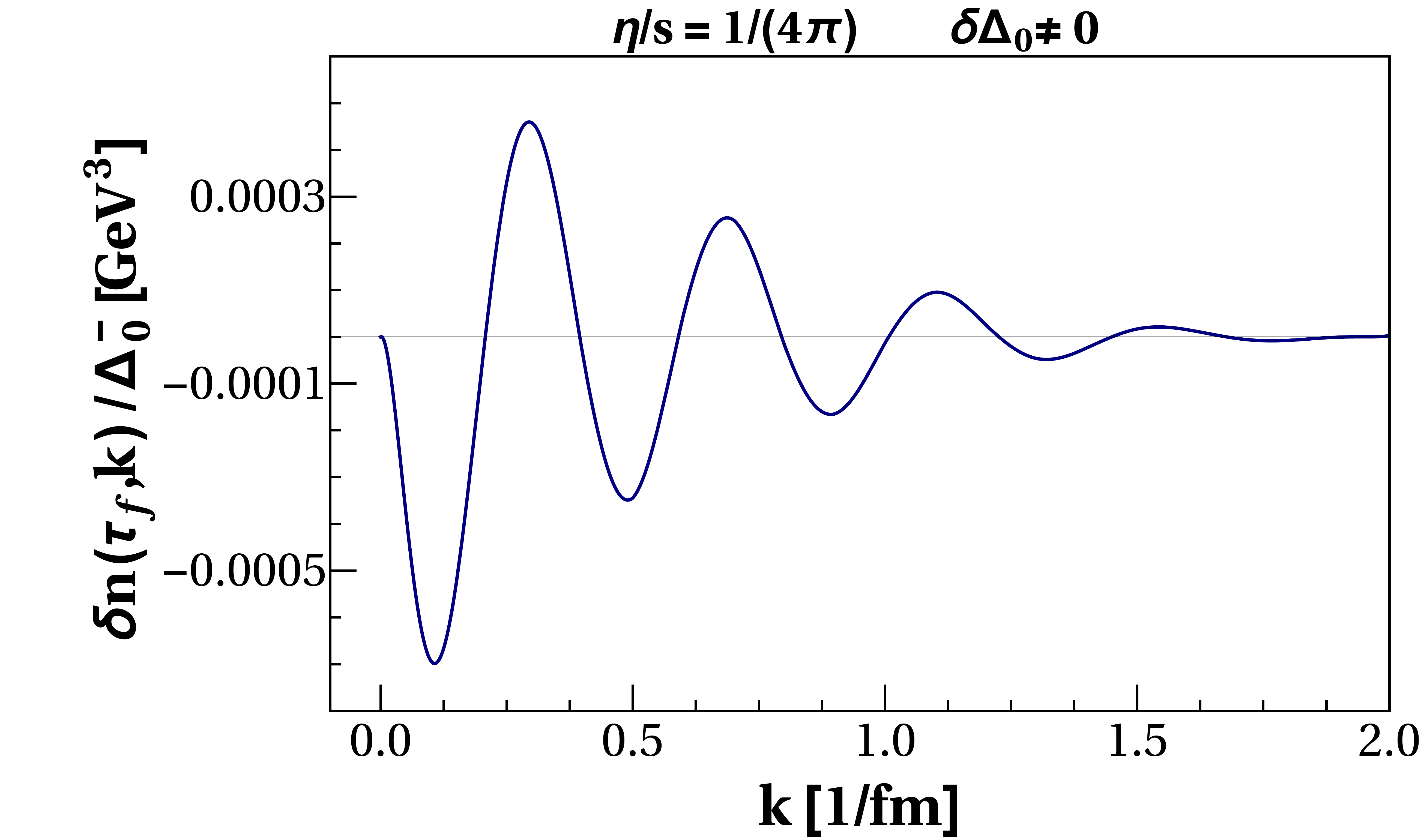}}
\end{picture} 
\end{tabular}
\end{centering}
\caption{(Color online) Amplitude of the perturbations at $\tau_f=10$ fm/c in units of the initial weight at $\tau=1$ fm/c as a function of the $k$-wave number. We choose $\eta/s=1/(4\pi)$, $T_0=$ 0.5 GeV and $\mu_0=$ 0.05 GeV. In each panel we choose a non-vanishing value initially for one of fluctuating fields while the remaining ones are set to zero.The top, middle and bottom panel corresponds to different initial conditions $\delta\epsilon_0 \neq 0$, $\delta n_0\neq 0$ and $\Delta^-_0=u^+_0-u^-_0 \neq 0$, respectively.} 
\label{F7}
\end{figure}
%----------------------------------------------------------------------------------------------------

The evolution equations for perturbations \eqref{eq:LinEvEqEnergyBesselFourier} - \eqref{eq:LinEvEqRapidityVelocityBesselFourier} simplify also in a situation where Bjorken boost invariance is realized as an exact symmetry instead of only on a statistical level. In that case one has $\delta u^\eta = 0$ and the perturbations $\delta \epsilon$, $\delta n$ etc. vanish except for $q=0$. Equation\ \eqref{eq:LinEvEqEnergyBesselFourier} becomes 
\begin{equation}
\begin{split}
\partial_\tau \delta \epsilon & + \left[ \frac{1}{\tau}+\frac{1}{\tau} \left( \frac{\partial p}{\partial \epsilon} \right)_n - \frac{1}{\tau^2} \left( \frac{\partial \zeta}{\partial \epsilon} \right)_n - \frac{4}{3\tau^2} \left( \frac{\partial \eta}{\partial \epsilon} \right)_n \right] \delta\epsilon\\
&+ \left[  \frac{1}{\tau} \left( \frac{\partial p}{\partial n} \right)_\epsilon - \frac{1}{\tau^2} \left( \frac{\partial \zeta}{\partial n} \right)_\epsilon - \frac{4}{3\tau^2} \left( \frac{\partial \eta}{\partial n} \right)_\epsilon \right] \delta n \\
&+ \left[ \bar\epsilon + \bar p - \frac{2}{\tau}\bar \zeta + \frac{4}{3\tau} \bar \eta \right] \frac{k}{\sqrt{2}} \left(\delta u^+ - \delta u^- \right)  = 0,
\end{split}
\label{eq:LinEvEqEnergyBesselFourierBjorkenBoost}
\end{equation}
and similarly Eq.\ \eqref{eq:LinEvEqDensityBesselFourier} becomes
\begin{equation}
\begin{split}
\partial_\tau \delta n  & + \frac{1}{\tau} \delta n + \left[ \bar n - \bar\kappa \left[\frac{\bar n\bar T}{\bar \epsilon + \bar p}\right]^2 \partial_\tau\left(\frac{\bar\mu}{\bar T}\right) \right] \frac{k}{\sqrt{2}} \left(\delta u^+ - \delta u^- \right) \\
& + \bar\kappa \left[\frac{\bar n\bar T}{\bar \epsilon + \bar p}\right]^2 \left( \frac{\partial(\mu/T)}{\partial\epsilon} \right)_n k^2 \, \delta\epsilon + \bar\kappa \left[\frac{\bar n\bar T}{\bar \epsilon + \bar p}\right]^2 \left( \frac{\partial(\mu/T)}{\partial n} \right)_\epsilon k^2 \, \delta n = 0.
\end{split}
\label{eq:LinEvEqDensityBesselFourierBjorken}
\end{equation}
One observes that \eqref{eq:LinEvEqEnergyBesselFourierBjorkenBoost} and \eqref{eq:LinEvEqDensityBesselFourierBjorken} depend on $\delta u^+$ and $\delta u^-$ only via the combination $\left(\delta u^+ - \delta u^-\right)/\sqrt{2}$, for which one obtains from Eq.\ \eqref{eq:LinEvEqPlusMinusVelocityBesselFourier},
 \begin{equation}
 \begin{split}
 & \left( \bar \epsilon + \bar p - \frac{1}{\tau}\bar \zeta + \frac{2 }{3\tau}\bar \eta \right) \partial_\tau \frac{1}{\sqrt{2}}\left(\delta u^+ - \delta u^- \right) \\
 & + \left[ \partial_\tau \bar p - \frac{1}{\tau}\partial_\tau\bar \zeta +\frac{1}{\tau^2} \bar \zeta + \frac{2}{3\tau} \partial_\tau \bar \eta + \frac{4}{3\tau^2} \bar \eta \right] \frac{1}{\sqrt{2}} \left( \delta u^+ - \delta u^- \right) \\
& - \left[ \left( \frac{\partial p}{\partial \epsilon} \right)_n - \frac{1}{\tau}\left( \frac{\partial \zeta}{\partial \epsilon} \right)_n + \frac{2}{3\tau} \left( \frac{\partial \eta}{\partial \epsilon} \right)_n \right] k \, \delta \epsilon  - \left[ \left( \frac{\partial p}{\partial n} \right)_\epsilon - \frac{1}{\tau}\left( \frac{\partial \zeta}{\partial n} \right)_\epsilon + \frac{2}{3\tau} \left( \frac{\partial \eta}{\partial n} \right)_\epsilon \right] k \, \delta n \\
& + \left[ \bar \zeta k^2 + \frac{4}{3} \bar \eta k^2 \right] \frac{1}{\sqrt{2}} \left(  \delta u^+ - \delta u^- \right)  = 0 .
 \end{split}
 \label{eq:LinEvEqPlusMinusVelocityBesselFourierBjorken}
 \end{equation}
Equations \eqref{eq:LinEvEqEnergyBesselFourierBjorkenBoost} - \eqref{eq:LinEvEqPlusMinusVelocityBesselFourierBjorken} together with the information about background quantities form a closed system that describes the analog of sound propagation and baryon number diffusion in the transverse plane of a longitudinally expanding fireball. The orthogonal combination of fluid velocity perturbations $\delta u^+ + \delta u^-$ is a shear mode with purely dissipative behavior (equation not shown).

It is interesting to compare these equations to the ones that govern perturbations in a static medium. In that case all terms that involve explicit factors $1/\tau$ or derivatives of background quantities with respect to $\tau$ vanish. For example, the analog of Eq.\ \eqref{eq:LinEvEqEnergyBesselFourierBjorkenBoost} is\begin{equation}
\begin{split}
\partial_\tau \delta \epsilon + \left( \bar\epsilon + \bar p \right) \frac{k}{\sqrt{2}} \left(\delta u^+ - \delta u^- \right)  = 0,
\end{split}
\label{eq:LinEvEqEnergyBesselFourierBjorkenBoostStatic}
\end{equation}
while the analog of Eq.\ \eqref{eq:LinEvEqDensityBesselFourierBjorken} is
\begin{equation}
\begin{split}
& \partial_\tau \delta n   + \bar n \frac{k}{\sqrt{2}} \left(\delta u^+ - \delta u^- \right) \\
& + \bar\kappa \left[\frac{\bar n\bar T}{\bar \epsilon + \bar p}\right]^2 \left( \frac{\partial(\mu/T)}{\partial\epsilon} \right)_n k^2 \, \delta\epsilon + \bar\kappa \left[\frac{\bar n\bar T}{\bar \epsilon + \bar p}\right]^2 \left( \frac{\partial(\mu/T)}{\partial n} \right)_\epsilon k^2 \, \delta n = 0,
\end{split}
\label{eq:LinEvEqDensityBesselFourierBjorkenStatic}
\end{equation}
and the analog of Eq.~\eqref{eq:LinEvEqPlusMinusVelocityBesselFourierBjorken} is
 \begin{equation}
 \begin{split}
 & \left( \bar \epsilon + \bar p \right) \partial_\tau \frac{1}{\sqrt{2}}\left(\delta u^+ - \delta u^- \right) - \left( \frac{\partial p}{\partial \epsilon} \right)_n k \, \delta \epsilon - \left( \frac{\partial p}{\partial n} \right)_\epsilon k \, \delta n \\
& + \left[ \bar \zeta  + \frac{4}{3} \bar \eta \right] k^2 \frac{1}{\sqrt{2}} \left(  \delta u^+ - \delta u^- \right)  = 0 .
 \end{split}
 \label{eq:LinEvEqPlusMinusVelocityBesselFourierBjorkenStatic}
 \end{equation}
 The set of equations \eqref{eq:LinEvEqEnergyBesselFourierBjorkenBoostStatic} - to \eqref{eq:LinEvEqPlusMinusVelocityBesselFourierBjorkenStatic} describes sound propagation in the presence of dissipation due to shear viscosity, bulk viscosity and heat conductivity. We observe that at least some of the additional terms in Eq.\ \eqref{eq:LinEvEqEnergyBesselFourierBjorkenBoost} compared to \eqref{eq:LinEvEqEnergyBesselFourierBjorkenBoostStatic} have the effect of an additional damping, in particular the square bracket in the first line of Eq.\ \eqref{eq:LinEvEqEnergyBesselFourierBjorkenBoost} is expected to be positive in the regime where fluid dynamics is applicable. Similarly, the leading additional term in Eq.\ \eqref{eq:LinEvEqDensityBesselFourierBjorken} compared to Eq.\ \eqref{eq:LinEvEqDensityBesselFourierBjorkenStatic} is the term $\frac{1}{\tau}\delta n$ that has a damping effect, as well. The situation is less clear for the additional terms in Eq.\ \eqref{eq:LinEvEqPlusMinusVelocityBesselFourierBjorken} compared to Eq.\ \eqref{eq:LinEvEqPlusMinusVelocityBesselFourierBjorkenStatic}, in particular the second line in Eq.\ \eqref{eq:LinEvEqPlusMinusVelocityBesselFourierBjorken} might actually conteract damping because $\partial_\tau \bar p$ is negative. However, at least for larger values of the wavenumber $k$ and non-zero viscosities the dissipative damping term in the last line of Eq.\ \eqref{eq:LinEvEqPlusMinusVelocityBesselFourierBjorken} is dominating.

Equations \eqref{eq:LinEvEqEnergyBesselFourierBjorkenBoost} - \eqref{eq:LinEvEqPlusMinusVelocityBesselFourierBjorken} simplify further if the background is symmetric under baryon number conjugation as discussed in Sect.~\ref{subsec:statbar}. In that case the perturbation in baryon number density $\delta n$ decouples from Eqs.\ \eqref{eq:LinEvEqEnergyBesselFourierBjorkenBoost} and \eqref{eq:LinEvEqPlusMinusVelocityBesselFourierBjorken} and is described by Eq.\ \eqref{eq:LinEvEqDensityBesselFourierBaryonConjugationSymmetry} (with $q=0$). Nevertheless, the remaining equations for $\delta \epsilon$ and $\left(\delta u^+ - \delta u^-\right)/\sqrt{2}$ remain coupled and have to be integrated numerically for a given background solution and wavenumber $k$. This has already been discussed in Ref.~\cite{Florchinger:2011qf}.

In Figs.\ \ref{F4}, \ref{F5} and \ref{F6} we show numerical solutions of the evolution equations \eqref{eq:LinEvEqEnergyBesselFourierBjorkenBoost}- \eqref{eq:LinEvEqPlusMinusVelocityBesselFourierBjorken} for the ideal EOS~\eqref{eq:idealEOS}. For the background fields we employ the scaling solution~\eqref{eq:scalingTmu}. We compare the numerical results for different initial conditions and two different values of the ratio of shear viscosity to entropy density and assume $\zeta=0$ for simplicity. More precisely, the left columns of Figs.\ \ref{F4}, \ref{F5} and \ref{F6} correspond to $\eta/s=1/(4\pi)$, the right columns to $\eta/s = 10/(4\pi)$. In all cases, the heat conductivity is taken to be related to the shear viscosity by \eqref{eq:thercond}. We also compare different values of the radial wavenumber $k=k^{(m)}_l = z^{(m)}_l / R$. We choose $R=10$ fm/c which corresponds to $k_1^{(1)}=0.38 \, \text{fm}^{-1}$ (black curves), $k_2^{(2)}=0.84 \, \text{fm}^{-1}$ (red curves) and $k_3^{(3)}=1.30 \, \text{fm}^{-1}$ (blue curves). In all cases, the modes with larger $k$ are damped more quickly as expected. In order to simplify the notation we use the abbreviation $\Delta^-\equiv\delta u^+ -\delta u^-$ in Figs.\ \ref{F4}, \ref{F5} and \ref{F6}.

In Fig.\ \ref{F4} we have chosen initial conditions with non-vanishing perturbations in energy density $\delta \epsilon(\tau) = \delta \epsilon_0$ while the perturbations in baryon number density $\delta n$ and fluid velocity $\delta u^+$, $\delta u^-$ vanish initially. The pressure gradients associated with $\delta \epsilon$ induce sound waves with the typical oscillating behavior between $\delta \epsilon$ and $\delta u^+ - \delta u^-$, modified by the longitudinal expansion and viscous damping. As expected, the oscillation frequency is larger for larger radial wave numbers $k$. The perturbation in energy density $\delta \epsilon$ induces also a small perturbation in baryon number density $\delta n$ at times $\tau>\tau_0$. This is due to the linear mixing between the different fluctuating fields ($\delta\epsilon, \delta u^+$ and $\delta u^-$) for non-vanishing background baryon chemical potential (we choose $\mu_0=0.05 \, \text{GeV}$). For $\bar \mu = \bar n = 0$, the evolution equation for $\delta n$ would decouple from the other fluctuating fields as we discussed in the previous section. Because we solve linearized equations for the perturbations, the solution scales linearly with the initial value $\delta \epsilon_0$.

In Fig. \ref{F5} we initialize with non-vanishing perturbation in the baryon number density $\delta n(\tau_0) = \delta n_0$ but set $\delta \epsilon(\tau_0) = \delta u^+(\tau_0) = \delta u^-(\tau_0) = 0$. The mode excited in this way has essentially diffusive behavior. This is most clearly seen in the intermediate panel which shows the temporal evolution of $\delta n/\delta n_0$. There are no oscillations seen but simply a decay in amplitude which is faster for large values of $k$. This decay is mainly a consequence of heat conductivity (or equivalently, baryon number diffusion). In addition to the baryon number density perturbation, also a (small) perturbation in $\delta \epsilon$ and $\delta u^+ - \delta u^-$ is excited for $\tau> \tau_0$. This is again a consequence of the non-vanishing baryon number density in the background. The behavior of these perturbations is oscillatory, i.e., of sound type. 

In Fig. \ref{F6} we choose initial conditions with $\Delta^-_0=\delta u^+_0 - \delta u^-_0 \neq 0$ while the perturbations $\delta \epsilon$ and $\delta n$ vanish initially. This results again in sound propagation of the typical oscillating type. In Fig. \ref{F6}
we also show the behavior of perturbations in the orthogonal combination $\Delta^+=\delta u^+ + \delta u^-$ which is a shear mode whose decay rate is determined by shear viscosity $\eta$. The shear viscosity dependence of the decay rate for this particular shear mode can be obtained directly from the corresponding evolution equation \eqref{eq:LinEvEqPlusMinusVelocityBesselFourier}.

In Fig. \ref{F7} we show the final amplitude of the perturbations in energy density (left column) and particle density (right panel) at $\tau_f=10$ fm/c as a function of the $k$-wave number in units of the initial weight at time $\tau_0$=1 fm/c for $\eta/s=1/(4\pi)$ and  different initial conditions of the perturbations of the fluctuating fields. This plot shows that some modes of the initial perturbations characterized by the $k$-wave number indeed survive the entire evolution of the system and at the same time, it also indicates the distribution of of the surviving modes at the scales of time relevant for the freeze-out surface.~\footnote{This can be understood directly when taking the Fourier transform of the fluctuating fields at $\tau=\tau_f$, e.g., Eq.~\eqref{eq:BesselFourierExpansionEnergy} The distribution of the fluctuating field as a function of its $k$-wave number determines the distribution of this field in coordinate space and determines the location of the maximum allowed correlation length in coordinate space.}. The uppermost panel corresponds to the non-zero value for the initial perturbation in energy density $\delta \epsilon(\tau_0) = \delta \epsilon_0$ while the remaining fluctuating fields, $\delta n_0, \delta u^+_0$ and $\delta u^-_0$, vanish exactly. The middle panel corresponds to the case where  $\delta n(\tau_0) = \delta n_0$ and $\delta \epsilon_0=\delta u^+_0= \delta u^-_0=0$. The bottom panel corresponds to $\Delta^-_0=\delta u^+_0-\delta u^-_0\neq 0$ and  $\delta \epsilon_0=\delta n_0=0$. For the sound wave type initial conditions ($\delta\epsilon_0\neq 0$ or $\Delta^-_0\neq 0$) the size of the amplitudes at $\tau_f$ present a damped oscillatory behavior while for the case when $\delta n_0\neq 0$ the fluctuation of the $\delta\epsilon$ and $\Delta^-$ present an oscillatory behavior while $\delta n$ shows a exponential type decay which is typical to diffusive processes. In all the cases we observe that essentially none of the modes survive for values of $k\geq 2\,\,\text{fm}^{-1}$. %The maximum of the amplitude of the fluctuating field at $\tau_f$ always occurs when $k\leq 1\,\,\text{fm}^{-1}$ and its location depends on the initial condition. This is a hint that correlations between different fluctuating fields in the transverse plane might get enhanced. In order to have a more quantitative answer, it would be required to implement a more realistic scenario than the Bjorken model for the hydrodynamical fields. In addition, the location of the maximum would depend on the initial distribution of the spectrum of fluctuations.    

We conclude this subsection by emphasizing again the observation that perturbations in baryon number density have a diffusive time evolution with a dissipation rate determined by the heat conductivity. For typical values corresponding to strong coupling behavior, the damping is rather strong but at least the modes with the smallest radial and azimuthal wave-numbers (small values of $m$ and $l$) are not dissipated completely and could have experimentally observables consequences.

%%%%%%%%%%%%%%%%%%%%%%%%%%%%%%%%%%%%%%%%%%%%%%%%%%%%%%%%%%%%%%%%%%%%%%%%%%%%%%%%%%%%%%
\subsection{Exact transverse translation and rotation symmetry} 
\label{subsec:extranrot}
%%%%%%%%%%%%%%%%%%%%%%%%%%%%%%%%%%%%%%%%%%%%%%%%%%%%%%%%%%%%%%%%%%%%%%%%%%%%%%%%%%%%%%
%--------------------------Fig. 8------------------------------------
\begin{figure}[h]
\begin{centering}
\begin{tabular}{c c}
\begin{picture}(190,150)
\put(0,8){\includegraphics[scale=0.195]{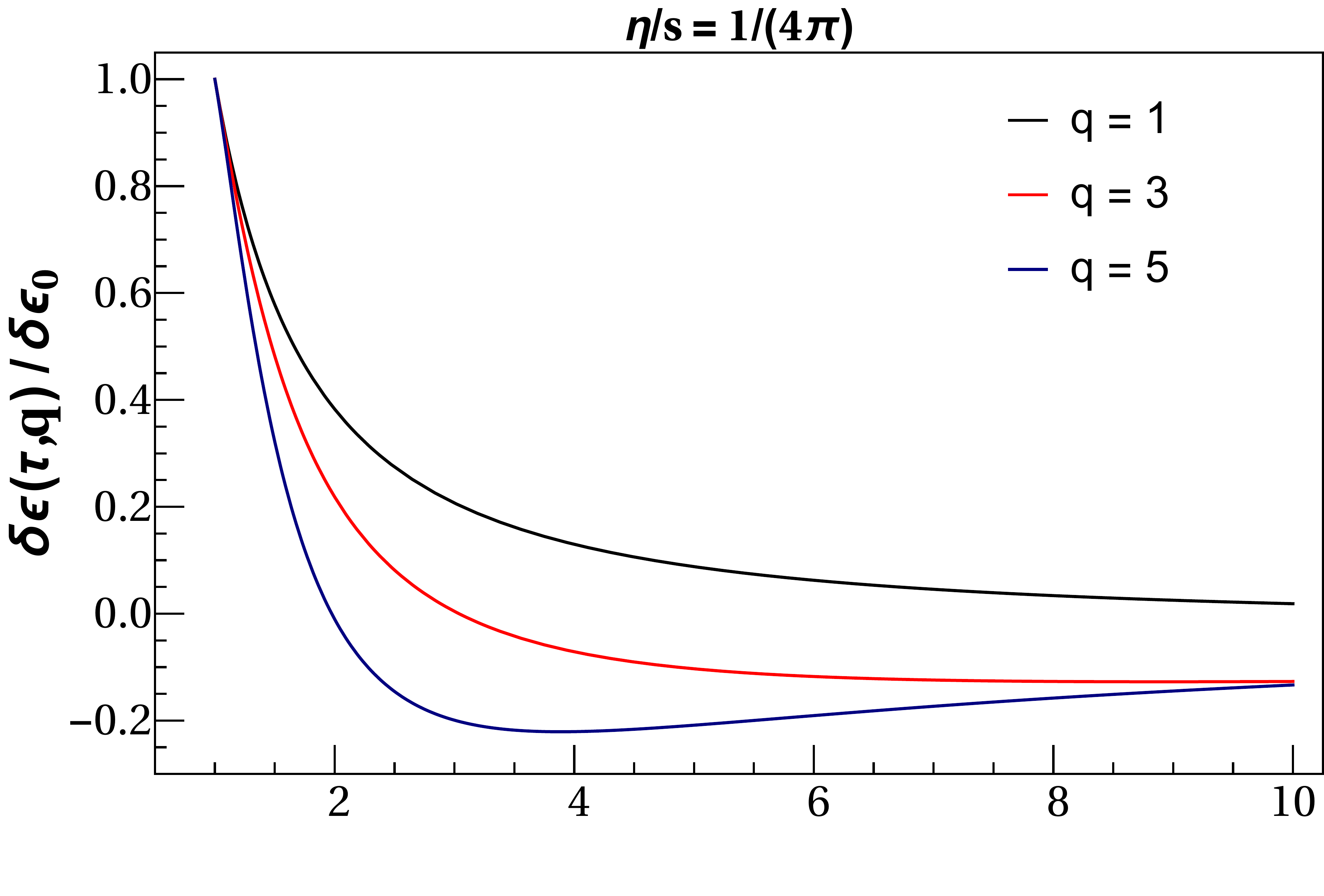}} 
\end{picture}
&
\begin{picture}(450,180)
\put(0,8){\includegraphics[scale=0.139]{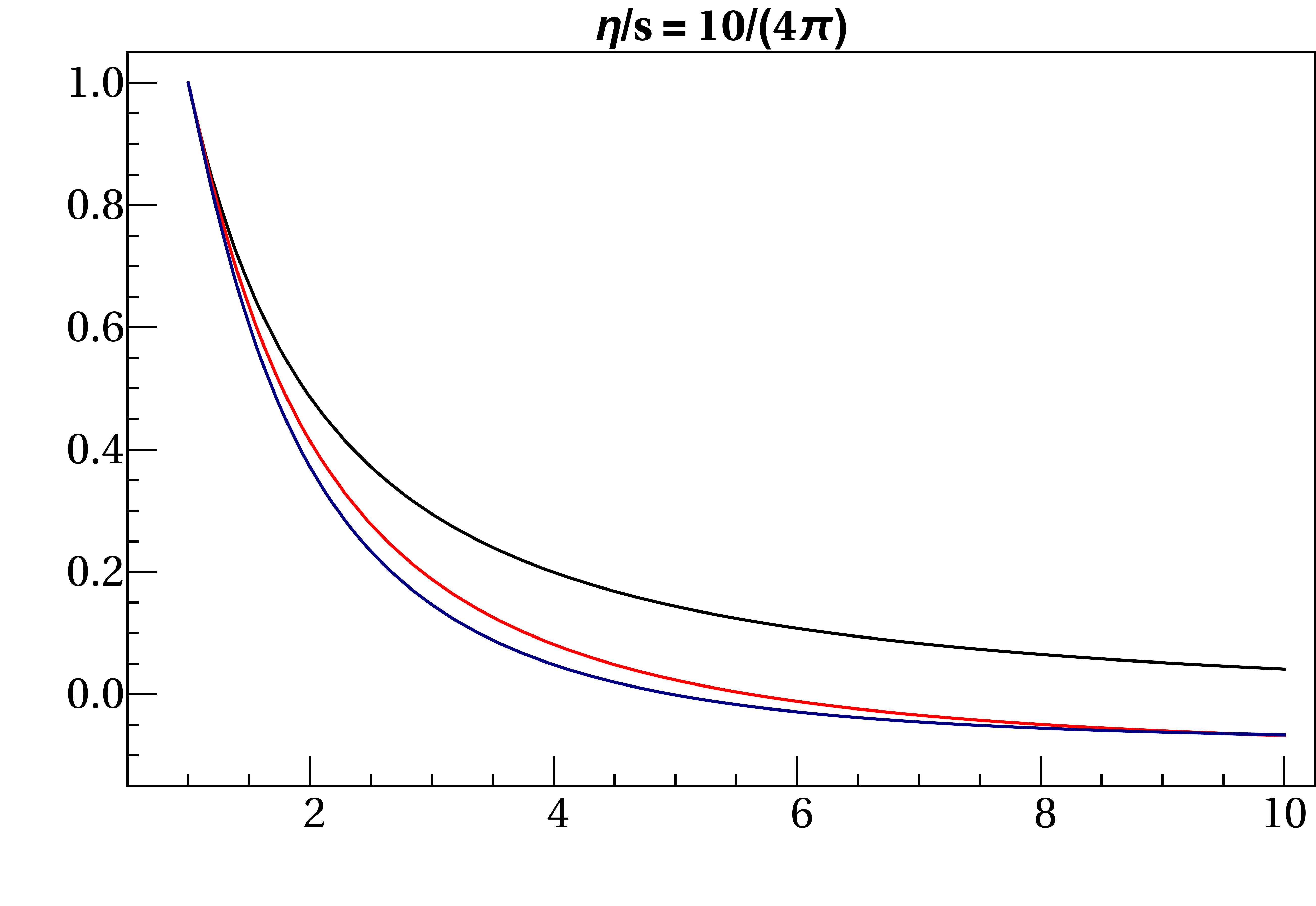}}
\end{picture}
 \\
\begin{picture}(247,140) 
\put(0,8){\includegraphics[scale=0.159]{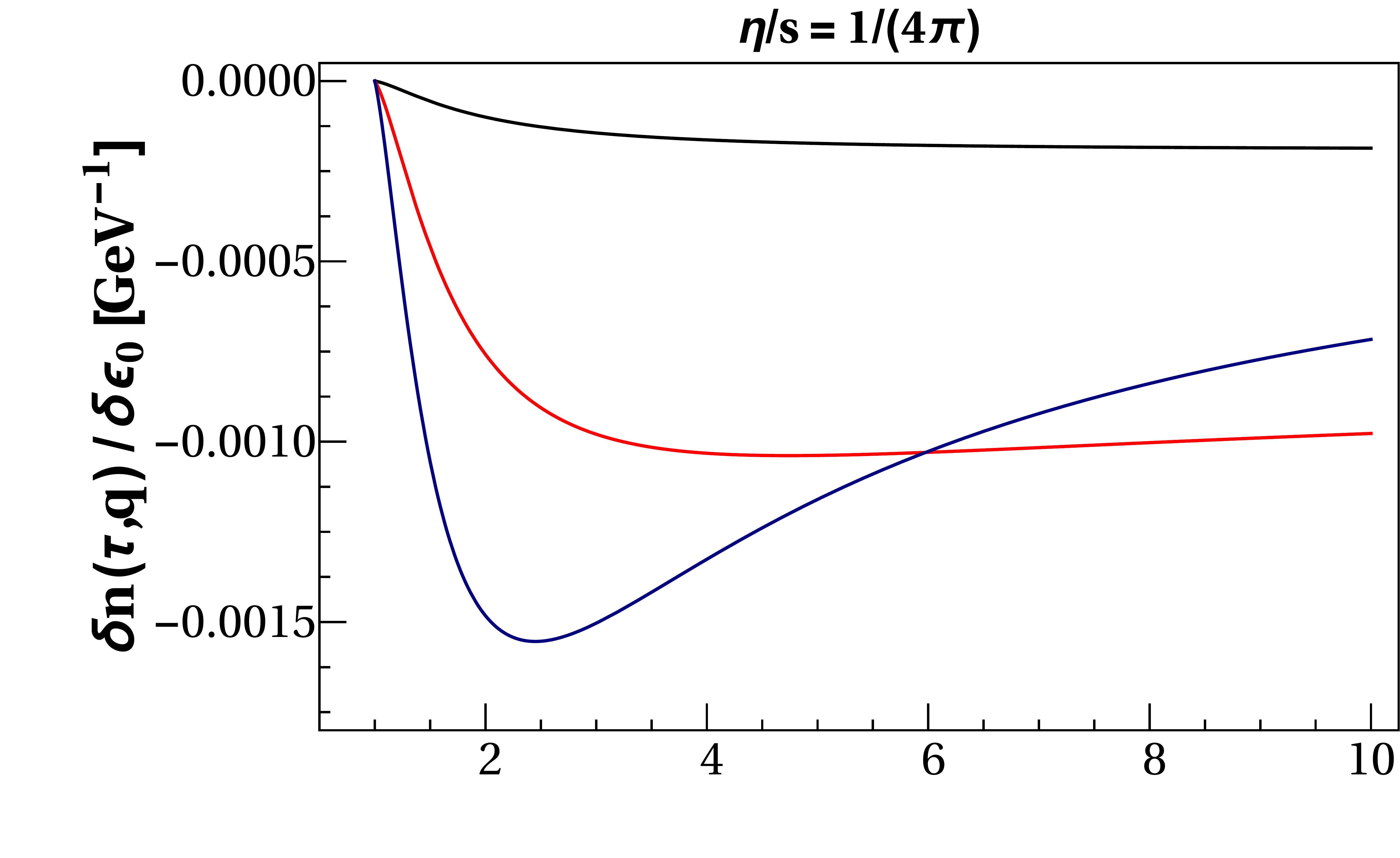} }
\end{picture}
&
\begin{picture}(485,130) 
\put(0,8){\includegraphics[scale=0.15]{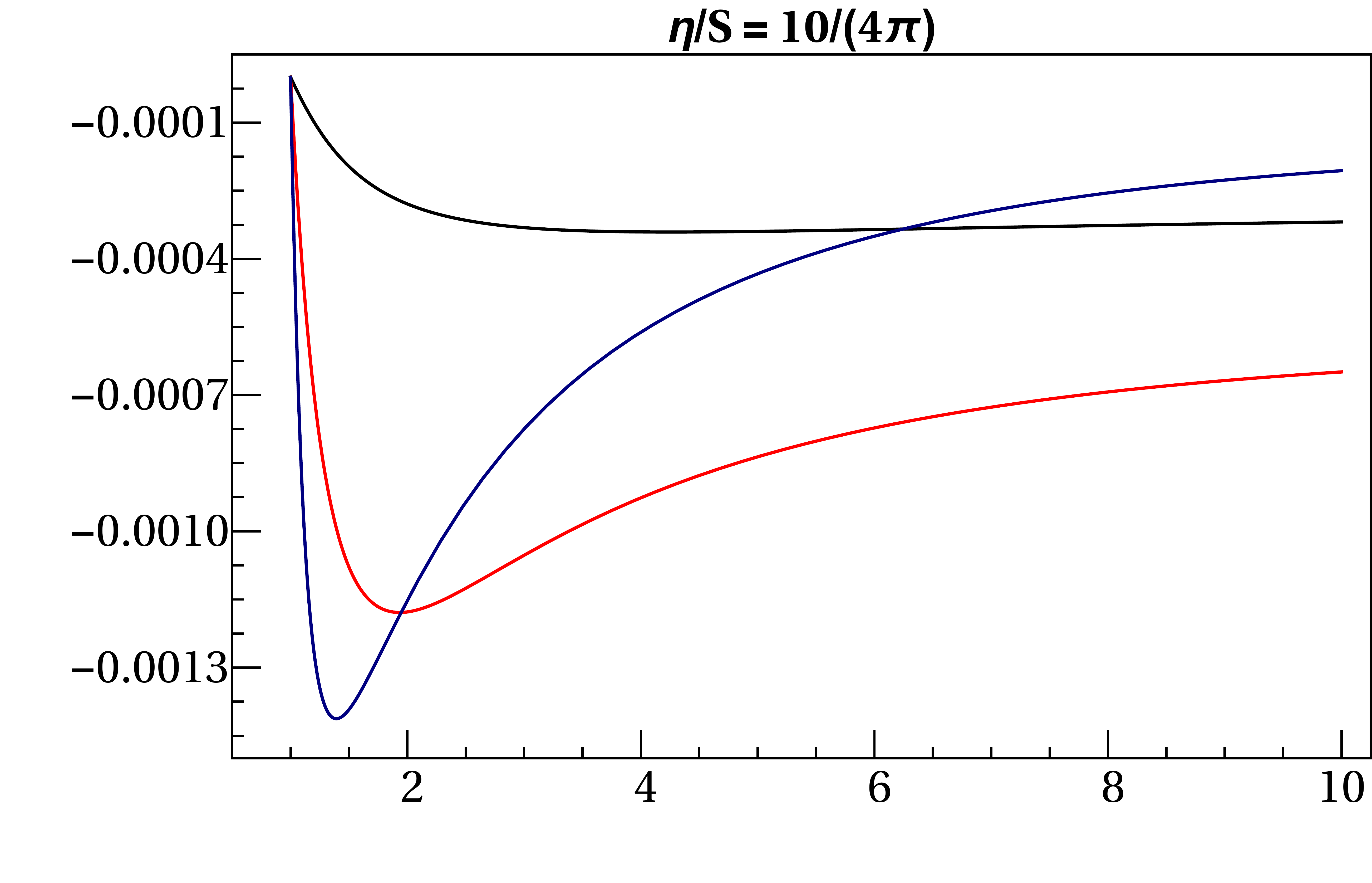}}
\end{picture}
\\
\begin{picture}(235,140) 
\put(0,8){\includegraphics[scale=0.155]{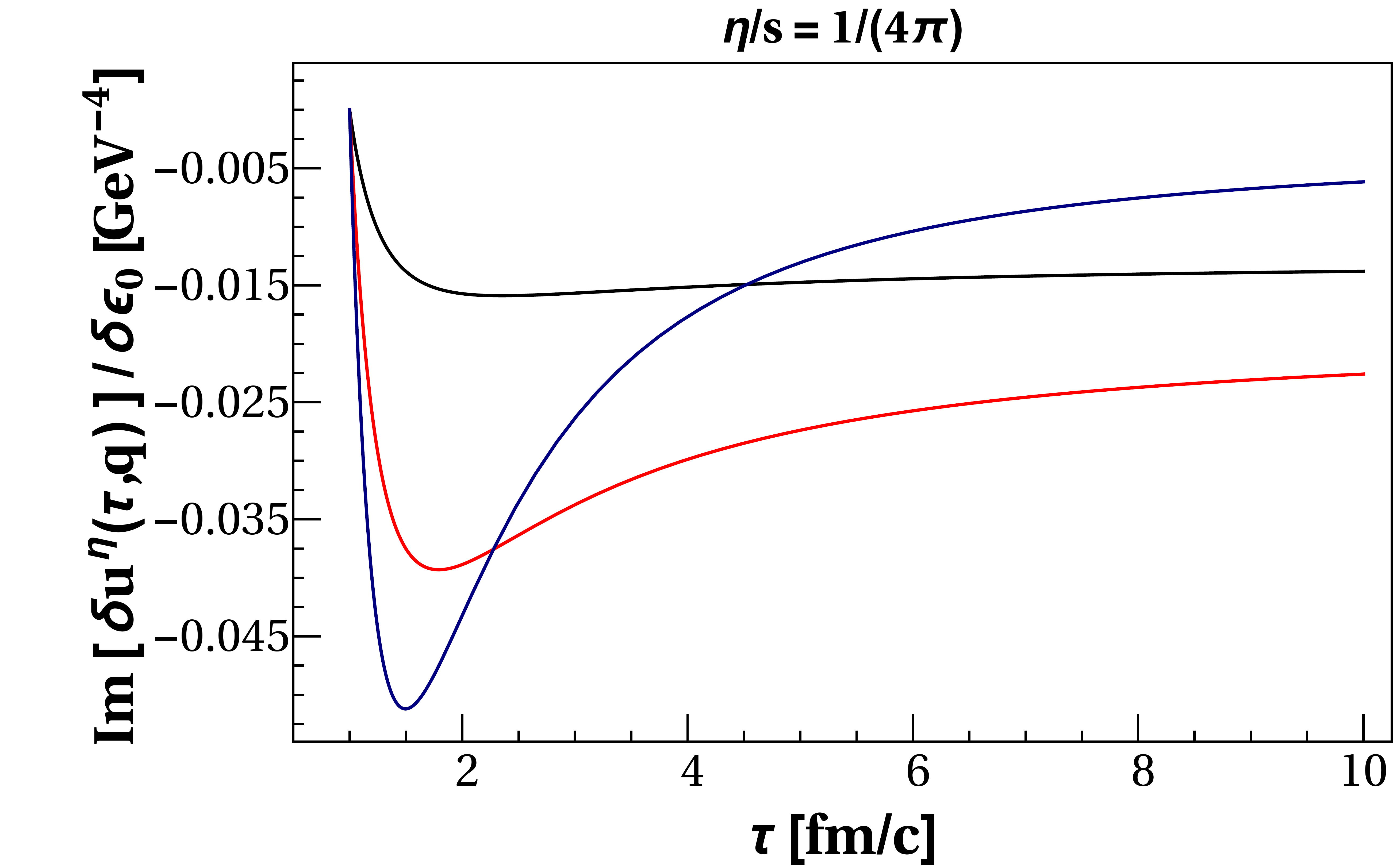}}
\end{picture}
&
\begin{picture}(465,140) 
\put(0,8){\includegraphics[scale=0.144]{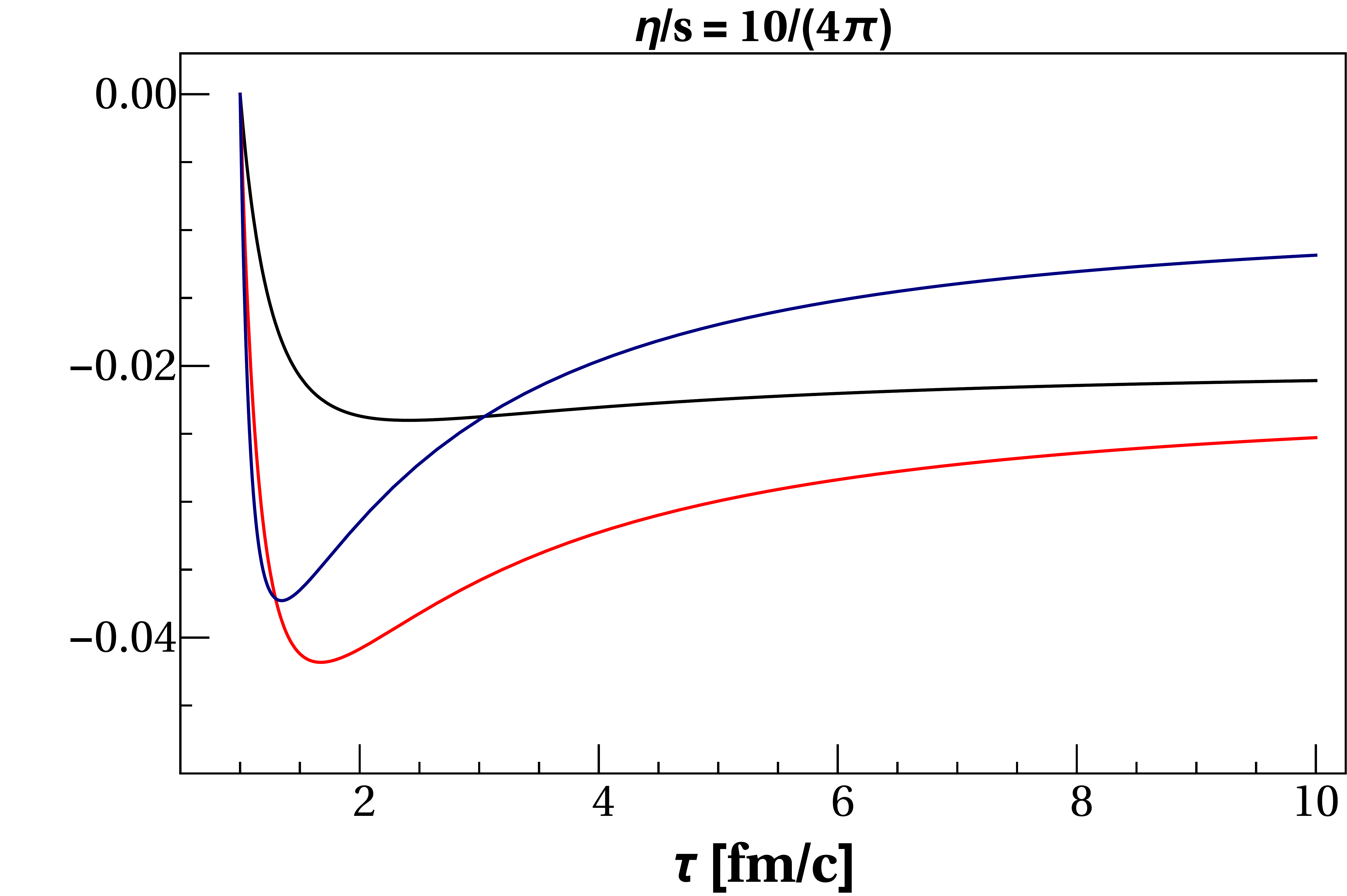}}
\end{picture} 
\end{tabular}
\end{centering}
\caption{(Color online) Evolution of perturbations in energy density, baryon number density and fluid velocity with exact transverse translation and rotation symmetry ($k = 0$, $\delta u^+=\delta u^-=0$) for different values of the rapidity wavenumber $q$: $q=1$ (black line), $q=3$ (red line) and $q=5$ (blue line). We compare two different values of the ratio of shear viscosity to entropy density $\eta/s=1/(4\pi)$ (left column) and $\eta/s=10/(4\pi)$ (right column). Heat conductivity is parametrized by Eq.\ \eqref{eq:thercond}. We use $T_0=$ 0.5 GeV, $\mu_0=$0.05 GeV, $\tau_0$= 1 fm/c, $\tau_f$=10 fm/c and for the initial values of the hydrodynamic fluctuations we choose $\delta\epsilon(\tau_0) \neq 0$, $\delta n(\tau_0)=\delta u^\eta(\tau_0)=0$. See text for further details.} 
\label{F8}
\end{figure}
%----------------------------------------------------------------------------------------------------
%--------------------------Fig. 9------------------------------------
\begin{figure}[h]
\begin{centering}
\begin{tabular}{c c}
\begin{picture}(190,150)
\put(0,8){\includegraphics[scale=0.195]{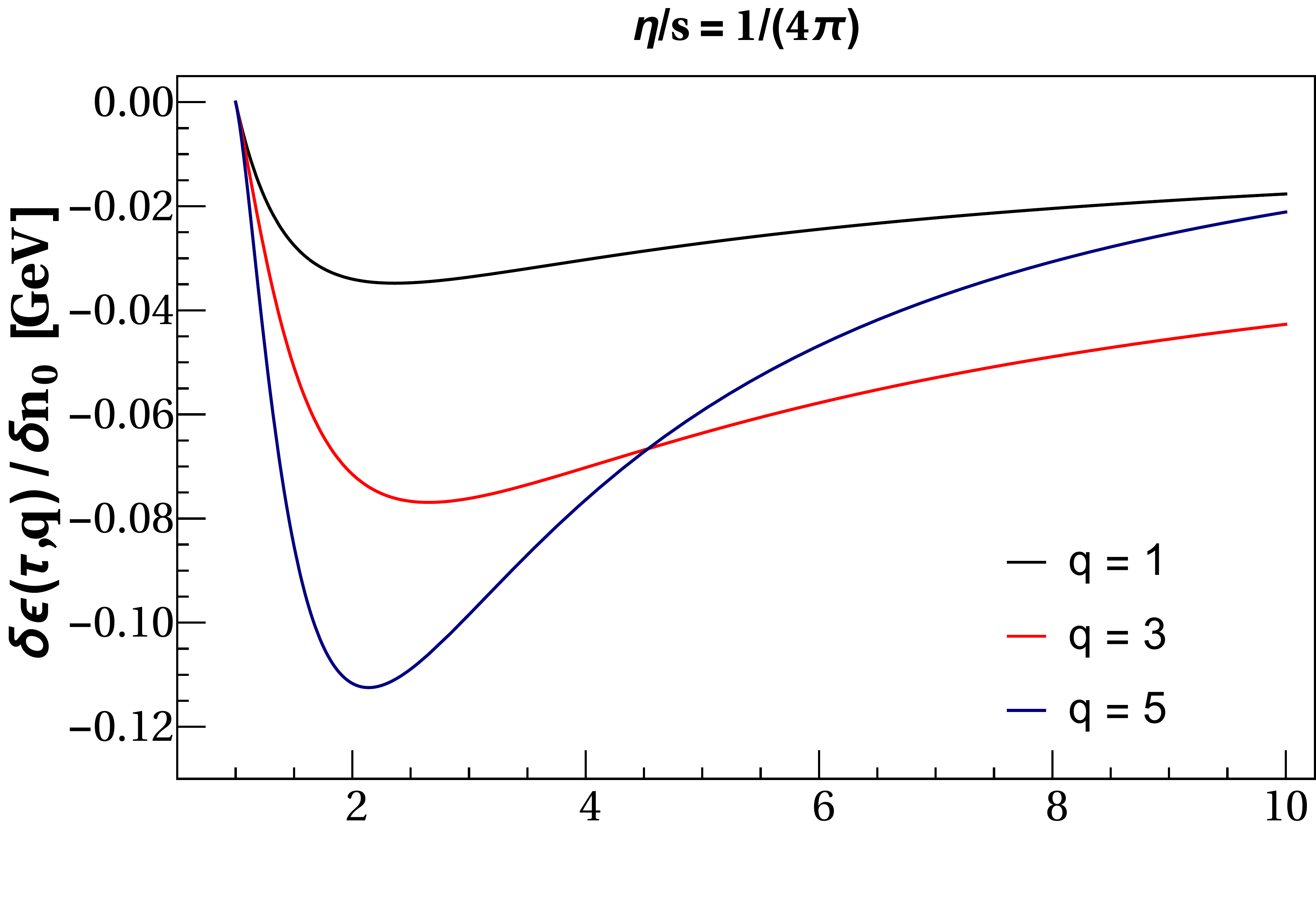}} 
\end{picture}
&
\begin{picture}(450,180)
\put(0,8){\includegraphics[scale=0.142]{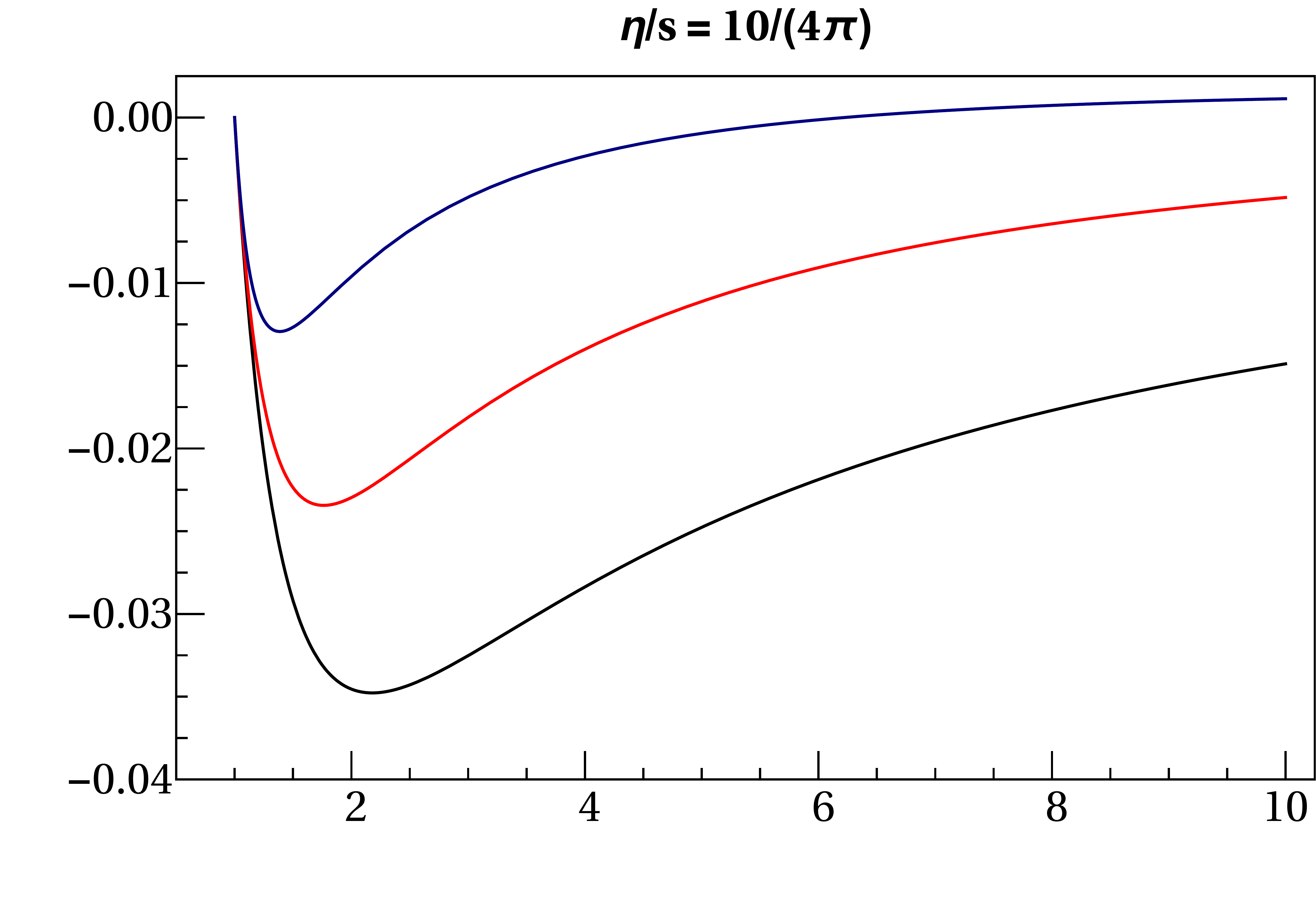}}
\end{picture}
 \\
\begin{picture}(177,140) 
\put(0,8){\includegraphics[scale=0.135]{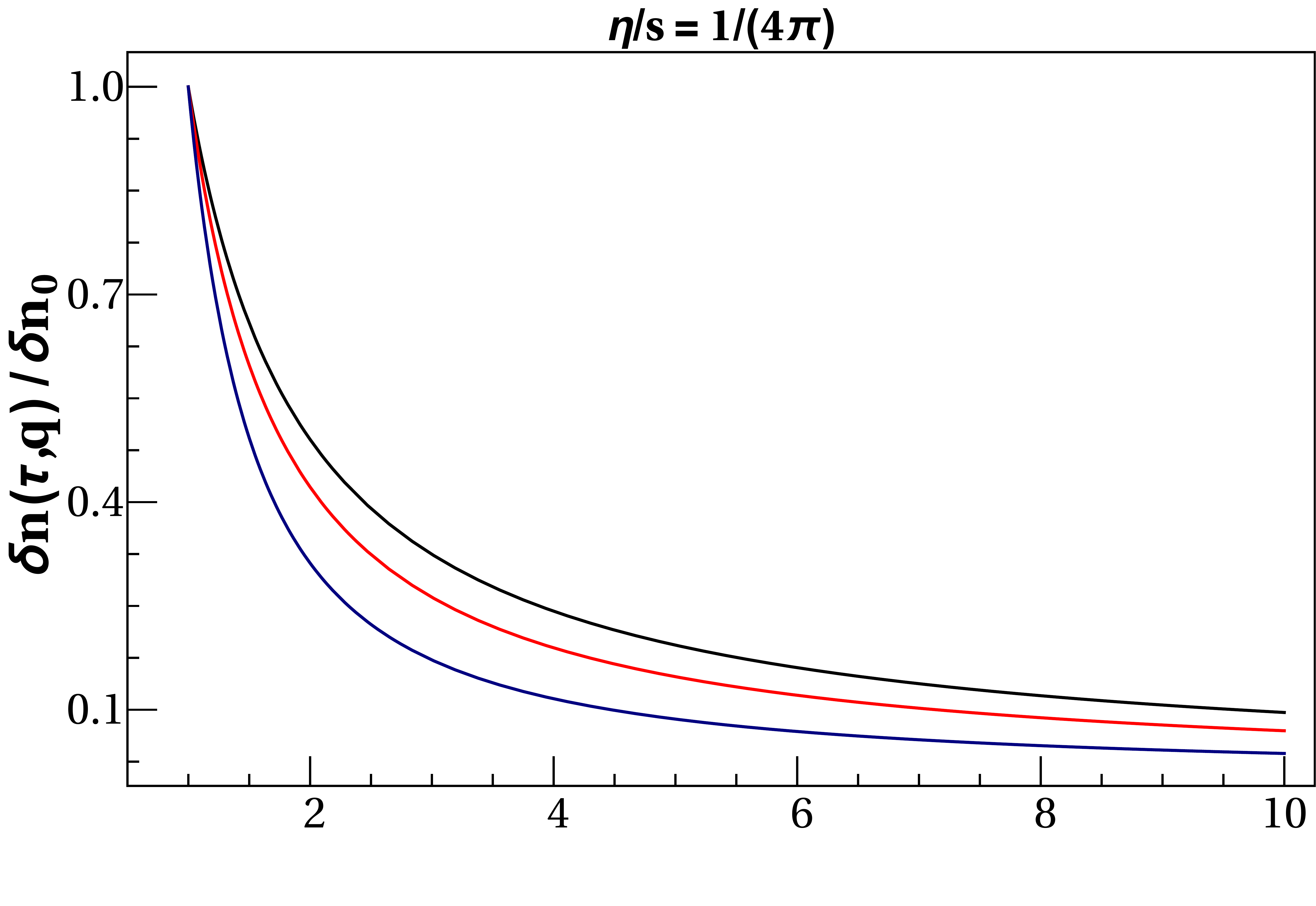} }
\end{picture}
&
\begin{picture}(433,130) 
\put(0,8){\includegraphics[scale=0.136]{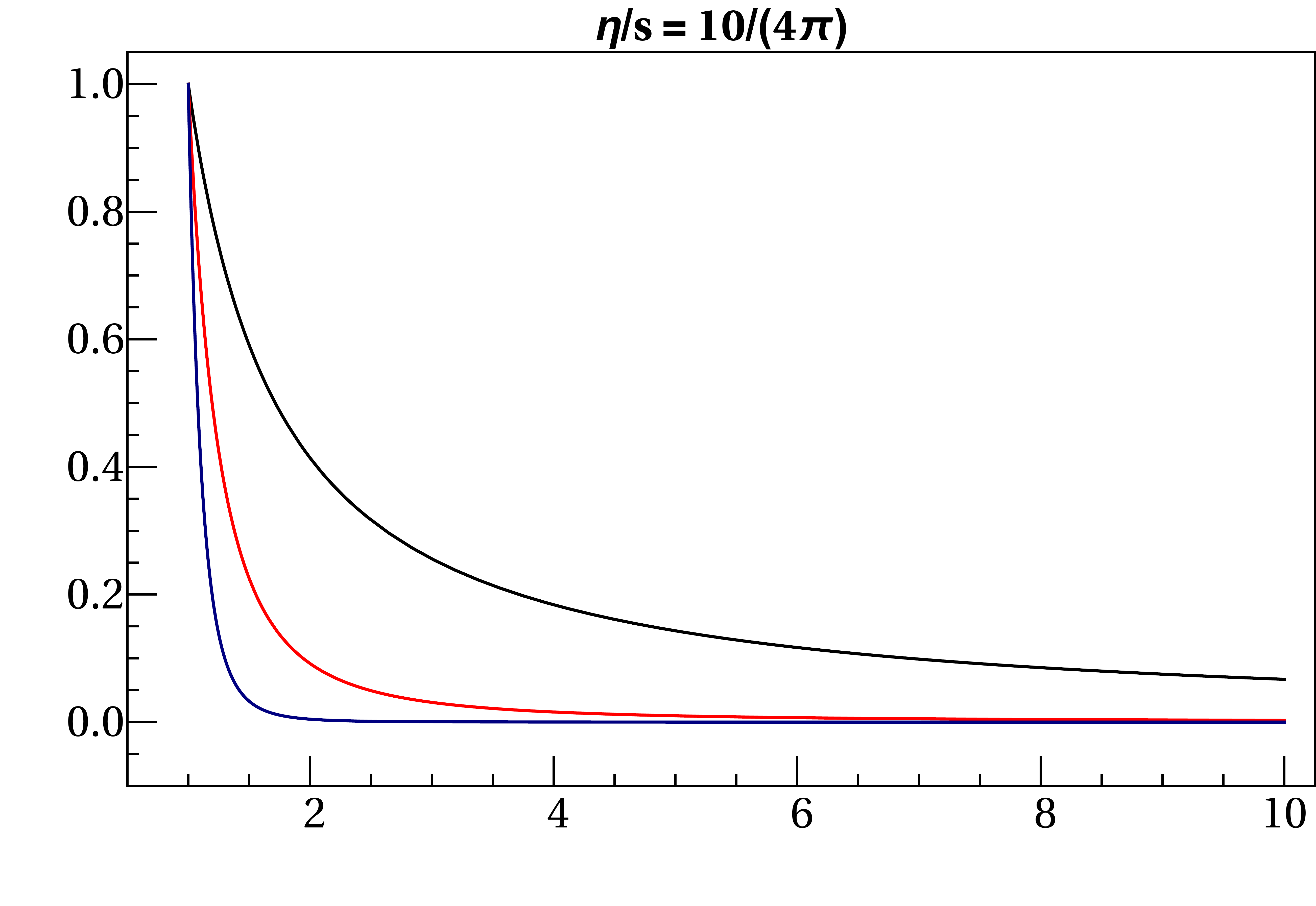}}
\end{picture}
\\
\begin{picture}(237,140) 
\put(0,8){\includegraphics[scale=0.155]{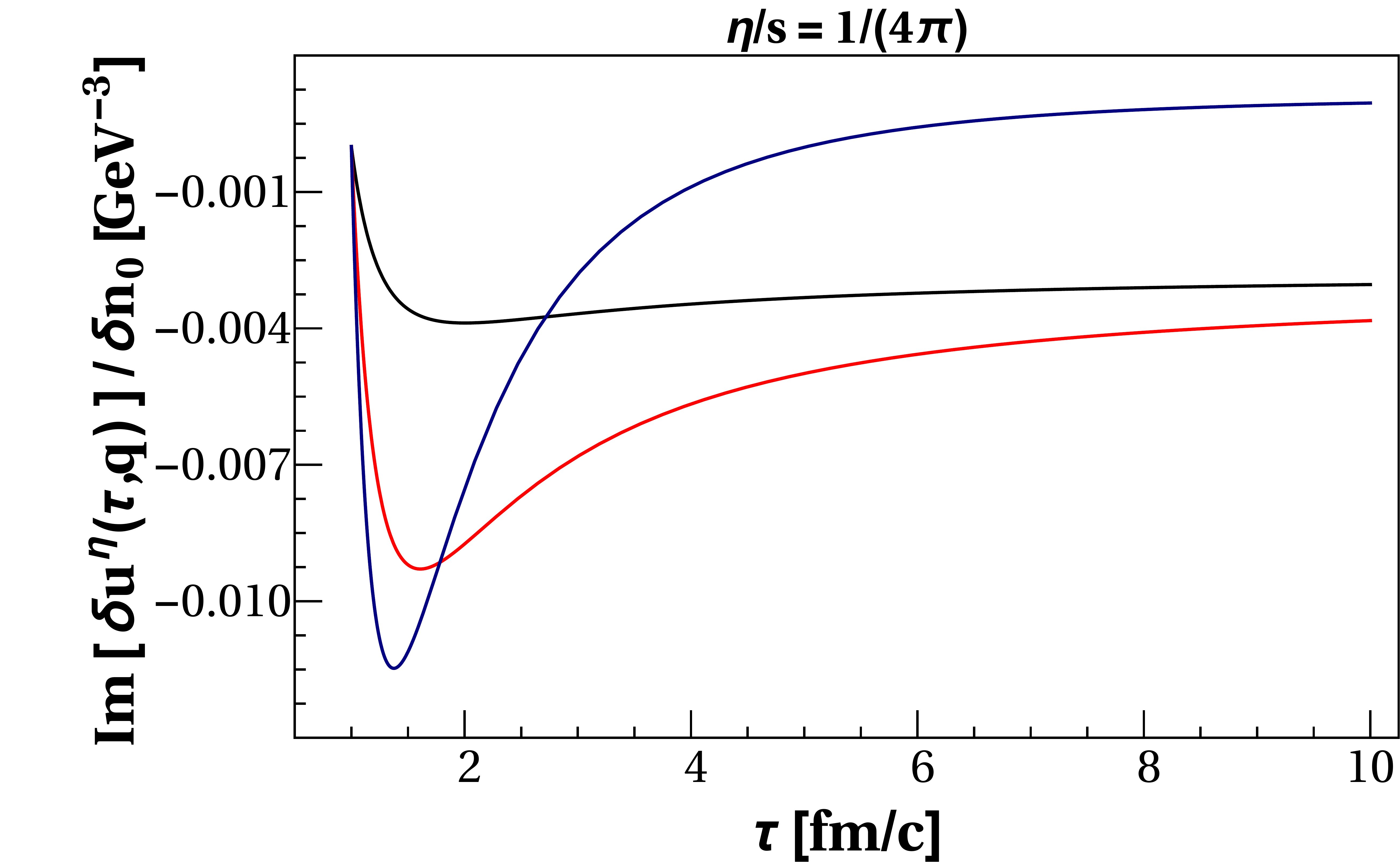}}
\end{picture}
&
\begin{picture}(460,140) 
\put(0,8){\includegraphics[scale=0.144]{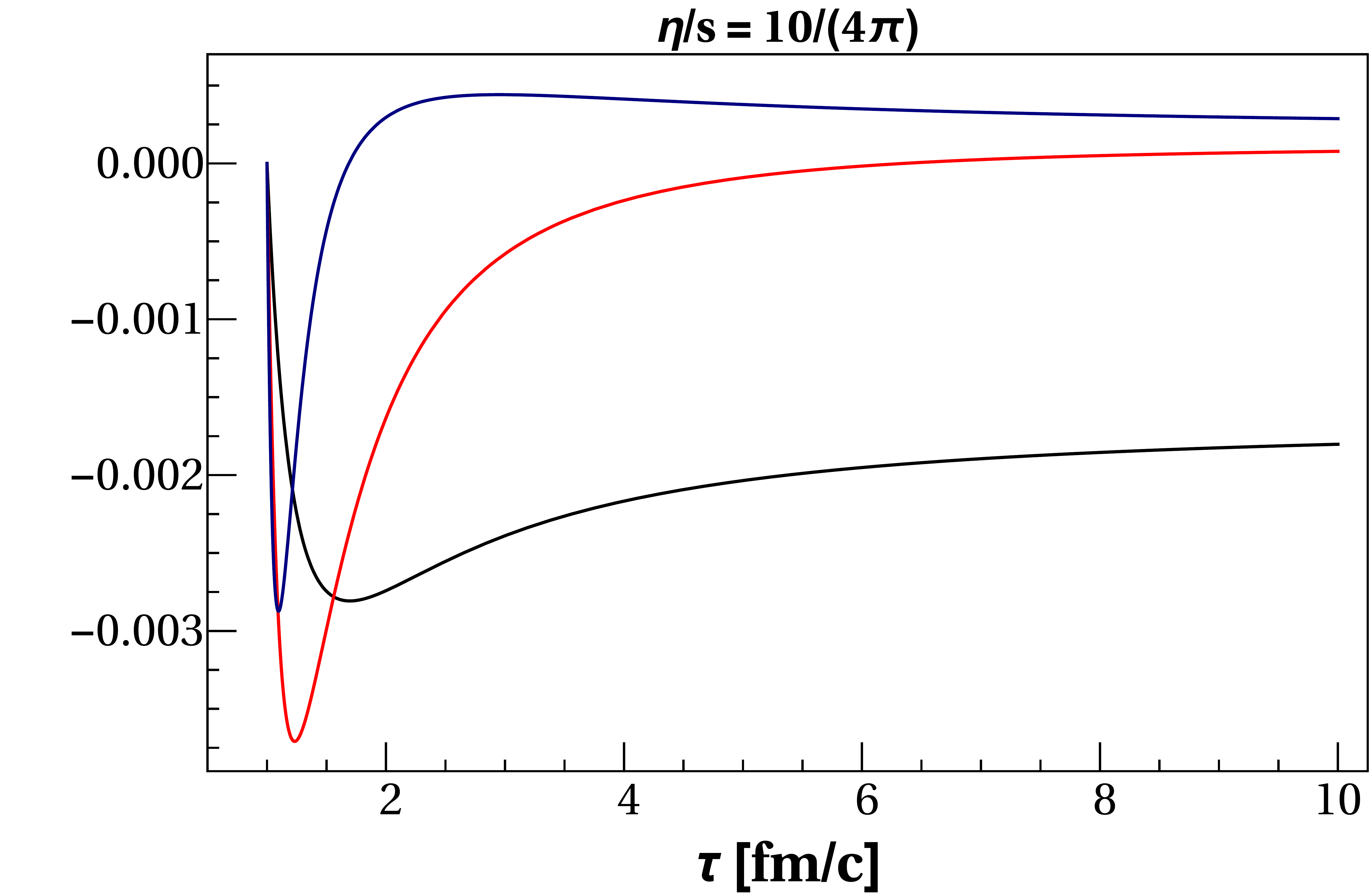}}
\end{picture} 
\end{tabular}
\end{centering}
\caption{(Color online) Same as Fig.\ \ref{F8} but for different initial values of the fluid perturbations:
$\delta n(\tau_0) \ne 0$, $\delta\epsilon(\tau_0)=\delta u^\eta(\tau_0)=0$. See text for further details.} 
\label{F9}
\end{figure}
%----------------------------------------------------------------------------------------------------
%--------------------------Fig. 10------------------------------------
\begin{figure}[h]
\begin{centering}
\begin{tabular}{c c}
\begin{picture}(190,150)
\put(0,8){\includegraphics[scale=0.195]{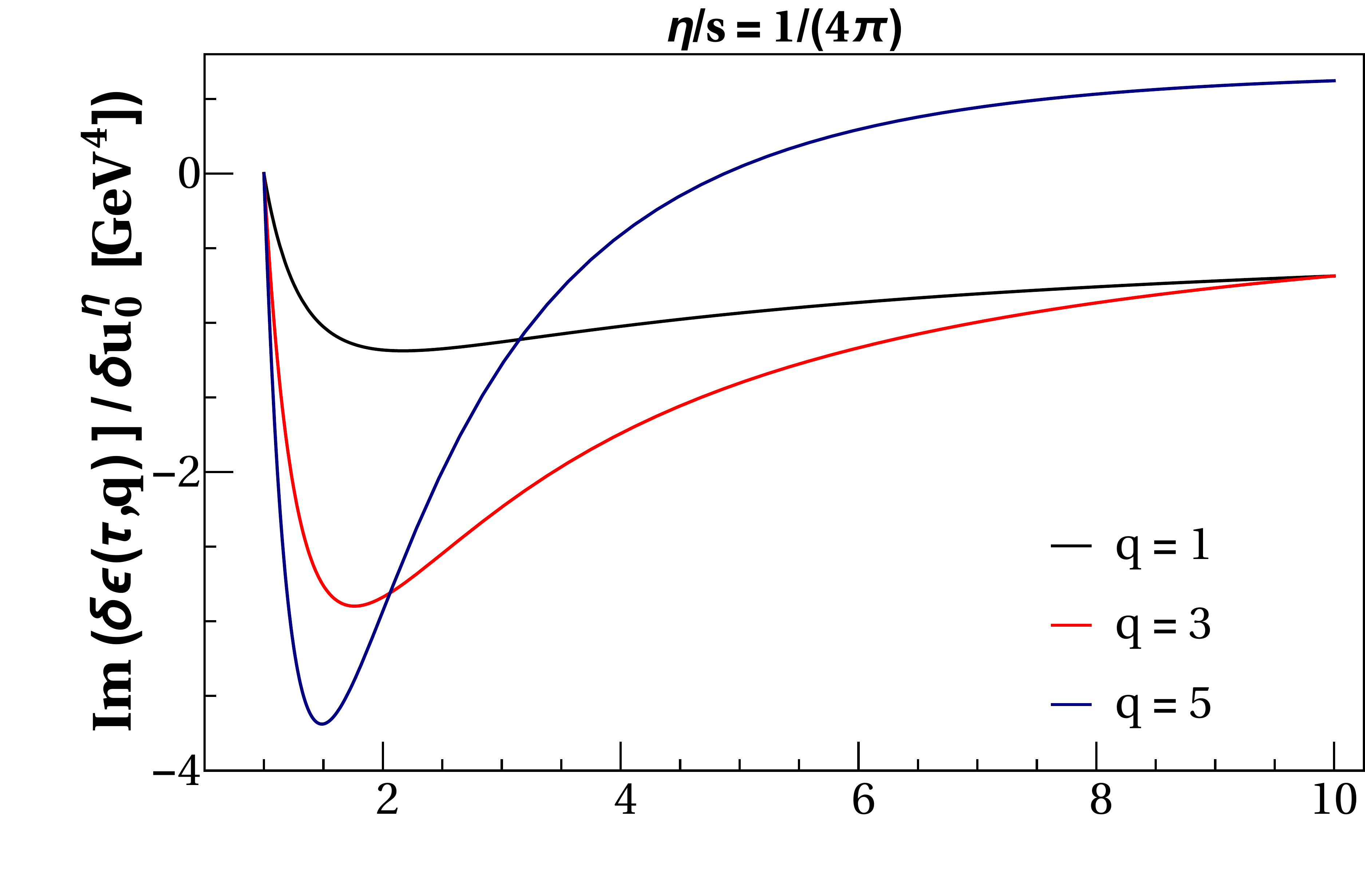}} 
\end{picture}
&
\begin{picture}(450,180)
\put(0,8){\includegraphics[scale=0.137]{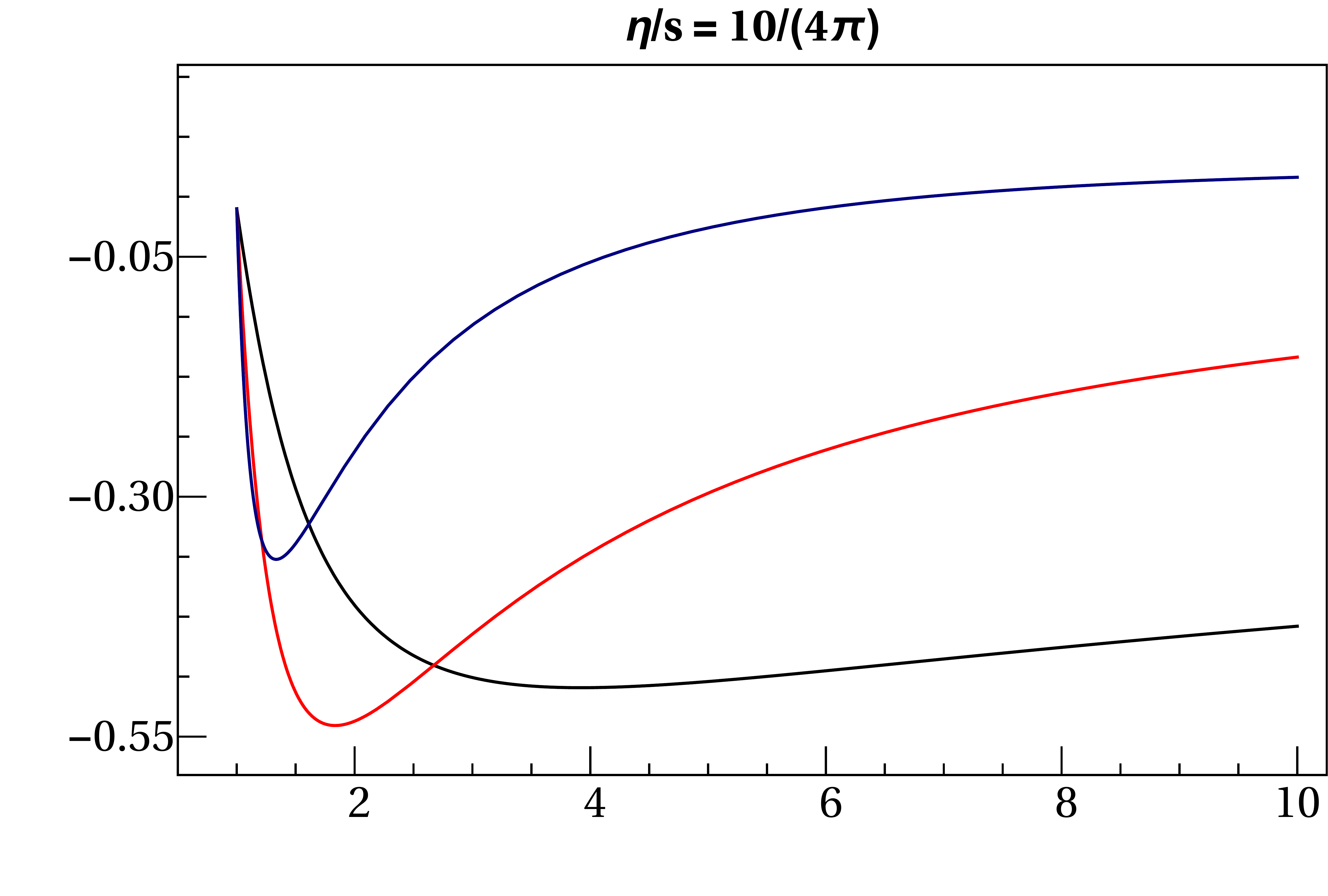}}
\end{picture}
 \\
\begin{picture}(223,140) 
\put(0,8){\includegraphics[scale=0.15]{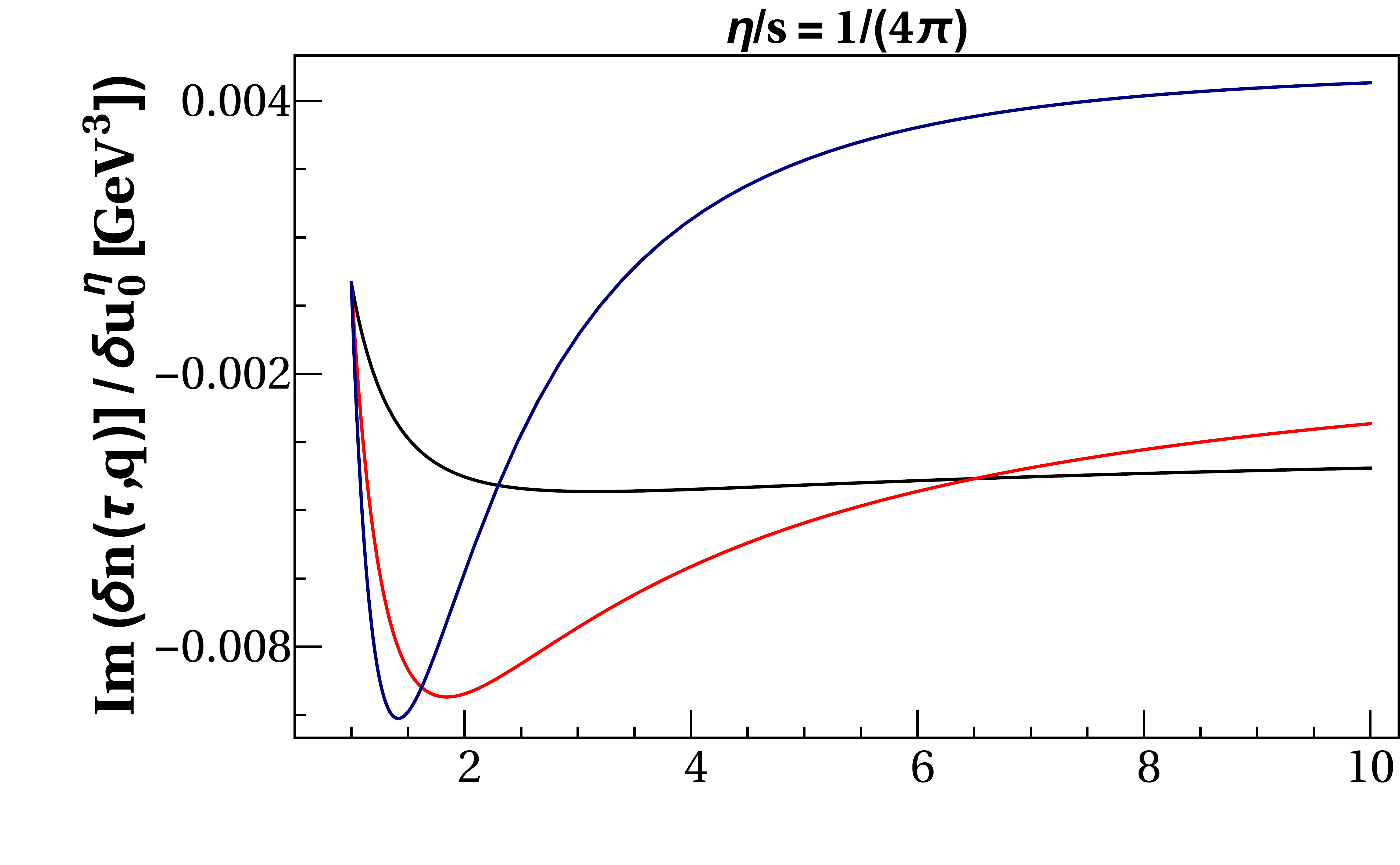} }
\end{picture}
&
\begin{picture}(467,130) 
\put(0,8){\includegraphics[scale=0.144]{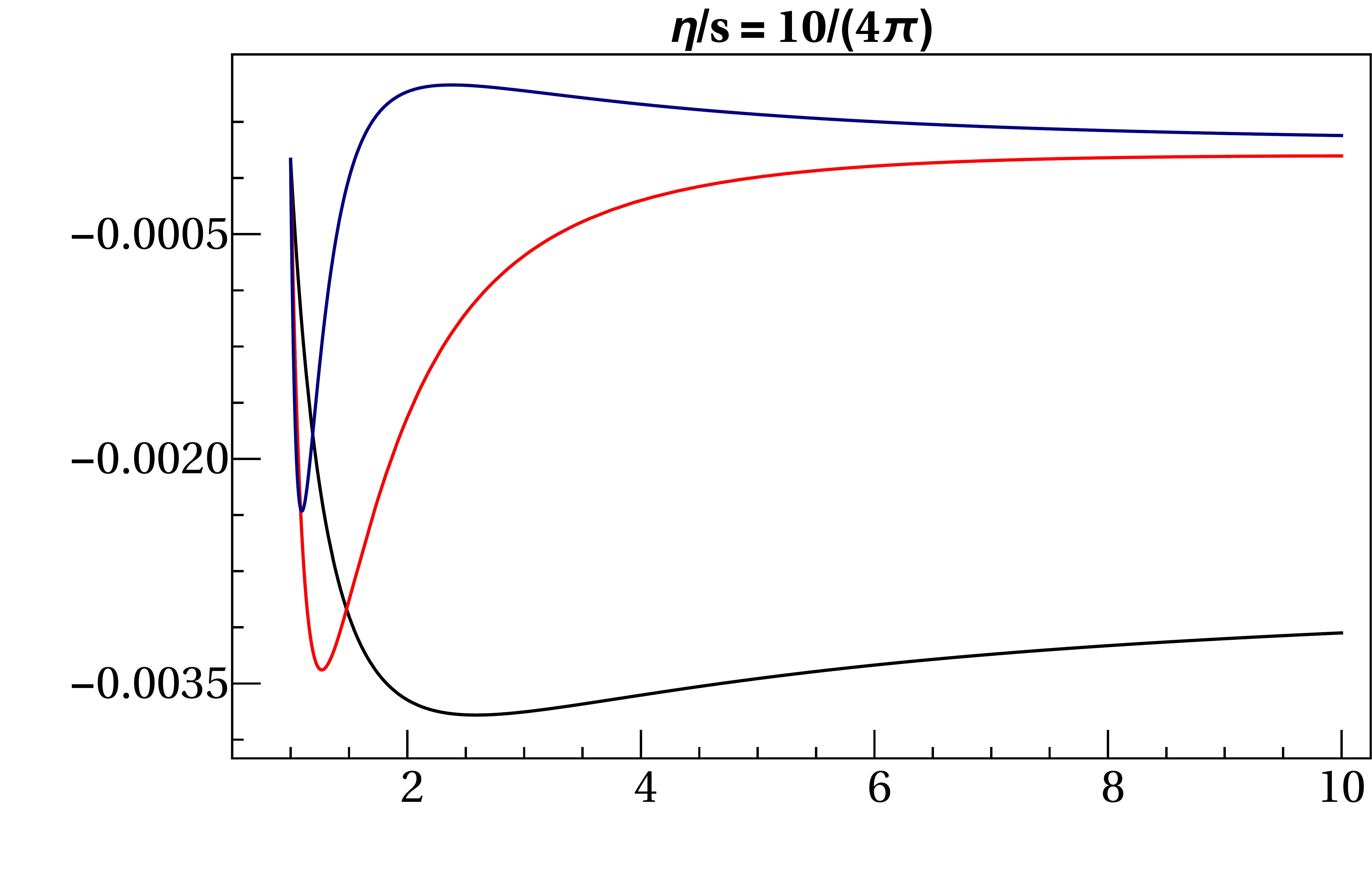}}
\end{picture}
\\
\begin{picture}(207,130) 
\put(0,8){\includegraphics[scale=0.145]{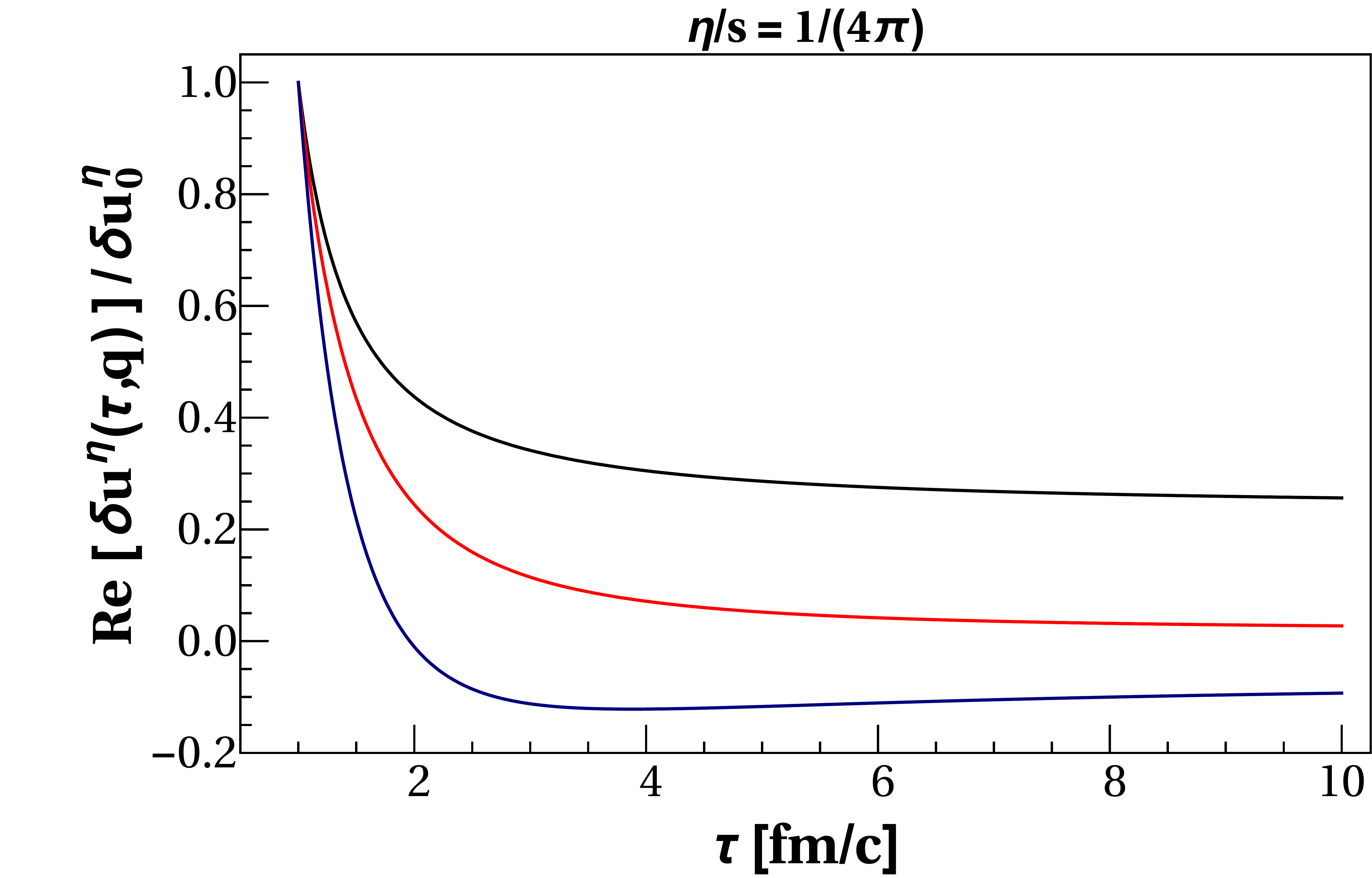}}
\end{picture}
&
\begin{picture}(447,130) 
\put(0,8){\includegraphics[scale=0.137]{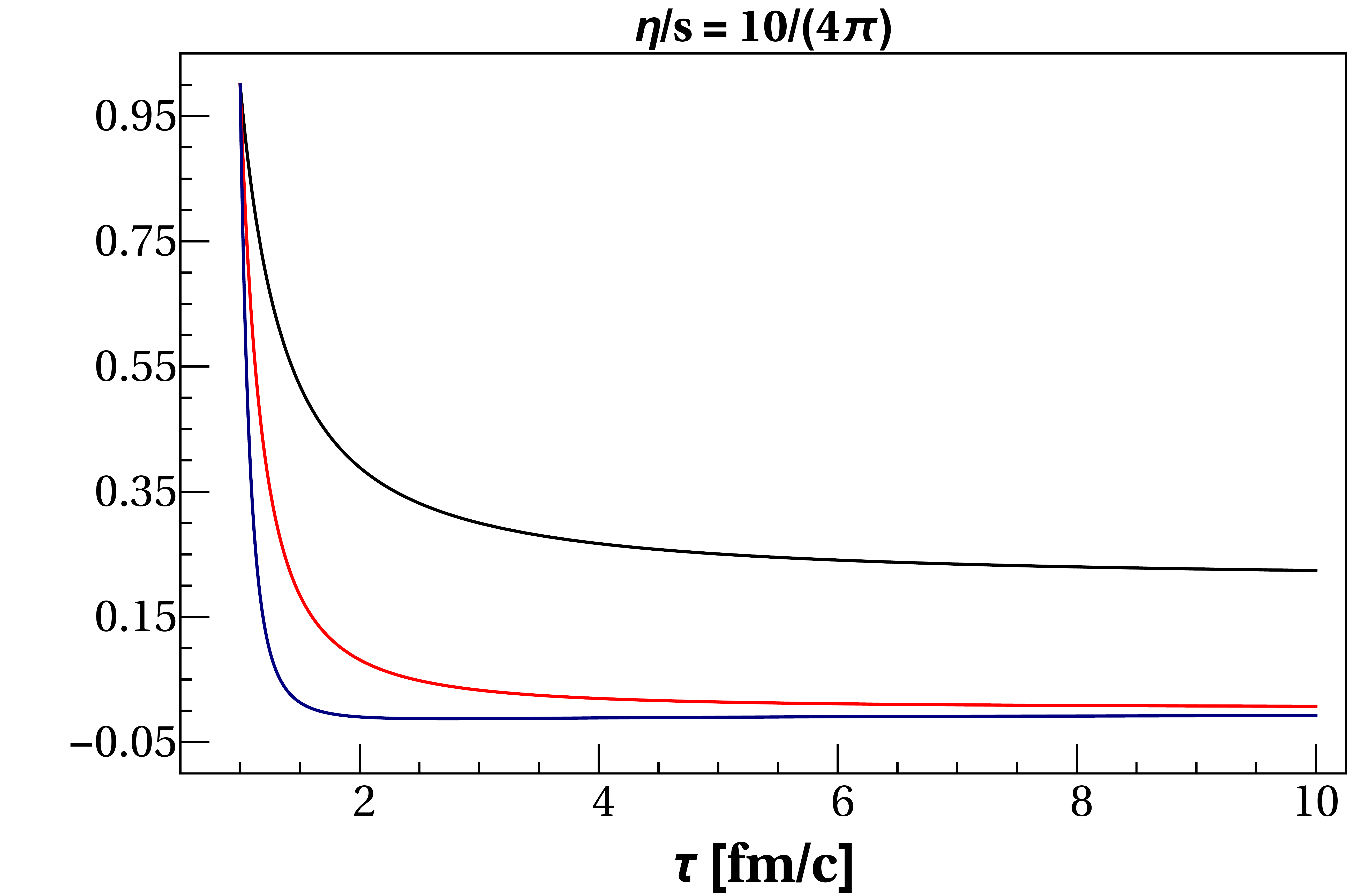}}
\end{picture} 
\end{tabular}
\end{centering}
\caption{(Color online) Same as Fig.\ \ref{F8} but for different initial values of the fluid perturbations:
$\delta u^\eta(\tau_0) \ne 0$, $\delta\epsilon(\tau_0)=\delta n(\tau_0)=0$. See text for further details.} 
\label{F10}
\end{figure}
%----------------------------------------------------------------------------------------------------
%--------------------------Fig. 11------------------------------------
\begin{figure}[h]
\begin{centering}
\begin{tabular}{c c }
\begin{picture}(215,150)
\put(0,8){\includegraphics[scale=0.128]{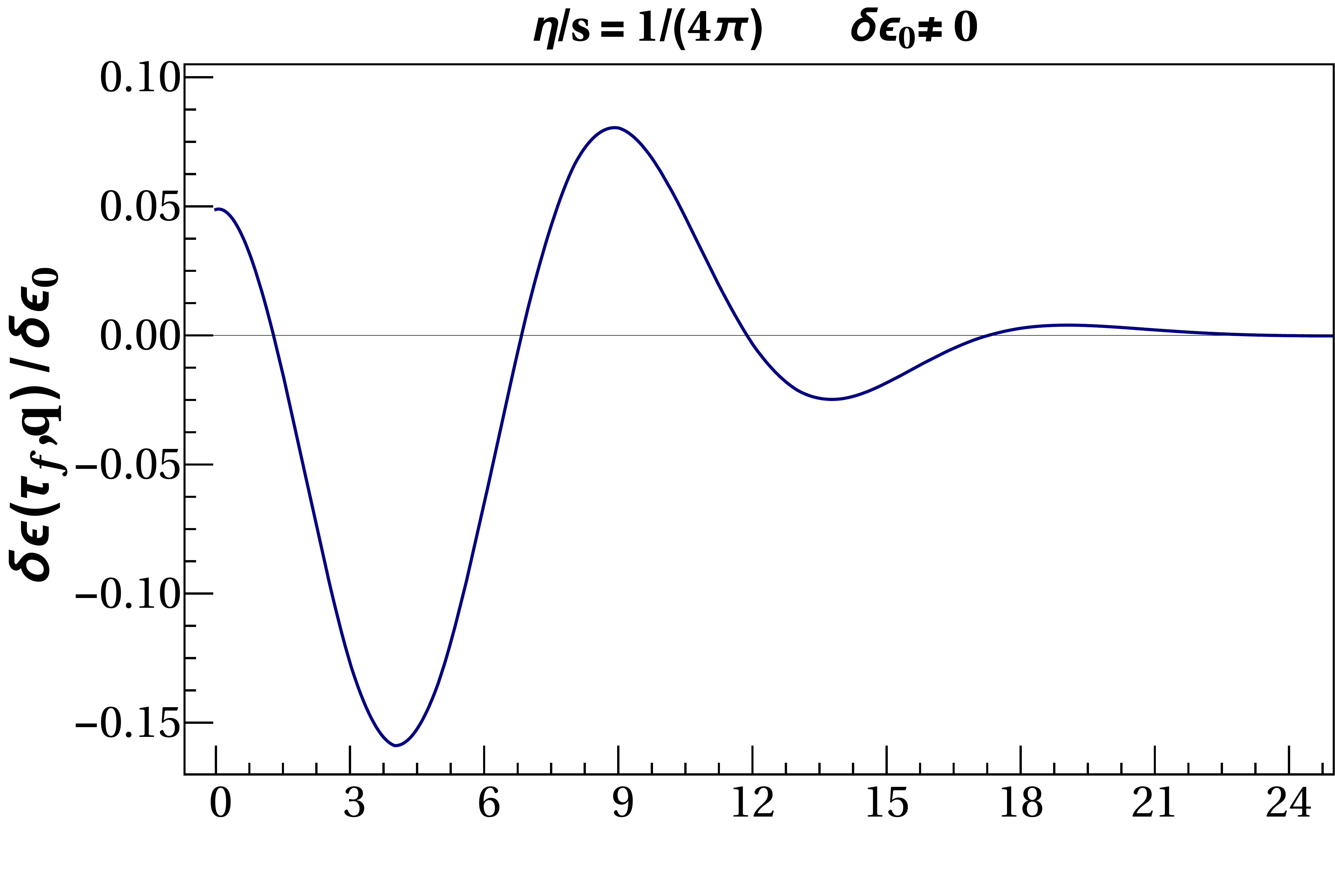}} 
\end{picture}
&
\begin{picture}(410,150)
\put(0,8){\includegraphics[scale=0.145]{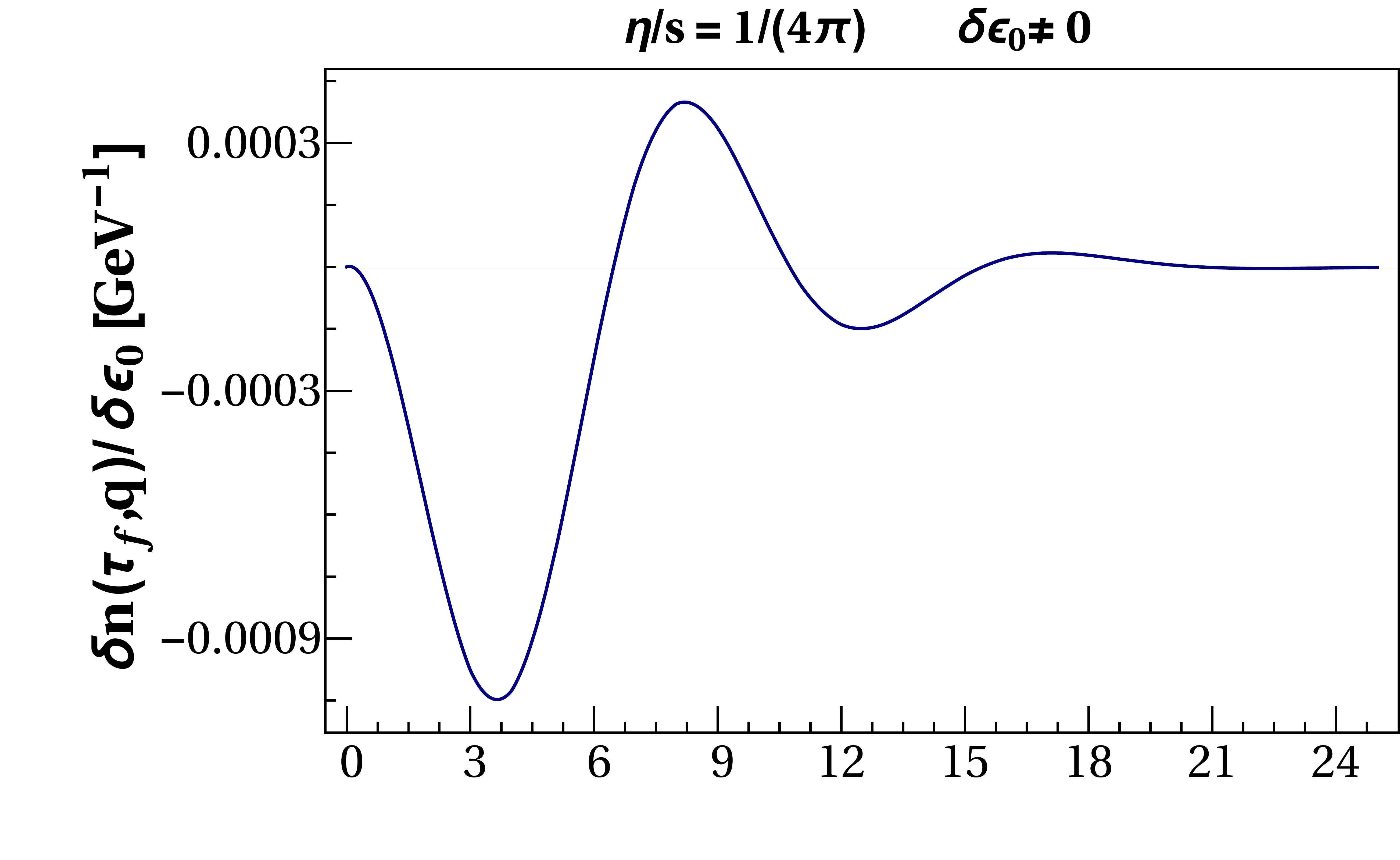}}
\end{picture}
 \\
\begin{picture}(245,130) 
\put(0,8){\includegraphics[scale=0.138]{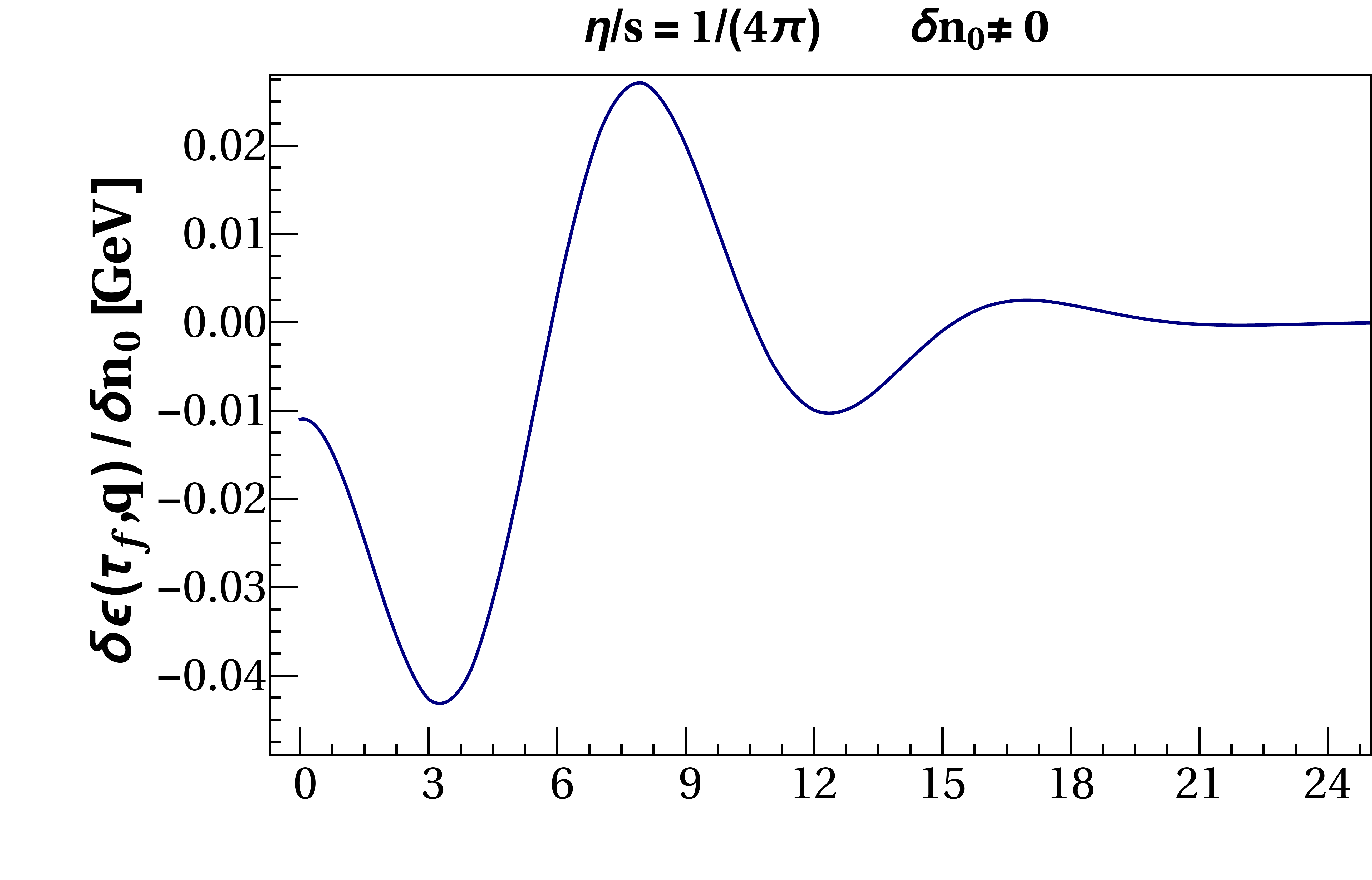} }
\end{picture}
&
\begin{picture}(355,130) 
\put(0,8){\includegraphics[scale=0.127]{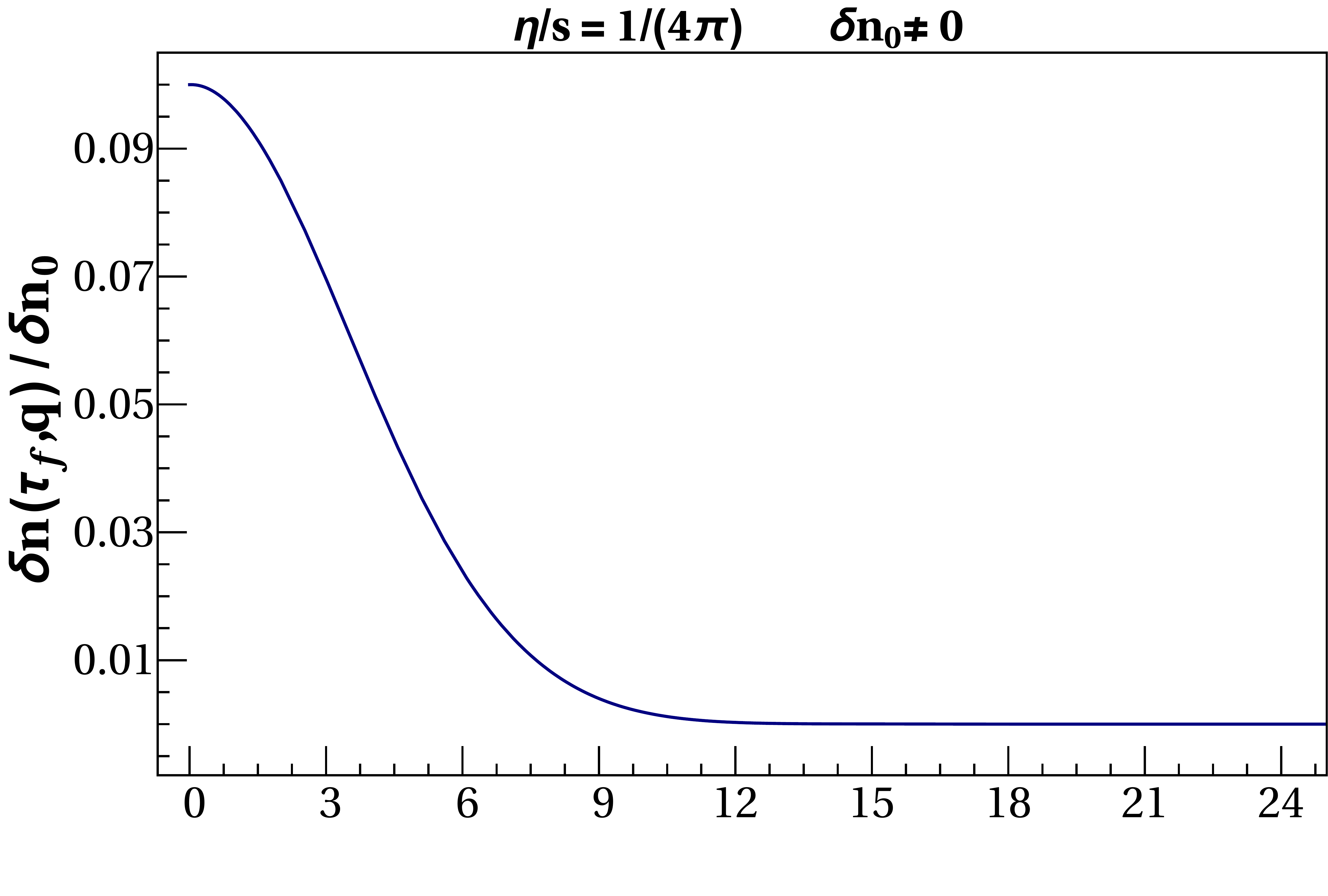}}
\end{picture}
\\
\begin{picture}(240,130) 
\put(0,8){\includegraphics[scale=0.135]{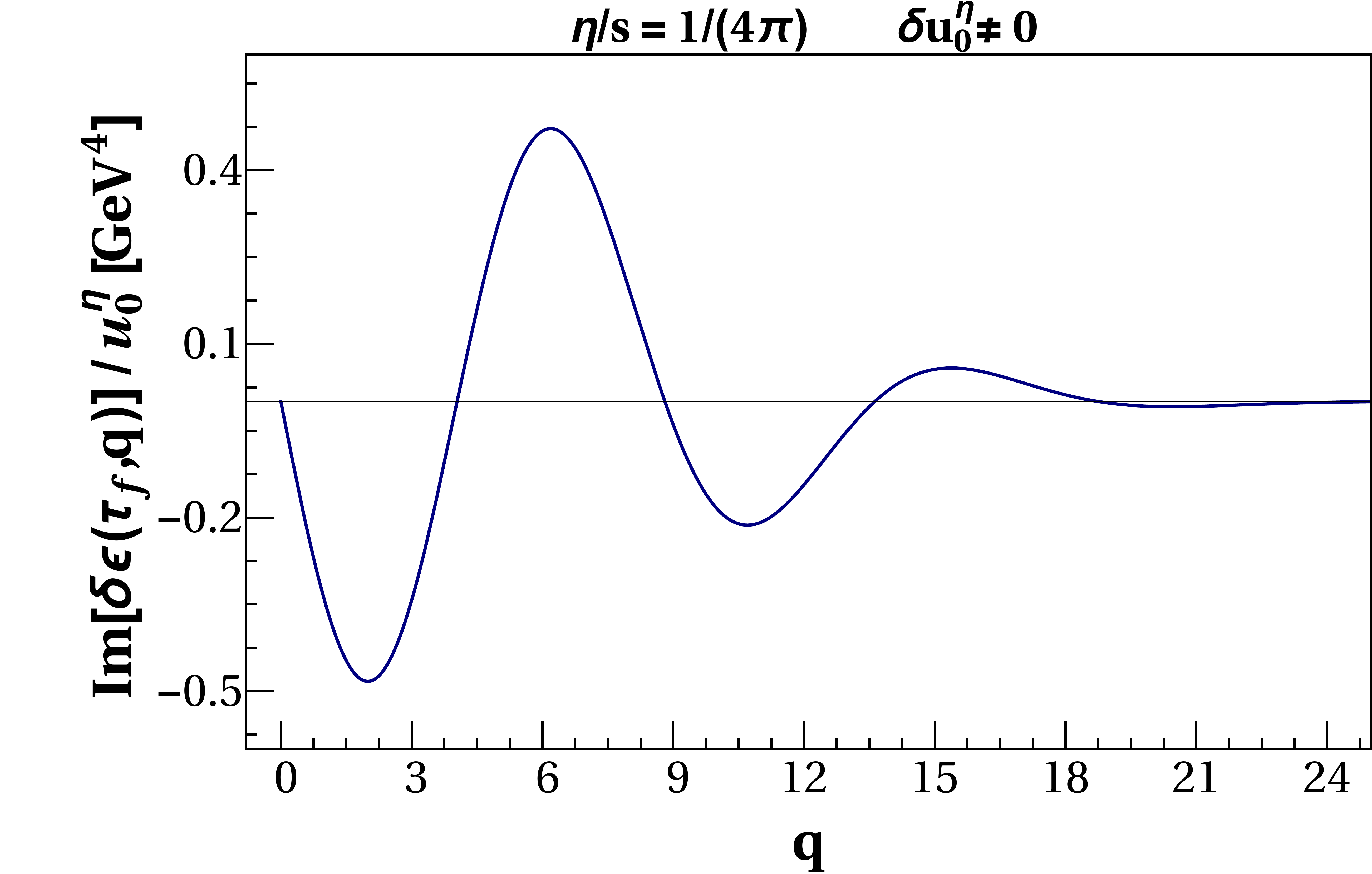}}
\end{picture}
&
\begin{picture}(410,130) 
\put(0,8){\includegraphics[scale=0.145]{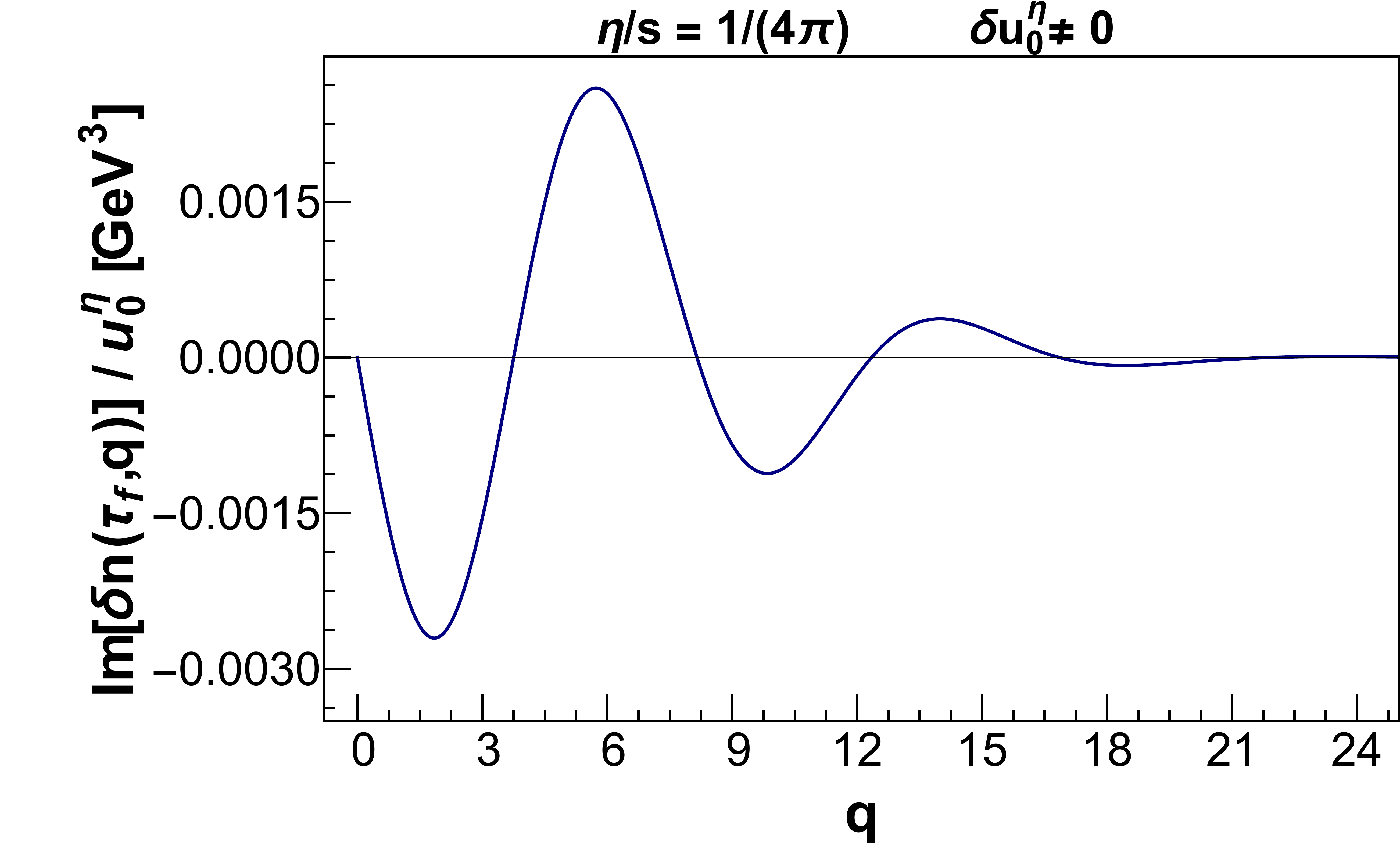}}
\end{picture} 
\end{tabular}
\end{centering}
\caption{(Color Online) Amplitude of the perturbations at $\tau_f=10$ fm/c in units of the initial weight at $\tau=1$ fm/c as a function of the $q$-wave number. We choose $\eta/s=1/(4\pi)$, $T_0=$ 0.5 GeV and $\mu_0=$ 0.05 GeV. In each panel we choose a non-vanishing value initially for one of the fluctuating fields while the remaining ones are set to zero. The top, middle and bottom panels correspond to different initial conditions $\delta\epsilon_0 \neq 0$, $\delta n_0\neq 0$ and $u^\eta_0\neq 0$, respectively.} 
\label{F11}
\end{figure}
%----------------------------------------------------------------------------------------------------

One can also consider a situation with exact symmetry under translations and rotations in the transverse plane. In that case only perturbations with $k=0$ are possible and the fluid velocities in transverse directions have to vanish, $\delta u^+ = \delta u^- = 0$. Again Eqs.~\eqref{eq:LinEvEqEnergyBesselFourier}-\eqref{eq:LinEvEqRapidityVelocityBesselFourier} simplify substantially, albeit not to a point where they can be integrated directly. Specifically, Eq.\ \eqref{eq:LinEvEqEnergyBesselFourier} becomes
\begin{equation}
\begin{split}
\partial_\tau \delta \epsilon & + \left[ \frac{1}{\tau}+\frac{1}{\tau} \left( \frac{\partial p}{\partial \epsilon} \right)_n - \frac{1}{\tau^2} \left( \frac{\partial \zeta}{\partial \epsilon} \right)_n - \frac{4}{3\tau^2} \left( \frac{\partial \eta}{\partial \epsilon} \right)_n \right] \delta\epsilon\\
&+ \left[  \frac{1}{\tau} \left( \frac{\partial p}{\partial n} \right)_\epsilon - \frac{1}{\tau^2} \left( \frac{\partial \zeta}{\partial n} \right)_\epsilon - \frac{4}{3\tau^2} \left( \frac{\partial \eta}{\partial n} \right)_\epsilon \right] \delta n 
+ \left[ \bar\epsilon + \bar p - \frac{2}{\tau}\bar \zeta - \frac{8}{3\tau} \bar \eta \right]   i q \, \delta u^\eta  = 0 ,
\end{split}
\label{eq:LinEvEqEnergyBesselFourierTransverseTranslationSymmetry}
\end{equation}
and the evolution equation for the perturbation in baryon number density \eqref{eq:LinEvEqDensityBesselFourier} becomes
\begin{equation}
\begin{split}
\partial_\tau \delta n  & + \frac{1}{\tau} \delta n + \left[ \bar n - \bar\kappa \left[\frac{\bar n\bar T}{\bar \epsilon + \bar p}\right]^2 \partial_\tau\left(\frac{\bar\mu}{\bar T}\right) \right] i q \, \delta u^\eta  \\
& + \bar\kappa \left[\frac{\bar n\bar T}{\bar \epsilon + \bar p}\right]^2 \left( \frac{\partial(\mu/T)}{\partial\epsilon} \right)_n \frac{q^2}{\tau^2} \delta\epsilon  + \bar\kappa \left[\frac{\bar n\bar T}{\bar \epsilon + \bar p}\right]^2 \left( \frac{\partial(\mu/T)}{\partial n} \right)_\epsilon   \frac{q^2}{\tau^2} \delta n = 0.
\end{split}
\label{eq:LinEvEqDensityBesselFourierTransverseTranslationSymmetry}
\end{equation}
Finally, the evolution equation for the rapidity component of the fluid velocity \eqref{eq:LinEvEqRapidityVelocityBesselFourier} becomes
\begin{equation}
\begin{split}
& \left( \bar \epsilon + \bar p - \frac{1}{\tau}\bar \zeta - \frac{4}{3\tau}\bar \eta \right) \partial_\tau \delta u^\eta 
+ \left[ \partial_\tau \bar p + \frac{2}{\tau}(\bar \epsilon + \bar p)  - \frac{1}{\tau}\partial_\tau\bar \zeta +\frac{1}{\tau^2} \bar \zeta - \frac{4}{3\tau} \partial_\tau \bar \eta - \frac{4}{3\tau^2} \bar \eta \right] \delta u^\eta \\
& + \left[ \left( \frac{\partial p}{\partial \epsilon} \right)_n - \frac{1}{\tau}\left( \frac{\partial \zeta}{\partial \epsilon} \right)_n + \frac{2}{3\tau} \left( \frac{\partial \eta}{\partial \epsilon} \right)_n \right] \frac{i q}{\tau^2} \delta \epsilon 
+ \left[ \left( \frac{\partial p}{\partial n} \right)_\epsilon - \frac{1}{\tau}\left( \frac{\partial \zeta}{\partial n} \right)_\epsilon + \frac{2}{3\tau} \left( \frac{\partial \eta}{\partial n} \right)_\epsilon \right] \frac{i q}{\tau^2} \delta n \\
& + \left( \bar \zeta + \frac{4}{3}\bar \eta \right) \frac{q^2}{\tau^2}  \delta u^\eta = 0.
\end{split}
\label{eq:LinEvEqRapidityVelocityBesselFourierTransverseTranslationSymmetry}
\end{equation}

These equations simplify further in a situation with vanishing baryon number density, the numerical solution for this situation has already been discussed in Ref.\ \cite{Florchinger:2011qf}. The solution of the fluctuating fields is in general complex but subject to the reality constraints $\delta\epsilon^*(\tau,q) = \delta\epsilon(\tau,-q)$ and similar for $\delta n$ and $\delta u^\eta$. 

In order to gain some qualitative insights let us consider a simple equation of state $\epsilon = 3 p$ while setting $\bar n = \bar \mu = 0$ and neglecting the effects of viscosities where they are sub-leading compared to other background terms. One can then derive for the variable $\delta = \delta \epsilon / \bar \epsilon$ the equation
\begin{equation}
\partial_\tau^2 \delta + \left[ \frac{5}{3\tau} + \left(\frac{\bar \zeta + 4 \bar \eta/3}{\bar \epsilon + \bar p} \right)\frac{q^2}{\tau^2} \right] \partial_\tau \delta + \frac{q^2}{3\tau^2} \delta = 0 .
\label{eq:longitudinalSoundWave}
\end{equation}
This equation describes sound propagation in the longitudinal direction on top of the expanding Bjorken background solution. Both the expansion and the viscosities have a damping effect as can be read of from the term $\sim \partial_\tau\delta$. The last term in Eq.\ \eqref{eq:longitudinalSoundWave} is due to pressure gradients and the actual driving term of sound propagation. It is somewhat different than in other situations because of the time dependence $\sim 1/\tau^2$.

More general, the set of equations \eqref{eq:LinEvEqEnergyBesselFourierTransverseTranslationSymmetry}, \eqref{eq:LinEvEqDensityBesselFourierTransverseTranslationSymmetry} and \eqref{eq:LinEvEqRapidityVelocityBesselFourierTransverseTranslationSymmetry} describe also baryon number density waves and diffusion in the longitudinal direction. We show numerical solutions to the evolution equations \eqref{eq:LinEvEqEnergyBesselFourierTransverseTranslationSymmetry} - \eqref{eq:LinEvEqRapidityVelocityBesselFourierTransverseTranslationSymmetry} in Figs.\ \ref{F8},~\ref{F9} and~\ref{F10} for different initial conditions. As we proceed in Sect.~\ref{subsec:exactbjo} we compare two different values of the ration of shear viscosity to entropy density $\eta/s = 1/ (4\pi)$ (left panel) and $\eta/ s = 10 / (4\pi)$ (right panel). For the background fields we use again the scaling solution~\eqref{eq:scalingTmu}.

For Fig. \ref{F8} we choose only $\delta \epsilon$ to be non-zero initially. Compared with the behavior of the transverse sound waves or sound waves in a static medium discussed in the previous section, the evolution of the resulting longitudinal sound waves is completely different. In particular, no proper oscillations are visible during the entire temporal evolution. Rather, one observes a decay in amplitude, in particular at early times. This effect of the longitudinal expansion is particularly strong for large values of $q$. At later times the damping actually weakens to the extent that amplitudes remain non-zero at the final time. Interestingly, the influence of viscosity on the time evolution of longitudinal perturbations is relatively weak. Some quantitative differences are of course visible between the left and right panels of Fig.\ \ref{F8} but qualitatively, the evolution is surprisingly similar. 

Figure\ \ref{F9} was obtained by selecting only $\delta n\neq 0$ at $\tau_0$. Again, we do not observe any proper oscillations of the fluctuating fields along the longitudinal direction. In this case the viscosity and heat conductivity have a somewhat larger effect. The amplitude of the fluctuating fields gets damped as one increases the value for the shear viscosity and heat conductivity (according to Eq.\ \eqref{eq:thercond}).

In Figure~\ \ref{F10} we choose $\delta u^\eta_0\neq 0$. As in the previous two situations, there is no proper oscillation visible for the time interval shown. 

Finally, Fig. \ref{F11} shows the final amplitude of perturbations at $\tau_f=10$ fm/c in units of the initial weight at time $\tau_0$ as a function of the longitudinal $q$-wavenumber. Figure \ref{F11} is obtained by choosing a non-vanishing value initially for one particular fluctuating field while the remaining fluctuating fields are initially set to zero. The top, middle and bottom panels of Fig. \ref{F11} corresponds to $\delta\epsilon_0\neq 0$, $\delta n_0\neq 0$ and $\delta u^-_0\neq 0$, respectively. We observe that the amplitude of the fluctuating modes goes asymptotically to zero for $q\geq 25$ which corresponds to a small window in the rapidity variable (i.e., $\Delta\eta\sim (\Delta q)^{-1}$). We expect that modes with intermediate and large $q$ would be damped stronger for earlier initialization time $\tau_0$.

Finally, in a situation where Bjorken boost invariance as well as translations and rotations in the transverse plane are realized exactly, i.\ e.,\ $\delta u^\eta = \delta u^+ = \delta u^- = k = q = 0$, Eqs.\ \eqref{eq:LinEvEqEnergyBesselFourier} and \eqref{eq:LinEvEqDensityBesselFourier} reduce simply to a linearized version of the Bjorken expansion in Eq.\ \eqref{eq:BjorkenExpEN} as it has to be.

%%%%%%%%%%%%%%%%%%%%%%%%%%%%%%%%%%%%%%%%%%%%%%%%%%%%%%%%%%%%%%%%%%%%%%%%%%%%%%%%
\section{The two point correlation function of baryonic particles}
\label{sec:corrFunct}
In this section we discuss the possibility to access the information about perturbations in the baryon number density experimentally by measuring a correlation function of the net number of baryons (baryons minus anti-baryons) as a function of the rapidity and azimuthal angle. We concentrate for simplicity on the case of vanishing background baryon number density as discussed in Sect.~\ref{subsec:statbar}.

Perturbations in baryon number density in position space as described by Eq.\ \eqref{eq:baryonfluctlongitudinal} are not directly accessible to experiments. However, a fluctuating baryon number density
and chemical potential on the kinetic freeze-out surface has an influence on the distribution
of particles with non-zero baryon number in momentum space. This concerns in particular
protons but also resonances with non-vanishing baryon number. Similar as for flow observables, there is a direct link between different harmonics in azimuthal angle and rapidity in the fluid dynamic description and the corresponding harmonics in the momentum space particle distribution. Thus, we can partly access the physical information contained in Eq.\ \eqref{eq:baryonfluctlongitudinal}. As an example, we consider a connected two-point correlation function of the type\footnote{The brackets $\langle\cdots\rangle$ in Eq.~\eqref{eq:twoparticleBaryoncorrelation} denote an event average
\begin{equation*}
\langle \mathcal{O}(x,y)\rangle=\lim_{N_\text{events}\to\infty}\,\frac{1}{N_\text{events}}\,\sum_{i=1}^{N_\text{events}}\,\mathcal{O}_i(x,y)\,.
\end{equation*}} 
\begin{equation}
C_\text{Baryon}(\phi_1-\phi_2, \eta_1-\eta_2) = \langle  n_\text{Baryons}(\phi_1,\eta_1) n_\text{Baryons}(\phi_2,\eta_2) \rangle_c ,
\label{eq:twoparticleBaryoncorrelation}
\end{equation}
which measures correlations of baryonic particles (i.e. the number of baryons minus anti-baryons) as a function of the difference between (particle momentum) azimuthal angles $\phi_1 - \phi_2$ and (particle momentum) rapidities $\eta_1 - \eta_2$. In Eq.\ \eqref{eq:twoparticleBaryoncorrelation}, $n_\text{Baryons}(\phi,\eta)$ is the number of baryons minus anti-baryons as found in the detector in a particular bin in azimuthal angle $\phi$ and rapidity $\eta$~\footnote{There is a complication due to the fact that neutrons cannot be measured experimentally. Further
studies are needed in order to quantify whether this presents a problem for observables as in ~\eqref{eq:twoparticleBaryoncorrelation} and if so,
how these can be overcome. Also, one should estimate possible contributions to Eq.~\eqref{eq:twoparticleBaryoncorrelation} from sources other
than fluid dynamics, such as resonance decays.}. We also introduce the Fourier representation
\begin{equation}
C_\text{Baryon}(\phi_1-\phi_2, \eta_1-\eta_2) = \sum_{m=-\infty}^\infty \int \frac{dq}{2\pi} \; \tilde C_\text{Baryon}(m,q) \, e^{im(\phi_1-\phi_2) + i q (\eta_1 - \eta_2)}.
\label{eq:twoparticleBaryoncorrelationFourier}
\end{equation}
The correlation function in Eqs.\ \eqref{eq:twoparticleBaryoncorrelation} and \eqref{eq:twoparticleBaryoncorrelationFourier} is determined by a combination of initial conditions (set at the time where a fluid dynamic description becomes valid) and response functions that describe how baryon number density perturbations propagate in the fluid dynamic regime and how they influence the particle distributions at freeze-out. 

In the following we discuss both parts in a bit more detail. First, the initial state after a heavy ion collision (and after the early non-equilibrium dynamics) at the time $\tau_0$ when a fluid dynamic description becomes valid, is characterized by a fluctuating baryon number density $\delta n(\tau_0, r, \phi, \eta)$ around some average or expectation value $\bar n(\tau_0,r)$. (The latter might be rather small at LHC and upper RHIC energies and we neglect it in the following.) For the fluctuating part we use a Bessel-Fourier decomposition
\begin{equation}
\delta n(\tau_0,r,\phi,\eta) = \sum_{m=-\infty}^\infty \sum_{l=1}^\infty \int \frac{d q}{2\pi} \; \delta n^{(m)}_{l}(q) \, e^{im\phi + i q \eta} J_{m}\left(z^{(m)}_l \rho(r)\right)
\label{eq:deltaNBesselFourierInitial}
\end{equation}
An event-by-event ensemble of initial conditions conditions for the baryon number density can be characterized in terms of the weights $\delta n^{(m)}_l(q)$. For example, the two-mode correlation function is
\begin{equation}
\langle \delta n^{(m_1)}_{l_1}(q_1) \; \delta n^{(m_2)}_{l_2}(q_2) \rangle =  2\pi \delta(q_1+q_2) \delta_{m_1+m_2,0} \; C^{(m)}_{\delta n \delta n; l_1,l_2} (q).
\label{eq:initialBaryonperturbationCorrelationBesselFourier}
\end{equation}
We have assumed here that the ensemble of initial conditions is statistically symmetric under azimuthal rotations and longitudinal boosts leading to the factors $\delta_{m_1+m_2,0} $ and $2\pi \delta(q_1+q_2) $ on the right hand side of Eq.\ \eqref{eq:initialBaryonperturbationCorrelationBesselFourier}.

For a single event with baryon number perturbation as in \eqref{eq:deltaNBesselFourierInitial}, the baryon number distribution in momentum space after kinetic freeze-out will be proportional to the weights $\delta n^{(m)}_l(q)$ within the linear response approximation. More specific, the Bjorken-boost and azimuthal rotation symmetries imply that one can write
\begin{equation}
n_\text{Baryons}^{(m)}(q) = \sum_l S_{\text{Baryons}; (m) l}(q) \delta n^{(m)}_l(q) ,
\label{eq:BaryonNumberLinearResponse}
\end{equation}
with linear baryon number response function $S_{\text{Baryons}; (m) l}(q)$. The object on the left hand side of Eq.\ \eqref{eq:BaryonNumberLinearResponse} is the Bessel-Fourier weight of the (momentum space) distribution of the number of baryons minus anti-baryons. The correlation function on the right hand side of \eqref{eq:twoparticleBaryoncorrelationFourier} can be written as 
\begin{equation}
\tilde C_\text{Baryon}(m,q) = \sum_{l_1,l_2=1}^\infty S_{\text{Baryon}; (m)l_1}(q) \, S_{\text{Baryon}; (-m)l_2}(-q) \, C^{(m)}_{\delta n \delta n; l_1,l_2} (q).
\end{equation}
For a more detailed discussion of the response function formalism briefly introduced here we refer to Ref.\ \cite{Floerchinger:2014fta}. 

The linear response functions $S_{\text{Baryon}; (m)l}(q)$ are in particular also affected by heat conductivity. More specific, the analog of the factor $\exp(-k^2 I_1 - q^2 I_2)$ in a situation with realistic transverse dependence and radial flow leads to a suppression of modes with $q^2 > 0$ and large values of $m$ and/or the radial wave number $l$. Qualitatively, one expects that the scale for the suppression in the transverse direction is set by the (time dependent) radius $R$ of the fireball. Moreover, the $l$'th zero crossings $z^{(m)}_l$ of the Bessel-functions $J_m(z)$ are for fixed $l$ approximately linear in $m$ (for the relevant values of $m$ and $l$, with prefactor of order unity) so that one expects qualitatively
\begin{equation}
\tilde C_\text{Baryon}(m,q)  \approx \exp(- 2 m^2 I_1^\prime - 2 q^2 I_2^\prime) \tilde C_\text{Baryon}^{\bar \kappa = 0}(m,q) ,
\label{eq:BaryonExpSurpression}
\end{equation}
where on the right hand side $\tilde C_\text{Baryon}^{\bar \kappa = 0}(m,q)$ would be the corresponding correlation function in the (somewhat hypothetical) situation of vanishing heat conductivity and the dissipative attenuation terms can be roughly estimated as
\begin{equation}
\begin{split}
I_1^\prime& \approx \int_{\tau_0}^{\tau_f}  d\tau \, \frac{1}{R^2} \, \bar\kappa \left[\frac{\bar n\bar T}{\bar \epsilon + \bar p}\right]^2 \left( \frac{\partial(\mu/T)}{\partial n} \right)_\epsilon, \\
I_2^\prime & \approx \int_{\tau_0}^{\tau_f}  d\tau \, \frac{1}{\tau^{2}} \, \bar\kappa \left[\frac{\bar n\bar T}{\bar \epsilon + \bar p}\right]^2 \left( \frac{\partial(\mu/T)}{\partial n} \right)_\epsilon .
\end{split}
\end{equation}
Going now back to the two-particle correlation function \eqref{eq:twoparticleBaryoncorrelation}, the exponential suppression factor in Eq.\ \eqref{eq:BaryonExpSurpression} implies for large $I_2^\prime$ long range correlations with respect to the rapidity difference $\eta_1 - \eta_2$, with a decay that is determined by the value of $I_2^\prime$ (except if $\tilde C_\text{Baryon}^{\bar \kappa = 0}(m,q)$ has a very strong decay with $q$ already) and a similar, although weaker, effect with respect to the azimuthal wavenumber $m$. In order to make our qualitative statements more precise, it is necessary to generalize the calculations described here to a more realistic background. A more realistic background would have a realistic transverse profile and expansion in addition to the longitudinal (boost-invariant) expansion. Moreover, one also has to perform more detailed studies of the initial conditions and kinetic freeze-out, that both affect $\tilde C_\text{Baryon}^{\bar \kappa = 0}(m,q)$.

%%%%%%%%%%%%%%%%%%%%%%%%%%%%%%%%%%%%%%%%%%%%%%%%%%%%%%%%%%%%%%%%%%%%%%%%%%%%%%%%
\section{Conclusions}
\label{sec:concl}

We have studied solutions of the fluid equations describing relativistic heavy ion collisions in the presence of a globally conserved quantum number (baryon number) using a background-fluctuating splitting. For the background we have assumed Bjorken boost and transverse translation and rotation invariance. This generalizes Bjorken's original solution to non-vanishing baryon number density as well as shear and bulk viscosities. Heat conductivity does not play a role on the background equations  since the diffusion current vanishes exactly due to the symmetries of the Bjorken flow. 

We derived evolution equations for the perturbations around this background solution. While the amplitude of these perturbations was assumed to  be small, such that linearized equations could be used, the formalism allows us to treat perturbations with arbitrary dependence on the transverse coordinates and rapidity. Technically, this is done by employing a Bessel-Fourier expansion. The partial differential equations of relativistic fluid dynamics become ordinary differential equations for the different modes that are characterized by radial, azimuthal and rapidity wave numbers. The evolution of these perturbations is governed by the thermodynamic properties encoded in the equation of state $p(T,\mu)$ as well as the transport properties (i.e., shear viscosity $\eta(T,\mu)$, bulk viscosity $\zeta(T,\mu)$ and heat conductivity $\kappa(T,\mu)$ in the first order formalism we use). 

Generically, one finds that perturbations with large wave numbers are damped more quickly by the dissipative processes, as expected. The dissipation of different modes depends on time in a different way and, in particular, deviations from Bjorken boost symmetry show a fast damping at early times. In principle, it might be possible to use these dependencies to probe transport and thermodynamic properties at different times in the evolution history and therefore for different temperatures of the quark-gluon plasma produced in a heavy ion collision.

In order to make more quantitative statements, one must take a realistic transverse density profile and expansion into account, of course. This has been done for perturbations with exact Bjorken boost symmetry and vanishing baryon number in Refs.\ \cite{Floerchinger:2013rya, Floerchinger:2013vua, Floerchinger:2013hza, Floerchinger:2014fta}. In the present paper we have concentrated mainly on the evolution of perturbations in baryon number density. They have diffusion-type evolution governed by the longitudinal expansion and heat conductivity. (In the Landau frame, heat conductivity can in fact be understood as baryon number diffusion.) There are characteristic differences in the dependencies on longitudinal and transverse wave numbers. More specific, baryon number perturbations are quickly ``flattened out'' in the longitudinal direction at early times.  

In principle, the information on baryon number perturbations is accessible experimentally via two-point (and higher order) correlation functions of particles with non-zero
baryon number, as a function of the difference in azimuthal angles and rapidities. Based on the evolution equations for perturbations, we expect long-range correlations in rapidity (a ``baryon number ridge''). For a more detailed theoretical picture one needs a better description of the local event-by-event fluctuations in baryon number density at the initial time when fluid dynamics becomes valid. Also, one should take a realistic transverse expansion into account and study the implications of baryon number perturbations at the kinetic freeze-out. (Formulas needed for this have already been derived in Ref.\ \cite{Floerchinger:2013hza}.) It would be very interesting to study net-baryon number correlations experimentally, as well as theoretically in more detail, and thereby constrain heat conductivity as another property of the quark-gluon plasma.

%%%%%%%%%%%%%%%%%%%%%%%%%%%%%%%%%%%%%%%%%%%%%%%%%%%%%%%%%%%%%%%%%%%%%%%%%%%%%%%%

%%%%%%%%%%%%%%%%%%%%%%%%%%%%%%%%%%%%%%%%%%%%%%%%%%%%%%%%%%%%%%%%%%%%%%%%%%%%%%%%
\section*{Acknowledgments}
%%%%%%%%%%%%%%%%%%%%%%%%%%%%%%%%%%%%%%%%%%%%%%%%%%%%%%%%%%%%%%%%%%%%%%%%%%%%%%%%

We thank A.~Beraudo, P.~Benincasa, M. Lisa, S. Voloshin, J.M. Torres-Rinc\'on, S. Gavin and U.~A.~Wiedemann for useful discussions. We thank to U. Heinz, J. Noronha, K. Rajagopal and D. Wertepny for their careful reading of our manuscript. MM thanks D.~Bazow for his help with some numerics. MM thanks the Physics Department of the Universidade de Santiago de Compostela and the CERN theory group for their hospitality during the initial stages of this project.  MM was supported by the U.S. Department of Energy, Office of Science, Office of Nuclear Physics under Awards No.
DE-SC0004286 and (within the framework of the JET Collaboration) No. DE-SC0004104.

%%%%%%%%%%%%%%%%%%%%%%%%%%%%%%%%%%%%%%%%%%%%%%%%%%%%%%%%%%%%%%%%%%%%%%%%%%%%%%%%
\begin{appendix}
%%%%%%%%%%%%%%%%%%%%%%%%%%%%%%%%%%%%%%%%%%%%%%%%%%%%%%%%%%%%%%%%%%%%%%%%%%%%%%%%
%%%%%%%%%%%%%%%%%%%%%%%%%%%%%%%%%%%%%%%%%%%%%%%%%%%%%%%%%%%%%%%%%%%%%%%%%%%%%%%%
\section{Thermodynamic relations in the grand canonical ensemble}
\label{app:ThermodynamicRelations}
%%%%%%%%%%%%%%%%%%%%%%%%%%%%%%%%%%%%%%%%%%%%%%%%%%%%%%%%%%%%%%%%%%%%%%%%%%%%%%%%
In this appendix we compile some thermodynamic relations in the grand canonical ensemble that we found useful in the context of relativistic fluid dynamics with a conserved charge. We start from the pressure $p(T,\mu)$, which is related to the thermodynamic potential of the grand canonical ensemble (the Landau potential) by $p=-\Omega/V$. The differential of pressure is
\beq
\label{eq:grandiff}
d p =  s dT + n d\mu \,.
\eeq
All thermodynamic quantities can be obtained from this and the Gibbs-Duhem relation $\epsilon+p = T s + \mu n$, for example
\beq
\label{eq:grandder}
s  =  \left(\frac{\partial p}{\partial T}\right)_\mu, \quad \quad n = \left(\frac{\partial p}{\partial \mu}\right)_T \,.
\eeq
In the following we will sometimes drop the subscripts with the convention that pressure is evaluated as a function of $T$ and $\mu$ unless indicated otherwise. Also we find it useful to express all susceptibilities in terms of the pressure and its derivatives. This avoids ambiguities and realizes Maxwell's relations automatically. For example, the energy density is obtained then as
\begin{equation}
\label{eq:grancan}
\epsilon  =  - p+ T \frac{\partial p}{\partial T}+\mu\frac{\partial p}{\partial \mu}.
\end{equation}
Its differential, as well as the one for density, are
\begin{subequations}
\label{eq:thermo}
\begin{align}
d\epsilon =&\left[ T \frac{\partial^2 p}{\partial T^2} + \mu \frac{\partial^2 p}{\partial T \partial \mu} \right]dT 
+ \left[ T \frac{\partial^2 p}{\partial T \partial \mu} + \mu \frac{\partial^2 p}{\partial \mu^2} \right] d\mu\,,
\\
dn = & \frac{\partial^2 p}{\partial T \partial \mu} dT + \frac{\partial^2 p}{\partial \mu^2} d\mu\,.
\end{align}
\end{subequations}
These linear relations can be inverted to yield $dT$ and $d\mu$ in terms of $d\epsilon$ and $d n$,
\begin{equation}
\begin{split}
dT = & \frac{\frac{\partial^2p}{\partial\mu^2}}{T \frac{\partial^2 p}{\partial T^2} \frac{\partial^2 p}{\partial \mu^2}- T \frac{\partial^2 p}{\partial T \partial \mu} \frac{\partial^2 p}{\partial T \partial \mu}} d\epsilon  -  
\frac{T\frac{\partial^2p}{\partial T \partial\mu} + \mu \frac{\partial^2 p}{\partial\mu^2}}{T \frac{\partial^2 p}{\partial T^2} \frac{\partial^2 p}{\partial \mu^2} - T \frac{\partial^2 p}{\partial T \partial \mu} \frac{\partial^2 p}{\partial T \partial \mu}}  d n, \\
d\mu = & - \frac{\frac{\partial^2p}{\partial T \partial\mu}}{T \frac{\partial^2 p}{\partial T^2} \frac{\partial^2 p}{\partial \mu^2} - T \frac{\partial^2 p}{\partial T \partial \mu} \frac{\partial^2 p}{\partial T \partial \mu}} d\epsilon  +  
\frac{T\frac{\partial^2p}{\partial T^2} + \mu \frac{\partial^2 p}{\partial T \partial \mu}}{T \frac{\partial^2 p}{\partial T^2} \frac{\partial^2 p}{\partial \mu^2} - T \frac{\partial^2 p}{\partial T \partial \mu} \frac{\partial^2 p}{\partial T \partial \mu}}  d n.
\end{split}
\end{equation}
Other useful quantities are the heat capacity densities
\begin{equation}
\begin{split}
c_V & = \frac{T}{V} \left(\frac{\partial S}{\partial T} \right)_{V,N} = T \left( \frac{\partial s}{\partial T} \right)_n = \frac{T\left( \frac{\partial^2p}{\partial T^2} \frac{\partial^2p}{\partial \mu^2} -  \frac{\partial^2p}{\partial T \partial \mu} \frac{\partial^2p}{\partial T \partial \mu}\right)}{ \frac{\partial^2 p}{\partial \mu^2}},\\
c_P & = \frac{T}{V}\left( \frac{\partial S}{\partial T} \right)_{P,N} = \frac{T}{s/n} \left( \frac{\partial (s/n)}{\partial T} \right)_P = \frac{T}{n^2} \left( n^2 \frac{\partial^2 p}{\partial T^2} - 2 s n \frac{\partial^2 p}{\partial T \partial \mu} + s^2 \frac{\partial^2 p}{\partial\mu^2} \right)
\end{split}
\end{equation}
the isothermal and adiabatic compressibilities
\begin{equation}
\begin{split}
\kappa_T = & - \frac{1}{V} \left( \frac{\partial V}{\partial p} \right)_{T,N} = \frac{1}{n} \left( \frac{\partial n}{\partial p} \right)_T= \frac{1}{n^2} \frac{\partial^2 p}{\partial\mu^2},\\
\kappa_S = & - \frac{1}{V} \left(\frac{\partial V}{\partial p}\right)_{S,N} = \frac{1}{n} \left( \frac{\partial n}{\partial p} \right)_{s/n}= \frac{ \frac{\partial^2 p}{\partial T^2} \frac{\partial^2 p}{\partial\mu^2} - \frac{\partial^2 p}{\partial T \partial\mu} \frac{\partial^2 p}{\partial T \partial\mu} }{n^2 \frac{\partial^2 p}{\partial T^2} - 2 s n \frac{\partial^2 p}{\partial T \partial \mu} + s^2 \frac{\partial^2 p}{\partial\mu^2}},
\end{split}
\end{equation}
the thermal expansion coefficient
\begin{equation}
\alpha = \frac{1}{V} \left( \frac{\partial V}{\partial T} \right)_{P,N} = - \frac{1}{n} \left( \frac{\partial n}{\partial T} \right)_P = \frac{1}{n^2} \left( s \frac{\partial^2 p}{\partial\mu^2} - n \frac{\partial^2 p}{\partial T\partial\mu} \right),
\end{equation}
the sound velocity at fixed entropy per particle
\begin{equation}
c_s^2 = \left( \frac{\partial p}{\partial \epsilon} \right)_{s/n} = \frac{n^2 \frac{\partial^2 p}{\partial T^2}-2 sn \frac{\partial^2 p}{\partial T \partial \mu}+ s^2 \frac{\partial^2 p}{\partial \mu^2}}{(\epsilon + p) \left( \frac{\partial^2 p}{\partial T^2} \frac{\partial^2 p}{\partial \mu^2} - \frac{\partial^2 p}{\partial T \partial\mu} \frac{\partial^2 p}{\partial T \partial\mu}\right)},
\end{equation}
and a modified sound velocity at fixed particle density
\begin{equation}
\tilde c_s^2 = \left( \frac{\partial p}{\partial \epsilon} \right)_n = \frac{s \frac{\partial^2 p}{\partial\mu^2}-n \frac{\partial^2 p}{\partial T \partial\mu}}{T\frac{\partial^2 p}{\partial T^2}\frac{\partial^2 p}{\partial\mu^2}-T \frac{\partial^2p}{\partial T\partial\mu}\frac{\partial^2 p}{\partial T \partial\mu}}.
\end{equation}
Both sound velocities agree for vanishing baryon number density, $n=0$.
Note that the usual relations 
\begin{equation}
\begin{split}
\frac{c_P}{c_V} = \frac{\kappa_T}{\kappa_S}, \quad\quad & \quad c_P-c_V = \frac{T \alpha^2}{\kappa_T},\\
\kappa_T-\kappa_S = \frac{T \alpha^2}{c_P}, \quad\quad & \quad c_s^2 = \frac{1}{\kappa_S (\epsilon+p)},
\end{split}
\end{equation}
are fulfilled. Moreover, one has
\begin{equation}
\tilde c_s^2 = \frac{\alpha}{c_V \kappa_T}, \quad\quad\quad \frac{1}{c_V}-\frac{1}{c_P} = \frac{T \tilde c_s^4}{c_s^2 (\epsilon+p)}
\end{equation}
%
%In terms of these susceptibilities one can write
%\begin{equation}
%\begin{split}
%dT = & \left[\frac{\epsilon + p}{c_V}\right] \frac{1}{\epsilon + p} d\epsilon + \left[\frac{T\alpha}{c_V \kappa_T} - \frac{\epsilon+p}{c_V}\right] \frac{1}{n} dn,\\
%d\mu = & \left[ -\frac{s(\epsilon+p)}{n c_V} + \frac{\alpha (\epsilon + p)}{n c_V \kappa_T}\right] \frac{1}{\epsilon + p} d\epsilon + \left[ \frac{1}{n\kappa_T} - \frac{(Ts+\epsilon+p)\alpha}{n c_V \kappa_T} + \frac{(\epsilon+p) s}{n c_V} \right] \frac{1}{n} dn,
%\end{split}
%\end{equation}
%which can be directly used to transform from the fluid dynamic equations from the variables $\epsilon$ and $n$ to $T$ and $\mu$.
%
%In general, for Navier-Stokes theory one needs also the variations of the shear viscosity $\eta(T,\mu)$, bulk viscosity $\zeta(T,\mu)$ and heat conductivity $\kappa(T,\mu)$,
%\begin{equation}
%\begin{split}
%d\eta  = & \left(\frac{\partial \eta}{\partial T}\right)_\mu dT + \left(\frac{\partial \eta}{\partial \mu}\right)_T d\mu \,, \\
%d\zeta = & \left(\frac{\partial \zeta}{\partial T}\right)_\mu dT+ \left(\frac{\partial \zeta}{\partial \mu}\right)_T d \mu \,, \\
%d\kappa = & \left(\frac{\partial \kappa}{\partial T}\right)_\mu dT+ \left(\frac{\partial \kappa}{\partial \mu}\right)_T d\mu \,.
%\end{split}
%\end{equation}
%
For the evolution equations of linear perturbations as discussed in Sec.~\ref{sec:fluc+Bjo} we need also
\begin{equation}
\begin{split}
\left( \frac{\partial p}{\partial \epsilon}\right)_n = & \frac{s\frac{\partial^2 p}{\partial\mu^2} - n\frac{\partial^2 p}{\partial T\partial\mu} }{ T \frac{\partial^2 p}{\partial T^2} \frac{\partial^2 p}{\partial\mu^2} - T \frac{\partial^2 p}{\partial T \partial \mu} \frac{\partial^2 p}{\partial T \partial\mu} },\\
\left( \frac{\partial p}{\partial n}\right)_\epsilon = & \frac{ T n \frac{\partial^2 p}{\partial T^2} + (Ts+\mu n) \frac{\partial^2 p}{\partial T \partial\mu} + \mu s \frac{\partial^2 p}{\partial \mu^2} }{ T \frac{\partial^2 p}{\partial T^2} \frac{\partial^2 p}{\partial\mu^2} - T \frac{\partial^2 p}{\partial T \partial \mu} \frac{\partial^2 p}{\partial T \partial\mu} }, \\
\left( \frac{\partial (\mu/T)}{\partial \epsilon}\right)_n = &  \frac{ \frac{1}{T} \frac{\partial^2 p}{\partial T \partial\mu} - \frac{\mu}{T^2}\frac{\partial^2 p}{\partial T\partial \mu} }{ T \frac{\partial^2 p}{\partial T^2} \frac{\partial^2 p}{\partial\mu^2} - T \frac{\partial^2 p}{\partial T \partial \mu} \frac{\partial^2 p}{\partial T \partial\mu} }, \\
\left( \frac{\partial (\mu/T)}{\partial n}\right)_\epsilon = &  \frac{  \frac{\partial^2 p}{\partial T^2} - \frac{\mu^2}{T^2} \frac{\partial^2 p}{\partial\mu^2}  }{ T \frac{\partial^2 p}{\partial T^2} \frac{\partial^2 p}{\partial\mu^2} - T \frac{\partial^2 p}{\partial T \partial \mu} \frac{\partial^2 p}{\partial T \partial\mu} },  \\
\left( \frac{\partial \zeta}{\partial \epsilon}\right)_n = & \frac{  \frac{\partial\zeta}{\partial T}\frac{\partial^2 p}{\partial \mu^2} - \frac{\partial \zeta}{\partial \mu} \frac{\partial^2 p}{\partial T \partial \mu}  }{ T \frac{\partial^2 p}{\partial T^2} \frac{\partial^2 p}{\partial\mu^2} - T \frac{\partial^2 p}{\partial T \partial \mu} \frac{\partial^2 p}{\partial T \partial\mu} },  \\
\left( \frac{\partial \zeta}{\partial n}\right)_\epsilon = &  \frac{  \frac{\partial\zeta}{\partial\mu} \left( T\frac{\partial^2 p}{\partial T^2} + \mu \frac{\partial^2 p}{\partial T \partial\mu} \right) - \frac{\partial \zeta}{\partial T} \left( T \frac{\partial^2 p}{\partial T \partial\mu} + \mu \frac{\partial^2 p}{\partial\mu^2} \right) }{ T \frac{\partial^2 p}{\partial T^2} \frac{\partial^2 p}{\partial\mu^2} - T \frac{\partial^2 p}{\partial T \partial \mu} \frac{\partial^2 p}{\partial T \partial\mu} },   \\
\left( \frac{\partial \eta}{\partial \epsilon}\right)_n = &  \frac{  \frac{\partial\eta}{\partial T}\frac{\partial^2 p}{\partial \mu^2} - \frac{\partial \eta}{\partial \mu} \frac{\partial^2 p}{\partial T \partial \mu}  }{ T \frac{\partial^2 p}{\partial T^2} \frac{\partial^2 p}{\partial\mu^2} - T \frac{\partial^2 p}{\partial T \partial \mu} \frac{\partial^2 p}{\partial T \partial\mu} },  \\
\left( \frac{\partial \eta}{\partial n}\right)_\epsilon = & \frac{  \frac{\partial\eta}{\partial\mu} \left( T\frac{\partial^2 p}{\partial T^2} + \mu \frac{\partial^2 p}{\partial T \partial\mu} \right) - \frac{\partial \eta}{\partial T} \left( T \frac{\partial^2 p}{\partial T \partial\mu} + \mu \frac{\partial^2 p}{\partial\mu^2} \right) }{ T \frac{\partial^2 p}{\partial T^2} \frac{\partial^2 p}{\partial\mu^2} - T \frac{\partial^2 p}{\partial T \partial \mu} \frac{\partial^2 p}{\partial T \partial\mu} },
\end{split}
\end{equation}
where, similarly to pressure $p(T,\mu)$, the bulk viscosity $\zeta(T,\mu)$ and shear viscosity $\eta(T,\mu)$ are functions of $T$ and $\mu$ on the right hand side.

\section{Linearized relativistic fluid dynamics}
\label{app:Linearizedrelativisticfluiddynamics}
In this appendix we discuss a background-fluctuation splitting for the fluid dynamic equations as it is used in Sec.~\ref{sec:fluc+Bjo}. We split the fluid dynamic fields into a background part and a perturbation according to
\begin{equation}
\begin{split}
u^\mu & = \bar u^\mu + \delta u^\mu,\\
\epsilon & = \bar \epsilon + \delta \epsilon,\\
n & = \bar n + \delta n,\\
\pi_\text{bulk} & = \bar \pi_\text{bulk} + \delta \pi_\text{bulk}, 
\end{split}
\end{equation}
and so on. The projector orthogonal to the fluid velocity is given by
\begin{equation}
\Delta^{\mu\nu} = \bar \Delta^{\mu\nu} + \delta \Delta^{\mu\nu},
\end{equation}
with $\delta \Delta^{\mu\nu} = \bar u^\mu \delta u^\nu + \delta u^\mu \bar u^\nu$. If one restricts to a linear treatment of perturbations, the equations of motion for the background are simply the full equations of motion \eqref{eq:eveqnsgeneral}. For the perturbations in energy and particle number density one obtains from Eq.\ \eqref{eq:eveqnsgeneral}
\begin{equation}
\begin{split}
\bar u^\mu \partial_\mu \delta \epsilon + \delta u^\mu \partial_\mu \bar \epsilon + (\bar \epsilon + \bar p + \bar \pi_\text{bulk}) \nabla_\mu \delta u^\mu + (\delta \epsilon + \delta p + \delta \pi_\text{bulk}) \nabla_\mu \bar u^\mu & \\
+ \bar \pi^{\mu\nu} \nabla_\mu \delta u_\nu + \delta \pi^{\mu\nu} \nabla_\mu \bar u_\nu & = 0,\\
\bar u^\mu \partial_\mu \delta n + \delta u^\mu \partial_\mu \bar n + \bar n \nabla_\mu \delta u^\mu + \delta n \nabla_\mu \bar u^\mu + \nabla_\mu \delta \nu^\mu & = 0,
\end{split}
\label{eq:B3}
\end{equation}
and for the fluid velocity
\begin{equation}
\begin{split}
(\bar \epsilon + \bar p + \bar \pi_\text{bulk}) \bar u^\mu \nabla_\mu \delta u^\nu + (\bar \epsilon + \bar p + \bar \pi_\text{bulk}) \delta u^\mu \nabla_\mu \bar u^\nu + (\delta \epsilon + \delta p + \delta \pi_\text{bulk}) \bar u^\mu \nabla_\mu \bar u^\nu & \\
+ \bar \Delta^{\nu\mu} \partial_\mu (\delta p + \delta \pi_\text{bulk}) + \delta \Delta^{\nu\mu} \partial_\mu (\bar p + \bar \pi_\text{bulk}) + \bar \Delta^\nu_{\;\;\,\alpha} \nabla_\mu \delta \pi^{\mu\alpha} + \delta\Delta^\nu_{\;\;\,\alpha} \nabla_\mu \bar \pi^{\mu\alpha} & = 0.
\end{split}
\label{eq:B4}
\end{equation}
In these equations one can see $\delta\epsilon$ and $\delta n$ as independent variables, to which other thermodynamic variables and the transport coefficients are related in the standard way, e.g.,
\begin{equation}
\begin{split}
\delta p = & \left( \frac{\partial p}{\partial \epsilon} \right)_n \delta \epsilon + \left( \frac{\partial p}{\partial n} \right)_\epsilon \delta n, \\
\partial_\mu \delta p = & \left( \frac{\partial p}{\partial \epsilon} \right)_n \partial_\mu\delta \epsilon + \left( \frac{\partial p}{\partial n} \right)_\epsilon \partial_\mu\delta n + \partial_\mu \left( \frac{\partial p}{\partial \epsilon}  \right)_n \delta \epsilon +   \partial_\mu \left( \frac{\partial p}{\partial n} \right)_\epsilon  \delta n.
\end{split}
\end{equation}
From the constitutive relation of first order fluid dynamics in Eq.\ \eqref{eq:NSshear} one finds
\begin{equation}
\delta \pi^{\mu\nu} = -2 \bar \eta \, \delta \sigma^{\mu\nu} - 2 \delta \eta \, \bar \sigma^{\mu\nu}
\end{equation}
with
\begin{equation}
\begin{split}
\delta \sigma^{\mu\nu} = \frac{1}{2} \bar \Delta^{\mu\alpha} \nabla_\alpha \delta u^\nu + \frac{1}{2} \bar \Delta^{\nu\alpha} \nabla_\alpha \delta u^\mu - \frac{1}{3}\bar\Delta^{\mu\nu} \nabla_\alpha \delta u^\alpha & \\
+ \frac{1}{2} \delta \Delta^{\mu\alpha} \nabla_\alpha \bar u^\nu + \frac{1}{2} \delta \Delta^{\nu\alpha} \nabla_\alpha \bar u^\mu - \frac{1}{3} \delta\Delta^{\mu\nu} \nabla_\alpha \bar u^\alpha.
\end{split}
\end{equation}
Similarly, for the bulk viscous pressure in Eq.\ \eqref{eq:NSbulk} one finds
\begin{equation}
\delta \pi_\text{bulk} = - \bar \zeta \, \delta \theta - \delta \zeta \, \bar\theta, 
\end{equation}
with 
\begin{equation}
\delta \theta = \nabla_\mu \delta u^\mu.
\end{equation}
Finally, the perturbation of the diffusion current is obtained from eq.\ \eqref{eq:dissNS} as
\begin{equation}
\delta \nu^\alpha = - \bar\kappa \left[ \frac{\bar n\bar T}{\bar \epsilon + \bar p} \right]^2 \delta\iota^\alpha - \delta \left( \kappa \left[ \frac{nT}{\epsilon + p} \right]^2 \right) \bar \iota^\alpha,
\end{equation}
with
\begin{equation}
\delta \iota^\alpha = \bar \Delta^{\alpha\beta}  \partial_\beta \, \delta \left( \mu/T\right) + \delta \Delta^{\alpha\beta} \partial_\beta \left( \bar \mu/ \bar T \right),
\end{equation}
and
\begin{equation}
\begin{split}
\partial_\beta \delta\left( \mu / T \right) = & \left( \frac{\partial(\mu/T)}{\partial \epsilon} \right)_n \partial_\beta \delta \epsilon + \left( \frac{\partial(\mu/T)}{\partial n} \right)_\epsilon \partial_\beta \delta n \\
& + \partial_\beta \left( \frac{\partial (\mu/T)}{\partial \epsilon} \right)_n \delta \epsilon + \partial_\beta \left( \frac{\partial(\mu/T)}{\partial n} \right)_\epsilon \delta n.
\end{split}
\end{equation}
Equations \eqref{eq:B3} and \eqref{eq:B4} also involve the following divergence of the shear stress perturbation
\begin{equation}
\begin{split}
\nabla_\mu \delta \pi^{\mu\nu} = & -2 (\partial_\mu \bar \eta) \delta\sigma^{\mu\nu} - 2 \bar \eta \nabla_\mu \delta\sigma^{\mu\nu}  -2 (\partial_\mu \delta \eta) \bar \sigma^{\mu\nu} - 2 \delta \eta \nabla_\mu \bar \sigma^{\mu\nu},
\end{split}
\end{equation}
with 
\begin{equation}
\begin{split}
\nabla_\mu \delta \sigma^{\mu\nu} = & \tfrac{1}{2} \bar \Delta^{\mu\alpha} \nabla_\mu \nabla_\alpha \delta u^\nu + \tfrac{1}{2} \bar \Delta^{\nu\alpha} \nabla_\mu \nabla_\alpha \delta u^\mu - \tfrac{1}{3} \bar\Delta^{\mu\nu} \nabla_\mu\nabla_\alpha\delta u^\alpha \\
& + \tfrac{1}{2} \delta\Delta^{\mu\alpha} \nabla_\mu\nabla_\alpha \bar u^\nu + \tfrac{1}{2} \delta \Delta^{\nu\alpha} \nabla_\mu \nabla_\alpha \bar u^\mu - \tfrac{1}{3} \delta\Delta^{\mu\nu} \nabla_\mu\nabla_\alpha \bar u^\alpha \\
& + \bar u^\mu (\nabla_\mu \bar u^\alpha) \nabla_\alpha \delta u^\nu + \tfrac{1}{6} (\nabla_\mu \bar u^\mu) \bar u^\alpha \nabla_\alpha \delta u^\nu\\
& + \bar u^\nu (\nabla_\mu \delta u^\alpha) \nabla_\alpha \bar u^\mu - \tfrac{2}{3} \bar u^\nu (\nabla_\mu \delta u^\mu) \nabla_\alpha \bar u^\alpha \\
& + \tfrac{1}{2} \delta u^\nu (\nabla_\mu \bar u^\alpha) \nabla_\alpha \bar u^\mu - \tfrac{1}{3} \delta u^\nu (\nabla_\mu \bar u^\mu) \nabla_\alpha \bar u^\alpha \\
& + \bar u^\mu (\nabla_\mu \delta u^\alpha) \nabla_\alpha \bar u^\nu + \tfrac{1}{6} (\nabla_\mu \bar u^\mu) \delta u^\alpha \nabla_\alpha \bar u^\nu  \\
& + \delta u^\mu (\nabla_\mu \bar u^\alpha) \nabla_\alpha \bar u^\nu + \tfrac{1}{6} (\nabla_\mu \delta u^\mu) \bar u^\alpha \nabla_\alpha \bar u^\nu,
\end{split}
\end{equation}
the derivative of the bulk viscous pressure perturbation
\begin{equation}
\begin{split}
\partial_\mu \delta \pi_\text{bulk} = - (\partial_\mu \bar \zeta) \delta \theta - \bar \theta \partial_\mu \delta \theta - (\partial_\mu \delta \theta) \bar \theta - \delta \zeta \partial_\mu \bar \theta,
\end{split}
\end{equation}
with
\begin{equation}
\partial_\mu \delta \theta = \nabla_\mu \nabla_\alpha \delta u^\alpha,
\end{equation}
and finally the divergence of the perturbation in the diffusion current
\begin{equation}
\begin{split}
\nabla_\alpha \nu^\alpha = & - \partial_\alpha \left[ \bar \kappa \left( \frac{\bar n \bar T}{\bar \epsilon + \bar p} \right)^2 \right] \delta \iota^\alpha - \bar \kappa \left( \frac{\bar n \bar T}{\bar \epsilon + \bar p} \right)^2 \nabla_\alpha \delta \iota^\alpha \\
& - \nabla_\alpha \delta \left[ \kappa \left( \frac{nT}{\epsilon + p} \right)^2 \right] \bar \iota^\alpha - \delta \left[ \kappa \left( \frac{nT}{\epsilon + p} \right)^2 \right] \nabla_\alpha \bar \iota^\alpha,
\end{split}
\end{equation}
with
\begin{equation}
\begin{split}
\nabla_\alpha \delta \iota^\alpha = & \bar \Delta^{\alpha\beta} \nabla_\alpha \partial_\beta \, \delta ( \mu/T) + \nabla_\alpha \bar \Delta^{\alpha\beta} \, \partial_\beta \, \delta ( \mu/T ) \\
& + \nabla_\alpha \delta \Delta^{\alpha\beta} \partial_\beta \left( \bar \mu/\bar T \right) + \delta\Delta^{\alpha\beta} \nabla_\alpha \partial_\beta \left(\bar \mu/ \bar T\right) .
\end{split}
\label{eq:B18}
\end{equation}
and
\begin{equation}
\begin{split}
\nabla_\alpha \partial_\beta \delta(\mu/T) = & \left( \frac{\partial (\mu/T)}{\partial \epsilon} \right)_n \nabla_\alpha \partial_\beta \delta \epsilon + \left( \frac{\partial (\mu/T)}{\partial n} \right)_\epsilon \nabla_\alpha \partial_\beta \delta n \\
& + \partial_\alpha \left( \frac{\partial (\mu/T)}{\partial \epsilon} \right)_n \partial_\beta \delta \epsilon + \partial_\alpha \left( \frac{\partial (\mu/T)}{\partial n} \right)_\epsilon \partial_\beta \delta n \\
& + \partial_\beta \left( \frac{\partial (\mu/T)}{\partial \epsilon} \right)_n \partial_\alpha \delta \epsilon + \partial_\beta \left( \frac{\partial (\mu/T)}{\partial n} \right)_\epsilon \partial_\alpha \delta n \\
& + \nabla_\alpha \partial_\beta \left( \frac{\partial (\mu/T)}{\partial \epsilon} \right)_n  \delta \epsilon + \nabla_\alpha \partial_\beta \left( \frac{\partial (\mu/T)}{\partial n} \right)_\epsilon \delta n .
\end{split}
\label{eq:B19}
\end{equation}
Note that the expression in Eq.\ \eqref{eq:B19} is contracted in Eq.\ \eqref{eq:B18} with the projector $\bar \Delta^{\alpha\beta}$. In many circumstances the background field changes only in the direction of $\bar u^\mu$ such that Eq.\ \eqref{eq:B19} simplifies substantially.

Note that the formulas compiled in this appendix allow us to obtain for a given background solution and thermodynamic equation of state linear evolution equations for the perturbations around this background solution. The independent variables of these linearized equations are the three independent components of $\delta u^\mu$ (one constraint is given by the condition $\bar u_\mu \delta u^\mu=0$) as well as $\delta \epsilon$ and $\delta n$. In the first order formalism of relativistic fluid dynamics, the equations for the perturbations are of parabolic type while they are expected to become of elliptic type when relaxation time terms are kept.

\end{appendix}
\bibliography{fluctuations}

\begin{thebibliography}{115}%
\makeatletter
\providecommand \@ifxundefined [1]{%
 \@ifx{#1\undefined}
}%
\providecommand \@ifnum [1]{%
 \ifnum #1\expandafter \@firstoftwo
 \else \expandafter \@secondoftwo
 \fi
}%
\providecommand \@ifx [1]{%
 \ifx #1\expandafter \@firstoftwo
 \else \expandafter \@secondoftwo
 \fi
}%
\providecommand \natexlab [1]{#1}%
\providecommand \enquote  [1]{``#1''}%
\providecommand \bibnamefont  [1]{#1}%
\providecommand \bibfnamefont [1]{#1}%
\providecommand \citenamefont [1]{#1}%
\providecommand \href@noop [0]{\@secondoftwo}%
\providecommand \href [0]{\begingroup \@sanitize@url \@href}%
\providecommand \@href[1]{\@@startlink{#1}\@@href}%
\providecommand \@@href[1]{\endgroup#1\@@endlink}%
\providecommand \@sanitize@url [0]{\catcode `\\12\catcode `\$12\catcode
  `\&12\catcode `\#12\catcode `\^12\catcode `\_12\catcode `\%12\relax}%
\providecommand \@@startlink[1]{}%
\providecommand \@@endlink[0]{}%
\providecommand \url  [0]{\begingroup\@sanitize@url \@url }%
\providecommand \@url [1]{\endgroup\@href {#1}{\urlprefix }}%
\providecommand \urlprefix  [0]{URL }%
\providecommand \Eprint [0]{\href }%
\providecommand \doibase [0]{http://dx.doi.org/}%
\providecommand \selectlanguage [0]{\@gobble}%
\providecommand \bibinfo  [0]{\@secondoftwo}%
\providecommand \bibfield  [0]{\@secondoftwo}%
\providecommand \translation [1]{[#1]}%
\providecommand \BibitemOpen [0]{}%
\providecommand \bibitemStop [0]{}%
\providecommand \bibitemNoStop [0]{.\EOS\space}%
\providecommand \EOS [0]{\spacefactor3000\relax}%
\providecommand \BibitemShut  [1]{\csname bibitem#1\endcsname}%
\let\auto@bib@innerbib\@empty
%</preamble>
\bibitem [{\citenamefont {Aamodt}\ \emph {et~al.}(2011)\citenamefont {Aamodt}
  \emph {et~al.}}]{ALICE:2011ab}%
  \BibitemOpen
  \bibfield  {author} {\bibinfo {author} {\bibfnamefont {K.}~\bibnamefont
  {Aamodt}} \emph {et~al.} (\bibinfo {collaboration} {ALICE}),\ }\bibfield
  {title} {\enquote {\bibinfo {title} {{Higher harmonic anisotropic flow
  measurements of charged particles in Pb-Pb collisions at $\sqrt{s_{NN}}$=2.76
  TeV}},}\ }\href {\doibase 10.1103/PhysRevLett.107.032301} {\bibfield
  {journal} {\bibinfo  {journal} {Phys.Rev.Lett.}\ }\textbf {\bibinfo {volume}
  {107}},\ \bibinfo {pages} {032301} (\bibinfo {year} {2011})},\ \Eprint
  {http://arxiv.org/abs/1105.3865} {arXiv:1105.3865 [nucl-ex]} \BibitemShut
  {NoStop}%
%%CITATION = ARXIV:1105.3865;%%
\bibitem [{\citenamefont {Aamodt}\ \emph {et~al.}(2012)\citenamefont {Aamodt}
  \emph {et~al.}}]{Aamodt:2011by}%
  \BibitemOpen
  \bibfield  {author} {\bibinfo {author} {\bibfnamefont {K.}~\bibnamefont
  {Aamodt}} \emph {et~al.} (\bibinfo {collaboration} {ALICE}),\ }\bibfield
  {title} {\enquote {\bibinfo {title} {{Harmonic decomposition of two-particle
  angular correlations in Pb-Pb collisions at $\sqrt{s_{NN}}=2.76$ TeV}},}\
  }\href {\doibase 10.1016/j.physletb.2012.01.060} {\bibfield  {journal}
  {\bibinfo  {journal} {Phys.Lett.}\ }\textbf {\bibinfo {volume} {B708}},\
  \bibinfo {pages} {249--264} (\bibinfo {year} {2012})},\ \Eprint
  {http://arxiv.org/abs/1109.2501} {arXiv:1109.2501 [nucl-ex]} \BibitemShut
  {NoStop}%
%%CITATION = ARXIV:1109.2501;%%
\bibitem [{\citenamefont {Chatrchyan}\ \emph {et~al.}(2012)\citenamefont
  {Chatrchyan} \emph {et~al.}}]{Chatrchyan:2012wg}%
  \BibitemOpen
  \bibfield  {author} {\bibinfo {author} {\bibfnamefont {Serguei}\ \bibnamefont
  {Chatrchyan}} \emph {et~al.} (\bibinfo {collaboration} {CMS}),\ }\bibfield
  {title} {\enquote {\bibinfo {title} {{Centrality dependence of dihadron
  correlations and azimuthal anisotropy harmonics in PbPb collisions at
  $\sqrt{s_{NN}}=2.76$ TeV}},}\ }\href {\doibase
  10.1140/epjc/s10052-012-2012-3} {\bibfield  {journal} {\bibinfo  {journal}
  {Eur.Phys.J.}\ }\textbf {\bibinfo {volume} {C72}},\ \bibinfo {pages} {2012}
  (\bibinfo {year} {2012})},\ \Eprint {http://arxiv.org/abs/1201.3158}
  {arXiv:1201.3158 [nucl-ex]} \BibitemShut {NoStop}%
%%CITATION = ARXIV:1201.3158;%%
\bibitem [{\citenamefont {Aad}\ \emph {et~al.}(2013)\citenamefont {Aad} \emph
  {et~al.}}]{Aad:2013xma}%
  \BibitemOpen
  \bibfield  {author} {\bibinfo {author} {\bibfnamefont {Georges}\ \bibnamefont
  {Aad}} \emph {et~al.} (\bibinfo {collaboration} {ATLAS}),\ }\bibfield
  {title} {\enquote {\bibinfo {title} {{Measurement of the distributions of
  event-by-event flow harmonics in lead-lead collisions at = 2.76 TeV with the
  ATLAS detector at the LHC}},}\ }\href {\doibase 10.1007/JHEP11(2013)183}
  {\bibfield  {journal} {\bibinfo  {journal} {JHEP}\ }\textbf {\bibinfo
  {volume} {1311}},\ \bibinfo {pages} {183} (\bibinfo {year} {2013})},\ \Eprint
  {http://arxiv.org/abs/1305.2942} {arXiv:1305.2942 [hep-ex]} \BibitemShut
  {NoStop}%
%%CITATION = ARXIV:1305.2942;%%
\bibitem [{\citenamefont {Adamczyk}\ \emph {et~al.}(2013)\citenamefont
  {Adamczyk} \emph {et~al.}}]{Adamczyk:2013waa}%
  \BibitemOpen
  \bibfield  {author} {\bibinfo {author} {\bibfnamefont {L.}~\bibnamefont
  {Adamczyk}} \emph {et~al.} (\bibinfo {collaboration} {STAR}),\ }\bibfield
  {title} {\enquote {\bibinfo {title} {{Third Harmonic Flow of Charged
  Particles in Au+Au Collisions at sqrtsNN = 200 GeV}},}\ }\href {\doibase
  10.1103/PhysRevC.88.014904} {\bibfield  {journal} {\bibinfo  {journal}
  {Phys.Rev.}\ }\textbf {\bibinfo {volume} {C88}},\ \bibinfo {pages} {014904}
  (\bibinfo {year} {2013})},\ \Eprint {http://arxiv.org/abs/1301.2187}
  {arXiv:1301.2187 [nucl-ex]} \BibitemShut {NoStop}%
%%CITATION = ARXIV:1301.2187;%%
\bibitem [{\citenamefont {Adare}\ \emph {et~al.}(2011)\citenamefont {Adare}
  \emph {et~al.}}]{Adare:2011tg}%
  \BibitemOpen
  \bibfield  {author} {\bibinfo {author} {\bibfnamefont {A.}~\bibnamefont
  {Adare}} \emph {et~al.} (\bibinfo {collaboration} {PHENIX}),\ }\bibfield
  {title} {\enquote {\bibinfo {title} {{Measurements of Higher-Order Flow
  Harmonics in Au+Au Collisions at $\sqrt{s_{NN}} = 200$ GeV}},}\ }\href
  {\doibase 10.1103/PhysRevLett.107.252301} {\bibfield  {journal} {\bibinfo
  {journal} {Phys.Rev.Lett.}\ }\textbf {\bibinfo {volume} {107}},\ \bibinfo
  {pages} {252301} (\bibinfo {year} {2011})},\ \Eprint
  {http://arxiv.org/abs/1105.3928} {arXiv:1105.3928 [nucl-ex]} \BibitemShut
  {NoStop}%
%%CITATION = ARXIV:1105.3928;%%
\bibitem [{\citenamefont {Alver}\ \emph {et~al.}(2008)\citenamefont {Alver},
  \citenamefont {Back}, \citenamefont {Baker}, \citenamefont {Ballintijn},
  \citenamefont {Barton} \emph {et~al.}}]{Alver:2008zza}%
  \BibitemOpen
  \bibfield  {author} {\bibinfo {author} {\bibfnamefont {B.}~\bibnamefont
  {Alver}}, \bibinfo {author} {\bibfnamefont {B.B.}\ \bibnamefont {Back}},
  \bibinfo {author} {\bibfnamefont {M.D.}\ \bibnamefont {Baker}}, \bibinfo
  {author} {\bibfnamefont {M.}~\bibnamefont {Ballintijn}}, \bibinfo {author}
  {\bibfnamefont {D.S.}\ \bibnamefont {Barton}},  \emph {et~al.},\ }\bibfield
  {title} {\enquote {\bibinfo {title} {{Importance of correlations and
  fluctuations on the initial source eccentricity in high-energy
  nucleus-nucleus collisions}},}\ }\href {\doibase 10.1103/PhysRevC.77.014906}
  {\bibfield  {journal} {\bibinfo  {journal} {Phys.Rev.}\ }\textbf {\bibinfo
  {volume} {C77}},\ \bibinfo {pages} {014906} (\bibinfo {year} {2008})},\
  \Eprint {http://arxiv.org/abs/0711.3724} {arXiv:0711.3724 [nucl-ex]}
  \BibitemShut {NoStop}%
%%CITATION = ARXIV:0711.3724;%%
\bibitem [{\citenamefont {Heinz}\ and\ \citenamefont
  {Snellings}(2013)}]{Heinz:2013th}%
  \BibitemOpen
  \bibfield  {author} {\bibinfo {author} {\bibfnamefont {Ulrich}\ \bibnamefont
  {Heinz}}\ and\ \bibinfo {author} {\bibfnamefont {Raimond}\ \bibnamefont
  {Snellings}},\ }\bibfield  {title} {\enquote {\bibinfo {title} {{Collective
  flow and viscosity in relativistic heavy-ion collisions}},}\ }\href {\doibase
  10.1146/annurev-nucl-102212-170540} {\bibfield  {journal} {\bibinfo
  {journal} {Ann.Rev.Nucl.Part.Sci.}\ }\textbf {\bibinfo {volume} {63}},\
  \bibinfo {pages} {123--151} (\bibinfo {year} {2013})},\ \Eprint
  {http://arxiv.org/abs/1301.2826} {arXiv:1301.2826 [nucl-th]} \BibitemShut
  {NoStop}%
%%CITATION = ARXIV:1301.2826;%%
\bibitem [{\citenamefont {Gale}\ \emph {et~al.}(2013)\citenamefont {Gale},
  \citenamefont {Jeon},\ and\ \citenamefont {Schenke}}]{Gale:2013da}%
  \BibitemOpen
  \bibfield  {author} {\bibinfo {author} {\bibfnamefont {Charles}\ \bibnamefont
  {Gale}}, \bibinfo {author} {\bibfnamefont {Sangyong}\ \bibnamefont {Jeon}}, \
  and\ \bibinfo {author} {\bibfnamefont {Bjoern}\ \bibnamefont {Schenke}},\
  }\bibfield  {title} {\enquote {\bibinfo {title} {{Hydrodynamic Modeling of
  Heavy-Ion Collisions}},}\ }\href {\doibase 10.1142/S0217751X13400113}
  {\bibfield  {journal} {\bibinfo  {journal} {Int.J.Mod.Phys.}\ }\textbf
  {\bibinfo {volume} {A28}},\ \bibinfo {pages} {1340011} (\bibinfo {year}
  {2013})},\ \Eprint {http://arxiv.org/abs/1301.5893} {arXiv:1301.5893
  [nucl-th]} \BibitemShut {NoStop}%
%%CITATION = ARXIV:1301.5893;%%
\bibitem [{\citenamefont {Fukushima}\ and\ \citenamefont
  {Hatsuda}(2011)}]{Fukushima:2010bq}%
  \BibitemOpen
  \bibfield  {author} {\bibinfo {author} {\bibfnamefont {Kenji}\ \bibnamefont
  {Fukushima}}\ and\ \bibinfo {author} {\bibfnamefont {Tetsuo}\ \bibnamefont
  {Hatsuda}},\ }\bibfield  {title} {\enquote {\bibinfo {title} {{The phase
  diagram of dense QCD}},}\ }\href {\doibase 10.1088/0034-4885/74/1/014001}
  {\bibfield  {journal} {\bibinfo  {journal} {Rept.Prog.Phys.}\ }\textbf
  {\bibinfo {volume} {74}},\ \bibinfo {pages} {014001} (\bibinfo {year}
  {2011})},\ \Eprint {http://arxiv.org/abs/1005.4814} {arXiv:1005.4814
  [hep-ph]} \BibitemShut {NoStop}%
%%CITATION = ARXIV:1005.4814;%%
\bibitem [{\citenamefont {Ding}\ \emph {et~al.}(2015)\citenamefont {Ding},
  \citenamefont {Karsch},\ and\ \citenamefont {Mukherjee}}]{Ding:2015ona}%
  \BibitemOpen
  \bibfield  {author} {\bibinfo {author} {\bibfnamefont {Heng-Tong}\
  \bibnamefont {Ding}}, \bibinfo {author} {\bibfnamefont {Frithjof}\
  \bibnamefont {Karsch}}, \ and\ \bibinfo {author} {\bibfnamefont {Swagato}\
  \bibnamefont {Mukherjee}},\ }\bibfield  {title} {\enquote {\bibinfo {title}
  {{Thermodynamics of strong-interaction matter from Lattice QCD}},}\ }\href
  {\doibase 10.1142/S0218301315300076} {\bibfield  {journal} {\bibinfo
  {journal} {Int. J. Mod. Phys.}\ }\textbf {\bibinfo {volume} {E24}},\ \bibinfo
  {pages} {1530007} (\bibinfo {year} {2015})},\ \Eprint
  {http://arxiv.org/abs/1504.05274} {arXiv:1504.05274 [hep-lat]} \BibitemShut
  {NoStop}%
%%CITATION = ARXIV:1504.05274;%%
\bibitem [{\citenamefont {Gazdzicki}\ and\ \citenamefont
  {Mrowczynski}(1992)}]{Gazdzicki:1992ri}%
  \BibitemOpen
  \bibfield  {author} {\bibinfo {author} {\bibfnamefont {M.}~\bibnamefont
  {Gazdzicki}}\ and\ \bibinfo {author} {\bibfnamefont {S.}~\bibnamefont
  {Mrowczynski}},\ }\bibfield  {title} {\enquote {\bibinfo {title} {{A Method
  to study 'equilibration' in nucleus-nucleus collisions}},}\ }\href {\doibase
  10.1007/BF01881715} {\bibfield  {journal} {\bibinfo  {journal} {Z.Phys.}\
  }\textbf {\bibinfo {volume} {C54}},\ \bibinfo {pages} {127--132} (\bibinfo
  {year} {1992})}\BibitemShut {NoStop}%
%%CITATION = ZEPYA,C54,127;%%
\bibitem [{\citenamefont {Stodolsky}(1995)}]{Stodolsky:1995ds}%
  \BibitemOpen
  \bibfield  {author} {\bibinfo {author} {\bibfnamefont {Leo}\ \bibnamefont
  {Stodolsky}},\ }\bibfield  {title} {\enquote {\bibinfo {title} {{Temperature
  fluctuations in multiparticle production}},}\ }\href {\doibase
  10.1103/PhysRevLett.75.1044} {\bibfield  {journal} {\bibinfo  {journal}
  {Phys.Rev.Lett.}\ }\textbf {\bibinfo {volume} {75}},\ \bibinfo {pages}
  {1044--1045} (\bibinfo {year} {1995})}\BibitemShut {NoStop}%
%%CITATION = PRLTA,75,1044;%%
\bibitem [{\citenamefont {Shuryak}(1998)}]{Shuryak:1997yj}%
  \BibitemOpen
  \bibfield  {author} {\bibinfo {author} {\bibfnamefont {Edward~V.}\
  \bibnamefont {Shuryak}},\ }\bibfield  {title} {\enquote {\bibinfo {title}
  {{Event per event analysis of heavy ion collisions and thermodynamical
  fluctuations}},}\ }\href {\doibase 10.1016/S0370-2693(98)00127-0} {\bibfield
  {journal} {\bibinfo  {journal} {Phys.Lett.}\ }\textbf {\bibinfo {volume}
  {B423}},\ \bibinfo {pages} {9--14} (\bibinfo {year} {1998})},\ \Eprint
  {http://arxiv.org/abs/hep-ph/9704456} {arXiv:hep-ph/9704456 [hep-ph]}
  \BibitemShut {NoStop}%
%%CITATION = HEP-PH/9704456;%%
\bibitem [{\citenamefont {Mrowczynski}(1998)}]{Mrowczynski:1997kz}%
  \BibitemOpen
  \bibfield  {author} {\bibinfo {author} {\bibfnamefont {Stanislaw}\
  \bibnamefont {Mrowczynski}},\ }\bibfield  {title} {\enquote {\bibinfo {title}
  {{Hadronic matter compressibility from event by event analysis of heavy ion
  collisions}},}\ }\href {\doibase 10.1016/S0370-2693(98)00492-4} {\bibfield
  {journal} {\bibinfo  {journal} {Phys.Lett.}\ }\textbf {\bibinfo {volume}
  {B430}},\ \bibinfo {pages} {9--14} (\bibinfo {year} {1998})},\ \Eprint
  {http://arxiv.org/abs/nucl-th/9712030} {arXiv:nucl-th/9712030 [nucl-th]}
  \BibitemShut {NoStop}%
%%CITATION = NUCL-TH/9712030;%%
\bibitem [{\citenamefont {Stephanov}\ \emph {et~al.}(1998)\citenamefont
  {Stephanov}, \citenamefont {Rajagopal},\ and\ \citenamefont
  {Shuryak}}]{Stephanov:1998dy}%
  \BibitemOpen
  \bibfield  {author} {\bibinfo {author} {\bibfnamefont {Misha~A.}\
  \bibnamefont {Stephanov}}, \bibinfo {author} {\bibfnamefont {K.}~\bibnamefont
  {Rajagopal}}, \ and\ \bibinfo {author} {\bibfnamefont {Edward~V.}\
  \bibnamefont {Shuryak}},\ }\bibfield  {title} {\enquote {\bibinfo {title}
  {{Signatures of the tricritical point in QCD}},}\ }\href {\doibase
  10.1103/PhysRevLett.81.4816} {\bibfield  {journal} {\bibinfo  {journal}
  {Phys.Rev.Lett.}\ }\textbf {\bibinfo {volume} {81}},\ \bibinfo {pages}
  {4816--4819} (\bibinfo {year} {1998})},\ \Eprint
  {http://arxiv.org/abs/hep-ph/9806219} {arXiv:hep-ph/9806219 [hep-ph]}
  \BibitemShut {NoStop}%
%%CITATION = HEP-PH/9806219;%%
\bibitem [{\citenamefont {Stephanov}\ \emph {et~al.}(1999)\citenamefont
  {Stephanov}, \citenamefont {Rajagopal},\ and\ \citenamefont
  {Shuryak}}]{Stephanov:1999zu}%
  \BibitemOpen
  \bibfield  {author} {\bibinfo {author} {\bibfnamefont {Misha~A.}\
  \bibnamefont {Stephanov}}, \bibinfo {author} {\bibfnamefont {K.}~\bibnamefont
  {Rajagopal}}, \ and\ \bibinfo {author} {\bibfnamefont {Edward~V.}\
  \bibnamefont {Shuryak}},\ }\bibfield  {title} {\enquote {\bibinfo {title}
  {{Event-by-event fluctuations in heavy ion collisions and the QCD critical
  point}},}\ }\href {\doibase 10.1103/PhysRevD.60.114028} {\bibfield  {journal}
  {\bibinfo  {journal} {Phys.Rev.}\ }\textbf {\bibinfo {volume} {D60}},\
  \bibinfo {pages} {114028} (\bibinfo {year} {1999})},\ \Eprint
  {http://arxiv.org/abs/hep-ph/9903292} {arXiv:hep-ph/9903292 [hep-ph]}
  \BibitemShut {NoStop}%
%%CITATION = HEP-PH/9903292;%%
\bibitem [{\citenamefont {Berdnikov}\ and\ \citenamefont
  {Rajagopal}(2000)}]{Berdnikov:1999ph}%
  \BibitemOpen
  \bibfield  {author} {\bibinfo {author} {\bibfnamefont {Boris}\ \bibnamefont
  {Berdnikov}}\ and\ \bibinfo {author} {\bibfnamefont {Krishna}\ \bibnamefont
  {Rajagopal}},\ }\bibfield  {title} {\enquote {\bibinfo {title} {{Slowing
  out-of-equilibrium near the QCD critical point}},}\ }\href {\doibase
  10.1103/PhysRevD.61.105017} {\bibfield  {journal} {\bibinfo  {journal}
  {Phys.Rev.}\ }\textbf {\bibinfo {volume} {D61}},\ \bibinfo {pages} {105017}
  (\bibinfo {year} {2000})},\ \Eprint {http://arxiv.org/abs/hep-ph/9912274}
  {arXiv:hep-ph/9912274 [hep-ph]} \BibitemShut {NoStop}%
%%CITATION = HEP-PH/9912274;%%
\bibitem [{\citenamefont {Jeon}\ and\ \citenamefont
  {Koch}(2000)}]{Jeon:2000wg}%
  \BibitemOpen
  \bibfield  {author} {\bibinfo {author} {\bibfnamefont {S.}~\bibnamefont
  {Jeon}}\ and\ \bibinfo {author} {\bibfnamefont {V.}~\bibnamefont {Koch}},\
  }\bibfield  {title} {\enquote {\bibinfo {title} {{Charged particle ratio
  fluctuation as a signal for QGP}},}\ }\href {\doibase
  10.1103/PhysRevLett.85.2076} {\bibfield  {journal} {\bibinfo  {journal}
  {Phys.Rev.Lett.}\ }\textbf {\bibinfo {volume} {85}},\ \bibinfo {pages}
  {2076--2079} (\bibinfo {year} {2000})},\ \Eprint
  {http://arxiv.org/abs/hep-ph/0003168} {arXiv:hep-ph/0003168 [hep-ph]}
  \BibitemShut {NoStop}%
%%CITATION = HEP-PH/0003168;%%
\bibitem [{\citenamefont {Asakawa}\ \emph {et~al.}(2000)\citenamefont
  {Asakawa}, \citenamefont {Heinz},\ and\ \citenamefont
  {Muller}}]{Asakawa:2000wh}%
  \BibitemOpen
  \bibfield  {author} {\bibinfo {author} {\bibfnamefont {Masayuki}\
  \bibnamefont {Asakawa}}, \bibinfo {author} {\bibfnamefont {Ulrich~W.}\
  \bibnamefont {Heinz}}, \ and\ \bibinfo {author} {\bibfnamefont {Berndt}\
  \bibnamefont {Muller}},\ }\bibfield  {title} {\enquote {\bibinfo {title}
  {{Fluctuation probes of quark deconfinement}},}\ }\href {\doibase
  10.1103/PhysRevLett.85.2072} {\bibfield  {journal} {\bibinfo  {journal}
  {Phys.Rev.Lett.}\ }\textbf {\bibinfo {volume} {85}},\ \bibinfo {pages}
  {2072--2075} (\bibinfo {year} {2000})},\ \Eprint
  {http://arxiv.org/abs/hep-ph/0003169} {arXiv:hep-ph/0003169 [hep-ph]}
  \BibitemShut {NoStop}%
%%CITATION = HEP-PH/0003169;%%
\bibitem [{\citenamefont {Voloshin}(2002)}]{Voloshin:2001ei}%
  \BibitemOpen
  \bibfield  {author} {\bibinfo {author} {\bibfnamefont {Sergei~A.}\
  \bibnamefont {Voloshin}} (\bibinfo {collaboration} {STAR}),\ }\bibfield
  {title} {\enquote {\bibinfo {title} {{Multiplicity and mean transverse
  momentum fluctuations in Au+Au collisions at RHIC}},}\ }\href {\doibase
  10.1063/1.1469997} {\bibfield  {journal} {\bibinfo  {journal} {AIP
  Conf.Proc.}\ }\textbf {\bibinfo {volume} {610}},\ \bibinfo {pages} {591--596}
  (\bibinfo {year} {2002})},\ \Eprint {http://arxiv.org/abs/nucl-ex/0109006}
  {arXiv:nucl-ex/0109006 [nucl-ex]} \BibitemShut {NoStop}%
%%CITATION = NUCL-EX/0109006;%%
\bibitem [{\citenamefont {Pruneau}\ \emph {et~al.}(2002)\citenamefont
  {Pruneau}, \citenamefont {Gavin},\ and\ \citenamefont
  {Voloshin}}]{Pruneau:2002yf}%
  \BibitemOpen
  \bibfield  {author} {\bibinfo {author} {\bibfnamefont {C.}~\bibnamefont
  {Pruneau}}, \bibinfo {author} {\bibfnamefont {S.}~\bibnamefont {Gavin}}, \
  and\ \bibinfo {author} {\bibfnamefont {S.}~\bibnamefont {Voloshin}},\
  }\bibfield  {title} {\enquote {\bibinfo {title} {{Methods for the study of
  particle production fluctuations}},}\ }\href {\doibase
  10.1103/PhysRevC.66.044904} {\bibfield  {journal} {\bibinfo  {journal}
  {Phys.Rev.}\ }\textbf {\bibinfo {volume} {C66}},\ \bibinfo {pages} {044904}
  (\bibinfo {year} {2002})},\ \Eprint {http://arxiv.org/abs/nucl-ex/0204011}
  {arXiv:nucl-ex/0204011 [nucl-ex]} \BibitemShut {NoStop}%
%%CITATION = NUCL-EX/0204011;%%
\bibitem [{\citenamefont {Hatta}\ and\ \citenamefont
  {Stephanov}(2003)}]{Hatta:2003wn}%
  \BibitemOpen
  \bibfield  {author} {\bibinfo {author} {\bibfnamefont {Y.}~\bibnamefont
  {Hatta}}\ and\ \bibinfo {author} {\bibfnamefont {M.A.}\ \bibnamefont
  {Stephanov}},\ }\bibfield  {title} {\enquote {\bibinfo {title} {{Proton
  number fluctuation as a signal of the QCD critical endpoint}},}\ }\href
  {\doibase 10.1103/PhysRevLett.91.102003} {\bibfield  {journal} {\bibinfo
  {journal} {Phys.Rev.Lett.}\ }\textbf {\bibinfo {volume} {91}},\ \bibinfo
  {pages} {102003} (\bibinfo {year} {2003})},\ \Eprint
  {http://arxiv.org/abs/hep-ph/0302002} {arXiv:hep-ph/0302002 [hep-ph]}
  \BibitemShut {NoStop}%
%%CITATION = HEP-PH/0302002;%%
\bibitem [{\citenamefont {Nonaka}\ and\ \citenamefont
  {Asakawa}(2005)}]{Nonaka:2004pg}%
  \BibitemOpen
  \bibfield  {author} {\bibinfo {author} {\bibfnamefont {Chiho}\ \bibnamefont
  {Nonaka}}\ and\ \bibinfo {author} {\bibfnamefont {Masayuki}\ \bibnamefont
  {Asakawa}},\ }\bibfield  {title} {\enquote {\bibinfo {title} {{Hydrodynamical
  evolution near the QCD critical end point}},}\ }\href {\doibase
  10.1103/PhysRevC.71.044904} {\bibfield  {journal} {\bibinfo  {journal}
  {Phys.Rev.}\ }\textbf {\bibinfo {volume} {C71}},\ \bibinfo {pages} {044904}
  (\bibinfo {year} {2005})},\ \Eprint {http://arxiv.org/abs/nucl-th/0410078}
  {arXiv:nucl-th/0410078 [nucl-th]} \BibitemShut {NoStop}%
%%CITATION = NUCL-TH/0410078;%%
\bibitem [{\citenamefont {Aziz}\ and\ \citenamefont
  {Gavin}(2004)}]{Aziz:2004qu}%
  \BibitemOpen
  \bibfield  {author} {\bibinfo {author} {\bibfnamefont {Mohamed~Abdel}\
  \bibnamefont {Aziz}}\ and\ \bibinfo {author} {\bibfnamefont {Sean}\
  \bibnamefont {Gavin}},\ }\bibfield  {title} {\enquote {\bibinfo {title}
  {{Causal diffusion and the survival of charge fluctuations in nuclear
  collisions}},}\ }\href {\doibase 10.1103/PhysRevC.70.034905} {\bibfield
  {journal} {\bibinfo  {journal} {Phys.Rev.}\ }\textbf {\bibinfo {volume}
  {C70}},\ \bibinfo {pages} {034905} (\bibinfo {year} {2004})},\ \Eprint
  {http://arxiv.org/abs/nucl-th/0404058} {arXiv:nucl-th/0404058 [nucl-th]}
  \BibitemShut {NoStop}%
%%CITATION = NUCL-TH/0404058;%%
\bibitem [{\citenamefont {Stephanov}(2009)}]{Stephanov:2008qz}%
  \BibitemOpen
  \bibfield  {author} {\bibinfo {author} {\bibfnamefont {M.A.}\ \bibnamefont
  {Stephanov}},\ }\bibfield  {title} {\enquote {\bibinfo {title} {{Non-Gaussian
  fluctuations near the QCD critical point}},}\ }\href {\doibase
  10.1103/PhysRevLett.102.032301} {\bibfield  {journal} {\bibinfo  {journal}
  {Phys.Rev.Lett.}\ }\textbf {\bibinfo {volume} {102}},\ \bibinfo {pages}
  {032301} (\bibinfo {year} {2009})},\ \Eprint {http://arxiv.org/abs/0809.3450}
  {arXiv:0809.3450 [hep-ph]} \BibitemShut {NoStop}%
%%CITATION = ARXIV:0809.3450;%%
\bibitem [{\citenamefont {Nahrgang}\ \emph {et~al.}(2012)\citenamefont
  {Nahrgang}, \citenamefont {Schuster}, \citenamefont {Mitrovski},
  \citenamefont {Stock},\ and\ \citenamefont {Bleicher}}]{Schuster:2009jv}%
  \BibitemOpen
  \bibfield  {author} {\bibinfo {author} {\bibfnamefont {Marlene}\ \bibnamefont
  {Nahrgang}}, \bibinfo {author} {\bibfnamefont {Tim}\ \bibnamefont
  {Schuster}}, \bibinfo {author} {\bibfnamefont {Michael}\ \bibnamefont
  {Mitrovski}}, \bibinfo {author} {\bibfnamefont {Reinhard}\ \bibnamefont
  {Stock}}, \ and\ \bibinfo {author} {\bibfnamefont {Marcus}\ \bibnamefont
  {Bleicher}},\ }\bibfield  {title} {\enquote {\bibinfo {title} {{Net-baryon-,
  net-proton-, and net-charge kurtosis in heavy-ion collisions within a
  relativistic transport approach}},}\ }\href {\doibase
  10.1140/epjc/s10052-012-2143-6} {\bibfield  {journal} {\bibinfo  {journal}
  {Eur.Phys.J.}\ }\textbf {\bibinfo {volume} {C72}},\ \bibinfo {pages} {2143}
  (\bibinfo {year} {2012})},\ \Eprint {http://arxiv.org/abs/0903.2911}
  {arXiv:0903.2911 [hep-ph]} \BibitemShut {NoStop}%
%%CITATION = ARXIV:0903.2911;%%
\bibitem [{\citenamefont {Athanasiou}\ \emph {et~al.}(2010)\citenamefont
  {Athanasiou}, \citenamefont {Rajagopal},\ and\ \citenamefont
  {Stephanov}}]{Athanasiou:2010kw}%
  \BibitemOpen
  \bibfield  {author} {\bibinfo {author} {\bibfnamefont {Christiana}\
  \bibnamefont {Athanasiou}}, \bibinfo {author} {\bibfnamefont {Krishna}\
  \bibnamefont {Rajagopal}}, \ and\ \bibinfo {author} {\bibfnamefont {Misha}\
  \bibnamefont {Stephanov}},\ }\bibfield  {title} {\enquote {\bibinfo {title}
  {{Using Higher Moments of Fluctuations and their Ratios in the Search for the
  QCD Critical Point}},}\ }\href {\doibase 10.1103/PhysRevD.82.074008}
  {\bibfield  {journal} {\bibinfo  {journal} {Phys.Rev.}\ }\textbf {\bibinfo
  {volume} {D82}},\ \bibinfo {pages} {074008} (\bibinfo {year} {2010})},\
  \Eprint {http://arxiv.org/abs/1006.4636} {arXiv:1006.4636 [hep-ph]}
  \BibitemShut {NoStop}%
%%CITATION = ARXIV:1006.4636;%%
\bibitem [{\citenamefont {Ling}\ \emph {et~al.}(2014)\citenamefont {Ling},
  \citenamefont {Springer},\ and\ \citenamefont {Stephanov}}]{Ling:2013ksb}%
  \BibitemOpen
  \bibfield  {author} {\bibinfo {author} {\bibfnamefont {Bo}~\bibnamefont
  {Ling}}, \bibinfo {author} {\bibfnamefont {Todd}\ \bibnamefont {Springer}}, \
  and\ \bibinfo {author} {\bibfnamefont {Mikhail}\ \bibnamefont {Stephanov}},\
  }\bibfield  {title} {\enquote {\bibinfo {title} {{Hydrodynamics of charge
  fluctuations and balance functions}},}\ }\href {\doibase
  10.1103/PhysRevC.89.064901} {\bibfield  {journal} {\bibinfo  {journal} {Phys.
  Rev.}\ }\textbf {\bibinfo {volume} {C89}},\ \bibinfo {pages} {064901}
  (\bibinfo {year} {2014})},\ \Eprint {http://arxiv.org/abs/1310.6036}
  {arXiv:1310.6036 [nucl-th]} \BibitemShut {NoStop}%
%%CITATION = ARXIV:1310.6036;%%
\bibitem [{\citenamefont {Albright}\ \emph {et~al.}(2015)\citenamefont
  {Albright}, \citenamefont {Kapusta},\ and\ \citenamefont
  {Young}}]{Albright:2015uua}%
  \BibitemOpen
  \bibfield  {author} {\bibinfo {author} {\bibfnamefont {M.}~\bibnamefont
  {Albright}}, \bibinfo {author} {\bibfnamefont {J.}~\bibnamefont {Kapusta}}, \
  and\ \bibinfo {author} {\bibfnamefont {C.}~\bibnamefont {Young}},\ }\bibfield
   {title} {\enquote {\bibinfo {title} {{Baryon Number Fluctuations from a
  Crossover Equation of State Compared to Heavy-Ion Collision Measurements in
  the Beam Energy Range $\sqrt{s_{NN}}$ = 7.7 to 200 GeV}},}\ }\href {\doibase
  10.1103/PhysRevC.92.044904} {\bibfield  {journal} {\bibinfo  {journal} {Phys.
  Rev.}\ }\textbf {\bibinfo {volume} {C92}},\ \bibinfo {pages} {044904}
  (\bibinfo {year} {2015})},\ \Eprint {http://arxiv.org/abs/1506.03408}
  {arXiv:1506.03408 [nucl-th]} \BibitemShut {NoStop}%
%%CITATION = ARXIV:1506.03408;%%
\bibitem [{\citenamefont {Mukherjee}\ \emph {et~al.}(2015)\citenamefont
  {Mukherjee}, \citenamefont {Venugopalan},\ and\ \citenamefont
  {Yin}}]{Mukherjee:2015swa}%
  \BibitemOpen
  \bibfield  {author} {\bibinfo {author} {\bibfnamefont {Swagato}\ \bibnamefont
  {Mukherjee}}, \bibinfo {author} {\bibfnamefont {Raju}\ \bibnamefont
  {Venugopalan}}, \ and\ \bibinfo {author} {\bibfnamefont {Yi}~\bibnamefont
  {Yin}},\ }\bibfield  {title} {\enquote {\bibinfo {title} {{Real time
  evolution of non-Gaussian cumulants in the QCD critical regime}},}\ }\href
  {\doibase 10.1103/PhysRevC.92.034912} {\bibfield  {journal} {\bibinfo
  {journal} {Phys. Rev.}\ }\textbf {\bibinfo {volume} {C92}},\ \bibinfo {pages}
  {034912} (\bibinfo {year} {2015})},\ \Eprint
  {http://arxiv.org/abs/1506.00645} {arXiv:1506.00645 [hep-ph]} \BibitemShut
  {NoStop}%
%%CITATION = ARXIV:1506.00645;%%
\bibitem [{\citenamefont {Stephanov}(2011)}]{Stephanov:2011pb}%
  \BibitemOpen
  \bibfield  {author} {\bibinfo {author} {\bibfnamefont {M.~A.}\ \bibnamefont
  {Stephanov}},\ }\bibfield  {title} {\enquote {\bibinfo {title} {{On the sign
  of kurtosis near the QCD critical point}},}\ }\href {\doibase
  10.1103/PhysRevLett.107.052301} {\bibfield  {journal} {\bibinfo  {journal}
  {Phys. Rev. Lett.}\ }\textbf {\bibinfo {volume} {107}},\ \bibinfo {pages}
  {052301} (\bibinfo {year} {2011})},\ \Eprint {http://arxiv.org/abs/1104.1627}
  {arXiv:1104.1627 [hep-ph]} \BibitemShut {NoStop}%
%%CITATION = ARXIV:1104.1627;%%
\bibitem [{\citenamefont {Adamczyk}\ \emph {et~al.}(2014)\citenamefont
  {Adamczyk} \emph {et~al.}}]{Adamczyk:2013dal}%
  \BibitemOpen
  \bibfield  {author} {\bibinfo {author} {\bibfnamefont {L.}~\bibnamefont
  {Adamczyk}} \emph {et~al.} (\bibinfo {collaboration} {STAR}),\ }\bibfield
  {title} {\enquote {\bibinfo {title} {{Energy Dependence of Moments of
  Net-proton Multiplicity Distributions at RHIC}},}\ }\href {\doibase
  10.1103/PhysRevLett.112.032302} {\bibfield  {journal} {\bibinfo  {journal}
  {Phys.Rev.Lett.}\ }\textbf {\bibinfo {volume} {112}},\ \bibinfo {pages}
  {032302} (\bibinfo {year} {2014})},\ \Eprint {http://arxiv.org/abs/1309.5681}
  {arXiv:1309.5681 [nucl-ex]} \BibitemShut {NoStop}%
%%CITATION = ARXIV:1309.5681;%%
\bibitem [{\citenamefont {Aggarwal}\ \emph {et~al.}(2010)\citenamefont
  {Aggarwal} \emph {et~al.}}]{Aggarwal:2010wy}%
  \BibitemOpen
  \bibfield  {author} {\bibinfo {author} {\bibfnamefont {M.M.}\ \bibnamefont
  {Aggarwal}} \emph {et~al.} (\bibinfo {collaboration} {STAR}),\ }\bibfield
  {title} {\enquote {\bibinfo {title} {{Higher Moments of Net-proton
  Multiplicity Distributions at RHIC}},}\ }\href {\doibase
  10.1103/PhysRevLett.105.022302} {\bibfield  {journal} {\bibinfo  {journal}
  {Phys.Rev.Lett.}\ }\textbf {\bibinfo {volume} {105}},\ \bibinfo {pages}
  {022302} (\bibinfo {year} {2010})},\ \Eprint {http://arxiv.org/abs/1004.4959}
  {arXiv:1004.4959 [nucl-ex]} \BibitemShut {NoStop}%
%%CITATION = ARXIV:1004.4959;%%
\bibitem [{\citenamefont {Abelev}\ \emph {et~al.}(2010)\citenamefont {Abelev}
  \emph {et~al.}}]{Abelev:2009bw}%
  \BibitemOpen
  \bibfield  {author} {\bibinfo {author} {\bibfnamefont {B.I.}\ \bibnamefont
  {Abelev}} \emph {et~al.} (\bibinfo {collaboration} {STAR}),\ }\bibfield
  {title} {\enquote {\bibinfo {title} {{Identified particle production,
  azimuthal anisotropy, and interferometry measurements in Au+Au collisions at
  s(NN)**(1/2) = 9.2- GeV}},}\ }\href {\doibase 10.1103/PhysRevC.81.024911}
  {\bibfield  {journal} {\bibinfo  {journal} {Phys.Rev.}\ }\textbf {\bibinfo
  {volume} {C81}},\ \bibinfo {pages} {024911} (\bibinfo {year} {2010})},\
  \Eprint {http://arxiv.org/abs/0909.4131} {arXiv:0909.4131 [nucl-ex]}
  \BibitemShut {NoStop}%
%%CITATION = ARXIV:0909.4131;%%
\bibitem [{\citenamefont {Sarkar}(2013)}]{Sarkar:2013iaa}%
  \BibitemOpen
  \bibfield  {author} {\bibinfo {author} {\bibfnamefont {Amal}\ \bibnamefont
  {Sarkar}} (\bibinfo {collaboration} {STAR}),\ }\bibfield  {title} {\enquote
  {\bibinfo {title} {{Higher moments of net-kaon multiplicity distributions at
  RHIC energies for the search of QCD critical point}},}\ }\bibfield
  {booktitle} {\emph {\bibinfo {booktitle} {{Proceedings, 8th International
  Workshop on Critical Point and Onset of Deconfinement (CPOD 2013)}}},\
  }\href@noop {} {\bibfield  {journal} {\bibinfo  {journal} {PoS}\ }\textbf
  {\bibinfo {volume} {CPOD2013}},\ \bibinfo {pages} {061} (\bibinfo {year}
  {2013})}\BibitemShut {NoStop}%
%%CITATION = POSCI,CPOD2013,061;%%
\bibitem [{\citenamefont {Landau}\ and\ \citenamefont
  {Lifshitz}(1980)}]{landau1980statistical}%
  \BibitemOpen
  \bibfield  {author} {\bibinfo {author} {\bibfnamefont {L.D.}\ \bibnamefont
  {Landau}}\ and\ \bibinfo {author} {\bibfnamefont {E.M.}\ \bibnamefont
  {Lifshitz}},\ }\href@noop {} {\emph {\bibinfo {title} {Statistical
  Physics,Part. 2}}},\ Course of theoretical physics\ (\bibinfo  {publisher}
  {Pergamon Press},\ \bibinfo {year} {1980})\BibitemShut {NoStop}%
\bibitem [{\citenamefont {Kovtun}(2012)}]{Kovtun:2012rj}%
  \BibitemOpen
  \bibfield  {author} {\bibinfo {author} {\bibfnamefont {Pavel}\ \bibnamefont
  {Kovtun}},\ }\bibfield  {title} {\enquote {\bibinfo {title} {{Lectures on
  hydrodynamic fluctuations in relativistic theories}},}\ }\href {\doibase
  10.1088/1751-8113/45/47/473001} {\bibfield  {journal} {\bibinfo  {journal}
  {J.Phys.}\ }\textbf {\bibinfo {volume} {A45}},\ \bibinfo {pages} {473001}
  (\bibinfo {year} {2012})},\ \Eprint {http://arxiv.org/abs/1205.5040}
  {arXiv:1205.5040 [hep-th]} \BibitemShut {NoStop}%
%%CITATION = ARXIV:1205.5040;%%
\bibitem [{\citenamefont {Kapusta}\ \emph {et~al.}(2012)\citenamefont
  {Kapusta}, \citenamefont {Muller},\ and\ \citenamefont
  {Stephanov}}]{Kapusta:2011gt}%
  \BibitemOpen
  \bibfield  {author} {\bibinfo {author} {\bibfnamefont {J.I.}\ \bibnamefont
  {Kapusta}}, \bibinfo {author} {\bibfnamefont {B.}~\bibnamefont {Muller}}, \
  and\ \bibinfo {author} {\bibfnamefont {M.}~\bibnamefont {Stephanov}},\
  }\bibfield  {title} {\enquote {\bibinfo {title} {{Relativistic Theory of
  Hydrodynamic Fluctuations with Applications to Heavy Ion Collisions}},}\
  }\href {\doibase 10.1103/PhysRevC.85.054906} {\bibfield  {journal} {\bibinfo
  {journal} {Phys.Rev.}\ }\textbf {\bibinfo {volume} {C85}},\ \bibinfo {pages}
  {054906} (\bibinfo {year} {2012})},\ \Eprint {http://arxiv.org/abs/1112.6405}
  {arXiv:1112.6405 [nucl-th]} \BibitemShut {NoStop}%
%%CITATION = ARXIV:1112.6405;%%
\bibitem [{\citenamefont {Hohenberg}\ and\ \citenamefont
  {Halperin}(1977)}]{Hohenberg:1977ym}%
  \BibitemOpen
  \bibfield  {author} {\bibinfo {author} {\bibfnamefont {P.~C.}\ \bibnamefont
  {Hohenberg}}\ and\ \bibinfo {author} {\bibfnamefont {B.~I.}\ \bibnamefont
  {Halperin}},\ }\bibfield  {title} {\enquote {\bibinfo {title} {{Theory of
  Dynamic Critical Phenomena}},}\ }\href {\doibase 10.1103/RevModPhys.49.435}
  {\bibfield  {journal} {\bibinfo  {journal} {Rev. Mod. Phys.}\ }\textbf
  {\bibinfo {volume} {49}},\ \bibinfo {pages} {435--479} (\bibinfo {year}
  {1977})}\BibitemShut {NoStop}%
%%CITATION = RMPHA,49,435;%%
\bibitem [{\citenamefont {Luettmer‐Strathmann}\ \emph
  {et~al.}(1995)\citenamefont {Luettmer‐Strathmann}, \citenamefont
  {Sengers},\ and\ \citenamefont {Olchowy}}]{strathmann1995}%
  \BibitemOpen
  \bibfield  {author} {\bibinfo {author} {\bibfnamefont {J.}~\bibnamefont
  {Luettmer‐Strathmann}}, \bibinfo {author} {\bibfnamefont {J.~V.}\
  \bibnamefont {Sengers}}, \ and\ \bibinfo {author} {\bibfnamefont {G.~A.}\
  \bibnamefont {Olchowy}},\ }\bibfield  {title} {\enquote {\bibinfo {title}
  {Non‐asymptotic critical behavior of the transport properties of fluids},}\
  }\href {\doibase http://dx.doi.org/10.1063/1.470718} {\bibfield  {journal}
  {\bibinfo  {journal} {The Journal of Chemical Physics}\ }\textbf {\bibinfo
  {volume} {103}},\ \bibinfo {pages} {7482--7501} (\bibinfo {year}
  {1995})}\BibitemShut {NoStop}%
\bibitem [{\citenamefont {Berg}\ and\ \citenamefont
  {Moldover}(1988)}]{Berg1988}%
  \BibitemOpen
  \bibfield  {author} {\bibinfo {author} {\bibfnamefont {Robert~F.}\
  \bibnamefont {Berg}}\ and\ \bibinfo {author} {\bibfnamefont {Michael~R.}\
  \bibnamefont {Moldover}},\ }\bibfield  {title} {\enquote {\bibinfo {title}
  {Critical exponent for the viscosity of four binary liquids},}\ }\href
  {\doibase http://dx.doi.org/10.1063/1.454890} {\bibfield  {journal} {\bibinfo
   {journal} {The Journal of Chemical Physics}\ }\textbf {\bibinfo {volume}
  {89}},\ \bibinfo {pages} {3694--3704} (\bibinfo {year} {1988})}\BibitemShut
  {NoStop}%
\bibitem [{\citenamefont {Kapusta}\ and\ \citenamefont
  {Torres-Rincon}(2012)}]{Kapusta:2012zb}%
  \BibitemOpen
  \bibfield  {author} {\bibinfo {author} {\bibfnamefont {Joseph~I.}\
  \bibnamefont {Kapusta}}\ and\ \bibinfo {author} {\bibfnamefont {Juan~M.}\
  \bibnamefont {Torres-Rincon}},\ }\bibfield  {title} {\enquote {\bibinfo
  {title} {{Thermal Conductivity and Chiral Critical Point in Heavy Ion
  Collisions}},}\ }\href {\doibase 10.1103/PhysRevC.86.054911} {\bibfield
  {journal} {\bibinfo  {journal} {Phys.Rev.}\ }\textbf {\bibinfo {volume}
  {C86}},\ \bibinfo {pages} {054911} (\bibinfo {year} {2012})},\ \Eprint
  {http://arxiv.org/abs/1209.0675} {arXiv:1209.0675 [nucl-th]} \BibitemShut
  {NoStop}%
%%CITATION = ARXIV:1209.0675;%%
\bibitem [{\citenamefont {Martinez}\ \emph {et~al.}()\citenamefont {Martinez},
  \citenamefont {Sievert},\ and\ \citenamefont {Wertepny}}]{DMS}%
  \BibitemOpen
  \bibfield  {author} {\bibinfo {author} {\bibfnamefont {M.}~\bibnamefont
  {Martinez}}, \bibinfo {author} {\bibfnamefont {Matthew~D.}\ \bibnamefont
  {Sievert}}, \ and\ \bibinfo {author} {\bibfnamefont {Douglas~E.}\
  \bibnamefont {Wertepny}},\ }\bibfield  {title} {\enquote {\bibinfo {title}
  {{Baryon Number Fluctuations and Quark Correlations in the CGC Framework}},}\
  }\href@noop {} {\ }\bibinfo {note} {Work in progress}\BibitemShut {NoStop}%
\bibitem [{\citenamefont {Bjorken}(1983)}]{Bjorken:1982qr}%
  \BibitemOpen
  \bibfield  {author} {\bibinfo {author} {\bibfnamefont {J.D.}\ \bibnamefont
  {Bjorken}},\ }\bibfield  {title} {\enquote {\bibinfo {title} {{Highly
  Relativistic Nucleus-Nucleus Collisions: The Central Rapidity Region}},}\
  }\href {\doibase 10.1103/PhysRevD.27.140} {\bibfield  {journal} {\bibinfo
  {journal} {Phys.Rev.}\ }\textbf {\bibinfo {volume} {D27}},\ \bibinfo {pages}
  {140--151} (\bibinfo {year} {1983})}\BibitemShut {NoStop}%
%%CITATION = PHRVA,D27,140;%%
\bibitem [{\citenamefont {Floerchinger}\ and\ \citenamefont
  {Wiedemann}(2011)}]{Florchinger:2011qf}%
  \BibitemOpen
  \bibfield  {author} {\bibinfo {author} {\bibfnamefont {Stefan}\ \bibnamefont
  {Floerchinger}}\ and\ \bibinfo {author} {\bibfnamefont {Urs~Achim}\
  \bibnamefont {Wiedemann}},\ }\bibfield  {title} {\enquote {\bibinfo {title}
  {{Fluctuations around Bjorken Flow and the onset of turbulent phenomena}},}\
  }\href {\doibase 10.1007/JHEP11(2011)100} {\bibfield  {journal} {\bibinfo
  {journal} {JHEP}\ }\textbf {\bibinfo {volume} {1111}},\ \bibinfo {pages}
  {100} (\bibinfo {year} {2011})},\ \Eprint {http://arxiv.org/abs/1108.5535}
  {arXiv:1108.5535 [nucl-th]} \BibitemShut {NoStop}%
%%CITATION = ARXIV:1108.5535;%%
\bibitem [{\citenamefont {Floerchinger}\ and\ \citenamefont
  {Wiedemann}(2014{\natexlab{a}})}]{Floerchinger:2013rya}%
  \BibitemOpen
  \bibfield  {author} {\bibinfo {author} {\bibfnamefont {Stefan}\ \bibnamefont
  {Floerchinger}}\ and\ \bibinfo {author} {\bibfnamefont {Urs~Achim}\
  \bibnamefont {Wiedemann}},\ }\bibfield  {title} {\enquote {\bibinfo {title}
  {{Mode-by-mode fluid dynamics for relativistic heavy ion collisions}},}\
  }\href {\doibase 10.1016/j.physletb.2013.12.025} {\bibfield  {journal}
  {\bibinfo  {journal} {Phys.Lett.}\ }\textbf {\bibinfo {volume} {B728}},\
  \bibinfo {pages} {407--411} (\bibinfo {year} {2014}{\natexlab{a}})},\ \Eprint
  {http://arxiv.org/abs/1307.3453} {arXiv:1307.3453 [hep-ph]} \BibitemShut
  {NoStop}%
%%CITATION = ARXIV:1307.3453;%%
\bibitem [{\citenamefont {Floerchinger}\ and\ \citenamefont
  {Wiedemann}(2013)}]{Floerchinger:2013vua}%
  \BibitemOpen
  \bibfield  {author} {\bibinfo {author} {\bibfnamefont {Stefan}\ \bibnamefont
  {Floerchinger}}\ and\ \bibinfo {author} {\bibfnamefont {Urs~Achim}\
  \bibnamefont {Wiedemann}},\ }\bibfield  {title} {\enquote {\bibinfo {title}
  {{Characterization of initial fluctuations for the hydrodynamical description
  of heavy ion collisions}},}\ }\href {\doibase 10.1103/PhysRevC.88.044906}
  {\bibfield  {journal} {\bibinfo  {journal} {Phys.Rev.}\ }\textbf {\bibinfo
  {volume} {C88}},\ \bibinfo {pages} {044906} (\bibinfo {year} {2013})},\
  \Eprint {http://arxiv.org/abs/1307.7611} {arXiv:1307.7611 [hep-ph]}
  \BibitemShut {NoStop}%
%%CITATION = ARXIV:1307.7611;%%
\bibitem [{\citenamefont {Floerchinger}\ \emph {et~al.}(2014)\citenamefont
  {Floerchinger}, \citenamefont {Wiedemann}, \citenamefont {Beraudo},
  \citenamefont {Del~Zanna}, \citenamefont {Inghirami} \emph
  {et~al.}}]{Floerchinger:2013tya}%
  \BibitemOpen
  \bibfield  {author} {\bibinfo {author} {\bibfnamefont {Stefan}\ \bibnamefont
  {Floerchinger}}, \bibinfo {author} {\bibfnamefont {Urs~Achim}\ \bibnamefont
  {Wiedemann}}, \bibinfo {author} {\bibfnamefont {Andrea}\ \bibnamefont
  {Beraudo}}, \bibinfo {author} {\bibfnamefont {Luca}\ \bibnamefont
  {Del~Zanna}}, \bibinfo {author} {\bibfnamefont {Gabriele}\ \bibnamefont
  {Inghirami}},  \emph {et~al.},\ }\bibfield  {title} {\enquote {\bibinfo
  {title} {{How (non-)linear is the hydrodynamics of heavy ion collisions?}}}\
  }\href {\doibase 10.1016/j.physletb.2014.06.049} {\bibfield  {journal}
  {\bibinfo  {journal} {Phys.Lett.}\ }\textbf {\bibinfo {volume} {B735}},\
  \bibinfo {pages} {305--310} (\bibinfo {year} {2014})},\ \Eprint
  {http://arxiv.org/abs/1312.5482} {arXiv:1312.5482 [hep-ph]} \BibitemShut
  {NoStop}%
%%CITATION = ARXIV:1312.5482;%%
\bibitem [{\citenamefont {Floerchinger}\ and\ \citenamefont
  {Wiedemann}(2014{\natexlab{b}})}]{Floerchinger:2013hza}%
  \BibitemOpen
  \bibfield  {author} {\bibinfo {author} {\bibfnamefont {Stefan}\ \bibnamefont
  {Floerchinger}}\ and\ \bibinfo {author} {\bibfnamefont {Urs~Achim}\
  \bibnamefont {Wiedemann}},\ }\bibfield  {title} {\enquote {\bibinfo {title}
  {{Kinetic freeze-out, particle spectra and harmonic flow coefficients from
  mode-by-mode hydrodynamics}},}\ }\href {\doibase 10.1103/PhysRevC.89.034914}
  {\bibfield  {journal} {\bibinfo  {journal} {Phys.Rev.}\ }\textbf {\bibinfo
  {volume} {C89}},\ \bibinfo {pages} {034914} (\bibinfo {year}
  {2014}{\natexlab{b}})},\ \Eprint {http://arxiv.org/abs/1311.7613}
  {arXiv:1311.7613 [hep-ph]} \BibitemShut {NoStop}%
%%CITATION = ARXIV:1311.7613;%%
\bibitem [{\citenamefont {Floerchinger}\ and\ \citenamefont
  {Wiedemann}(2014{\natexlab{c}})}]{Floerchinger:2014fta}%
  \BibitemOpen
  \bibfield  {author} {\bibinfo {author} {\bibfnamefont {Stefan}\ \bibnamefont
  {Floerchinger}}\ and\ \bibinfo {author} {\bibfnamefont {Urs~Achim}\
  \bibnamefont {Wiedemann}},\ }\bibfield  {title} {\enquote {\bibinfo {title}
  {{Statistics of initial density perturbations in heavy ion collisions and
  their fluid dynamic response}},}\ }\href {\doibase 10.1007/JHEP08(2014)005}
  {\bibfield  {journal} {\bibinfo  {journal} {JHEP}\ }\textbf {\bibinfo
  {volume} {1408}},\ \bibinfo {pages} {005} (\bibinfo {year}
  {2014}{\natexlab{c}})},\ \Eprint {http://arxiv.org/abs/1405.4393}
  {arXiv:1405.4393 [hep-ph]} \BibitemShut {NoStop}%
%%CITATION = ARXIV:1405.4393;%%
\bibitem [{\citenamefont {Brouzakis}\ \emph {et~al.}(2015)\citenamefont
  {Brouzakis}, \citenamefont {Floerchinger}, \citenamefont {Tetradis},\ and\
  \citenamefont {Wiedemann}}]{Brouzakis:2014gka}%
  \BibitemOpen
  \bibfield  {author} {\bibinfo {author} {\bibfnamefont {Nikolaos}\
  \bibnamefont {Brouzakis}}, \bibinfo {author} {\bibfnamefont {Stefan}\
  \bibnamefont {Floerchinger}}, \bibinfo {author} {\bibfnamefont {Nikolaos}\
  \bibnamefont {Tetradis}}, \ and\ \bibinfo {author} {\bibfnamefont
  {Urs~Achim}\ \bibnamefont {Wiedemann}},\ }\bibfield  {title} {\enquote
  {\bibinfo {title} {{Nonlinear evolution of density and flow perturbations on
  a Bjorken background}},}\ }\href {\doibase 10.1103/PhysRevD.91.065007}
  {\bibfield  {journal} {\bibinfo  {journal} {Phys.Rev.}\ }\textbf {\bibinfo
  {volume} {D91}},\ \bibinfo {pages} {065007} (\bibinfo {year} {2015})},\
  \Eprint {http://arxiv.org/abs/1411.2912} {arXiv:1411.2912 [hep-ph]}
  \BibitemShut {NoStop}%
%%CITATION = ARXIV:1411.2912;%%
\bibitem [{\citenamefont {Floerchinger}\ and\ \citenamefont
  {Zapp}(2014)}]{Floerchinger:2014yqa}%
  \BibitemOpen
  \bibfield  {author} {\bibinfo {author} {\bibfnamefont {Stefan}\ \bibnamefont
  {Floerchinger}}\ and\ \bibinfo {author} {\bibfnamefont {Korinna~C.}\
  \bibnamefont {Zapp}},\ }\bibfield  {title} {\enquote {\bibinfo {title}
  {{Hydrodynamics and Jets in Dialogue}},}\ }\href {\doibase
  10.1140/epjc/s10052-014-3189-4} {\bibfield  {journal} {\bibinfo  {journal}
  {Eur.Phys.J.}\ }\textbf {\bibinfo {volume} {C74}},\ \bibinfo {pages} {3189}
  (\bibinfo {year} {2014})},\ \Eprint {http://arxiv.org/abs/1407.1782}
  {arXiv:1407.1782 [hep-ph]} \BibitemShut {NoStop}%
%%CITATION = ARXIV:1407.1782;%%
\bibitem [{\citenamefont {Hiscock}\ and\ \citenamefont
  {Lindblom}(1983)}]{Hiscock:1983zz}%
  \BibitemOpen
  \bibfield  {author} {\bibinfo {author} {\bibfnamefont {W.A.}\ \bibnamefont
  {Hiscock}}\ and\ \bibinfo {author} {\bibfnamefont {L.}~\bibnamefont
  {Lindblom}},\ }\bibfield  {title} {\enquote {\bibinfo {title} {{Stability and
  causality in dissipative relativistic fluids}},}\ }\href {\doibase
  10.1016/0003-4916(83)90288-9} {\bibfield  {journal} {\bibinfo  {journal}
  {Annals Phys.}\ }\textbf {\bibinfo {volume} {151}},\ \bibinfo {pages}
  {466--496} (\bibinfo {year} {1983})}\BibitemShut {NoStop}%
%%CITATION = APNYA,151,466;%%
\bibitem [{\citenamefont {Pu}\ \emph {et~al.}(2010)\citenamefont {Pu},
  \citenamefont {Koide},\ and\ \citenamefont {Rischke}}]{Pu:2009fj}%
  \BibitemOpen
  \bibfield  {author} {\bibinfo {author} {\bibfnamefont {Shi}\ \bibnamefont
  {Pu}}, \bibinfo {author} {\bibfnamefont {Tomoi}\ \bibnamefont {Koide}}, \
  and\ \bibinfo {author} {\bibfnamefont {Dirk~H.}\ \bibnamefont {Rischke}},\
  }\bibfield  {title} {\enquote {\bibinfo {title} {{Does stability of
  relativistic dissipative fluid dynamics imply causality?}}}\ }\href {\doibase
  10.1103/PhysRevD.81.114039} {\bibfield  {journal} {\bibinfo  {journal}
  {Phys.Rev.}\ }\textbf {\bibinfo {volume} {D81}},\ \bibinfo {pages} {114039}
  (\bibinfo {year} {2010})},\ \Eprint {http://arxiv.org/abs/0907.3906}
  {arXiv:0907.3906 [hep-ph]} \BibitemShut {NoStop}%
%%CITATION = ARXIV:0907.3906;%%
\bibitem [{\citenamefont {Bazavov}\ \emph {et~al.}(2014)\citenamefont {Bazavov}
  \emph {et~al.}}]{Bazavov:2014pvz}%
  \BibitemOpen
  \bibfield  {author} {\bibinfo {author} {\bibfnamefont {A.}~\bibnamefont
  {Bazavov}} \emph {et~al.} (\bibinfo {collaboration} {HotQCD Collaboration}),\
  }\bibfield  {title} {\enquote {\bibinfo {title} {{Equation of state in ( 2+1
  )-flavor QCD}},}\ }\href {\doibase 10.1103/PhysRevD.90.094503} {\bibfield
  {journal} {\bibinfo  {journal} {Phys.Rev.}\ }\textbf {\bibinfo {volume}
  {D90}},\ \bibinfo {pages} {094503} (\bibinfo {year} {2014})},\ \Eprint
  {http://arxiv.org/abs/1407.6387} {arXiv:1407.6387 [hep-lat]} \BibitemShut
  {NoStop}%
%%CITATION = ARXIV:1407.6387;%%
\bibitem [{\citenamefont {Borsanyi}\ \emph {et~al.}(2014)\citenamefont
  {Borsanyi}, \citenamefont {Fodor}, \citenamefont {Hoelbling}, \citenamefont
  {Katz}, \citenamefont {Krieg} \emph {et~al.}}]{Borsanyi:2013bia}%
  \BibitemOpen
  \bibfield  {author} {\bibinfo {author} {\bibfnamefont {Szabocls}\
  \bibnamefont {Borsanyi}}, \bibinfo {author} {\bibfnamefont {Zoltan}\
  \bibnamefont {Fodor}}, \bibinfo {author} {\bibfnamefont {Christian}\
  \bibnamefont {Hoelbling}}, \bibinfo {author} {\bibfnamefont {Sandor~D.}\
  \bibnamefont {Katz}}, \bibinfo {author} {\bibfnamefont {Stefan}\ \bibnamefont
  {Krieg}},  \emph {et~al.},\ }\bibfield  {title} {\enquote {\bibinfo {title}
  {{Full result for the QCD equation of state with 2+1 flavors}},}\ }\href
  {\doibase 10.1016/j.physletb.2014.01.007} {\bibfield  {journal} {\bibinfo
  {journal} {Phys.Lett.}\ }\textbf {\bibinfo {volume} {B730}},\ \bibinfo
  {pages} {99--104} (\bibinfo {year} {2014})},\ \Eprint
  {http://arxiv.org/abs/1309.5258} {arXiv:1309.5258 [hep-lat]} \BibitemShut
  {NoStop}%
%%CITATION = ARXIV:1309.5258;%%
\bibitem [{\citenamefont {Haque}\ \emph {et~al.}(2013)\citenamefont {Haque},
  \citenamefont {Mustafa},\ and\ \citenamefont {Strickland}}]{Haque:2012my}%
  \BibitemOpen
  \bibfield  {author} {\bibinfo {author} {\bibfnamefont {Najmul}\ \bibnamefont
  {Haque}}, \bibinfo {author} {\bibfnamefont {Munshi~G.}\ \bibnamefont
  {Mustafa}}, \ and\ \bibinfo {author} {\bibfnamefont {Michael}\ \bibnamefont
  {Strickland}},\ }\bibfield  {title} {\enquote {\bibinfo {title} {{Two-loop
  HTL pressure at finite temperature and chemical potential}},}\ }\href
  {\doibase 10.1103/PhysRevD.87.105007} {\bibfield  {journal} {\bibinfo
  {journal} {Phys.Rev.}\ }\textbf {\bibinfo {volume} {D87}},\ \bibinfo {pages}
  {105007} (\bibinfo {year} {2013})},\ \Eprint {http://arxiv.org/abs/1212.1797}
  {arXiv:1212.1797 [hep-ph]} \BibitemShut {NoStop}%
%%CITATION = ARXIV:1212.1797;%%
\bibitem [{\citenamefont {Haque}\ \emph
  {et~al.}(2014{\natexlab{a}})\citenamefont {Haque}, \citenamefont
  {Bandyopadhyay}, \citenamefont {Andersen}, \citenamefont {Mustafa},
  \citenamefont {Strickland},\ and\ \citenamefont {Su}}]{Haque:2014rua}%
  \BibitemOpen
  \bibfield  {author} {\bibinfo {author} {\bibfnamefont {Najmul}\ \bibnamefont
  {Haque}}, \bibinfo {author} {\bibfnamefont {Aritra}\ \bibnamefont
  {Bandyopadhyay}}, \bibinfo {author} {\bibfnamefont {Jens~O.}\ \bibnamefont
  {Andersen}}, \bibinfo {author} {\bibfnamefont {Munshi~G.}\ \bibnamefont
  {Mustafa}}, \bibinfo {author} {\bibfnamefont {Michael}\ \bibnamefont
  {Strickland}}, \ and\ \bibinfo {author} {\bibfnamefont {Nan}\ \bibnamefont
  {Su}},\ }\bibfield  {title} {\enquote {\bibinfo {title} {{Three-loop HTLpt
  thermodynamics at finite temperature and chemical potential}},}\ }\href
  {\doibase 10.1007/JHEP05(2014)027} {\bibfield  {journal} {\bibinfo  {journal}
  {JHEP}\ }\textbf {\bibinfo {volume} {05}},\ \bibinfo {pages} {027} (\bibinfo
  {year} {2014}{\natexlab{a}})},\ \Eprint {http://arxiv.org/abs/1402.6907}
  {arXiv:1402.6907 [hep-ph]} \BibitemShut {NoStop}%
%%CITATION = ARXIV:1402.6907;%%
\bibitem [{\citenamefont {Haque}\ \emph
  {et~al.}(2014{\natexlab{b}})\citenamefont {Haque}, \citenamefont {Andersen},
  \citenamefont {Mustafa}, \citenamefont {Strickland},\ and\ \citenamefont
  {Su}}]{Haque:2013sja}%
  \BibitemOpen
  \bibfield  {author} {\bibinfo {author} {\bibfnamefont {Najmul}\ \bibnamefont
  {Haque}}, \bibinfo {author} {\bibfnamefont {Jens~O.}\ \bibnamefont
  {Andersen}}, \bibinfo {author} {\bibfnamefont {Munshi~G.}\ \bibnamefont
  {Mustafa}}, \bibinfo {author} {\bibfnamefont {Michael}\ \bibnamefont
  {Strickland}}, \ and\ \bibinfo {author} {\bibfnamefont {Nan}\ \bibnamefont
  {Su}},\ }\bibfield  {title} {\enquote {\bibinfo {title} {{Three-loop pressure
  and susceptibility at finite temperature and density from hard-thermal-loop
  perturbation theory}},}\ }\href {\doibase 10.1103/PhysRevD.89.061701}
  {\bibfield  {journal} {\bibinfo  {journal} {Phys. Rev.}\ }\textbf {\bibinfo
  {volume} {D89}},\ \bibinfo {pages} {061701} (\bibinfo {year}
  {2014}{\natexlab{b}})},\ \Eprint {http://arxiv.org/abs/1309.3968}
  {arXiv:1309.3968 [hep-ph]} \BibitemShut {NoStop}%
%%CITATION = ARXIV:1309.3968;%%
\bibitem [{\citenamefont {Vuorinen}(2003)}]{Vuorinen:2003fs}%
  \BibitemOpen
  \bibfield  {author} {\bibinfo {author} {\bibfnamefont {A.}~\bibnamefont
  {Vuorinen}},\ }\bibfield  {title} {\enquote {\bibinfo {title} {{The Pressure
  of QCD at finite temperatures and chemical potentials}},}\ }\href {\doibase
  10.1103/PhysRevD.68.054017} {\bibfield  {journal} {\bibinfo  {journal}
  {Phys.Rev.}\ }\textbf {\bibinfo {volume} {D68}},\ \bibinfo {pages} {054017}
  (\bibinfo {year} {2003})},\ \Eprint {http://arxiv.org/abs/hep-ph/0305183}
  {arXiv:hep-ph/0305183 [hep-ph]} \BibitemShut {NoStop}%
%%CITATION = HEP-PH/0305183;%%
\bibitem [{\citenamefont {de~Forcrand}(2009)}]{deForcrand:2010ys}%
  \BibitemOpen
  \bibfield  {author} {\bibinfo {author} {\bibfnamefont {Philippe}\
  \bibnamefont {de~Forcrand}},\ }\bibfield  {title} {\enquote {\bibinfo {title}
  {{Simulating QCD at finite density}},}\ }\href@noop {} {\bibfield  {journal}
  {\bibinfo  {journal} {PoS}\ }\textbf {\bibinfo {volume} {LAT2009}},\ \bibinfo
  {pages} {010} (\bibinfo {year} {2009})},\ \Eprint
  {http://arxiv.org/abs/1005.0539} {arXiv:1005.0539 [hep-lat]} \BibitemShut
  {NoStop}%
%%CITATION = ARXIV:1005.0539;%%
\bibitem [{\citenamefont {Huovinen}\ and\ \citenamefont
  {Petreczky}(2010)}]{Huovinen:2009yb}%
  \BibitemOpen
  \bibfield  {author} {\bibinfo {author} {\bibfnamefont {Pasi}\ \bibnamefont
  {Huovinen}}\ and\ \bibinfo {author} {\bibfnamefont {Pter}\ \bibnamefont
  {Petreczky}},\ }\bibfield  {title} {\enquote {\bibinfo {title} {{QCD Equation
  of State and Hadron Resonance Gas}},}\ }\href {\doibase
  10.1016/j.nuclphysa.2010.02.015} {\bibfield  {journal} {\bibinfo  {journal}
  {Nucl.Phys.}\ }\textbf {\bibinfo {volume} {A837}},\ \bibinfo {pages} {26--53}
  (\bibinfo {year} {2010})},\ \Eprint {http://arxiv.org/abs/0912.2541}
  {arXiv:0912.2541 [hep-ph]} \BibitemShut {NoStop}%
%%CITATION = ARXIV:0912.2541;%%
\bibitem [{\citenamefont {Csikor}\ \emph {et~al.}(2004)\citenamefont {Csikor},
  \citenamefont {Egri}, \citenamefont {Fodor}, \citenamefont {Katz},
  \citenamefont {Szabo},\ and\ \citenamefont {Toth}}]{Csikor:2004ik}%
  \BibitemOpen
  \bibfield  {author} {\bibinfo {author} {\bibfnamefont {F.}~\bibnamefont
  {Csikor}}, \bibinfo {author} {\bibfnamefont {G.~I.}\ \bibnamefont {Egri}},
  \bibinfo {author} {\bibfnamefont {Z.}~\bibnamefont {Fodor}}, \bibinfo
  {author} {\bibfnamefont {S.~D.}\ \bibnamefont {Katz}}, \bibinfo {author}
  {\bibfnamefont {K.~K.}\ \bibnamefont {Szabo}}, \ and\ \bibinfo {author}
  {\bibfnamefont {A.~I.}\ \bibnamefont {Toth}},\ }\bibfield  {title} {\enquote
  {\bibinfo {title} {{Equation of state at finite temperature and chemical
  potential, lattice QCD results}},}\ }\href {\doibase
  10.1088/1126-6708/2004/05/046} {\bibfield  {journal} {\bibinfo  {journal}
  {JHEP}\ }\textbf {\bibinfo {volume} {05}},\ \bibinfo {pages} {046} (\bibinfo
  {year} {2004})},\ \Eprint {http://arxiv.org/abs/hep-lat/0401016}
  {arXiv:hep-lat/0401016 [hep-lat]} \BibitemShut {NoStop}%
%%CITATION = HEP-LAT/0401016;%%
\bibitem [{\citenamefont {Fodor}\ \emph {et~al.}(2003)\citenamefont {Fodor},
  \citenamefont {Katz},\ and\ \citenamefont {Szabo}}]{Fodor:2002km}%
  \BibitemOpen
  \bibfield  {author} {\bibinfo {author} {\bibfnamefont {Z.}~\bibnamefont
  {Fodor}}, \bibinfo {author} {\bibfnamefont {S.~D.}\ \bibnamefont {Katz}}, \
  and\ \bibinfo {author} {\bibfnamefont {K.~K.}\ \bibnamefont {Szabo}},\
  }\bibfield  {title} {\enquote {\bibinfo {title} {{The QCD equation of state
  at nonzero densities: Lattice result}},}\ }\href {\doibase
  10.1016/j.physletb.2003.06.011} {\bibfield  {journal} {\bibinfo  {journal}
  {Phys. Lett.}\ }\textbf {\bibinfo {volume} {B568}},\ \bibinfo {pages}
  {73--77} (\bibinfo {year} {2003})},\ \Eprint
  {http://arxiv.org/abs/hep-lat/0208078} {arXiv:hep-lat/0208078 [hep-lat]}
  \BibitemShut {NoStop}%
%%CITATION = HEP-LAT/0208078;%%
\bibitem [{\citenamefont {Allton}\ \emph {et~al.}(2003)\citenamefont {Allton},
  \citenamefont {Ejiri}, \citenamefont {Hands}, \citenamefont {Kaczmarek},
  \citenamefont {Karsch}, \citenamefont {Laermann},\ and\ \citenamefont
  {Schmidt}}]{Allton:2003vx}%
  \BibitemOpen
  \bibfield  {author} {\bibinfo {author} {\bibfnamefont {C.~R.}\ \bibnamefont
  {Allton}}, \bibinfo {author} {\bibfnamefont {S.}~\bibnamefont {Ejiri}},
  \bibinfo {author} {\bibfnamefont {S.~J.}\ \bibnamefont {Hands}}, \bibinfo
  {author} {\bibfnamefont {O.}~\bibnamefont {Kaczmarek}}, \bibinfo {author}
  {\bibfnamefont {F.}~\bibnamefont {Karsch}}, \bibinfo {author} {\bibfnamefont
  {E.}~\bibnamefont {Laermann}}, \ and\ \bibinfo {author} {\bibfnamefont
  {C.}~\bibnamefont {Schmidt}},\ }\bibfield  {title} {\enquote {\bibinfo
  {title} {{The Equation of state for two flavor QCD at nonzero chemical
  potential}},}\ }\href {\doibase 10.1103/PhysRevD.68.014507} {\bibfield
  {journal} {\bibinfo  {journal} {Phys. Rev.}\ }\textbf {\bibinfo {volume}
  {D68}},\ \bibinfo {pages} {014507} (\bibinfo {year} {2003})},\ \Eprint
  {http://arxiv.org/abs/hep-lat/0305007} {arXiv:hep-lat/0305007 [hep-lat]}
  \BibitemShut {NoStop}%
%%CITATION = HEP-LAT/0305007;%%
\bibitem [{\citenamefont {Allton}\ \emph {et~al.}(2005)\citenamefont {Allton},
  \citenamefont {Doring}, \citenamefont {Ejiri}, \citenamefont {Hands},
  \citenamefont {Kaczmarek}, \citenamefont {Karsch}, \citenamefont {Laermann},\
  and\ \citenamefont {Redlich}}]{Allton:2005gk}%
  \BibitemOpen
  \bibfield  {author} {\bibinfo {author} {\bibfnamefont {C.~R.}\ \bibnamefont
  {Allton}}, \bibinfo {author} {\bibfnamefont {M.}~\bibnamefont {Doring}},
  \bibinfo {author} {\bibfnamefont {S.}~\bibnamefont {Ejiri}}, \bibinfo
  {author} {\bibfnamefont {S.~J.}\ \bibnamefont {Hands}}, \bibinfo {author}
  {\bibfnamefont {O.}~\bibnamefont {Kaczmarek}}, \bibinfo {author}
  {\bibfnamefont {F.}~\bibnamefont {Karsch}}, \bibinfo {author} {\bibfnamefont
  {E.}~\bibnamefont {Laermann}}, \ and\ \bibinfo {author} {\bibfnamefont
  {K.}~\bibnamefont {Redlich}},\ }\bibfield  {title} {\enquote {\bibinfo
  {title} {{Thermodynamics of two flavor QCD to sixth order in quark chemical
  potential}},}\ }\href {\doibase 10.1103/PhysRevD.71.054508} {\bibfield
  {journal} {\bibinfo  {journal} {Phys. Rev.}\ }\textbf {\bibinfo {volume}
  {D71}},\ \bibinfo {pages} {054508} (\bibinfo {year} {2005})},\ \Eprint
  {http://arxiv.org/abs/hep-lat/0501030} {arXiv:hep-lat/0501030 [hep-lat]}
  \BibitemShut {NoStop}%
%%CITATION = HEP-LAT/0501030;%%
\bibitem [{\citenamefont {Bernard}\ \emph {et~al.}(2008)\citenamefont
  {Bernard}, \citenamefont {DeTar}, \citenamefont {Levkova}, \citenamefont
  {Gottlieb}, \citenamefont {Heller}, \citenamefont {Hetrick}, \citenamefont
  {Sugar},\ and\ \citenamefont {Toussaint}}]{Bernard:2007nm}%
  \BibitemOpen
  \bibfield  {author} {\bibinfo {author} {\bibfnamefont {C.}~\bibnamefont
  {Bernard}}, \bibinfo {author} {\bibfnamefont {Carleton~E.}\ \bibnamefont
  {DeTar}}, \bibinfo {author} {\bibfnamefont {L.}~\bibnamefont {Levkova}},
  \bibinfo {author} {\bibfnamefont {Steven}\ \bibnamefont {Gottlieb}}, \bibinfo
  {author} {\bibfnamefont {U.~M.}\ \bibnamefont {Heller}}, \bibinfo {author}
  {\bibfnamefont {J.~E.}\ \bibnamefont {Hetrick}}, \bibinfo {author}
  {\bibfnamefont {R.}~\bibnamefont {Sugar}}, \ and\ \bibinfo {author}
  {\bibfnamefont {D.}~\bibnamefont {Toussaint}},\ }\bibfield  {title} {\enquote
  {\bibinfo {title} {{QCD thermodynamics with 2+1 flavors at nonzero chemical
  potential}},}\ }\href {\doibase 10.1103/PhysRevD.77.014503} {\bibfield
  {journal} {\bibinfo  {journal} {Phys. Rev.}\ }\textbf {\bibinfo {volume}
  {D77}},\ \bibinfo {pages} {014503} (\bibinfo {year} {2008})},\ \Eprint
  {http://arxiv.org/abs/0710.1330} {arXiv:0710.1330 [hep-lat]} \BibitemShut
  {NoStop}%
%%CITATION = ARXIV:0710.1330;%%
\bibitem [{\citenamefont {Basak}\ \emph {et~al.}(2008)\citenamefont {Basak}
  \emph {et~al.}}]{Basak:2009uv}%
  \BibitemOpen
  \bibfield  {author} {\bibinfo {author} {\bibfnamefont {S.}~\bibnamefont
  {Basak}} \emph {et~al.} (\bibinfo {collaboration} {MILC}),\ }\bibfield
  {title} {\enquote {\bibinfo {title} {{QCD equation of state at non-zero
  chemical potential}},}\ }\bibfield  {booktitle} {\emph {\bibinfo {booktitle}
  {{Proceedings, 26th International Symposium on Lattice field theory (Lattice
  2008)}}},\ }\href@noop {} {\bibfield  {journal} {\bibinfo  {journal} {PoS}\
  }\textbf {\bibinfo {volume} {LATTICE2008}},\ \bibinfo {pages} {171} (\bibinfo
  {year} {2008})},\ \Eprint {http://arxiv.org/abs/0910.0276} {arXiv:0910.0276
  [hep-lat]} \BibitemShut {NoStop}%
%%CITATION = ARXIV:0910.0276;%%
\bibitem [{\citenamefont {DeTar}\ \emph {et~al.}(2010)\citenamefont {DeTar},
  \citenamefont {Levkova}, \citenamefont {Gottlieb}, \citenamefont {Heller},
  \citenamefont {Hetrick}, \citenamefont {Sugar},\ and\ \citenamefont
  {Toussaint}}]{DeTar:2010xm}%
  \BibitemOpen
  \bibfield  {author} {\bibinfo {author} {\bibfnamefont {C.}~\bibnamefont
  {DeTar}}, \bibinfo {author} {\bibfnamefont {L.}~\bibnamefont {Levkova}},
  \bibinfo {author} {\bibfnamefont {Steven}\ \bibnamefont {Gottlieb}}, \bibinfo
  {author} {\bibfnamefont {U.~M.}\ \bibnamefont {Heller}}, \bibinfo {author}
  {\bibfnamefont {J.~E.}\ \bibnamefont {Hetrick}}, \bibinfo {author}
  {\bibfnamefont {R.}~\bibnamefont {Sugar}}, \ and\ \bibinfo {author}
  {\bibfnamefont {D.}~\bibnamefont {Toussaint}},\ }\bibfield  {title} {\enquote
  {\bibinfo {title} {{QCD thermodynamics with nonzero chemical potential at
  $N\_{t}=6$ and effects from heavy quarks}},}\ }\href {\doibase
  10.1103/PhysRevD.81.114504} {\bibfield  {journal} {\bibinfo  {journal} {Phys.
  Rev.}\ }\textbf {\bibinfo {volume} {D81}},\ \bibinfo {pages} {114504}
  (\bibinfo {year} {2010})},\ \Eprint {http://arxiv.org/abs/1003.5682}
  {arXiv:1003.5682 [hep-lat]} \BibitemShut {NoStop}%
%%CITATION = ARXIV:1003.5682;%%
\bibitem [{\citenamefont {Borsanyi}\ \emph {et~al.}(2012)\citenamefont
  {Borsanyi}, \citenamefont {Endrodi}, \citenamefont {Fodor}, \citenamefont
  {Katz}, \citenamefont {Krieg}, \citenamefont {Ratti},\ and\ \citenamefont
  {Szabo}}]{Borsanyi:2012cr}%
  \BibitemOpen
  \bibfield  {author} {\bibinfo {author} {\bibfnamefont {Sz.}\ \bibnamefont
  {Borsanyi}}, \bibinfo {author} {\bibfnamefont {G.}~\bibnamefont {Endrodi}},
  \bibinfo {author} {\bibfnamefont {Z.}~\bibnamefont {Fodor}}, \bibinfo
  {author} {\bibfnamefont {S.~D.}\ \bibnamefont {Katz}}, \bibinfo {author}
  {\bibfnamefont {S.}~\bibnamefont {Krieg}}, \bibinfo {author} {\bibfnamefont
  {C.}~\bibnamefont {Ratti}}, \ and\ \bibinfo {author} {\bibfnamefont {K.~K.}\
  \bibnamefont {Szabo}},\ }\bibfield  {title} {\enquote {\bibinfo {title} {{QCD
  equation of state at nonzero chemical potential: continuum results with
  physical quark masses at order $mu^2$}},}\ }\href {\doibase
  10.1007/JHEP08(2012)053} {\bibfield  {journal} {\bibinfo  {journal} {JHEP}\
  }\textbf {\bibinfo {volume} {08}},\ \bibinfo {pages} {053} (\bibinfo {year}
  {2012})},\ \Eprint {http://arxiv.org/abs/1204.6710} {arXiv:1204.6710
  [hep-lat]} \BibitemShut {NoStop}%
%%CITATION = ARXIV:1204.6710;%%
\bibitem [{\citenamefont {Hegde}(2014)}]{Hegde:2014wga}%
  \BibitemOpen
  \bibfield  {author} {\bibinfo {author} {\bibfnamefont {Prasad}\ \bibnamefont
  {Hegde}} (\bibinfo {collaboration} {BNL-Bielefeld-CCNU}),\ }\bibfield
  {title} {\enquote {\bibinfo {title} {{The QCD equation of state to
  $\mathcal{O}(\mu\_B^4)$}},}\ }\bibfield  {booktitle} {\emph {\bibinfo
  {booktitle} {{Proceedings, 32nd International Symposium on Lattice Field
  Theory (Lattice 2014)}}},\ }\href@noop {} {\bibfield  {journal} {\bibinfo
  {journal} {PoS}\ }\textbf {\bibinfo {volume} {LATTICE2014}},\ \bibinfo
  {pages} {226} (\bibinfo {year} {2014})},\ \Eprint
  {http://arxiv.org/abs/1412.6727} {arXiv:1412.6727 [hep-lat]} \BibitemShut
  {NoStop}%
%%CITATION = ARXIV:1412.6727;%%
\bibitem [{\citenamefont {Takaishi}\ \emph {et~al.}(2009)\citenamefont
  {Takaishi}, \citenamefont {de~Forcrand},\ and\ \citenamefont
  {Nakamura}}]{Takaishi:2010kc}%
  \BibitemOpen
  \bibfield  {author} {\bibinfo {author} {\bibfnamefont {Tetsuya}\ \bibnamefont
  {Takaishi}}, \bibinfo {author} {\bibfnamefont {Philippe}\ \bibnamefont
  {de~Forcrand}}, \ and\ \bibinfo {author} {\bibfnamefont {Atsushi}\
  \bibnamefont {Nakamura}},\ }\bibfield  {title} {\enquote {\bibinfo {title}
  {{Equation of State at Finite Density from Imaginary Chemical Potential}},}\
  }\bibfield  {booktitle} {\emph {\bibinfo {booktitle} {{Proceedings, 27th
  International Symposium on Lattice field theory (Lattice 2009)}}},\
  }\href@noop {} {\bibfield  {journal} {\bibinfo  {journal} {PoS}\ }\textbf
  {\bibinfo {volume} {LAT2009}},\ \bibinfo {pages} {198} (\bibinfo {year}
  {2009})},\ \Eprint {http://arxiv.org/abs/1002.0890} {arXiv:1002.0890
  [hep-lat]} \BibitemShut {NoStop}%
%%CITATION = ARXIV:1002.0890;%%
\bibitem [{\citenamefont {Borsanyi}\ \emph {et~al.}(2010)\citenamefont
  {Borsanyi}, \citenamefont {Endrodi}, \citenamefont {Fodor}, \citenamefont
  {Jakovac}, \citenamefont {Katz}, \citenamefont {Krieg}, \citenamefont
  {Ratti},\ and\ \citenamefont {Szabo}}]{Borsanyi:2010cj}%
  \BibitemOpen
  \bibfield  {author} {\bibinfo {author} {\bibfnamefont {Szabolcs}\
  \bibnamefont {Borsanyi}}, \bibinfo {author} {\bibfnamefont {Gergely}\
  \bibnamefont {Endrodi}}, \bibinfo {author} {\bibfnamefont {Zoltan}\
  \bibnamefont {Fodor}}, \bibinfo {author} {\bibfnamefont {Antal}\ \bibnamefont
  {Jakovac}}, \bibinfo {author} {\bibfnamefont {Sandor~D.}\ \bibnamefont
  {Katz}}, \bibinfo {author} {\bibfnamefont {Stefan}\ \bibnamefont {Krieg}},
  \bibinfo {author} {\bibfnamefont {Claudia}\ \bibnamefont {Ratti}}, \ and\
  \bibinfo {author} {\bibfnamefont {Kalman~K.}\ \bibnamefont {Szabo}},\
  }\bibfield  {title} {\enquote {\bibinfo {title} {{The QCD equation of state
  with dynamical quarks}},}\ }\href {\doibase 10.1007/JHEP11(2010)077}
  {\bibfield  {journal} {\bibinfo  {journal} {JHEP}\ }\textbf {\bibinfo
  {volume} {11}},\ \bibinfo {pages} {077} (\bibinfo {year} {2010})},\ \Eprint
  {http://arxiv.org/abs/1007.2580} {arXiv:1007.2580 [hep-lat]} \BibitemShut
  {NoStop}%
%%CITATION = ARXIV:1007.2580;%%
\bibitem [{\citenamefont {Arnold}\ \emph {et~al.}(2000)\citenamefont {Arnold},
  \citenamefont {Moore},\ and\ \citenamefont {Yaffe}}]{Arnold:2000dr}%
  \BibitemOpen
  \bibfield  {author} {\bibinfo {author} {\bibfnamefont {Peter~Brockway}\
  \bibnamefont {Arnold}}, \bibinfo {author} {\bibfnamefont {Guy~D.}\
  \bibnamefont {Moore}}, \ and\ \bibinfo {author} {\bibfnamefont {Laurence~G.}\
  \bibnamefont {Yaffe}},\ }\bibfield  {title} {\enquote {\bibinfo {title}
  {{Transport coefficients in high temperature gauge theories. 1. Leading log
  results}},}\ }\href {\doibase 10.1088/1126-6708/2000/11/001} {\bibfield
  {journal} {\bibinfo  {journal} {JHEP}\ }\textbf {\bibinfo {volume} {0011}},\
  \bibinfo {pages} {001} (\bibinfo {year} {2000})},\ \Eprint
  {http://arxiv.org/abs/hep-ph/0010177} {arXiv:hep-ph/0010177 [hep-ph]}
  \BibitemShut {NoStop}%
%%CITATION = HEP-PH/0010177;%%
\bibitem [{\citenamefont {Arnold}\ \emph {et~al.}(2003)\citenamefont {Arnold},
  \citenamefont {Moore},\ and\ \citenamefont {Yaffe}}]{Arnold:2003zc}%
  \BibitemOpen
  \bibfield  {author} {\bibinfo {author} {\bibfnamefont {Peter~Brockway}\
  \bibnamefont {Arnold}}, \bibinfo {author} {\bibfnamefont {Guy~D}\
  \bibnamefont {Moore}}, \ and\ \bibinfo {author} {\bibfnamefont {Laurence~G.}\
  \bibnamefont {Yaffe}},\ }\bibfield  {title} {\enquote {\bibinfo {title}
  {{Transport coefficients in high temperature gauge theories. 2. Beyond
  leading log}},}\ }\href {\doibase 10.1088/1126-6708/2003/05/051} {\bibfield
  {journal} {\bibinfo  {journal} {JHEP}\ }\textbf {\bibinfo {volume} {0305}},\
  \bibinfo {pages} {051} (\bibinfo {year} {2003})},\ \Eprint
  {http://arxiv.org/abs/hep-ph/0302165} {arXiv:hep-ph/0302165 [hep-ph]}
  \BibitemShut {NoStop}%
%%CITATION = HEP-PH/0302165;%%
\bibitem [{\citenamefont {Arnold}\ \emph {et~al.}(2006)\citenamefont {Arnold},
  \citenamefont {Dogan},\ and\ \citenamefont {Moore}}]{Arnold:2006fz}%
  \BibitemOpen
  \bibfield  {author} {\bibinfo {author} {\bibfnamefont {Peter~Brockway}\
  \bibnamefont {Arnold}}, \bibinfo {author} {\bibfnamefont {Caglar}\
  \bibnamefont {Dogan}}, \ and\ \bibinfo {author} {\bibfnamefont {Guy~D.}\
  \bibnamefont {Moore}},\ }\bibfield  {title} {\enquote {\bibinfo {title} {{The
  Bulk Viscosity of High-Temperature QCD}},}\ }\href {\doibase
  10.1103/PhysRevD.74.085021} {\bibfield  {journal} {\bibinfo  {journal}
  {Phys.Rev.}\ }\textbf {\bibinfo {volume} {D74}},\ \bibinfo {pages} {085021}
  (\bibinfo {year} {2006})},\ \Eprint {http://arxiv.org/abs/hep-ph/0608012}
  {arXiv:hep-ph/0608012 [hep-ph]} \BibitemShut {NoStop}%
%%CITATION = HEP-PH/0608012;%%
\bibitem [{\citenamefont {Heiselberg}\ and\ \citenamefont
  {Pethick}(1993)}]{Heiselberg:1993cr}%
  \BibitemOpen
  \bibfield  {author} {\bibinfo {author} {\bibfnamefont {H.}~\bibnamefont
  {Heiselberg}}\ and\ \bibinfo {author} {\bibfnamefont {C.J.}\ \bibnamefont
  {Pethick}},\ }\bibfield  {title} {\enquote {\bibinfo {title} {{Transport and
  relaxation in degenerate quark plasmas}},}\ }\href {\doibase
  10.1103/PhysRevD.48.2916} {\bibfield  {journal} {\bibinfo  {journal}
  {Phys.Rev.}\ }\textbf {\bibinfo {volume} {D48}},\ \bibinfo {pages}
  {2916--2928} (\bibinfo {year} {1993})}\BibitemShut {NoStop}%
%%CITATION = PHRVA,D48,2916;%%
\bibitem [{\citenamefont {Danielewicz}\ and\ \citenamefont
  {Gyulassy}(1985)}]{Danielewicz:1984ww}%
  \BibitemOpen
  \bibfield  {author} {\bibinfo {author} {\bibfnamefont {P.}~\bibnamefont
  {Danielewicz}}\ and\ \bibinfo {author} {\bibfnamefont {M.}~\bibnamefont
  {Gyulassy}},\ }\bibfield  {title} {\enquote {\bibinfo {title} {{Dissipative
  Phenomena in Quark Gluon Plasmas}},}\ }\href {\doibase
  10.1103/PhysRevD.31.53} {\bibfield  {journal} {\bibinfo  {journal}
  {Phys.Rev.}\ }\textbf {\bibinfo {volume} {D31}},\ \bibinfo {pages} {53--62}
  (\bibinfo {year} {1985})}\BibitemShut {NoStop}%
%%CITATION = PHRVA,D31,53;%%
\bibitem [{\citenamefont {Policastro}\ \emph {et~al.}(2001)\citenamefont
  {Policastro}, \citenamefont {Son},\ and\ \citenamefont
  {Starinets}}]{Policastro:2001yc}%
  \BibitemOpen
  \bibfield  {author} {\bibinfo {author} {\bibfnamefont {G.}~\bibnamefont
  {Policastro}}, \bibinfo {author} {\bibfnamefont {Dan~T.}\ \bibnamefont
  {Son}}, \ and\ \bibinfo {author} {\bibfnamefont {Andrei~O.}\ \bibnamefont
  {Starinets}},\ }\bibfield  {title} {\enquote {\bibinfo {title} {{The Shear
  viscosity of strongly coupled N=4 supersymmetric Yang-Mills plasma}},}\
  }\href {\doibase 10.1103/PhysRevLett.87.081601} {\bibfield  {journal}
  {\bibinfo  {journal} {Phys.Rev.Lett.}\ }\textbf {\bibinfo {volume} {87}},\
  \bibinfo {pages} {081601} (\bibinfo {year} {2001})},\ \Eprint
  {http://arxiv.org/abs/hep-th/0104066} {arXiv:hep-th/0104066 [hep-th]}
  \BibitemShut {NoStop}%
%%CITATION = HEP-TH/0104066;%%
\bibitem [{\citenamefont {Son}\ and\ \citenamefont
  {Starinets}(2006)}]{Son:2006em}%
  \BibitemOpen
  \bibfield  {author} {\bibinfo {author} {\bibfnamefont {Dam~T.}\ \bibnamefont
  {Son}}\ and\ \bibinfo {author} {\bibfnamefont {Andrei~O.}\ \bibnamefont
  {Starinets}},\ }\bibfield  {title} {\enquote {\bibinfo {title}
  {{Hydrodynamics of r-charged black holes}},}\ }\href {\doibase
  10.1088/1126-6708/2006/03/052} {\bibfield  {journal} {\bibinfo  {journal}
  {JHEP}\ }\textbf {\bibinfo {volume} {0603}},\ \bibinfo {pages} {052}
  (\bibinfo {year} {2006})},\ \Eprint {http://arxiv.org/abs/hep-th/0601157}
  {arXiv:hep-th/0601157 [hep-th]} \BibitemShut {NoStop}%
%%CITATION = HEP-TH/0601157;%%
\bibitem [{\citenamefont {Buchel}(2005)}]{Buchel:2005cv}%
  \BibitemOpen
  \bibfield  {author} {\bibinfo {author} {\bibfnamefont {Alex}\ \bibnamefont
  {Buchel}},\ }\bibfield  {title} {\enquote {\bibinfo {title} {{Transport
  properties of cascading gauge theories}},}\ }\href {\doibase
  10.1103/PhysRevD.72.106002} {\bibfield  {journal} {\bibinfo  {journal}
  {Phys.Rev.}\ }\textbf {\bibinfo {volume} {D72}},\ \bibinfo {pages} {106002}
  (\bibinfo {year} {2005})},\ \Eprint {http://arxiv.org/abs/hep-th/0509083}
  {arXiv:hep-th/0509083 [hep-th]} \BibitemShut {NoStop}%
%%CITATION = HEP-TH/0509083;%%
\bibitem [{\citenamefont {Schafer}\ and\ \citenamefont
  {Teaney}(2009)}]{Schafer:2009dj}%
  \BibitemOpen
  \bibfield  {author} {\bibinfo {author} {\bibfnamefont {Thomas}\ \bibnamefont
  {Schafer}}\ and\ \bibinfo {author} {\bibfnamefont {Derek}\ \bibnamefont
  {Teaney}},\ }\bibfield  {title} {\enquote {\bibinfo {title} {{Nearly Perfect
  Fluidity: From Cold Atomic Gases to Hot Quark Gluon Plasmas}},}\ }\href
  {\doibase 10.1088/0034-4885/72/12/126001} {\bibfield  {journal} {\bibinfo
  {journal} {Rept.Prog.Phys.}\ }\textbf {\bibinfo {volume} {72}},\ \bibinfo
  {pages} {126001} (\bibinfo {year} {2009})},\ \Eprint
  {http://arxiv.org/abs/0904.3107} {arXiv:0904.3107 [hep-ph]} \BibitemShut
  {NoStop}%
%%CITATION = ARXIV:0904.3107;%%
\bibitem [{\citenamefont {Weinberg}(1971)}]{Weinberg:1971mx}%
  \BibitemOpen
  \bibfield  {author} {\bibinfo {author} {\bibfnamefont {Steven}\ \bibnamefont
  {Weinberg}},\ }\bibfield  {title} {\enquote {\bibinfo {title} {{Entropy
  generation and the survival of protogalaxies in an expanding universe}},}\
  }\href {\doibase 10.1086/151073} {\bibfield  {journal} {\bibinfo  {journal}
  {Astrophys.J.}\ }\textbf {\bibinfo {volume} {168}},\ \bibinfo {pages} {175}
  (\bibinfo {year} {1971})}\BibitemShut {NoStop}%
%%CITATION = ASJOA,168,175;%%
\bibitem [{\citenamefont {Denicol}\ \emph
  {et~al.}(2012{\natexlab{a}})\citenamefont {Denicol}, \citenamefont {Molnár},
  \citenamefont {Niemi},\ and\ \citenamefont {Rischke}}]{Denicol:2012es}%
  \BibitemOpen
  \bibfield  {author} {\bibinfo {author} {\bibfnamefont {G.S.}\ \bibnamefont
  {Denicol}}, \bibinfo {author} {\bibfnamefont {E.}~\bibnamefont {Molnár}},
  \bibinfo {author} {\bibfnamefont {H.}~\bibnamefont {Niemi}}, \ and\ \bibinfo
  {author} {\bibfnamefont {D.H.}\ \bibnamefont {Rischke}},\ }\bibfield  {title}
  {\enquote {\bibinfo {title} {{Derivation of fluid dynamics from kinetic
  theory with the 14-moment approximation}},}\ }\href {\doibase
  10.1140/epja/i2012-12170-x} {\bibfield  {journal} {\bibinfo  {journal}
  {Eur.Phys.J.}\ }\textbf {\bibinfo {volume} {A48}},\ \bibinfo {pages} {170}
  (\bibinfo {year} {2012}{\natexlab{a}})},\ \Eprint
  {http://arxiv.org/abs/1206.1554} {arXiv:1206.1554 [nucl-th]} \BibitemShut
  {NoStop}%
%%CITATION = ARXIV:1206.1554;%%
\bibitem [{\citenamefont {Denicol}\ \emph
  {et~al.}(2012{\natexlab{b}})\citenamefont {Denicol}, \citenamefont {Niemi},
  \citenamefont {Molnar},\ and\ \citenamefont {Rischke}}]{Denicol:2012cn}%
  \BibitemOpen
  \bibfield  {author} {\bibinfo {author} {\bibfnamefont {G.S.}\ \bibnamefont
  {Denicol}}, \bibinfo {author} {\bibfnamefont {H.}~\bibnamefont {Niemi}},
  \bibinfo {author} {\bibfnamefont {E.}~\bibnamefont {Molnar}}, \ and\ \bibinfo
  {author} {\bibfnamefont {D.H.}\ \bibnamefont {Rischke}},\ }\bibfield  {title}
  {\enquote {\bibinfo {title} {{Derivation of transient relativistic fluid
  dynamics from the Boltzmann equation}},}\ }\href {\doibase
  10.1103/PhysRevD.85.114047, 10.1103/PhysRevD.91.039902} {\bibfield  {journal}
  {\bibinfo  {journal} {Phys.Rev.}\ }\textbf {\bibinfo {volume} {D85}},\
  \bibinfo {pages} {114047} (\bibinfo {year} {2012}{\natexlab{b}})},\ \Eprint
  {http://arxiv.org/abs/1202.4551} {arXiv:1202.4551 [nucl-th]} \BibitemShut
  {NoStop}%
%%CITATION = ARXIV:1202.4551;%%
\bibitem [{\citenamefont {Molnár}\ \emph {et~al.}(2014)\citenamefont
  {Molnár}, \citenamefont {Niemi}, \citenamefont {Denicol},\ and\
  \citenamefont {Rischke}}]{Molnar:2013lta}%
  \BibitemOpen
  \bibfield  {author} {\bibinfo {author} {\bibfnamefont {E.}~\bibnamefont
  {Molnár}}, \bibinfo {author} {\bibfnamefont {H.}~\bibnamefont {Niemi}},
  \bibinfo {author} {\bibfnamefont {G.S.}\ \bibnamefont {Denicol}}, \ and\
  \bibinfo {author} {\bibfnamefont {D.H.}\ \bibnamefont {Rischke}},\ }\bibfield
   {title} {\enquote {\bibinfo {title} {{Relative importance of second-order
  terms in relativistic dissipative fluid dynamics}},}\ }\href {\doibase
  10.1103/PhysRevD.89.074010} {\bibfield  {journal} {\bibinfo  {journal}
  {Phys.Rev.}\ }\textbf {\bibinfo {volume} {D89}},\ \bibinfo {pages} {074010}
  (\bibinfo {year} {2014})},\ \Eprint {http://arxiv.org/abs/1308.0785}
  {arXiv:1308.0785 [nucl-th]} \BibitemShut {NoStop}%
%%CITATION = ARXIV:1308.0785;%%
\bibitem [{\citenamefont {Denicol}\ \emph {et~al.}(2014)\citenamefont
  {Denicol}, \citenamefont {Jeon},\ and\ \citenamefont
  {Gale}}]{Denicol:2014vaa}%
  \BibitemOpen
  \bibfield  {author} {\bibinfo {author} {\bibfnamefont {G.S.}\ \bibnamefont
  {Denicol}}, \bibinfo {author} {\bibfnamefont {S.}~\bibnamefont {Jeon}}, \
  and\ \bibinfo {author} {\bibfnamefont {C.}~\bibnamefont {Gale}},\ }\bibfield
  {title} {\enquote {\bibinfo {title} {{Transport Coefficients of Bulk Viscous
  Pressure in the 14-moment approximation}},}\ }\href {\doibase
  10.1103/PhysRevC.90.024912} {\bibfield  {journal} {\bibinfo  {journal}
  {Phys.Rev.}\ }\textbf {\bibinfo {volume} {C90}},\ \bibinfo {pages} {024912}
  (\bibinfo {year} {2014})},\ \Eprint {http://arxiv.org/abs/1403.0962}
  {arXiv:1403.0962 [nucl-th]} \BibitemShut {NoStop}%
%%CITATION = ARXIV:1403.0962;%%
\bibitem [{\citenamefont {Jaiswal}\ \emph {et~al.}(2014)\citenamefont
  {Jaiswal}, \citenamefont {Ryblewski},\ and\ \citenamefont
  {Strickland}}]{Jaiswal:2014isa}%
  \BibitemOpen
  \bibfield  {author} {\bibinfo {author} {\bibfnamefont {Amaresh}\ \bibnamefont
  {Jaiswal}}, \bibinfo {author} {\bibfnamefont {Radoslaw}\ \bibnamefont
  {Ryblewski}}, \ and\ \bibinfo {author} {\bibfnamefont {Michael}\ \bibnamefont
  {Strickland}},\ }\bibfield  {title} {\enquote {\bibinfo {title} {{Transport
  coefficients for bulk viscous evolution in the relaxation time
  approximation}},}\ }\href {\doibase 10.1103/PhysRevC.90.044908} {\bibfield
  {journal} {\bibinfo  {journal} {Phys.Rev.}\ }\textbf {\bibinfo {volume}
  {C90}},\ \bibinfo {pages} {044908} (\bibinfo {year} {2014})},\ \Eprint
  {http://arxiv.org/abs/1407.7231} {arXiv:1407.7231 [hep-ph]} \BibitemShut
  {NoStop}%
%%CITATION = ARXIV:1407.7231;%%
\bibitem [{\citenamefont {Florkowski}\ \emph {et~al.}(2015)\citenamefont
  {Florkowski}, \citenamefont {Jaiswal}, \citenamefont {Maksymiuk},
  \citenamefont {Ryblewski},\ and\ \citenamefont
  {Strickland}}]{Florkowski:2015lra}%
  \BibitemOpen
  \bibfield  {author} {\bibinfo {author} {\bibfnamefont {Wojciech}\
  \bibnamefont {Florkowski}}, \bibinfo {author} {\bibfnamefont {Amaresh}\
  \bibnamefont {Jaiswal}}, \bibinfo {author} {\bibfnamefont {Ewa}\ \bibnamefont
  {Maksymiuk}}, \bibinfo {author} {\bibfnamefont {Radoslaw}\ \bibnamefont
  {Ryblewski}}, \ and\ \bibinfo {author} {\bibfnamefont {Michael}\ \bibnamefont
  {Strickland}},\ }\bibfield  {title} {\enquote {\bibinfo {title}
  {{Relativistic quantum transport coefficients for second-order viscous
  hydrodynamics}},}\ }\href {\doibase 10.1103/PhysRevC.91.054907} {\bibfield
  {journal} {\bibinfo  {journal} {Phys.Rev.}\ }\textbf {\bibinfo {volume}
  {C91}},\ \bibinfo {pages} {054907} (\bibinfo {year} {2015})}\BibitemShut
  {NoStop}%
%%CITATION = ARXIV:1503.03226;%%
\bibitem [{\citenamefont {Noronha-Hostler}\ \emph {et~al.}(2013)\citenamefont
  {Noronha-Hostler}, \citenamefont {Denicol}, \citenamefont {Noronha},
  \citenamefont {Andrade},\ and\ \citenamefont
  {Grassi}}]{Noronha-Hostler:2013gga}%
  \BibitemOpen
  \bibfield  {author} {\bibinfo {author} {\bibfnamefont {Jacquelyn}\
  \bibnamefont {Noronha-Hostler}}, \bibinfo {author} {\bibfnamefont
  {Gabriel~S.}\ \bibnamefont {Denicol}}, \bibinfo {author} {\bibfnamefont
  {Jorge}\ \bibnamefont {Noronha}}, \bibinfo {author} {\bibfnamefont {Rone
  P.~G.}\ \bibnamefont {Andrade}}, \ and\ \bibinfo {author} {\bibfnamefont
  {Frederique}\ \bibnamefont {Grassi}},\ }\bibfield  {title} {\enquote
  {\bibinfo {title} {{Bulk Viscosity Effects in Event-by-Event Relativistic
  Hydrodynamics}},}\ }\href {\doibase 10.1103/PhysRevC.88.044916} {\bibfield
  {journal} {\bibinfo  {journal} {Phys.Rev.}\ }\textbf {\bibinfo {volume}
  {C88}},\ \bibinfo {pages} {044916} (\bibinfo {year} {2013})},\ \Eprint
  {http://arxiv.org/abs/1305.1981} {arXiv:1305.1981 [nucl-th]} \BibitemShut
  {NoStop}%
%%CITATION = ARXIV:1305.1981;%%
\bibitem [{\citenamefont {Brigante}\ \emph
  {et~al.}(2008{\natexlab{a}})\citenamefont {Brigante}, \citenamefont {Liu},
  \citenamefont {Myers}, \citenamefont {Shenker},\ and\ \citenamefont
  {Yaida}}]{Brigante:2007nu}%
  \BibitemOpen
  \bibfield  {author} {\bibinfo {author} {\bibfnamefont {Mauro}\ \bibnamefont
  {Brigante}}, \bibinfo {author} {\bibfnamefont {Hong}\ \bibnamefont {Liu}},
  \bibinfo {author} {\bibfnamefont {Robert~C.}\ \bibnamefont {Myers}}, \bibinfo
  {author} {\bibfnamefont {Stephen}\ \bibnamefont {Shenker}}, \ and\ \bibinfo
  {author} {\bibfnamefont {Sho}\ \bibnamefont {Yaida}},\ }\bibfield  {title}
  {\enquote {\bibinfo {title} {{Viscosity Bound Violation in Higher Derivative
  Gravity}},}\ }\href {\doibase 10.1103/PhysRevD.77.126006} {\bibfield
  {journal} {\bibinfo  {journal} {Phys.Rev.}\ }\textbf {\bibinfo {volume}
  {D77}},\ \bibinfo {pages} {126006} (\bibinfo {year} {2008}{\natexlab{a}})},\
  \Eprint {http://arxiv.org/abs/0712.0805} {arXiv:0712.0805 [hep-th]}
  \BibitemShut {NoStop}%
%%CITATION = ARXIV:0712.0805;%%
\bibitem [{\citenamefont {Brigante}\ \emph
  {et~al.}(2008{\natexlab{b}})\citenamefont {Brigante}, \citenamefont {Liu},
  \citenamefont {Myers}, \citenamefont {Shenker},\ and\ \citenamefont
  {Yaida}}]{Brigante:2008gz}%
  \BibitemOpen
  \bibfield  {author} {\bibinfo {author} {\bibfnamefont {Mauro}\ \bibnamefont
  {Brigante}}, \bibinfo {author} {\bibfnamefont {Hong}\ \bibnamefont {Liu}},
  \bibinfo {author} {\bibfnamefont {Robert~C.}\ \bibnamefont {Myers}}, \bibinfo
  {author} {\bibfnamefont {Stephen}\ \bibnamefont {Shenker}}, \ and\ \bibinfo
  {author} {\bibfnamefont {Sho}\ \bibnamefont {Yaida}},\ }\bibfield  {title}
  {\enquote {\bibinfo {title} {{The Viscosity Bound and Causality
  Violation}},}\ }\href {\doibase 10.1103/PhysRevLett.100.191601} {\bibfield
  {journal} {\bibinfo  {journal} {Phys.Rev.Lett.}\ }\textbf {\bibinfo {volume}
  {100}},\ \bibinfo {pages} {191601} (\bibinfo {year} {2008}{\natexlab{b}})},\
  \Eprint {http://arxiv.org/abs/0802.3318} {arXiv:0802.3318 [hep-th]}
  \BibitemShut {NoStop}%
%%CITATION = ARXIV:0802.3318;%%
\bibitem [{\citenamefont {Kats}\ and\ \citenamefont
  {Petrov}(2009)}]{Kats:2007mq}%
  \BibitemOpen
  \bibfield  {author} {\bibinfo {author} {\bibfnamefont {Yevgeny}\ \bibnamefont
  {Kats}}\ and\ \bibinfo {author} {\bibfnamefont {Pavel}\ \bibnamefont
  {Petrov}},\ }\bibfield  {title} {\enquote {\bibinfo {title} {{Effect of
  curvature squared corrections in AdS on the viscosity of the dual gauge
  theory}},}\ }\href {\doibase 10.1088/1126-6708/2009/01/044} {\bibfield
  {journal} {\bibinfo  {journal} {JHEP}\ }\textbf {\bibinfo {volume} {0901}},\
  \bibinfo {pages} {044} (\bibinfo {year} {2009})},\ \Eprint
  {http://arxiv.org/abs/0712.0743} {arXiv:0712.0743 [hep-th]} \BibitemShut
  {NoStop}%
%%CITATION = ARXIV:0712.0743;%%
\bibitem [{\citenamefont {Natsuume}\ and\ \citenamefont
  {Okamura}(2008)}]{Natsuume:2007ty}%
  \BibitemOpen
  \bibfield  {author} {\bibinfo {author} {\bibfnamefont {Makoto}\ \bibnamefont
  {Natsuume}}\ and\ \bibinfo {author} {\bibfnamefont {Takashi}\ \bibnamefont
  {Okamura}},\ }\bibfield  {title} {\enquote {\bibinfo {title} {{Causal
  hydrodynamics of gauge theory plasmas from AdS/CFT duality}},}\ }\href
  {\doibase 10.1103/PhysRevD.78.089902, 10.1103/PhysRevD.77.066014} {\bibfield
  {journal} {\bibinfo  {journal} {Phys.Rev.}\ }\textbf {\bibinfo {volume}
  {D77}},\ \bibinfo {pages} {066014} (\bibinfo {year} {2008})},\ \Eprint
  {http://arxiv.org/abs/0712.2916} {arXiv:0712.2916 [hep-th]} \BibitemShut
  {NoStop}%
%%CITATION = ARXIV:0712.2916;%%
\bibitem [{\citenamefont {Buchel}\ \emph {et~al.}(2009)\citenamefont {Buchel},
  \citenamefont {Myers},\ and\ \citenamefont {Sinha}}]{Buchel:2008vz}%
  \BibitemOpen
  \bibfield  {author} {\bibinfo {author} {\bibfnamefont {Alex}\ \bibnamefont
  {Buchel}}, \bibinfo {author} {\bibfnamefont {Robert~C.}\ \bibnamefont
  {Myers}}, \ and\ \bibinfo {author} {\bibfnamefont {Aninda}\ \bibnamefont
  {Sinha}},\ }\bibfield  {title} {\enquote {\bibinfo {title} {{Beyond eta/s =
  1/4 pi}},}\ }\href {\doibase 10.1088/1126-6708/2009/03/084} {\bibfield
  {journal} {\bibinfo  {journal} {JHEP}\ }\textbf {\bibinfo {volume} {0903}},\
  \bibinfo {pages} {084} (\bibinfo {year} {2009})},\ \Eprint
  {http://arxiv.org/abs/0812.2521} {arXiv:0812.2521 [hep-th]} \BibitemShut
  {NoStop}%
%%CITATION = ARXIV:0812.2521;%%
\bibitem [{\citenamefont {Erdmenger}\ \emph {et~al.}(2011)\citenamefont
  {Erdmenger}, \citenamefont {Kerner},\ and\ \citenamefont
  {Zeller}}]{Erdmenger:2010xm}%
  \BibitemOpen
  \bibfield  {author} {\bibinfo {author} {\bibfnamefont {Johanna}\ \bibnamefont
  {Erdmenger}}, \bibinfo {author} {\bibfnamefont {Patrick}\ \bibnamefont
  {Kerner}}, \ and\ \bibinfo {author} {\bibfnamefont {Hansjorg}\ \bibnamefont
  {Zeller}},\ }\bibfield  {title} {\enquote {\bibinfo {title} {{Non-universal
  shear viscosity from Einstein gravity}},}\ }\href {\doibase
  10.1016/j.physletb.2011.04.009} {\bibfield  {journal} {\bibinfo  {journal}
  {Phys. Lett.}\ }\textbf {\bibinfo {volume} {B699}},\ \bibinfo {pages}
  {301--304} (\bibinfo {year} {2011})},\ \Eprint
  {http://arxiv.org/abs/1011.5912} {arXiv:1011.5912 [hep-th]} \BibitemShut
  {NoStop}%
%%CITATION = ARXIV:1011.5912;%%
\bibitem [{\citenamefont {Rebhan}\ and\ \citenamefont
  {Steineder}(2012)}]{Rebhan:2011vd}%
  \BibitemOpen
  \bibfield  {author} {\bibinfo {author} {\bibfnamefont {Anton}\ \bibnamefont
  {Rebhan}}\ and\ \bibinfo {author} {\bibfnamefont {Dominik}\ \bibnamefont
  {Steineder}},\ }\bibfield  {title} {\enquote {\bibinfo {title} {{Violation of
  the Holographic Viscosity Bound in a Strongly Coupled Anisotropic Plasma}},}\
  }\href {\doibase 10.1103/PhysRevLett.108.021601} {\bibfield  {journal}
  {\bibinfo  {journal} {Phys. Rev. Lett.}\ }\textbf {\bibinfo {volume} {108}},\
  \bibinfo {pages} {021601} (\bibinfo {year} {2012})},\ \Eprint
  {http://arxiv.org/abs/1110.6825} {arXiv:1110.6825 [hep-th]} \BibitemShut
  {NoStop}%
%%CITATION = ARXIV:1110.6825;%%
\bibitem [{\citenamefont {Critelli}\ \emph {et~al.}(2014)\citenamefont
  {Critelli}, \citenamefont {Finazzo}, \citenamefont {Zaniboni},\ and\
  \citenamefont {Noronha}}]{Critelli:2014kra}%
  \BibitemOpen
  \bibfield  {author} {\bibinfo {author} {\bibfnamefont {R.}~\bibnamefont
  {Critelli}}, \bibinfo {author} {\bibfnamefont {S.~I.}\ \bibnamefont
  {Finazzo}}, \bibinfo {author} {\bibfnamefont {M.}~\bibnamefont {Zaniboni}}, \
  and\ \bibinfo {author} {\bibfnamefont {J.}~\bibnamefont {Noronha}},\
  }\bibfield  {title} {\enquote {\bibinfo {title} {{Anisotropic shear viscosity
  of a strongly coupled non-Abelian plasma from magnetic branes}},}\ }\href
  {\doibase 10.1103/PhysRevD.90.066006} {\bibfield  {journal} {\bibinfo
  {journal} {Phys. Rev.}\ }\textbf {\bibinfo {volume} {D90}},\ \bibinfo {pages}
  {066006} (\bibinfo {year} {2014})},\ \Eprint {http://arxiv.org/abs/1406.6019}
  {arXiv:1406.6019 [hep-th]} \BibitemShut {NoStop}%
%%CITATION = ARXIV:1406.6019;%%
\bibitem [{\citenamefont {Meyer}(2007)}]{Meyer:2007ic}%
  \BibitemOpen
  \bibfield  {author} {\bibinfo {author} {\bibfnamefont {Harvey~B.}\
  \bibnamefont {Meyer}},\ }\bibfield  {title} {\enquote {\bibinfo {title} {{A
  Calculation of the shear viscosity in SU(3) gluodynamics}},}\ }\href
  {\doibase 10.1103/PhysRevD.76.101701} {\bibfield  {journal} {\bibinfo
  {journal} {Phys. Rev.}\ }\textbf {\bibinfo {volume} {D76}},\ \bibinfo {pages}
  {101701} (\bibinfo {year} {2007})},\ \Eprint {http://arxiv.org/abs/0704.1801}
  {arXiv:0704.1801 [hep-lat]} \BibitemShut {NoStop}%
%%CITATION = ARXIV:0704.1801;%%
\bibitem [{\citenamefont {Meyer}(2009)}]{Meyer:2009jp}%
  \BibitemOpen
  \bibfield  {author} {\bibinfo {author} {\bibfnamefont {Harvey~B.}\
  \bibnamefont {Meyer}},\ }\bibfield  {title} {\enquote {\bibinfo {title}
  {{Transport properties of the quark-gluon plasma from lattice QCD}},}\
  }\bibfield  {booktitle} {\emph {\bibinfo {booktitle} {{Proceedings, 21st
  International Conference on Ultra-Relativistic nucleus nucleus collisions
  (Quark matter 2009)}}},\ }\href {\doibase 10.1016/j.nuclphysa.2009.09.053}
  {\bibfield  {journal} {\bibinfo  {journal} {Nucl. Phys.}\ }\textbf {\bibinfo
  {volume} {A830}},\ \bibinfo {pages} {641C--648C} (\bibinfo {year} {2009})},\
  \Eprint {http://arxiv.org/abs/0907.4095} {arXiv:0907.4095 [hep-lat]}
  \BibitemShut {NoStop}%
%%CITATION = ARXIV:0907.4095;%%
\bibitem [{\citenamefont {Christiansen}\ \emph {et~al.}(2015)\citenamefont
  {Christiansen}, \citenamefont {Haas}, \citenamefont {Pawlowski},\ and\
  \citenamefont {Strodthoff}}]{Christiansen:2014ypa}%
  \BibitemOpen
  \bibfield  {author} {\bibinfo {author} {\bibfnamefont {Nicolai}\ \bibnamefont
  {Christiansen}}, \bibinfo {author} {\bibfnamefont {Michael}\ \bibnamefont
  {Haas}}, \bibinfo {author} {\bibfnamefont {Jan~M.}\ \bibnamefont
  {Pawlowski}}, \ and\ \bibinfo {author} {\bibfnamefont {Nils}\ \bibnamefont
  {Strodthoff}},\ }\bibfield  {title} {\enquote {\bibinfo {title} {{Transport
  Coefficients in Yang--Mills Theory and QCD}},}\ }\href {\doibase
  10.1103/PhysRevLett.115.112002} {\bibfield  {journal} {\bibinfo  {journal}
  {Phys. Rev. Lett.}\ }\textbf {\bibinfo {volume} {115}},\ \bibinfo {pages}
  {112002} (\bibinfo {year} {2015})},\ \Eprint {http://arxiv.org/abs/1411.7986}
  {arXiv:1411.7986 [hep-ph]} \BibitemShut {NoStop}%
%%CITATION = ARXIV:1411.7986;%%
\bibitem [{\citenamefont {Haas}\ \emph {et~al.}(2014)\citenamefont {Haas},
  \citenamefont {Fister},\ and\ \citenamefont {Pawlowski}}]{Haas:2013hpa}%
  \BibitemOpen
  \bibfield  {author} {\bibinfo {author} {\bibfnamefont {Michael}\ \bibnamefont
  {Haas}}, \bibinfo {author} {\bibfnamefont {Leonard}\ \bibnamefont {Fister}},
  \ and\ \bibinfo {author} {\bibfnamefont {Jan~M.}\ \bibnamefont {Pawlowski}},\
  }\bibfield  {title} {\enquote {\bibinfo {title} {{Gluon spectral functions
  and transport coefficients in Yang--Mills theory}},}\ }\href {\doibase
  10.1103/PhysRevD.90.091501} {\bibfield  {journal} {\bibinfo  {journal}
  {Phys.Rev.}\ }\textbf {\bibinfo {volume} {D90}},\ \bibinfo {pages} {091501}
  (\bibinfo {year} {2014})},\ \Eprint {http://arxiv.org/abs/1308.4960}
  {arXiv:1308.4960 [hep-ph]} \BibitemShut {NoStop}%
%%CITATION = ARXIV:1308.4960;%%
\bibitem [{\citenamefont {Benincasa}\ \emph {et~al.}(2006)\citenamefont
  {Benincasa}, \citenamefont {Buchel},\ and\ \citenamefont
  {Starinets}}]{Benincasa:2005iv}%
  \BibitemOpen
  \bibfield  {author} {\bibinfo {author} {\bibfnamefont {Paolo}\ \bibnamefont
  {Benincasa}}, \bibinfo {author} {\bibfnamefont {Alex}\ \bibnamefont
  {Buchel}}, \ and\ \bibinfo {author} {\bibfnamefont {Andrei~O.}\ \bibnamefont
  {Starinets}},\ }\bibfield  {title} {\enquote {\bibinfo {title} {{Sound waves
  in strongly coupled non-conformal gauge theory plasma}},}\ }\href {\doibase
  10.1016/j.nuclphysb.2005.11.005} {\bibfield  {journal} {\bibinfo  {journal}
  {Nucl.Phys.}\ }\textbf {\bibinfo {volume} {B733}},\ \bibinfo {pages}
  {160--187} (\bibinfo {year} {2006})},\ \Eprint
  {http://arxiv.org/abs/hep-th/0507026} {arXiv:hep-th/0507026 [hep-th]}
  \BibitemShut {NoStop}%
%%CITATION = HEP-TH/0507026;%%
\bibitem [{\citenamefont {Buchel}(2010)}]{Buchel:2010gd}%
  \BibitemOpen
  \bibfield  {author} {\bibinfo {author} {\bibfnamefont {Alex}\ \bibnamefont
  {Buchel}},\ }\bibfield  {title} {\enquote {\bibinfo {title} {{Critical
  phenomena in N=4 SYM plasma}},}\ }\href {\doibase
  10.1016/j.nuclphysb.2010.07.017} {\bibfield  {journal} {\bibinfo  {journal}
  {Nucl.Phys.}\ }\textbf {\bibinfo {volume} {B841}},\ \bibinfo {pages} {59--99}
  (\bibinfo {year} {2010})},\ \Eprint {http://arxiv.org/abs/1005.0819}
  {arXiv:1005.0819 [hep-th]} \BibitemShut {NoStop}%
%%CITATION = ARXIV:1005.0819;%%
\bibitem [{\citenamefont {Buchel}(2012)}]{Buchel:2011uj}%
  \BibitemOpen
  \bibfield  {author} {\bibinfo {author} {\bibfnamefont {Alex}\ \bibnamefont
  {Buchel}},\ }\bibfield  {title} {\enquote {\bibinfo {title} {{Violation of
  the holographic bulk viscosity bound}},}\ }\href {\doibase
  10.1103/PhysRevD.85.066004} {\bibfield  {journal} {\bibinfo  {journal}
  {Phys.Rev.}\ }\textbf {\bibinfo {volume} {D85}},\ \bibinfo {pages} {066004}
  (\bibinfo {year} {2012})},\ \Eprint {http://arxiv.org/abs/1110.0063}
  {arXiv:1110.0063 [hep-th]} \BibitemShut {NoStop}%
%%CITATION = ARXIV:1110.0063;%%
\bibitem [{\citenamefont {Pitaevskii}\ and\ \citenamefont
  {Lifshitz}(1981)}]{Landau}%
  \BibitemOpen
  \bibfield  {author} {\bibinfo {author} {\bibfnamefont {L.~P.}\ \bibnamefont
  {Pitaevskii}}\ and\ \bibinfo {author} {\bibfnamefont {E.~M.}\ \bibnamefont
  {Lifshitz}},\ }\bibfield  {title} {\enquote {\bibinfo {title} {{Physical
  kinetics}},}\ }\href@noop {} {\bibfield  {journal} {\bibinfo  {journal}
  {Pergamon Press, Oxford}\ } (\bibinfo {year} {1981})}\BibitemShut {NoStop}%
\bibitem [{\citenamefont {Rougemont}\ \emph {et~al.}(2015)\citenamefont
  {Rougemont}, \citenamefont {Noronha},\ and\ \citenamefont
  {Noronha-Hostler}}]{Rougemont:2015ona}%
  \BibitemOpen
  \bibfield  {author} {\bibinfo {author} {\bibfnamefont {Romulo}\ \bibnamefont
  {Rougemont}}, \bibinfo {author} {\bibfnamefont {Jorge}\ \bibnamefont
  {Noronha}}, \ and\ \bibinfo {author} {\bibfnamefont {Jacquelyn}\ \bibnamefont
  {Noronha-Hostler}},\ }\bibfield  {title} {\enquote {\bibinfo {title}
  {{Suppression of baryon diffusion and transport in a baryon rich strongly
  coupled quark-gluon plasma}},}\ }\href {\doibase
  10.1103/PhysRevLett.115.202301} {\bibfield  {journal} {\bibinfo  {journal}
  {Phys. Rev. Lett.}\ }\textbf {\bibinfo {volume} {115}},\ \bibinfo {pages}
  {202301} (\bibinfo {year} {2015})},\ \Eprint
  {http://arxiv.org/abs/1507.06972} {arXiv:1507.06972 [hep-ph]} \BibitemShut
  {NoStop}%
%%CITATION = ARXIV:1507.06972;%%
\bibitem [{\citenamefont {Finazzo}\ \emph {et~al.}(2015)\citenamefont
  {Finazzo}, \citenamefont {Rougemont}, \citenamefont {Marrochio},\ and\
  \citenamefont {Noronha}}]{Finazzo:2014cna}%
  \BibitemOpen
  \bibfield  {author} {\bibinfo {author} {\bibfnamefont {Stefano~I.}\
  \bibnamefont {Finazzo}}, \bibinfo {author} {\bibfnamefont {Romulo}\
  \bibnamefont {Rougemont}}, \bibinfo {author} {\bibfnamefont {Hugo}\
  \bibnamefont {Marrochio}}, \ and\ \bibinfo {author} {\bibfnamefont {Jorge}\
  \bibnamefont {Noronha}},\ }\bibfield  {title} {\enquote {\bibinfo {title}
  {{Hydrodynamic transport coefficients for the non-conformal quark-gluon
  plasma from holography}},}\ }\href {\doibase 10.1007/JHEP02(2015)051}
  {\bibfield  {journal} {\bibinfo  {journal} {JHEP}\ }\textbf {\bibinfo
  {volume} {1502}},\ \bibinfo {pages} {051} (\bibinfo {year} {2015})},\ \Eprint
  {http://arxiv.org/abs/1412.2968} {arXiv:1412.2968 [hep-ph]} \BibitemShut
  {NoStop}%
%%CITATION = ARXIV:1412.2968;%%
\bibitem [{\citenamefont {Gubser}(2010)}]{Gubser:2010ze}%
  \BibitemOpen
  \bibfield  {author} {\bibinfo {author} {\bibfnamefont {Steven~S.}\
  \bibnamefont {Gubser}},\ }\bibfield  {title} {\enquote {\bibinfo {title}
  {{Symmetry constraints on generalizations of Bjorken flow}},}\ }\href
  {\doibase 10.1103/PhysRevD.82.085027} {\bibfield  {journal} {\bibinfo
  {journal} {Phys.Rev.}\ }\textbf {\bibinfo {volume} {D82}},\ \bibinfo {pages}
  {085027} (\bibinfo {year} {2010})},\ \Eprint {http://arxiv.org/abs/1006.0006}
  {arXiv:1006.0006 [hep-th]} \BibitemShut {NoStop}%
%%CITATION = ARXIV:1006.0006;%%
\bibitem [{\citenamefont {Kouno}\ \emph {et~al.}(1990)\citenamefont {Kouno},
  \citenamefont {Maruyama}, \citenamefont {Takagi},\ and\ \citenamefont
  {Saito}}]{Kouno:1989ps}%
  \BibitemOpen
  \bibfield  {author} {\bibinfo {author} {\bibfnamefont {Hiroaki}\ \bibnamefont
  {Kouno}}, \bibinfo {author} {\bibfnamefont {Masahiro}\ \bibnamefont
  {Maruyama}}, \bibinfo {author} {\bibfnamefont {Fujio}\ \bibnamefont
  {Takagi}}, \ and\ \bibinfo {author} {\bibfnamefont {Koichi}\ \bibnamefont
  {Saito}},\ }\bibfield  {title} {\enquote {\bibinfo {title} {{Relativistic
  Hydrodynamics of Quark - Gluon Plasma and Stability of Scaling Solutions}},}\
  }\href {\doibase 10.1103/PhysRevD.41.2903} {\bibfield  {journal} {\bibinfo
  {journal} {Phys.Rev.}\ }\textbf {\bibinfo {volume} {D41}},\ \bibinfo {pages}
  {2903} (\bibinfo {year} {1990})}\BibitemShut {NoStop}%
%%CITATION = PHRVA,D41,2903;%%
\bibitem [{\citenamefont {Muronga}(2002)}]{Muronga:2001zk}%
  \BibitemOpen
  \bibfield  {author} {\bibinfo {author} {\bibfnamefont {Azwinndini}\
  \bibnamefont {Muronga}},\ }\bibfield  {title} {\enquote {\bibinfo {title}
  {{Second order dissipative fluid dynamics for ultrarelativistic nuclear
  collisions}},}\ }\href {\doibase 10.1103/PhysRevLett.88.062302} {\bibfield
  {journal} {\bibinfo  {journal} {Phys.Rev.Lett.}\ }\textbf {\bibinfo {volume}
  {88}},\ \bibinfo {pages} {062302} (\bibinfo {year} {2002})},\ \Eprint
  {http://arxiv.org/abs/nucl-th/0104064} {arXiv:nucl-th/0104064 [nucl-th]}
  \BibitemShut {NoStop}%
%%CITATION = NUCL-TH/0104064;%%
\bibitem [{\citenamefont {Muronga}(2004)}]{Muronga:2003ta}%
  \BibitemOpen
  \bibfield  {author} {\bibinfo {author} {\bibfnamefont {Azwinndini}\
  \bibnamefont {Muronga}},\ }\bibfield  {title} {\enquote {\bibinfo {title}
  {{Causal theories of dissipative relativistic fluid dynamics for nuclear
  collisions}},}\ }\href {\doibase 10.1103/PhysRevC.69.034903} {\bibfield
  {journal} {\bibinfo  {journal} {Phys.Rev.}\ }\textbf {\bibinfo {volume}
  {C69}},\ \bibinfo {pages} {034903} (\bibinfo {year} {2004})},\ \Eprint
  {http://arxiv.org/abs/nucl-th/0309055} {arXiv:nucl-th/0309055 [nucl-th]}
  \BibitemShut {NoStop}%
%%CITATION = NUCL-TH/0309055;%%
\bibitem [{\citenamefont {Baier}\ \emph {et~al.}(2006)\citenamefont {Baier},
  \citenamefont {Romatschke},\ and\ \citenamefont {Wiedemann}}]{Baier:2006um}%
  \BibitemOpen
  \bibfield  {author} {\bibinfo {author} {\bibfnamefont {Rudolf}\ \bibnamefont
  {Baier}}, \bibinfo {author} {\bibfnamefont {Paul}\ \bibnamefont
  {Romatschke}}, \ and\ \bibinfo {author} {\bibfnamefont {Urs~Achim}\
  \bibnamefont {Wiedemann}},\ }\bibfield  {title} {\enquote {\bibinfo {title}
  {{Dissipative hydrodynamics and heavy ion collisions}},}\ }\href {\doibase
  10.1103/PhysRevC.73.064903} {\bibfield  {journal} {\bibinfo  {journal}
  {Phys.Rev.}\ }\textbf {\bibinfo {volume} {C73}},\ \bibinfo {pages} {064903}
  (\bibinfo {year} {2006})},\ \Eprint {http://arxiv.org/abs/hep-ph/0602249}
  {arXiv:hep-ph/0602249 [hep-ph]} \BibitemShut {NoStop}%
%%CITATION = HEP-PH/0602249;%%
\bibitem [{\citenamefont {Song}\ and\ \citenamefont
  {Heinz}(2008)}]{Song:2007ux}%
  \BibitemOpen
  \bibfield  {author} {\bibinfo {author} {\bibfnamefont {Huichao}\ \bibnamefont
  {Song}}\ and\ \bibinfo {author} {\bibfnamefont {Ulrich~W.}\ \bibnamefont
  {Heinz}},\ }\bibfield  {title} {\enquote {\bibinfo {title} {{Causal viscous
  hydrodynamics in 2+1 dimensions for relativistic heavy-ion collisions}},}\
  }\href {\doibase 10.1103/PhysRevC.77.064901} {\bibfield  {journal} {\bibinfo
  {journal} {Phys. Rev.}\ }\textbf {\bibinfo {volume} {C77}},\ \bibinfo {pages}
  {064901} (\bibinfo {year} {2008})},\ \Eprint {http://arxiv.org/abs/0712.3715}
  {arXiv:0712.3715 [nucl-th]} \BibitemShut {NoStop}%
%%CITATION = ARXIV:0712.3715;%%
\end{thebibliography}%
\end{document}